\documentclass{pasa}%

\title[\textsc{Phantom}]{{\sc Phantom}: A smoothed particle hydrodynamics and magnetohydrodynamics code for astrophysics}
\author[Price et al.]{Daniel J. Price$^1$\thanks{daniel.price@monash.edu}, James Wurster$^{2,1}$,  Terrence S. Tricco$^{4, 1}$, Chris Nixon$^{3}$, St\'even Toupin$^{5}$, Alex Pettitt$^6$, Conrad Chan$^{1}$, Daniel Mentiplay$^{1}$, Guillaume Laibe$^7$, Simon Glover$^{8}$, Clare Dobbs$^{2}$, Rebecca Nealon$^{3,1}$, David Liptai$^{1}$, Hauke Worpel$^{9,1}$, Cl\'ement Bonnerot$^{10}$, Giovanni Dipierro$^{3}$, Giulia Ballabio$^3$, Enrico Ragusa$^{11}$, Christoph Federrath$^{12}$, Roberto Iaconi$^{13}$, Thomas Reichardt$^{13}$, Duncan Forgan$^{14}$, Mark Hutchison$^{15,16,1}$, Thomas Constantino$^2$, Ben Ayliffe$^{17,1}$, Kieran Hirsh$^{1}$ \and Giuseppe Lodato$^{11}$\\
\affil{$^1$Monash Centre for Astrophysics (MoCA) and School of Physics and Astronomy, Monash University, Vic. 3800, Australia}%
\affil{$^2$School of Physics, University of Exeter, Stocker Rd., Exeter EX4 4QL, UK}
\affil{$^{3}$Theoretical Astrophysics Group, Department of Physics \& Astronomy, University of Leicester, Leicester LE1 7RH, UK}
\affil{$^{4}$Canadian Institute for Theoretical Astrophysics (CITA), University of Toronto, 60 St. George Street, Toronto, ON M5S 3H8, Canada}
\affil{$^5$Institut d'Astronomie et d'Astrophysique (IAA), Universit\'e Libre de Bruxelles (ULB), CP226, Boulevard du Triomphe B1050 Brussels, Belgium}
\affil{$^6$Department of Cosmosciences, Hokkaido University, Sapporo 060-0810, Japan}
\affil{$^7$Univ Lyon, ENS de Lyon, CNRS, Centre de Recherche Astrophysique de Lyon UMR5574, F-69230, Saint-Genis-Laval, France}
\affil{$^8$Zentrum f\"ur Astronomie der Universit\"at Heidelberg, Institut f\"ur Theoretische Astrophysik, Albert-Ueberle-Str 2, D-69120 Heidelberg, Germany}
\affil{$^{9}$AIP Potsdam, An der Sternwarte 16, 14482 Potsdam, Germany}
\affil{$^{10}$Leiden Observatory, Leiden University, PO Box 9513, NL-2300 RA Leiden, the Netherlands}
\affil{$^{11}$Dipartimento di Fisica, Universit\`a Degli Studi di Milano, Via Celoria 16, Milano, 20133, Italy}
\affil{$^{12}$Research School of Astronomy and Astrophysics, Australian National University, Canberra, ACT 2611, Australia}
\affil{$^{13}$Department of Physics and Astronomy, Macquarie University, Sydney, Australia}
\affil{$^{14}$St Andrews Centre for Exoplanet Science and School of Physics and Astronomy, University of St. Andrews, North Haugh, St. Andrews, Fife KY16 9SS, UK}
\affil{$^{15}$Physikalisches Institut, Universit\"at Bern, Gesellschaftstrasse 6, 3012 Bern, Switzerland}
\affil{$^{16}$Institute for Computational Science, University of Zurich, Winterthurerstrasse 190, CH-8057 Z\"urich, Switzerland}
\affil{$^{17}$Met Office, FitzRoy Road, Exeter, EX1 3PB, UK}}
\jid{PASA}
\doi{10.1017/pas.\the\year.xxx}
\jyear{\the\year}
\usepackage{bm}

\def\vg{{\bm v}_{\rm g}}
\def\vd{{\bm v}_{\rm d}}

\def\vca{\bm a}
\def\vcr{\bm r}
\def\vcv{\bm v}

\def\vcT{\bm T}

\newcommand\mion[2]{\mbox{#1\,{\footnotesize #2}}} 

\usepackage{etoolbox}
\makeatletter
\patchcmd\@combinedblfloats{\box\@outputbox}{\unvbox\@outputbox}{}{%
  \errmessage{\noexpand\@combinedblfloats could not be patched}%
}%
\makeatother

\usepackage{pstricks}

\usepackage{empheq}

\usepackage{float}
\newfloat{listing}{tbhp}{lst}
\floatname{listing}{Listing}

\usepackage{natbib,url,aas_macros}
\usepackage{hyperref}
\hypersetup{colorlinks,citecolor=blue,linkcolor=blue,urlcolor=blue}

\defcitealias{pricelaibe15}{PL15}
\defcitealias{balsarakim04}{BK04}

\begin{document}
\begin{abstract}
We present \textsc{Phantom}, a fast, parallel, modular and low-memory smoothed particle hydrodynamics and magnetohydrodynamics code developed over the last decade for astrophysical applications in three dimensions. The code has been developed with a focus on stellar, galactic, planetary and high energy astrophysics and has already been used widely for studies of accretion discs and turbulence, from the birth of planets to how black holes accrete. Here we describe and test the core algorithms as well as modules for magnetohydrodynamics, self-gravity, sink particles, dust-gas mixtures, H$_{2}$ chemistry, physical viscosity, external forces including numerous galactic potentials, Lense-Thirring precession, Poynting-Robertson drag and stochastic turbulent driving. \textsc{Phantom} is hereby made publicly available.
\end{abstract}

\maketitle

\section{Introduction}
 Numerical simulations are the `third pillar' of astrophysics, standing alongside observations and analytic theory. Since it is difficult to perform laboratory experiments in the relevant physical regimes and over the correct range of length and time-scales involved in most astrophysical problems, we turn instead to `numerical experiments' in the computer for understanding and insight. As algorithms and simulation codes become ever more sophisticated, the public availability of simulation codes has become crucial to ensure that these experiments can be both verified and reproduced.

 \textsc{Phantom} is a smoothed particle hydrodynamics (SPH) code developed over the last decade. It has been used widely for studies of turbulence \citep[e.g.][]{kitsionasetal09,pricefederrath10,pfb11}, accretion \citep[e.g.][]{lodatoprice10,nkp12,rlp12}, star formation including non-ideal magnetohydrodynamics \citep[e.g.][]{wpb16,wpb17}, star cluster formation \citep{liptaietal17}, and for studies of the Galaxy \citep{pettittetal14,dobbsetal16} as well as for simulating dust-gas mixtures \citep[e.g.][]{dipierroetal15,ragusaetal17,tpl17}. Although the initial applications and some details of the basic algorithm were described in \citet{pricefederrath10}, \citet{lodatoprice10} and \citet{price12}, the code itself has never been described in detail and, until now, has remained closed-source.

One of the initial design goals of \textsc{Phantom} was to have a low memory footprint. 
 A secondary motivation was the need for a public SPH code that is not primarily focused on cosmology, as in the highly successful \textsc{Gadget} code \citep{syw01,springel05}. The needs of different communities produce rather different outcomes in the code. For cosmology, the main focus is on simulating the gravitational collapse of dark matter in large volumes of the universe, with gas having only a secondary effect. This is reflected in the ability of the public \textsc{Gadget}-2 code to scale to exceedingly large numbers of dark matter particles, yet neglecting elements of the core SPH algorithm that are essential for stellar and planetary problems, such as the \citet{morrismonaghan97} artificial viscosity switch (c.f. the debate between \citealt{bauerspringel12} and \citealt{price12a}), the ability to use a spatially variable gravitational force softening \citep{bateburkert97,pricemonaghan07} or any kind of artificial conductivity, necessary for the correct treatment of contact discontinuities \citep{chowmonaghan97,pricemonaghan05,rosswogprice07,price08}. Almost all of these have since been implemented in development versions of \textsc{Gadget}-3 (e.g.~\citealt{iannuzzidolag11,becketal16}; see recent comparison project by \citealt{sembolinietal16}) but remain unavailable or unused in the public version. Likewise, the implementation of dust, non-ideal MHD and other physics relevant to star and planet formation is unlikely to be high priority in a code designed for studying cosmology or galaxy formation.
 
 Similarly, the \textsc{sphng} code \citep{benzetal90,bate95} has been a workhorse for our group for simulations of star formation \citep[e.g.][]{pricebate07,pricebate09,ptb12,lbp15} and accretion discs \citep[e.g.][]{lodatorice04,clc09}, contains a rich array of physics necessary for star and planet formation including all of the above algorithms, but the legacy nature of the code makes it difficult to modify or debug and there are no plans to make it public.
 
  \textsc{Gasoline} \citep{wsq04} is another code that has been widely and successfully used for galaxy formation simulations, with its successor, \textsc{Gasoline 2} \citep{wkq17}, recently publicly released. \citet{hubberetal11} have developed {\sc Seren} with similar goals to \textsc{Phantom}, focused on star cluster simulations. {\sc Seren} thus presents more advanced $N$-body algorithms compared to what is in \textsc{Phantom} but does not yet include magnetic fields, dust or H$_2$ chemistry.

  A third motivation was the need to distinguish between the `high performance' code used for 3D simulations from simpler codes used to develop and test algorithms, such as our already-public \textsc{ndspmhd} code \citep{price12}. \textsc{Phantom} is designed to `take what works and make it fast', rather than containing options for every possible variation on the SPH algorithm. Obsolete options are actively deleted.

 The initial release of \textsc{Phantom} has been developed with a focus on stellar, planetary and Galactic astrophysics as well as accretion discs. In this first paper, coinciding with the first stable public release, we describe and validate the core algorithms as well as some example applications. Various novel aspects and optimisation strategies are also presented. This paper is an attempt to document in detail what is currently available in the code, which include modules for magnetohydrodynamics, dust-gas mixtures, self-gravity and a range of other physics. The paper is also designed to serve as guide to the correct use of the various algorithms. Stable releases of \textsc{Phantom} are posted on the web\footnote{\url{https://phantomsph.bitbucket.io/}}, while the development version and wiki documentation are available on the \textsc{Bitbucket} platform\footnote{\url{https://bitbucket.org/danielprice/phantom}}.
 
 The paper is organised as follows: We describe the numerical methods in Section~\ref{sec:methods} with corresponding numerical tests in Section~\ref{sec:tests}. We cover SPH basics (\S\ref{sec:fundamentals}), our implementation of hydrodynamics (\S\ref{sec:hydro}; \S\ref{sec:hydrotest}), the timestepping algorithm (\S\ref{sec:timeint}), external forces (\S\ref{sec:extf}, \S\ref{sec:extftest}), turbulent forcing (\S\ref{sec:turbforcing}, \S\ref{sec:turbulence}), accretion disc viscosity (\S\ref{sec:discviscosity}, \S\ref{sec:disctest}), Navier-Stokes viscosity (\S\ref{sec:viscosity}, \S\ref{sec:visctest}), sink particles (\S\ref{sec:sinks}, \S\ref{sec:sinktest}), stellar physics (\S\ref{sec:stellar}), MHD (\S\ref{sec:mhd}, \S\ref{sec:mhdtests}), non-ideal MHD (\S\ref{sec:nonideal}, \S\ref{sec:nimhdtest}), self-gravity (\S\ref{sec:gravity}, \S\ref{sec:sg}), dust-gas mixtures (\S\ref{sec:dust}, \S\ref{sec:dusttest}), ISM chemistry and cooling (\S\ref{sec:ism}, \S\ref{sec:ismtest}) and particle injection (\S\ref{sec:inject}). We present the algorithms for generating initial conditions in Section~\ref{sec:initial}. Our approach to software engineering is described in Section~\ref{sec:software}. We give five examples of recent applications highlighting different aspects of {\sc Phantom} in Section~\ref{sec:apps}. We summarise in Section~\ref{sec:summary}.
 



\section{Numerical method}
\label{sec:methods}
 \textsc{Phantom} is based on the Smoothed Particle Hydrodynamics (SPH) technique, invented by \citet{lucy77} and \citet{gingoldmonaghan77} and the subject of numerous reviews \citep{benz90,monaghan92,monaghan05,monaghan12,rosswog09,springel10,price12}. 
 
 In the following we adopt the convention that $a$, $b$ and $c$ refer to particle indices; $i$, $j$ and $k$ refer to vector or tensor indices and $n$ and $m$ refer to indexing of nodes in the treecode.
 
\subsection{Fundamentals}
\label{sec:fundamentals}
\subsubsection{Lagrangian hydrodynamics}
  SPH solves the equations of hydrodynamics in Lagrangian form. The fluid is discretised onto a set of `particles' of mass $m$ that are moved with the local fluid velocity ${\bm v}$. Hence the two basic equations common to all physics in \textsc{Phantom} are
\begin{align}
\frac{{\rm d}{\bm r}}{{\rm d}t} & = {\bm v}, \label{eq:dxdt}\\
\frac{{\rm d}\rho}{{\rm d}t} & = -\rho (\nabla\cdot{\bm v}), \label{eq:cty}
\end{align}
where ${\bm r}$ is the particle position and $\rho$ is the density. These equations use the Lagrangian time derivative, ${\rm d}/{\rm d}t \equiv \partial / \partial t + {\bm v} \cdot \nabla$, and are the Lagrangian update of the particle position and the continuity equation (expressing the conservation of mass), respectively.

\subsubsection{Conservation of mass in SPH}
\label{sec:density}
The density is computed in \textsc{Phantom} using the usual SPH density sum,
\begin{equation}
\rho_{a} = \sum_{b} m_{b} W(\vert{\bm r}_{a} - {\bm r}_{b}\vert , h_{a}), \label{eq:rhosum}
\end{equation}
where $a$ and $b$ are particle labels, $m$ is the mass of the particle, $W$ is the smoothing kernel, $h$ is the smoothing length and the sum is over neighbouring particles (i.e. those within $R_{\rm kern} h$, where $R_{\rm kern}$ is the dimensionless cutoff radius of the smoothing kernel). Taking the Lagrangian time derivative of (\ref{eq:rhosum}), one obtains the discrete form of (\ref{eq:cty}) in SPH
\begin{equation}
\frac{{\rm d}\rho_{a}}{{\rm d}t} = \frac{1}{\Omega_{a}} \sum_{b} m_{b} ({\bm v}_{a} - {\bm v}_{b}) \cdot \nabla_{a} W_{ab} (h_{a}), \label{eq:sphcty}
\end{equation}
where $W_{ab}(h_{a})\equiv W(\vert{\bm r}_{a} - {\bm r}_{b}\vert , h_{a})$ and $\Omega_{a}$ is a term related to the gradient of the smoothing length  \citep{springelhernquist02,monaghan02} given by
\begin{equation}
\Omega_{a} \equiv 1 - \frac{\partial h_{a}}{\partial \rho_{a}}\sum_{b} m_{b} \frac{\partial W_{ab} (h_{a})}{\partial h_{a}}. \label{eq:omega}
\end{equation}
Equation (\ref{eq:sphcty}) is not used directly to compute the density in \textsc{Phantom}, since evaluating (\ref{eq:rhosum}) provides a time-independent solution to (\ref{eq:cty}) (see e.g. \citealt{monaghan92, price12} for details). The time-dependent version (\ref{eq:sphcty}) is equivalent to (\ref{eq:rhosum}) up to a boundary term \citep[see][]{price08} but is only used in \textsc{Phantom} to predict the smoothing length at the next timestep in order to reduce the number of iterations required to evaluate the density (see below).

 Since (\ref{eq:rhosum}), (\ref{eq:sphcty}) and (\ref{eq:omega}) all depend on the kernel evaluated on neighbours within $R_{\rm kern}$ times $h_{a}$, all three of these summations may be computed simultaneously using a single loop over the same set of neighbours. Details of the neighbour finding procedure are given in Section~\ref{sec:neighb}, below.

\subsubsection{Setting the smoothing length}
\label{sec:hset}
 The smoothing length itself is specified as a function of the particle number density, $n$, via
\begin{equation}
h_{a} = h_{\rm fact} n_{a}^{-1/3} = h_{\rm fact} \left( \frac{m_{a}}{\rho_{a}} \right)^{1/3}, \label{eq:hrho}
\end{equation}
where $h_{\rm fact}$ is the proportionality factor specifying the smoothing length in terms of the mean local particle spacing and the second equality holds only for equal mass particles, which are enforced in \textsc{Phantom}. The restriction to equal mass particles means that the resolution strictly follows mass, which may be restrictive for  problems involving large density contrasts \citep[e.g.][]{hutchisonetal16}. However, our view is that the potential pitfalls of unequal mass particles \citep[see e.g.][]{monaghanprice06} are currently too great to allow for a robust implementation in a public code.

As described in \citet{price12}, the proportionality constant $h_{\rm fact}$ can be related to the mean neighbour number according to
\begin{equation}
\overline{N}_{\rm neigh} = \frac43 \pi (R_{\rm kern} h_{\rm fact})^{3}, \label{eq:neighbnum}
\end{equation}
however this is only equal to the \emph{actual} neighbour number for particles in a uniform density distribution (more specifically, for a density distribution with no second derivative), meaning that the actual neighbour number varies. The default setting for $h_{\rm fact}$ is 1.2, corresponding to an average of 57.9 neighbours for a kernel truncated at $2h$ (i.e. for $R_{\rm kern} = 2$) in three dimensions. Table~\ref{tab:kernels} lists the settings recommended for different choices of kernel. The derivative required in (\ref{eq:omega}) is given by
\begin{equation}
\frac{\partial h_{a}}{\partial \rho_{a}} = -\frac{3 h_{a}}{\rho_{a}}.
\end{equation}

\subsubsection{Iterations for h and $\rho$}
\label{sec:hrho}
The mutual dependence of $\rho$ and $h$ means that a rootfinding procedure is necessary to solve both (\ref{eq:rhosum}) and (\ref{eq:hrho}) simultaneously. The procedure implemented in \textsc{Phantom} follows \citet{pricemonaghan04a} and \citet{pricemonaghan07}, solving, for each particle, the equation
\begin{equation}
f(h_{a}) = \rho_{\rm sum} (h_{a}) - \rho(h_{a}) = 0, \label{eq:fh}
\end{equation}
where $\rho_{\rm sum}$ is the density computed from (\ref{eq:rhosum}) and 
\begin{equation}
\rho(h_{a}) = m_{a} (h_{\rm fact}/h_{a})^{3},
\end{equation}
from (\ref{eq:hrho}). Equation (\ref{eq:fh}) is solved with Newton-Raphson iterations,
\begin{equation}
h_{a, {\rm new}} = h_{a} - \frac{f(h_{a})}{f'(h_{a})},
\end{equation}
where the derivative is given by
\begin{align}
f'(h_{a}) = \sum_{b} m_{b}  \frac{\partial W_{ab} (h_{a})}{\partial h_{a}} - \frac{\partial \rho_{a}}{\partial h_{a}} = -\frac{3\rho_{a}}{h_{a}} \Omega_{a}.
\end{align}
The iterations proceed until $\vert h_{a, {\rm new}} - h_{a}\vert/h_{a, 0} < \epsilon_h$, where $h_{a,0}$ is the smoothing length of particle $a$ at the start of the iteration procedure and $\epsilon_h$ is the tolerance. The convergence with Newton-Raphson is fast, with a quadratic reduction in the error at each iteration, meaning that no more than 2--3 iterations are required even with a rapidly changing density field. We avoid further iterations by predicting the smoothing length from the previous timestep according to
\begin{equation}
h^{0}_{a} = h_{a} + \Delta t \frac{{\rm d}h_{a}}{{\rm d}t} = h_{a} + \Delta t \frac{\partial h_{a}}{\partial \rho_{a}} \frac{{\rm d}\rho_{a}}{{\rm d} t},
\end{equation}
where ${\rm d}\rho_{a}/{\rm d}t$ is evaluated from (\ref{eq:sphcty}).

Since $h$ and $\rho$ are mutually dependent, we store only the smoothing length, from which the density can be obtained at any time via a function call evaluating $\rho(h)$. The default value of $\epsilon_h$ is $10^{-4}$ so that $h$ and $\rho$ can be used interchangeably. Setting a small tolerance does not significantly change the computational cost, as the iterations quickly fall below a tolerance of `one neighbour' according to (\ref{eq:neighbnum}), so any iterations beyond this refer to loops over the same set of neighbours which can be efficiently cached. However, it is important that the tolerance may be enforced to arbitrary precision rather than being an integer as implemented in the public version of \textsc{Gadget}, since (\ref{eq:fh}) expresses a mathematical relationship between $h$ and $\rho$ that is assumed throughout the derivation of the SPH algorithm \citep[see discussion in][]{price12}. The precision to which this is enforced places a lower limit on the total energy conservation. Fortunately floating point neighbour numbers are now default in most \textsc{Gadget}-3 variants also.

\subsubsection{Kernel functions}
\label{sec:kfunc}
We write the kernel function in the form
\begin{equation}
W_{ab}(r,h) \equiv \frac{C_{\rm norm}}{h^{3}} f(q),
\end{equation}
where $C_{\rm norm}$ is a normalisation constant, the factor of $h^{3}$ gives the dimensions of inverse volume and $f(q)$ is a dimensionless function of $q \equiv \vert {\bm r}_{a} - {\bm r}_{b} \vert / h$. Various relations for kernels in this form are given in \citet{morrisphd} and in Appendix~B of \citet{price10}. Those used in \textsc{Phantom} are the kernel gradient
\begin{equation}
\nabla_{a} W_{ab} = \hat{\bm r}_{ab} F_{ab},\textrm{ where } F_{ab} \equiv \frac{C_{\rm norm}}{h^{4}} f'(q), \label{eq:fab}
\end{equation}
and the derivative of the kernel with respect to $h$,
\begin{equation}
\frac{\partial W_{ab}(r,h)}{\partial h} = - \frac{C_{\rm norm}}{h^{4}} \left[ 3 f(q) + qf'(q) \right].
\end{equation}
 Notice that the ${\partial W}/{\partial h}$ term in particular can be evaluated simply from the functions needed to compute the density and kernel gradient and hence does not need to be derived separately if a different kernel is used.

\subsubsection{Choice of smoothing kernel}
\label{sec:kdefs}
 The default kernel function in SPH for the last 30 years (since \citealt{monaghanlattanzio85}) has been the $M_{4}$ cubic spline from the \citet{schoenberg46a} B-spline family, given by
\begin{equation}
f(q) = \left\{ \begin{array}{ll}
1 - \frac32 q^{2} + \frac34 q^{3}, & 0 \le q < 1; \\
\frac{1}{4}(2-q)^3, & 1 \le q < 2; \\
0. & q \ge 2, \end{array} \right. \label{eq:cubicspline}
\end{equation}
where the normalisation constant $C_{\rm norm} = 1/\pi$ in 3D and the compact support of the function implies that $R_{\rm kern} = 2$. While the cubic spline kernel is satisfactory for many applications, it is not always the best choice. Most SPH kernels are based on approximating the Gaussian, but with compact support to avoid the $\mathcal{O}(N^{2})$ computational cost. Convergence in SPH is guaranteed to be second order ($\propto h^{2}$) to the degree that the finite summations over neighbouring particles approximate integrals \citep[e.g.][]{monaghan92,monaghan05,price12}. Hence the choice of kernel and the effect that a given kernel has on the particle distribution are important considerations.

 In general, more accurate results will be obtained with a kernel with a larger compact support radius, since it will better approximate the Gaussian which has excellent convergence and stability properties \citep{morrisphd,price12,dehnenaly12}. However, care is required. One should not simply increase $h_{\rm fact}$ for the cubic spline kernel because even though this implies more neighbours [via~(\ref{eq:neighbnum})], it increases the resolution length. For the B-splines it also leads to the onset of the `pairing instability' where the particle distribution becomes unstable to transverse modes, leading to particles forming close pairs \citep{thomascouchman92,morrisphd,morris96,bot04,price12,dehnenaly12}. This is the motivation of our default choice of $h_{\rm fact} = 1.2$ for the cubic spline kernel, since it is just short of the maximum neighbour number that can be used while remaining stable to the pairing instability.
 
\begin{figure}
   \centering
   \includegraphics[width=\columnwidth]{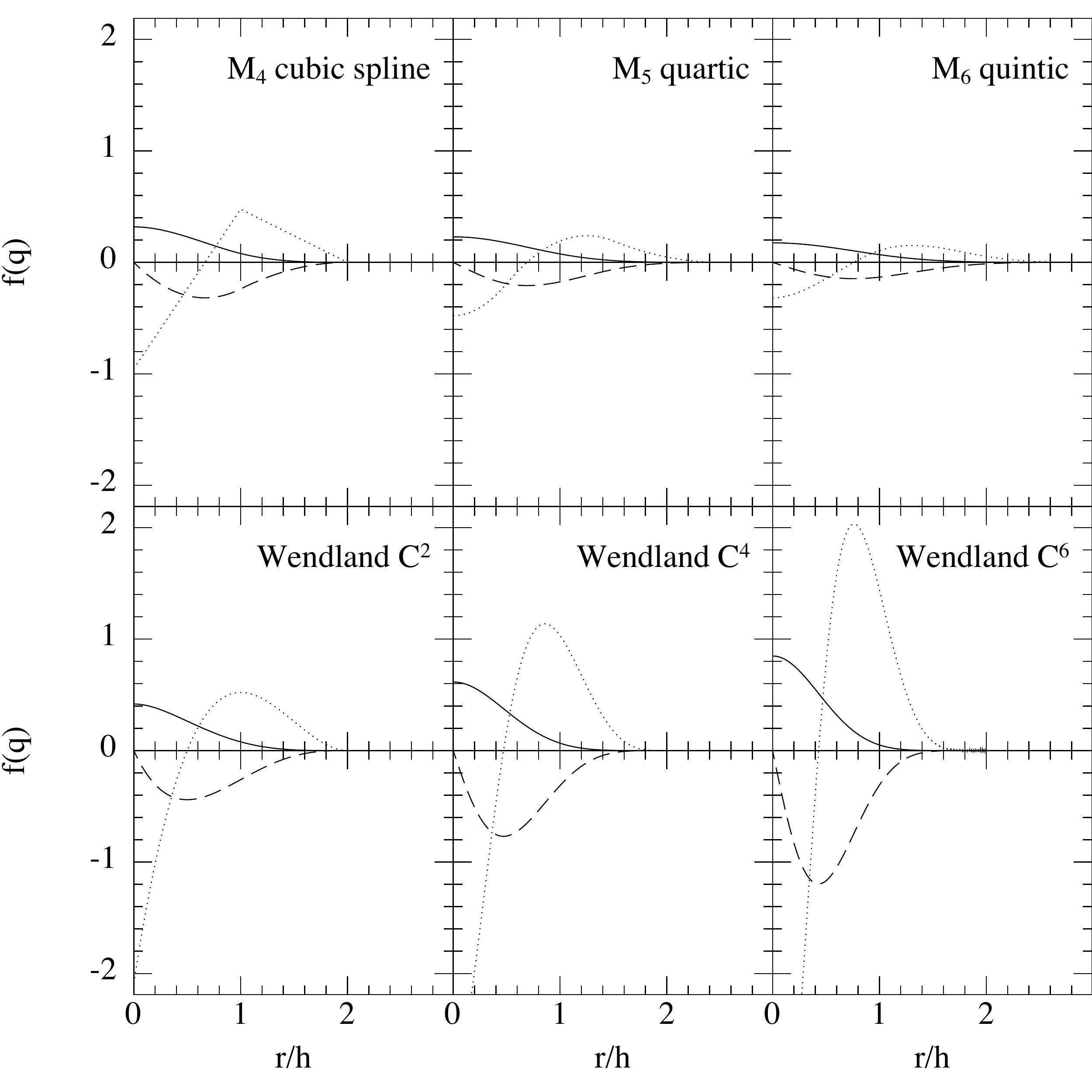} 
   \caption{Smoothing kernels available in \textsc{Phantom} (solid lines) together with their first (dashed lines) and second (dotted lines) derivatives. Wendland kernels in \textsc{Phantom} (bottom row) are given compact support radii of 2, whereas the B-spline kernels (top row) adopt the traditional practice where the support radius increases by 0.5. Thus, use of alternative kernels requires adjustment of $h_{\rm fact}$, the ratio of smoothing length to particle spacing (see Table~\ref{tab:kernels}).}
   \label{fig:kernels}
\end{figure}
\begin{table}
 \setlength{\tabcolsep}{5pt}
\begin{tabular}{llccccc}
\hline
\hline
Kernel & $R_{\rm kern}$ & $\sigma^{2}/h^2$ & $\sigma/h$ & $h_{\rm fact}$ & $h^{\rm d}_{\rm fact}$ & $N_{\rm neigh}$\\
\hline
$M_{4}$ & 2.0 & 9/10 & 0.95 & 1.0--1.2  & 1.2 & 57.9 \\
$M_{5}$ & 2.5 & 23/20 & 1.07 & 1.0--1.2 & 1.2 & 113 \\
$M_{6}$ & 3.0 & 7/5 & 1.18 & 1.0--1.1 & 1.0 & 113 \\
$C^{2}$ & 2.0 & 4/5 & 0.89 & $\geq 1.35$  & 1.4 & 92 \\
$C^{4}$ & 2.0 & 8/13 & 0.78 & $\geq 1.55$  & 1.6 & 137 \\
$C^{6}$ & 2.0 & 1/2 & 0.71 & $\geq 1.7$  & 2.2 & 356 \\
\hline
\hline
\end{tabular}
\caption{Compact support radii, variance, standard deviation, recommended ranges of $h_{\rm fact}$ and recommended default $h_{\rm fact}$ settings ($h_{\rm fact}^{d}$) for the kernel functions available in \textsc{Phantom}}
\label{tab:kernels}
\end{table}

 A better approach to reducing kernel bias is to keep the same resolution length\footnote{This leads to the question of what is the appropriate definition of the `smoothing length' to use when comparing kernels with different compact support radii. Recently it has been shown convincingly by \citet{dehnenaly12} and \citet{violeauleroy14} that the resolution length in SPH is proportional to the standard deviation of $W$. Hence the Gaussian has the same resolution length as the $M_{6}$ quintic with compact support radius of $3h$ with $h_{\rm fact} = 1.2$.  Setting the number of neighbours, though related, is not a good way of specifying the resolution length.} but to use a kernel that has a larger compact support radius. The traditional approach \citep[e.g.][]{morrisphd,morris96,bot04,price12} has been to use the higher kernels in the B-spline series, i.e. the $M_{5}$ quartic which extends to $2.5h$
\begin{equation}
f(q) = \left\{ \begin{array}{ll}
\left(\frac52 -q\right)^4 - 5\left(\frac32 -q\right)^4 + 10\left(\frac12-q\right)^4, & 0 \le q < \frac12, \\
\left(\frac52 -q\right)^4 - 5\left(\frac32 -q\right)^4, & \frac12 \le q < \frac32, \\
\left(\frac52 -q\right)^4, & \frac32 \le q < \frac52, \\
0, & q \ge \frac52, \end{array} \right. \label{eq:quarticspline} 
\end{equation}
where $C_{\rm norm} = 1/(20\pi)$, and the $M_{6}$ quintic extending to $3h$,
\begin{equation}
f(q) = \left\{ \begin{array}{ll}
(3-q)^5 - 6(2-q)^5 + 15(1-q)^5, & 0 \le q < 1, \\
(3-q)^5 - 6(2-q)^5, & 1 \le q < 2, \\
(3-q)^5, & 2 \le q < 3, \\
0, & q \ge 3, \end{array} \right. \label{eq:quinticspline}
\end{equation}
where $C_{\rm norm} = 1/(120\pi)$ in 3D. The quintic in particular gives results virtually indistinguishable from the Gaussian for most problems. 

 Recently, there has been tremendous interest in the use of the Wendland kernels \citep{wendland95}, particularly since \citet{dehnenaly12} showed that they are stable to the pairing instability at all neighbour numbers despite having a Gaussian-like shape and compact support. These functions are constructed as the unique polynomial functions with compact support but with a positive Fourier transform, which turns out to be a necessary condition for stability against the pairing instability \citep{dehnenaly12}. The three dimensional Wendland kernels scaled to a radius of $2h$ are given by $C^{2}$,
\begin{equation}
f(q) = \begin{cases} \left(1 - \frac{q}{2} \right)^{4} \left(2 q + 1\right), & q < 2, \\0, & q \geq 2, \end{cases}
\end{equation}
where $C_{\rm norm} = 21/(16\pi)$; the $C^{4}$ kernel,
\begin{equation}
f(q) = \begin{cases} \left(1 - \frac{q}{2}\right)^{6} \left(\frac{35 q^{2}}{12} + 3 q + 1\right), & q < 2, \\0, & q\geq 2, \end{cases}
\end{equation}
where $C_{\rm norm} = 495/(256\pi)$, and the $C^{6}$ kernel,
\begin{equation}
f(q) = \begin{cases} \left(1 - \frac{q}{2}\right)^{8} \left(4 q^{3} + \frac{25 q^{2}}{4} + 4 q + 1\right), & q < 2, \\0, & q \geq 2, \end{cases}
\end{equation}
where $C_{\rm norm} = 1365/(512\pi)$. Figure~\ref{fig:kernels} graphs $f(q)$ and its first and second derivative for each of the kernels available in \textsc{Phantom}.

 Several authors have argued for use of the Wendland kernels by default. For example, \citet{rosswog15} found best results on simple test problems using the $C^{6}$ Wendland kernel. However `best' in that case implied using an average of 356 neighbours in 3D (i.e. $h_{\rm fact} = 2.2$ with $R_{\rm kern} = 2.0$) which is a factor of $6$ more expensive than the standard approach. Similarly, \citet{huetal14} recommend the $C^{4}$ kernel with 200 neighbours which is $3.5\times$ more expensive. The large number of neighbours are needed because the Wendland kernels are always worse than the B-splines for a given number of neighbours due to the positive Fourier transform, meaning that the kernel bias (related to the Fourier transform) is always positive where the B-spline errors oscillate around zero \citep{dehnenaly12}. Hence whether or not this additional cost is worthwhile depends on the application. A more comprehensive analysis would be valuable here, as the `best' choice of kernel remains an open question (see also the kernels proposed by \citealt{cgr08,garcia-senzetal14}). An even broader question regards the kernel used for dissipation terms, for gravitational force softening and for drag in two-fluid applications (discussed further in Section~\ref{sec:dust}; \citealt{laibeprice12} found that double-hump shaped kernels led to more than an order of magnitude improvement in accuracy when used for drag terms).
 
 A simple and practical approach to checking that kernel bias does not affect the solution that we have used and advocate when using \textsc{Phantom} is to first attempt a simulation with the cubic spline, but then to check the results with a low resolution calculation using the quintic kernel. If the results are identical then it indicates that the kernel bias is not important, but if not then use of smoother but costlier kernels such as $M_{6}$ or $C^{6}$ may be warranted. Wendland kernels are mainly useful for preventing the pairing instability and are necessary if one desires to employ a large number of neighbours.

\subsubsection{Neighbour finding}
\label{sec:neighb}
\begin{figure}
   \centering
   \includegraphics[width=0.45\textwidth]{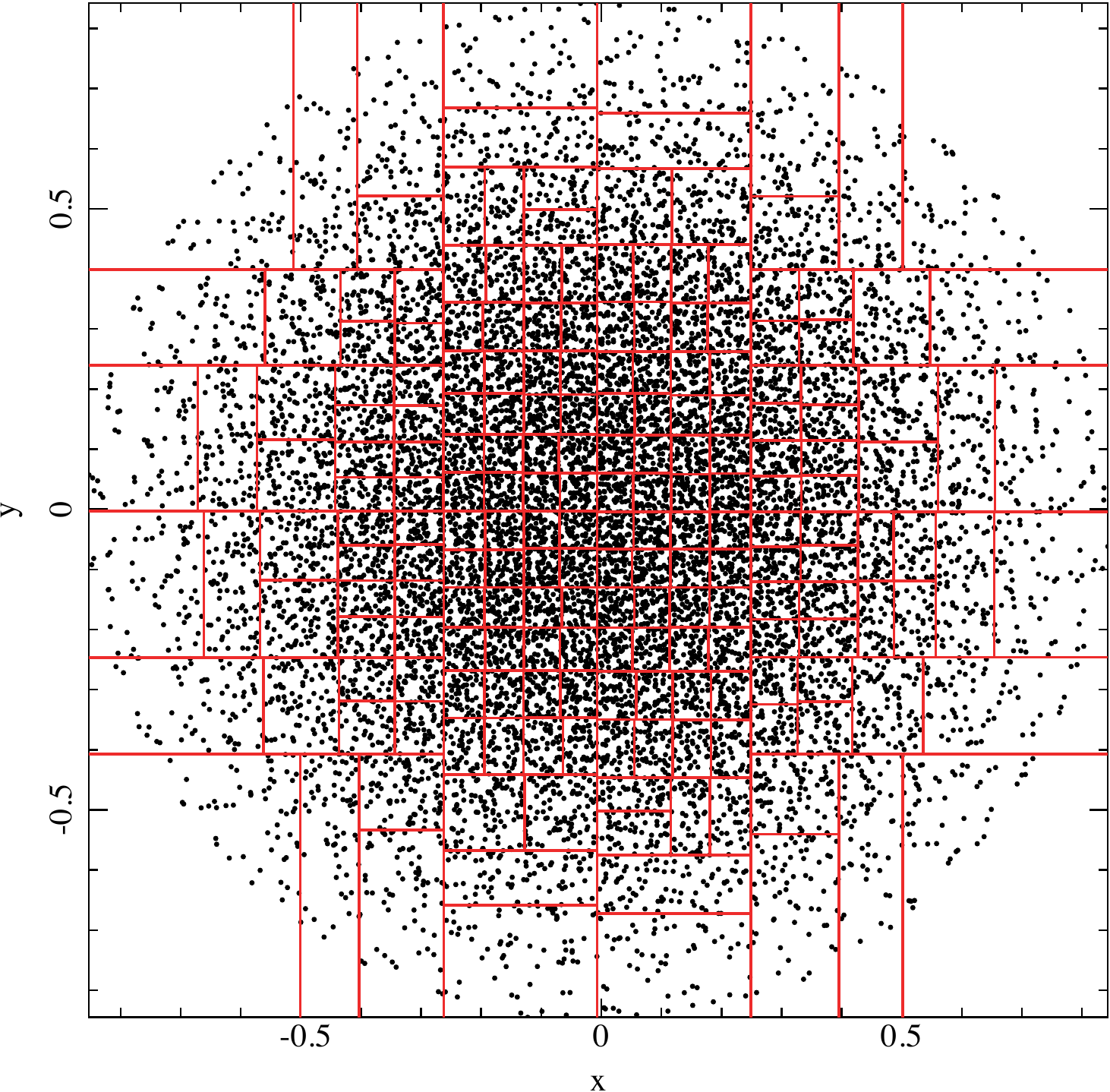} 
   \caption{Example of the $k$d-tree build. For illustrative purposes only we have constructed a two dimensional version of the tree on the projected particle distribution in the x-y plane of the particle distribution from a polytrope test with 13,115 particles. Each level of the tree recursively splits the particle distribution in half, bisecting the longest axis at the centre of mass until the number of particles in a given cell is $< N_{\rm min}$. For clarity we have used $N_{\rm min}=100$ in the above example, while $N_{\rm min}=10$ by default.}
\label{fig:kdtree}
\end{figure}

 Finding neighbours is the main computational expense to any SPH code. Earlier versions of \textsc{Phantom} contained three different options for neighbour-finding: A Cartesian grid, a cylindrical grid and a $k$d-tree. This was because we wrote the code originally with non-self-gravitating problems in mind, for which the overhead associated with a treecode is unnecessary. Since the implementation of self-gravity in \textsc{Phantom} the $k$d-tree has become the default, and is now sufficiently well optimised that the fixed-grid modules are more efficient only for simulations that do not employ either self-gravity or individual particle timesteps, which are rare in astrophysics.

 A key optimisation strategy employed in \textsc{Phantom} is to perform the neighbour search for groups of particles. The results of this search (i.e. positions of all trial neighbours) are then cached and used to check for neighbours for individual particles in the group. Our $k$d-tree algorithm closely follows \citet{gaftonrosswog11}, splitting the particles recursively based on the centre of mass and bisecting the longest axis at each level (Figure~\ref{fig:kdtree}). The tree build is refined until a cell contains less than $N_{\rm min}$ particles, which is then referred to as a `leaf node'. By default, $N_{\rm min} = 10$. The neighbour search is then performed once for each leaf node. Further details are given in Appendix~\ref{sec:maketree}.

\subsection{Hydrodynamics}
\label{sec:hydro}

\subsubsection{Compressible hydrodynamics}
The equations of compressible hydrodynamics are solved in the form
\begin{align}
\frac{{\rm d}{\bm v}}{{\rm d}t} = & -\frac{\nabla P}{\rho} + \Pi_{\rm shock} + {\bm a}_{\rm ext}({\bm r}, t) \nonumber\\
&  + {\bm a}_{\rm sink-gas} + {\bm a}_{\rm selfgrav}, \label{eq:mom} \\
\frac{{\rm d}u}{{\rm d}t} = & -\frac{P}{\rho} \left(\nabla\cdot{\bm v}\right) + \Lambda_{\rm shock} - \frac{\Lambda_{\rm cool}}{\rho}, \label{eq:dudt}
\end{align}
where $P$ is the pressure, $u$ is the specific internal energy, ${\bm a}_{\rm ext}$, ${\bm a}_{\rm sink-gas}$ and ${\bm a}_{\rm selfgrav}$ refer to (optional) accelerations from `external' or `body' forces (Section~\ref{sec:extf}), sink particles (Section~\ref{sec:sinks}) and self-gravity (Section~\ref{sec:gravity}), respectively. $\Pi_{\rm shock}$ and $\Lambda_{\rm shock}$ are dissipation terms required to give the correct entropy increase at a shock front, and $\Lambda_{\rm cool}$ is a cooling term.

\subsubsection{Equation of state}
\label{sec:eos}
 The equation set is closed by an equation of state relating the pressure to the density and/or internal energy. For an ideal gas this is given by
\begin{equation}
P = (\gamma - 1)\rho u, \label{eq:eos}
\end{equation}
where $\gamma$ is the adiabatic index and the sound speed $c_{\rm s}$ is given by 
\begin{equation}
c_{\rm s} = \sqrt{\frac{\gamma P}{\rho}}.
\end{equation} 
The internal energy, $u$, can be related to the gas temperature, $T$, using
\begin{equation}
P = \frac{\rho k_{\rm B} T}{\mu m_{\rm H}},
\end{equation}
giving
\begin{equation}
T = \frac{\mu m_{\rm H}}{k_{\rm B}} (\gamma -1) u,
\label{eq:tempu}
\end{equation}
where $k_{\rm B}$ is Boltzmann's constant, $\mu$ is the mean molecular weight and $m_{\rm H}$ is the mass of a hydrogen atom. Thus to infer the temperature one needs to specify a composition, but only the internal energy affects the gas dynamics. Equation~(\ref{eq:eos}) with $\gamma = 5/3$ is the default equation of state in \textsc{Phantom}.

 In the case where shocks are assumed to radiate away all of the heat generated at the shock front (i.e. $\Lambda_{\rm shock} = 0$) and there is no cooling ($\Lambda_{\rm cool} = 0$), (\ref{eq:dudt}) becomes simply, using (\ref{eq:cty})
\begin{equation}
\frac{{\rm d} u}{{\rm d} t} = \frac{P}{\rho^{2}} \frac{{\rm d}\rho}{{\rm d}t}, \label{eq:du}
\end{equation}
which, using (\ref{eq:eos}) can be integrated to give
\begin{equation}
P =  K \rho^{\gamma}, \label{eq:polyk}
\end{equation}
where $K$ is the polytropic constant. Even more simply, in the case where the temperature is assumed constant, or prescribed as a function of position, the equation of state is simply
\begin{equation}
P = c_{\rm s}^{2} \rho. \label{eq:isoeos}
\end{equation}
In both of these cases, (\ref{eq:polyk}) and (\ref{eq:isoeos}), the internal energy does not need to be stored. In this case the temperature is effectively set by the value of $K$ (and the density if $\gamma \neq 1$). Specifically,
\begin{equation}
T = \frac{\mu m_{\rm H}}{k_{\rm B}} K \rho^{\gamma - 1}.
\label{eq:tempk}
\end{equation}

\subsubsection{Code units}
\label{sec:units}
 For pure hydrodynamics physical units are irrelevant to the numerical results since (\ref{eq:dxdt})--(\ref{eq:cty}) and (\ref{eq:mom})--(\ref{eq:dudt}) are scale free to all but the Mach number. Hence setting physical units is only useful when comparing simulations with nature, when physical heating or cooling rates are applied via (\ref{eq:dudt}), or when one wishes to interpret the results in terms of temperature using (\ref{eq:tempu}) or (\ref{eq:tempk}).

 In the case where gravitational forces are applied, either using an external force (Section~\ref{sec:extf}) or using self-gravity (Section~\ref{sec:gravity}), we adopt the standard procedure of transforming units such that $G=1$ in code units, i.e.
\begin{equation}
u_{\rm time} = \sqrt{\frac{u_{\rm dist}^{3}}{{\rm G}u_{\rm mass}}},
\end{equation}
where $u_{\rm time}$, $u_{\rm dist}$ and $u_{\rm mass}$ are the units of time, length and mass, respectively. Additional constraints apply when using relativistic terms (Section~\ref{sec:lt}) or magnetic fields (Section~\ref{sec:magunits}).

\subsubsection{Equation of motion in SPH}
We follow the variable smoothing length formulation described by \citet{price12}, \citet{pricefederrath10} and \citet{lodatoprice10}. We discretise (\ref{eq:mom}) using
\begin{align}
\frac{{\rm d}{\bm v}_{a}}{{\rm d}t} = & - \sum_{b} m_{b} \left[ \frac{P_{a} + q^{a}_{ab}}{\rho_{a}^{2}\Omega_{a}}\nabla_a W_{ab}(h_{a})  +  \frac{P_{b}+q^{b}_{ab}}{\rho_{b}^{2}\Omega_{b}}\nabla_a W_{ab}(h_{b}) \right] \nonumber \\
& + {\bm a}_{\rm ext} ({\bm x}_{a}, t) + {\bm a}^{a}_{\rm sink-gas} + {\bm a}^{a}_{\rm selfgrav}, \label{eq:sphmom}
\end{align}
where the $q^{a}_{ab}$ and $q^{b}_{ab}$ terms represent the artificial viscosity (discussed in Section~\ref{sec:av}, below).
 
\subsubsection{Internal energy equation}
\label{sec:dudt}
 The internal energy equation (\ref{eq:dudt}) is discretised using the time derivative of the density sum (c.f. \ref{eq:du}), which from (\ref{eq:sphcty}) gives
\begin{equation}
\frac{{\rm d} u_{a}}{{\rm d} t} = \frac{P_{a}}{\rho_{a}^{2}\Omega_{a}} \sum_{b} m_{b} {\bm v}_{ab} \cdot \nabla_{a} W_{ab} (h_{a}) + \Lambda_{\rm shock} - \frac{\Lambda_{\rm cool}}{\rho}. \label{eq:dudtsph}
\end{equation}
where ${\bm v}_{ab} \equiv {\bm v}_a - {\bm v}_b$. Indeed, in the variational formulation of SPH \citep[e.g.][]{price12}, this expression is used as a constraint to derive (\ref{eq:sphmom}), which guarantees both the conservation of energy and entropy (the latter in the absence of dissipation terms). The shock capturing terms in the internal energy equation are discussed below. 

By default we assume an adiabatic gas, meaning that $P{\rm d}V$ work and shock heating terms contribute to the thermal energy of the gas, no energy is radiated to the environment, and total energy is conserved. To approximate a radiative gas one may set one or both of these terms to zero. Neglecting the shock heating term, $\Lambda_{\rm shock}$, gives an approximation equivalent to a polytropic equation of state (\ref{eq:polyk}), as described in Section~\ref{sec:eos}. Setting both shock and work contributions to zero implies that ${\rm d}u/{\rm d}t = 0$, meaning that each particle will simply retain its initial temperature.

\subsubsection{Conservation of energy in SPH}
\label{sec:energyconservation}
 Does evolving the internal energy equation imply that total energy is not conserved? Wrong! Total energy in SPH, for the case of hydrodynamics, is given by
\begin{equation}
E = \sum_a m_a \left(\frac12 v_a^2 + u_a\right).
\end{equation}
Taking the (Lagrangian) time derivative, we find that conservation of energy corresponds to
\begin{equation}
\frac{{\rm d}E}{{\rm d}t} = \sum_{a} m_{a} \left( {\bm v}_{a} \cdot \frac{{\rm d} {\bm v}_{a}}{{\rm d} t} + \frac{{\rm d} u_{a}}{{\rm d} t} \right) = 0.
\label{eq:dedt}
\end{equation}
Inserting our expressions (\ref{eq:sphmom}) and (\ref{eq:dudtsph}), and neglecting for the moment dissipative terms and external forces, we find
\begin{align}
\frac{{\rm d}E}{{\rm d}t} = -\sum_{a} \sum_{b} m_{a}  m_{b} & \left[ \frac{P_{a}\bm{v}_b}{\rho_{a}^{2}\Omega_{a}}\cdot\nabla_a W_{ab}(h_{a})  \right. \nonumber \\
& \left. +  \frac{P_{b}\bm{v}_a}{\rho_{b}^{2}\Omega_{b}}\cdot \nabla_a W_{ab}(h_{b}) \right] = 0.
\label{eq:dedtsph}
\end{align}
The double summation on the right hand side equals zero because the kernel gradient, and hence the overall sum, is antisymmetric. That is, $\nabla_a W_{ab} = -\nabla_b W_{ba}$. This means one can relabel the summation indices arbitrarily in one half of the sum, and add it to one half of the original sum to give zero. One may straightforwardly verify that this remains true when one includes the dissipative terms (see below).

 This means that even though we employ the internal energy equation, total energy remains conserved to machine precision in the spatial discretisation. That is, energy is conserved irrespective of the number of particles, the number of neighbours or the choice of smoothing kernel. The only non-conservation of energy arises from the ordinary differential equation solver one employs to solve the left hand side of the equations. We thus employ a symplectic time integration scheme in order to preserve the conservation properties as accurately as possible (Section~\ref{sec:timestepping}).

\subsubsection{Shock-capturing: momentum equation}
\label{sec:av}
The shock capturing dissipation terms are implemented following \citet{monaghan97}, derived by analogy with Riemann solvers from the special relativistic dissipation terms proposed by \citet{chowmonaghan97}. These were extended by \citet{pricemonaghan04a,pricemonaghan05} to magnetohydrodynamics (MHD) and recently to dust-gas mixtures by \citet{laibeprice14a}. In a recent paper, \citet{puriramachandran14} found this approach, along with the \citet{morrismonaghan97} switch (which they referred to as the `Monaghan-Price-Morris' formulation) to be the most accurate and robust method for shock-capturing in SPH when compared to several other approaches, including Godunov SPH \citep[e.g.][]{inutsuka02,chawhitworth03}.

 The formulation in \textsc{Phantom} differs from that given in \citet{price12} only by the way that the density and signal speed in the $q$ terms are averaged, as described in \citet{pricefederrath10} and \citet{lodatoprice10}. That is, we use
\begin{equation}
\Pi_{\rm shock}^{a} \equiv -\sum_{b} m_{b} \left[ \frac{q^{a}_{ab}}{\rho_{a}^{2}\Omega_{a}}\nabla_a W_{ab}(h_{a})  +  \frac{q^{b}_{ab}}{\rho_{b}^{2}\Omega_{b}}\nabla_a W_{ab}(h_{b}) \right], \label{eq:pishock} \\
\end{equation}
where
\begin{equation}
q^{a}_{ab} = \begin{cases}
-\frac12 \rho_{a} v_{{\rm sig}, a} {\bm v}_{ab} \cdot \hat{\bm r}_{ab}, &  {\bm v}_{ab} \cdot \hat{\bm r}_{ab} < 0 \\
0 & \text{otherwise} \end{cases}\label{eq:qvisc}
\end{equation}
where ${\bm v}_{ab} \equiv {\bm v}_{a} - {\bm v}_{b}$, $\hat{\bm r}_{ab} \equiv ({\bm r}_{a} - {\bm r}_{b})/\vert{\bm r}_{a} - {\bm r}_{b} \vert$ is the unit vector along the line of sight between the particles, and $v_{\rm sig}$ is the maximum signal speed, which depends on the physics implemented. For hydrodynamics this is given by
\begin{equation}
v_{{\rm sig},a} = \alpha^{\rm AV}_{a} c_{{\rm s},a} + \beta^{\rm AV} \vert {\bm v}_{ab} \cdot \hat{\bm r}_{ab} \vert, 
\label{eq:vsig}
\end{equation}
where in general $\alpha^{\rm AV}_{a} \in [0,1]$ is controlled by a switch (see Section~\ref{sec:switches}, below), while $\beta^{\rm AV} = 2$ by default. 

Importantly, $\alpha$ does \emph{not} multiply the $\beta^{\rm AV}$ term. The $\beta^{\rm AV}$ term provides a second order \citeauthor{vonneumannrichtmyer50}-like term that prevents particle interpenetration \citep[e.g.][]{lattanzioetal86,monaghan89} and thus $\beta^{\rm AV} \geq 2$ is needed wherever particle penetration may occur. This is important in accretion disc simulations where use of a low $\alpha$ may be acceptable in the absence of strong shocks, but a low $\beta$ will lead to particle penetration of the disc midplane, which is the cause of a number of convergence issues \citep{merubate11,merubate12}. \citet{pricefederrath10} found that $\beta^{\rm AV} = 4$ was necessary at high Mach number ($M \gtrsim 5$) to maintain a sharp shock structure, which despite nominally increasing the viscosity was found to give less dissipation overall because particle penetration no longer occurred at shock fronts.

\subsubsection{Shock-capturing: internal energy equation}
\label{sec:ac}
 The key insight from \citet{chowmonaghan97} was that shock capturing not only involves a viscosity term but involves dissipating the jump in each component of the energy, implying a conductivity term in hydrodynamics and resistive dissipation in MHD (see Section~\ref{sec:mhddiss}). The resulting contribution to the internal energy equation is given by \citep[e.g.][]{price12}
\begin{align}
\Lambda_{\rm shock} & \equiv - \frac{1}{\Omega_{a}\rho_a} \sum_{b} m_{b} v_{{\rm sig}, a} \frac12 ({\bm v}_{ab} \cdot \hat{\bm r}_{ab})^{2} F_{ab}(h_{a}) \nonumber \\
& + \sum_{b} m_{b} \alpha_{u} v^{u}_{\rm sig} (u_{a} - u_{b}) \frac12 \left[ \frac{F_{ab} (h_{a})}{\Omega_{a} \rho_{a}} + \frac{F_{ab} (h_{b})}{\Omega_{b} \rho_{b}}   \right] \nonumber \\
& + \Lambda_{\rm artres}, \label{eq:shockheating}
\end{align}
where the first term provides the viscous shock heating, the second term provides an artificial thermal conductivity and $F_{ab}$ is defined as in (\ref{eq:fab}) and $\Lambda_{\rm artres}$ is the heating due to artificial resistivity (Equation~\ref{eq:resheating}). The signal speed we use for conductivity term differs from the one used for viscosity, as discussed by \citet{price08} and \citet{price12}. In \textsc{Phantom} we use
\begin{equation}
v^{u}_{\rm sig} = \sqrt{\frac{\vert P_{a} - P_{b} \vert}{\overline{\rho}_{ab}}}, \label{eq:vsigupr}
\end{equation}
for simulations that do not involve self-gravity or external body forces \citep{price08}, and
\begin{equation}
v^{u}_{\rm sig} = \vert {\bm v}_{ab} \cdot \hat{\bm r}_{ab} \vert, \label{eq:vsigu}
\end{equation}
for simulations that do \citep{wvc08}. The importance of the conductivity term for treating contact discontinuities was highlighted by \citet{price08}, explaining the poor results found by \citet{agertzetal07} in SPH simulations of Kelvin-Helmholtz instabilities run across contact discontinuities (discussed further in Section~\ref{sec:kh}). With (\ref{eq:vsigu}), we have found there is no need for further switches to reduce conductivity (e.g. \citealt{price04,pricemonaghan05,valdarnini16}), since the effective thermal conductivity $\kappa$ is second order in the smoothing length ($\propto h^{2}$). \textsc{Phantom} therefore uses $\alpha_{u} = 1$ by default in (\ref{eq:shockheating}) and we have not yet found a situation where this leads to excess smoothing.

 It may be readily shown that the total energy remains conserved in the presence of dissipation by combining (\ref{eq:shockheating}) with the corresponding dissipative terms in (\ref{eq:sphmom}). The contribution to the entropy from both viscosity and conductivity is also positive definite (see the appendix in \citealt{pricemonaghan04a} for the mathematical proof in the case of conductivity).

\subsubsection{Shock detection}
\label{sec:switches}
 The standard approach to reducing dissipation in SPH away from shocks for the last 15 years has been the switch proposed by \citet{morrismonaghan97}, where the dimensionless viscosity parameter $\alpha$ is evolved for each particle $a$ according to
\begin{equation}
\frac{{\rm d}\alpha_{a}}{{\rm d}t} = \max(-\nabla\cdot{\bm v}_{a}, 0) - \frac{(\alpha_{a} - \alpha_{\rm min})}{\tau_{a}},
\end{equation}
where $\tau \equiv h / \left(\sigma_{\rm decay} v_{\rm sig}\right)$ and $\sigma_{\rm decay} = 0.1$ by default. We set $v_{\rm sig}$ in the decay time equal to the sound speed to avoid the need to store ${\rm d}\alpha/{\rm d}t$, since $\nabla\cdot{\bm v}$ is already stored in order to compute (\ref{eq:sphcty}). This is the switch used for numerous turbulence applications with \textsc{Phantom} \citep[e.g.][]{pricefederrath10,pfb11,tpf16} where it is important to minimise numerical dissipation in order to maximise the Reynolds number \citep[e.g.][]{valdarnini11,price12a}.

 More recently, \citet{cullendehnen10} proposed a more advanced switch using the time derivative of the velocity divergence. A modified version based on the gradient of the velocity divergence was also proposed by \citet{readhayfield12}. We implement a variation on the \citet{cullendehnen10} switch, using a shock indicator of the form
\begin{equation}
A_{a} = \xi_{a} \max \left[-\frac{{\rm d}}{{\rm d}t}(\nabla\cdot{\bm v}_{a}), 0 \right] ,
\label{eq:avsource}
\end{equation}
where
\begin{equation}
\xi = \frac{\vert \nabla \cdot {\bm v} \vert^{2}}{\vert \nabla \cdot {\bm v} \vert^{2} + \vert \nabla \times {\bm v} \vert^{2}}
\end{equation}
is a modification of the \citet{balsara95} viscosity limiter for shear flows. We use this to set $\alpha$ according to
\begin{equation}
\alpha_{{\rm loc}, a} =\min\left( \frac{10 h_{a}^{2} A_{a}}{c_{{\rm s}, a}^{2}}, \alpha_{\rm max} \right),
\end{equation}
where $c_{\rm s}$ is the sound speed and $\alpha_{\rm max}~=~1$. We use $c_{\rm s}$ in the expression for $\alpha_{\rm loc}$ also for magnetohydrodynamics (Section~\ref{sec:mhd}) since we found using the magnetosonic speed led to a poor treatment of MHD shocks. If $\alpha_{{\rm loc}, a}~>~\alpha_{a}$ we set $\alpha_{a} = \alpha_{{\rm loc}, a}$, otherwise we evolve $\alpha_{a}$ according to
\begin{equation}
\frac{{\rm d}\alpha_{a}}{{\rm d}t} = - \frac{(\alpha_{a} - \alpha_{{\rm loc}, a})}{\tau_{a}},
\end{equation}
where $\tau$ is defined as in the \citet{morrismonaghan97} version, above. We evolve $\alpha$ in the predictor part of the integrator only, i.e. with a first order time integration, to avoid complications in the corrector step. However, we perform the predictor step implicitly using a backward Euler method, i.e.
\begin{equation}
\alpha^{n+1}_{a} = \frac{\alpha_{a}^{n} +  \alpha_{{\rm loc}, a} \Delta t  / \tau_{a}}{1 + \Delta t / \tau_{a}},
\end{equation}
which ensures that the decay is stable regardless of the timestep (we do this for the Morris \& Monaghan method also).

We use the method outlined in Appendix B3 of \citet{cullendehnen10} to compute ${\rm d}(\nabla\cdot{\bm v}_{a})/{{\rm d}t}$. That is, we first compute the gradient tensors of the velocity, ${\bm v}$, and acceleration, ${\bm a}$ (used from the previous timestep), during the density loop using an SPH derivative operator that is exact to linear order, that is, with the matrix correction outlined in \citet{price04,price12}, namely
\begin{equation}
R^{ij}_{a} \frac{\partial v^{k}_{a}}{\partial x^{j}_{a}} = \sum_{b} m_{b} \left(v^{k}_{b} - v^{k}_{a}\right) \frac{\partial W_{ab}(h_{a})}{\partial x^{i}} ,
\label{eq:dvdxlin}
\end{equation}
where
\begin{equation}
R^{ij}_{a} = \sum_{b} m_{b} \left(x^{i}_{b} - x^{i}_{a}\right) \frac{\partial W_{ab}(h_{a})}{\partial x^{j}}  \approx \delta^{ij},
\label{eq:rmatrix}
\end{equation}
and repeated tensor indices imply summation. Finally, we construct the time derivative of the velocity divergence according to
\begin{equation}
\frac{{\rm d}}{{\rm d}t}\left(\frac{\partial v_{a}^{i}}{\partial x_{a}^{i}}\right) = \frac{\partial a_{a}^{i}}{\partial x_{a}^{i}} - \frac{\partial v_{a}^{i}}{\partial x_{a}^{j}} \frac{\partial v_{a}^{j}}{\partial x_{a}^{i}},
\label{eq:ddivvdt}
\end{equation}
where, as previously, repeated $i$ and $j$ indices imply summation. In Cartesian coordinates the last term in (\ref{eq:ddivvdt}) can be written out explicitly using
\begin{align}
\frac{\partial v_{a}^{i}}{\partial x_{a}^{j}} \frac{\partial v_{a}^{j}}{\partial x_{a}^{i}} & = \left(\frac{\partial v^{x}}{\partial x} \right)^{2} + \left(\frac{\partial v^{y}}{\partial y} \right)^{2} + \left(\frac{\partial v^{z}}{\partial z} \right)^{2} \nonumber \\
& + 2 \left[\frac{\partial v^{x}}{\partial y} \frac{\partial v^{y}}{\partial x} + \frac{\partial v^{x}}{\partial z} \frac{\partial v^{z}}{\partial x} + \frac{\partial v^{z}}{\partial y} \frac{\partial v^{y}}{\partial z}\right].
\end{align}

\subsubsection{Cooling}
 The cooling term $\Lambda_{\rm cool}$ can be set either from detailed chemical calculations (Section~\ref{sec:ismcooling}) or, for discs, by the simple `$\beta$-cooling' prescription of \citet{gammie01}, namely
\begin{equation}
\Lambda_{\rm cool} = \frac{\rho u}{t_{\rm cool}},
\end{equation}
where
\begin{equation}
t_{\rm cool} = \frac{\Omega(R)}{\beta_{\rm cool}},
\label{eq:tcool}
\end{equation}
with $\beta_{\rm cool}$ an input parameter to the code specifying the cooling timescale in terms of the local orbital time. We compute $\Omega$ in (\ref{eq:tcool}) using $\Omega \equiv 1/(x^2 + y^2 + z^2)^{3/2}$, i.e. assuming Keplerian rotation around a central object with mass equal to unity, with $G=1$ in code units.

\subsubsection{Conservation of linear and angular momentum}
\label{sec:momconservation}
 The total linear momentum is given by
\begin{equation}
\bm{P} = \sum_{a} m_{a} {\bm v}_{a},
\end{equation}
such that conservation of momentum corresponds to
\begin{equation}
\frac{{\rm d}\bm{P}}{{\rm d}t} = \sum_{a} m_{a} \frac{{\rm d}{\bm v}_{a}}{{\rm d}t} = 0.
\end{equation}
Inserting our discrete equation (\ref{eq:sphmom}), we find
\begin{align}
\frac{{\rm d}\bm{P}}{{\rm d}t} = \sum_{a} \sum_{b} m_a m_{b} & \left[ \frac{P_{a} + q^{a}_{ab}}{\rho_{a}^{2}\Omega_{a}}\nabla_a W_{ab}(h_{a}) \right. \nonumber \\
& \left. +  \frac{P_{b}+q^{b}_{ab}}{\rho_{b}^{2}\Omega_{b}}\nabla_a W_{ab}(h_{b}) \right] = 0.
\end{align}
where, as for the total energy (Section~\ref{sec:energyconservation}), the double summation is zero because of the antisymmetry of the kernel gradient. The same argument applies to the conservation of angular momentum,
\begin{equation}
\sum_{a} m_{a} {\bm r}_{a} \times {\bm v}_{a},
\end{equation}
(see e.g. \citealt{price12} for a detailed proof). As with total energy, this means linear and angular momentum are exactly conserved by our SPH scheme to the accuracy with which they are conserved by the timestepping scheme. 

In \textsc{Phantom}, linear and angular momentum are both conserved to round-off error (typically $\sim 10^{-16}$ in double precision) with global timestepping, but exact conservation is violated when using individual particle timesteps or when using the $k$d-tree to compute gravitational forces. The magnitude of these quantities, as well as the total energy and the individual components of energy (kinetic, internal, potential and magnetic), should thus be monitored by the user at runtime. Typically with individual timesteps one should expect energy conservation to $\Delta E /E \sim 10^{-3}$ and linear and angular momentum conservation to $\sim 10^{-6}$ with default code settings. The code execution is aborted if conservation errors exceed 10\%.

\subsection{Time integration}
\label{sec:timeint}

\subsubsection{Timestepping algorithm}
\label{sec:timestepping}
We integrate the equations of motion using a generalisation of the Leapfrog integrator which is reversible in the case of both velocity dependent forces and derivatives which depend on the velocity field. The basic integrator is the Leapfrog method in `Kick-Drift-Kick' or `Velocity Verlet' form \citep{verlet67}, where the positions and velocities of particles are updated from time $t^{n}$ to $t^{n+1}$ according to
\begin{align}
{\bm v}^{n+\frac12} & = {\bm v}^{n} + \frac12 \Delta t {\bm a}^{n}, \\
{\bm r}^{n+1} & = {\bm r}^{n} + \Delta t {\bm v}^{n+\frac12}, \\
{\bm a}^{n + 1} & = {\bm a}({\bm r}^{n+1}), \\
{\bm v}^{n + 1} & = {\bm v}^{n+\frac12} + \frac12 \Delta t {\bm a}^{n+1}, \label{eq:lfcorr}
\end{align}
where $\Delta t \equiv t^{n+1} - t^{n}$. This is identical to the formulation of Leapfrog used in other astrophysical SPH codes \citep[e.g.][]{springel05,wsq04}. The Verlet scheme, being both reversible and symplectic \citep[e.g.][]{haireretal03}, preserves the Hamiltonian nature of the SPH algorithm \citep[e.g.][]{gingoldmonaghan82a,monaghanprice01}. In particular, both linear and angular momentum are exactly conserved, there is no long-term energy drift, and phase space volume is conserved (e.g. for orbital dynamics). In SPH this is complicated by velocity-dependent terms in the acceleration from the shock-capturing dissipation terms. In this case the corrector step, (\ref{eq:lfcorr}), becomes implicit. The approach we take is to notice that these terms are not usually dominant over the position-dependent terms. Hence we use a first-order prediction of the velocity, as follows
\begin{align}
{\bm v}^{n+\frac12} & = {\bm v}^{n} + \frac12 \Delta t {\bm a}^{n}, \label{eq:lfvhalf} \\
{\bm r}^{n+1} & = {\bm r}^{n} + \Delta t {\bm v}^{n+\frac12}, \label{eq:xpred} \\
{\bm v}^{*} & = {\bm v}^{n+\frac12} + \frac12 \Delta t {\bm a}^{n},\label{eq:vpred} \\ 
{\bm a}^{n + 1} & = {\bm a}({\bm r}^{n+1}, {\bm v}^{*}),\label{eq:arv} \\
{\bm v}^{n + 1} & = {\bm v}^{*} + \frac12 \Delta t  \left[ {\bm a}^{n+1} - {\bm a}^{n} \right]. \label{eq:lfcorrv}
\end{align}
At the end of the step we then check if the error in the first order prediction is less than some tolerance $\epsilon$ according to
\begin{equation}
e = \frac{ \vert {\bm v}^{n+1} - {\bm v}^{*} \vert}{ \vert {\bm v}^{\rm mag} \vert} < \epsilon_{\rm v}, \label{eq:lfconvergencecriterion}
\end{equation}
where ${\bm v}^{\rm mag}$ is the mean velocity on all SPH particles (we set the error to zero if $\vert{\bm v}^{\rm mag}\vert = 0$) and by default $\epsilon_{\rm v} = 10^{-2}$. If this criterion is violated, then we recompute the accelerations by replacing ${\bm v}^{*}$ with ${\bm v}^{n+1}$ and iterating (\ref{eq:arv}) and (\ref{eq:lfcorrv}) until the criterion in (\ref{eq:lfconvergencecriterion}) is satisfied. In practice this happens rarely, but occurs for example in the first few steps of the Sedov problem where the initial conditions are discontinuous (Section~\ref{sec:sedov}). As each iteration is as expensive as halving the timestep, we also constrain the subsequent timestep such that iterations should not occur, i.e.
\begin{equation}
\Delta t = \min \left( \Delta t, \frac{\Delta t}{\sqrt{e_{\rm max} / \epsilon}} \right),
\end{equation}
where $e_{\rm max} = \max(e)$ over all particles. A caveat to the above is that velocity iterations are not currently implemented when using individual particle timesteps.

Additional variables such as the internal energy, $u$, and the magnetic field, ${\bm B}$, are timestepped with a predictor and trapezoidal corrector step in the same manner as the velocity, following (\ref{eq:vpred}) and (\ref{eq:lfcorrv}).

 Velocity-dependent external forces are treated separately, as described in Section~\ref{sec:extf}, below.

\subsubsection{Timestep constraints}
\label{sec:timestep}
 The timestep itself is determined at the end of each step, and is constrained to be less than the maximum stable timestep. For a given particle, $a$, this is given by \citep[e.g.][]{lattanzioetal86,monaghan97},
\begin{equation}
\Delta t_{{\rm C},a} \equiv C_{\rm cour} \frac{h_a}{v^{\rm dt}_{{\rm sig},a}}, \label{eq:dtcour}
\end{equation}
where $C_{\rm cour} = 0.3$ by default \citep{lattanzioetal86} and $v^{\rm dt}_{\rm sig}$ is taken as the maximum of (\ref{eq:vsig}) over the particle's neighbours assuming $\alpha^{\rm AV} = \max(\alpha^{\rm AV},1)$. The criterion above differs from the usual Courant-Friedrichs-Lewy condition used in Eulerian codes \citep{cfl28} because it depends only on the difference in velocity between neighbouring particles, not the absolute value.

An additional constraint is applied from the accelerations (the `force condition'), where
\begin{equation}
\Delta t_{{\rm f}, a} \equiv C_{\rm force} \sqrt{\frac{h_{a}}{\vert {\bm a}_{a} \vert}}, \label{eq:dtforce}
\end{equation}
where $C_{\rm force} = 0.25$ by default. A separate timestep constraint is applied for external forces
\begin{equation}
\Delta t_{{\rm ext},a} \equiv C_{\rm force} \sqrt{\frac{h}{\vert {\bm a}_{{\rm ext},a} \vert}}, \label{eq:dtext}
\end{equation}
and for accelerations to SPH particles to/from sink particles (Section~\ref{sec:sinks}, below)
\begin{equation}
\Delta t_{{\rm sink-gas},a} \equiv C_{\rm force} \sqrt{\frac{h_{a}}{\vert {\bm a}_{{\rm sink-gas},a} \vert}}. \label{eq:dtsinkgas}
\end{equation}
For external forces with potentials defined such that $\Phi \to 0$ as $r\to \infty$ an additional constraint  is applied using \citep{dehnenread11}
\begin{equation}
\Delta t_{{\Phi},a} \equiv C_{\rm force} \eta_\Phi \sqrt{\frac{\vert\Phi_{a}\vert}{\vert \nabla \Phi \vert_{a}^{2}}}, \label{eq:dtphi}
\end{equation}
where $\eta_\Phi = 0.05$ (see Section~\ref{sec:dtsinks}).

 The timestep for particle $a$ is then taken to be the minimum of all of the above constraints, i.e.
\begin{align}
\Delta t_{a} = \min & \left(\Delta t_{\rm C}, \Delta t_{\rm f}, \Delta t_{\rm ext}, \Delta t_{\rm sink-gas}, \Delta t_{\Phi} \right)_a,
\label{eq:dtmin}
\end{align}
with possible other constraints arising from additional physics as described in their respective sections. With global timestepping the resulting timestep is the minimum over all particles,
\begin{equation}
\Delta t = \min_{a} ( \Delta t_a).
\end{equation}

\subsubsection{Substepping of external forces}
\label{sec:respa}
In the case where the timestep is dominated by any of the external force timesteps, i.e.\ (\ref{eq:dtext})--(\ref{eq:dtphi}), we implement an operator splitting approach implemented according to the reversible reference system propagator algorithm (RESPA) derived by \citet{tbm92} for molecular dynamics. RESPA splits the acceleration into `long range' and `short range' contributions, which in \textsc{Phantom} are defined to be the SPH and external/point-mass accelerations, respectively. 

Our implementation follows \citet{tbm92} (see their Appendix~B), where the velocity is first predicted to the half step using the `long range' forces, followed by an inner loop where the positions are updated with the current velocity and the velocities are updated with the `short range' accelerations. Thus the timestepping proceeds according to
\begin{empheq}{align}
{\bm v} & = {\bm v} + \frac{\Delta t_{\rm sph}}{2} {\bm a}_{\rm sph}^{n},
\end{empheq}
\begin{empheq}[left=\textrm{over substeps} \empheqlbrace ]{align}
{\bm v} & = {\bm v} + \frac{\Delta t_{\rm ext}}{2} {\bm a}_{\rm ext}^{m}, & \label{eq:respa1} \\
{\bm r} & = {\bm r} + \Delta t_{\rm ext} {\bm v},  & \\
& \textrm{get } {\bm a}_{\rm ext}({\bm r}), &  \\
{\bm v} & = {\bm v} + \frac{\Delta t_{\rm ext}}{2} {\bm a}_{\rm ext}^{m+1}, \label{eq:respa4}
\end{empheq}
\vspace{-1em}
\begin{empheq}{align}
& \textrm{get } {\bm a}_{\rm sph}({\bm r}),\\
{\bm v} & = {\bm v} + \frac{\Delta t_{\rm sph}}{2} {\bm a}_{\rm sph}^{n}.
\end{empheq}
where ${\bm a}_{\rm SPH}$ indicates the SPH acceleration evaluated from (\ref{eq:sphmom}) and ${\bm a}_{\rm ext}$ indicates the external forces. The SPH and external accelerations are stored separately to enable this. $\Delta t_{\rm ext}$ is the minimum of all timesteps relating to sink-gas and external forces (equations~\ref{eq:dtext}--\ref{eq:dtphi}) while $\Delta t_{\rm sph}$ is the timestep relating to the SPH forces (equations \ref{eq:dtcour}, \ref{eq:dtforce} and \ref{eq:dtcool}).  $\Delta t_{\rm ext}$ is allowed to vary on each substep, so we take as many steps as required such that $\sum_j^{m-1} \Delta t_{{\rm ext},j} + \Delta t_{{\rm ext},f}= \Delta t_{\rm SPH}$, where $\Delta t_{{\rm ext},f} <  \Delta t_{{\rm ext},j}$ is chosen so that the sum will identically equal $\Delta t_{\rm SPH}$.  The number of substeps is $m \approx {\rm int}(\Delta t_{\rm ext, min}/\Delta t_{\rm SPH, min}) + 1$, where the minimum is taken over all particles.

\subsubsection{Individual particle timesteps}
\label{sec:indtimesteps}
 For simulations of stiff problems with a large range in timestep over the domain, it is more efficient to allow each particle to evolve on its own timestep independently \citep{bate95,springel05,saitohmakino10}. This violates all of the conservation properties of the Leapfrog integrator (see \citealt{makinoetal06} for an attempt to solve this), but can speed up the calculation by an order of magnitude or more. We implement this in the usual block-stepped manner by assigning timesteps in factor-of-two decrements from some maximum timestep $\Delta t_{\rm max}$, which for convenience is set equal to the time between output files.
 
 We implement a timestep limiter where the timestep for an active particle is constrained to be within a factor of 2 of its neighbours, similar to condition employed by \citet{saitohmakino09}.  Additionally, inactive particles will be woken up as required to ensure that their timestep is within a factor of 2 of its neighbours.

 The practical side of individual timestepping is described in Appendix~\ref{sec:indtimestepprac}.
 
\subsection{External forces}
\label{sec:extf}

\subsubsection{Point mass potential}
\label{sec:extptmass}
The simplest external force describes a point mass, $M$, at the origin, which yields gravitational potential and acceleration,
\begin{equation}
\Phi_{a} = -\frac{GM}{r_{a}}; \hspace{1cm} {\bm a}_{{\rm ext},a} = -\nabla\Phi_{a} = -\frac{GM}{\vert {\bm r}_{a} \vert^{3}} {\bm r}_{a}, \label{eq:externptmass}
\end{equation}
where $r_{a} \equiv \vert {\bm r}_{a} \vert \equiv \sqrt{{\bm r}_{a}\cdot{\bm r}_{a}}$. When this potential is used, we allow for particles within a certain radius, $R_{\rm acc}$, from the origin to be accreted. This allows for a simple treatment of accretion discs where the mass of the disc is assumed to be negligible compared to the mass of the central object. The accreted mass in this case is recorded but not added to the central mass. For more massive discs, or when the accreted mass is significant with respect to the central mass, it is better to model the central star using a sink particle (Section~\ref{sec:sinks}) where there are mutual gravitational forces between the star and the disc, and any mass accreted is added to the point mass (Section~\ref{sec:accrete}).

\subsubsection{Binary potential}
\label{sec:extbinary}
We provide the option to model motion in binary systems where the mass of the disc is negligible. In this case the binary motion is prescribed using
\begin{align}
{\bm r}_{1} & = [(1 - M) \cos(t),(1- M) \sin(t), 0], \\
{\bm r}_{2} & = [- M \cos(t), - M \sin(t), 0],
\end{align}
where $M$ is the mass ratio in units of the total mass (which is therefore unity). For this potential, $G$ and $\Omega$ are set to unity in computational units, where $\Omega$ is the angular velocity of the binary. Thus only $M$ needs to be specified to fix both $m_1$ and $m_{2}$. Hence the binary remains fixed on a circular orbit at $r=1$. The binary potential is therefore
\begin{equation}
\Phi_{a} = -\frac{M}{\vert {\bm r}_{a} - {\bm r}_{1} \vert} - \frac{(1- M) }{\vert {\bm r}_{a} - {\bm r}_{2} \vert}, \label{eq:extbinarypotential}
\end{equation}
such that the external acceleration is given by
\begin{equation}
{\bm a}_{{\rm ext},a} = -\nabla\Phi_{a} =  - M \frac{({\bm r}_{a} - {\bm r}_{1})}{\vert {\bm r}_{a} - {\bm r}_{1} \vert^3} - (1- M) \frac{({\bm r}_{a} - {\bm r}_{2})}{\vert {\bm r}_{a} - {\bm r}_{2} \vert^3}. \label{eq:extbinary}
\end{equation}
Again, there is an option to accrete particles that fall within a certain radius from either star ($R_{{\rm acc}, 1}$ or $R_{{\rm acc}, 2}$, respectively). For most binary accretion disc simulations (e.g. planet migration) it is better to use `live' sink particles to represent the binary so that there is feedback between the binary and the disc (we have used a live binary in all of our simulations to date, e.g. \citealt{nkp13,flp13,martinetal14,martinetal14a,doganetal15,rlp16,ragusaetal17}), but the binary potential remains useful under limited circumstances --- in particular when one wishes to turn off the feedback between the disc and the binary.

Given that the binary potential is time-dependent, for efficiency we compute the position of the binary only once at the start of each timestep, and use these stored positions to compute the accelerations of the SPH particles via (\ref{eq:extbinary}).

\subsubsection{Binary potential with gravitational wave decay}
An alternative binary potential including the effects of gravitational wave decay was used by \citet*{clp16} to study the squeezing of discs during the merger of supermassive black holes. Here the motion of the binary is prescribed according to
\begin{align}
{\bm r}_{1} & = \left[-\frac{m_{2}}{m_{1} + m_{2}} a \cos(\theta), -\frac{m_{2}}{m_{1} + m_{2}} a \sin(\theta), 0 \right], \nonumber \\
{\bm r}_{2} & = \left[\frac{m_{1}}{m_{1} + m_{2}} a \cos(\theta), \frac{m_{1}}{m_{1} + m_{2}} a \sin(\theta), 0 \right],
\end{align}
where the semi-major axis, $a$, decays according to
\begin{equation}
a(t) = a_{0} \left(1 - \frac{t}{\tau} \right)^{\frac14} .
\end{equation}
The initial separation is $a_{0}$, with $\tau$ defined as the time to merger, given by the usual expression \citep[e.g.][]{lodatoetal09}
\begin{equation}
\tau \equiv \frac{5}{256} \frac{a_{0}^{4}}{\mu_{12} (m_{1} + m_{2})^{2}},
\end{equation}
where 
\begin{equation}
\mu_{12} \equiv \frac{m_1 m_2}{m_1 + m_2}.
\end{equation}
The angle $\theta$ is defined using
\begin{equation}
\Omega \equiv \frac{{\rm d}\theta}{{\rm d}t} = \sqrt{\frac{G(m_{1} + m_{2})}{a^{3}}}.
\end{equation}
Inserting the expression for $a$ and integrating gives \citep{clp16}
\begin{equation}
\theta(t) = -\frac{8\tau}{5} \sqrt{\frac{G(m_{1} + m_{2})}{a_{0}^{3}}} \left( 1 - \frac{t}{\tau} \right).
\end{equation}
The positions of the binary, ${\bm r}_1$ and ${\bm r}_2$, can be inserted into (\ref{eq:extbinarypotential}) to obtain the binary potential, with the acceleration as given in (\ref{eq:extbinary}). The above can be used as a simple example of a time-dependent external potential.

\subsubsection{Galactic potentials}
\label{sec:galdisc}
We implement a range of external forces representing various galactic potentials, as used in \citet{pettittetal14}. These include arm, bar, halo, disc and spheroidal components. We refer the reader to the paper above for the actual forms of the potentials.

For the non-axisymmetric potentials a few important parameters that determine the morphology can be changed at run-time rather than compile time. These include the pattern speed, arm number, arm pitch angle and bar axis lengths (where applicable). In the case of non-axisymmetric components, the user should be aware that some will add mass to the system, whereas others simply perturb the galactic disc. These potentials can be used for any galactic system, but the various default scale lengths and masses are chosen to match the Milky Way's rotation curve \citep{2012PASJ...64...75S}.

The most basic potential in \textsc{Phantom} is a simple logarithmic potential from \citet{1987gady.book.....B}, which allows for the reproduction of a purely flat rotation curve with steep decrease at the galactic centre, and approximates the halo, bulge and disc contributions. Also included is the standard flattened disc potential of Miyamoto-Nagai \citep{1975PASJ...27..533M} and an exponential profile disc, specifically the form from \citet{2013MNRAS.428.2311K}. Several spheroidal components are available, including the potentials of \citet{1911MNRAS..71..460P}, \citet{1990ApJ...356..359H} and \citet{1930ApJ....71..231H}. These can be used generally for bulges and halos if given suitable mass and scale-lengths. We also include a few halo-specific profiles; the NFW \citep{1996ApJ...462..563N}, \citet{1991MNRAS.249..523B}, \citet{1981ApJ...251...61C} and the \citet{1991RMxAA..22..255A} potentials.

The arm potentials include some of the more complicated profiles. The first is the potential of \citet{2002ApJS..142..261C}, which is a relatively straightforward superposition of three sinusoidal-based spiral components to damp the potential ``troughs" in the inter-arm minima. The other spiral potential is from \citet{2003ApJ...582..230P}, and is more complicated. Here the arms are constructed from a superposition of oblate spheroids whose loci are placed along a standard logarithmic spiral. As the force from this potential is computationally expensive it is prudent to pre-compute a grid of potential/force and read it at run time. The python code to generate the appropriate grid files is distributed with the code.

Finally, the bar components: We include the bar potentials of \citet{2000AJ....119..800D}, \citet{2001PASJ...53.1163W}, the ``S" shaped bar of \citet{MNR:MNR17857}, both biaxial and triaxial versions provided in \citet{1992ApJ...397...44L}, and the boxy-bulge bar of \citet{2012MNRAS.427.1429W}. This final bar is contains both a small inner non-axisymmetric bulge and longer bar component, with the forces calculated by use of Hernquist-Ostriker expansion coefficients of the bar density field. \textsc{Phantom} contains the coefficients for several different forms of this bar potential.

\subsubsection{Lense-Thirring precession}
\label{sec:lt}
Lense-Thirring precession \citep{lensethirring18} from a spinning black hole is implemented in a Post-Newtonian approximation following \citet{nelsonpapaloizou00}, which has been used in \citet{nixonetal12} and \citet{npn15,nealonetal16}. In this case the external acceleration consists of a point mass potential (Section~\ref{sec:extptmass}) and the Lense-Thirring term,
\begin{equation}
{\bm a}_{{\rm ext},a} = -\nabla\Phi_a + {\bm v}_a \times {\bm \Omega}_{p,a},
\end{equation}
where $\Phi_a$ is given by (\ref{eq:externptmass}) and ${\bm v}_a \times {\bm \Omega}_{p,a}$ is the gravitomagnetic acceleration. A dipole approximation is used, yielding
\begin{equation}
{\bm \Omega}_{p,a} \equiv \frac{2{\bm S}}{\vert {\bm r}_a \vert^{3}} - \frac{6({\bm S}\cdot{\bm r}_a) {\bm r}_a}{\vert {\bm r}_a \vert^{5}},
\end{equation}
with ${\bm S} = a_{\rm spin} (GM)^{2} {\bm k}/c^{3}$, where ${\bm k}$ is a unit vector in the direction of the black hole spin. When using the Lense-Thirring force, geometric units are assumed such that $G=M=c=1$, as described in Section~\ref{sec:units}, but with the additional constraints on the unit system from $M$ and $c$.

Since in this case the external force depends on velocity, it cannot be implemented directly into Leapfrog. The method we employ to achieve this is simpler than those proposed elsewhere (c.f. attempts by \citealt{quinnetal10} and \citealt{reintremaine11} to adapt the Leapfrog integrator to Hill's equations). Our approach is to split the acceleration into position and velocity-dependent parts, i.e.
\begin{equation}
{\bm a}_{\rm ext} = {\bm a}_{\rm ext,x}({\bm r}) + {\bm a}_{\rm ext,v}({\bm r}, {\bm v}).
\end{equation}
The position dependent part (i.e. $-\nabla\Phi({\bm r})$) is integrated as normal. The velocity dependent Lense-Thirring term is added to the predictor step, (\ref{eq:xpred})--(\ref{eq:vpred}), as usual, but the corrector step, (\ref{eq:lfcorrv}), is written in the form
\begin{equation}
{\bm v}^{n + 1} = {\bm v}^{n + \frac12} + \frac12 \Delta t  \left[ {\bm a}_{\rm sph}^{n+1} + {\bm a}_{\rm ext,x}^{n+1} +  {\bm a}_{\rm ext,v}({\bm r}^{n+1}, {\bm v}^{n+1})\right]. \label{eq:lfcorrextv}
\end{equation}
where ${\bm v}^{n + \frac12} \equiv {\bm v}^{n} + \frac12 \Delta t {\bm a}^{n}$ as in (\ref{eq:lfvhalf}). This equation is implicit but the trick is to notice that it can be solved analytically for simple forces\footnote{The procedure for Hill's equations would be identical to our method for Lense-Thirring precession. The method we use is both simpler and more direct than any of the schemes proposed by \citet{quinnetal10} and \citet{reintremaine11}, and is time-reversible unlike the methods proposed in those papers.}. In the case of Lense-Thirring precession, we have
\begin{equation}
{\bm v}^{n + 1} = \tilde{\bm v} +  \frac12 \Delta t  \left[{\bm v}^{n+1} \times {\bm \Omega}_{p}({\bm r}^{n+1}) \right], \label{eq:ltcorr}
\end{equation}
where $\tilde{\bm v} \equiv {\bm v}^{n + \frac12} + \frac12 \Delta t  ({\bm a}_{\rm sph}^{n+1} + {\bm a}_{\rm ext,x}^{n+1})$. We therefore have a matrix equation in the form
\begin{equation}
{\bm R} {\bm v}^{n+1} = \tilde{\bm v}, \label{eq:rmatv}
\end{equation}
where ${\bm R}$ is the $3\times 3$ matrix given by
\begin{equation}
{\bm R} \equiv \left[ \begin{array}{ccc}
1 & -  \frac{\Delta t}{2} \Omega^{z}_{p} & \frac{\Delta t}{2} \Omega^{y}_{p} \\
 \frac{\Delta t}{2}  \Omega^{z}_{p} &  1 & -\frac{\Delta t}{2}  \Omega^{x}_{p} \\
 -\frac{\Delta t}{2} \Omega^{y}_{p} & \frac{\Delta t}{2}  \Omega^{x}_{p} & 1 \\
\end{array} \right]. \label{eq:rmat}
\end{equation}
Rearranging (\ref{eq:rmatv}), ${\bm v}^{n+1}$ is obtained by using
\begin{equation}
 {\bm v}^{n+1} = {\bm R}^{-1}\tilde{\bm v},
\end{equation}
where ${\bm R}^{-1}$ is the inverse of ${\bm R}$ and can be computed analytically.

\subsubsection{Generalised Newtonian potential}
\label{sec:gnewton}

The generalised Newtonian potential described by \citet{tejedarosswog13} is implemented, where the acceleration terms are given by
\begin{equation}
{\bm a}_{{\rm ext},a} = -\frac{GM{\bm r}_a}{\vert {\bm r}_a \vert^{3}}f^{2} + \frac{2R_{\rm g}  {\bm v}_a ({\bm v}_a \cdot{\bm r}_a)}{\vert {\bm r}_a \vert^{3} f} - \frac{3 R_{\rm g} {\bm r}_a ({\bm v}_a \times {\bm r}_a)^{2}}{\vert {\bm r}_a \vert^{5}}, 
\end{equation}
with $R_{\rm g} \equiv GM/c^{3}$ and $f \equiv \left(1 - 2R_{\rm g}/\vert {\bm r}_a \vert \right)$. See \citet{bonnerotetal16} for a recent application. This potential reproduces several features of the \citet{schwarzschild16} spacetime, in particular reproducing the orbital and epicyclic frequencies to better than 7 per cent \citep{tejedarosswog13}. As the acceleration involves velocity-dependent terms, it requires a semi-implicit solution like Lense-Thirring precession. Since the matrix equation is rather involved for this case, the corrector step is iterated using fixed point iterations until the velocity of each particle is converged to a tolerance of 1 per cent. 

\subsubsection{Poynting-Robertson drag}
\label{sec:prdrag}
The radiation drag from a central point-like, gravitating, radiating, and non-rotating object may be applied as an external force. The implementation is intended to be as general as possible. The acceleration of a particle subject to these external forces is
\begin{align}
{\bm a}_{{\rm ext},a} =& \dfrac{(k_0\beta_{\rm PR}-1)GM}{\vert {\bm r}_a \vert^3} \vcr_a \nonumber \\
& -\beta_{\rm PR} \left(k_1\dfrac{GM}{\vert {\bm r}_a \vert^3}\dfrac{v_r}{c}\vcr_a -k_2\dfrac{GM}{\vert {\bm r}_a \vert^2}\dfrac{\vcv_a}{c}\right),
\label{eq:prdrag}
\end{align}
where $v_r$ is the component of the velocity in the radial direction. The parameter $\beta_{\rm PR}$ is the ratio of radiation to gravitational forces, supplied by a separate user-written module. Relativistic effects are neglected because these are thought to be less important than radiation forces for low ($\beta_{\rm PR} < 0.01$) luminosities, even in accreting neutron star systems where a strong gravitational field is present (e.g., \citealt{millerlamb93}).

The three terms on the right side of (\ref{eq:prdrag}) correspond respectively to gravity (reduced by outward radiation pressure), redshift-related modification to radiation pressure caused by radial motion, and Poynting-Robertson drag against the direction of motion. These three terms can be scaled independently by changing the three parameters $k_0$, $k_1$ and $k_2$, whose default values are unity. Rotation of the central object can be crudely emulated by changing $k_2$.

As for Lense-Thirring precession, the ${\bm a}^{n+1}$ term of the Leapfrog integration scheme can be expanded into velocity-dependent and non velocity-dependent component. We obtain, after some algebra,
\begin{equation}
\vcv^{n+1} =-\dfrac{\vcT-Qk_1(\vcv^{n+1}\cdot\hat{\vcr})\hat{\vcr}}{1+Qk_2},
\label{eqn:prd_updatedvel}
\end{equation}
where
\begin{equation}
\vcT=\vcv^n+\frac{1}{2}\Delta t\vca^n-\dfrac{(1-k_0\beta_{\rm PR})GM\Delta t}{2r^3}\vcr
\end{equation}
and 
\begin{equation}
Q=\dfrac{GM\beta_{\rm PR}\Delta t}{2cr^2}.
\end{equation}
Equation (\ref{eqn:prd_updatedvel}) yields a set of simultaneous equations for the three vector components that can be solved analytically. A detailed derivation is given in \cite{worpel15}.

\subsubsection{Coriolis and centrifugal forces}
\label{sec:coriolis}
 Under certain circumstances it is useful to perform calculations in a corotating reference frame (e.g. for damping binary stars into equilibrium with each other). 
The resulting acceleration terms are given by
\begin{equation}
{\bm a}_{{\rm ext},a} = -{\bm \Omega} \times ({\bm \Omega} \times {\bm r}_a) - 2 ({\bm \Omega} \times {\bm v}_a),
\end{equation}
which are the centrifugal and Coriolis terms, respectively, with ${\bm \Omega}$ the angular rotation vector. The timestepping algorithm is as described above for Lense-Thirring precession, with the velocity dependent term handled by solving the $3\times 3$ matrix in the Leapfrog corrector step.

\subsection{Driven turbulence}
\label{sec:turbforcing}
\textsc{Phantom} implements turbulence driving in periodic domains via an Ornstein-Uhlenbeck stochastic driving of the acceleration field, as first suggested by \citet{eswaranpope88}. This is an SPH adaptation of the module used in the grid-based simulations by \citet{shn06} and \citet{fks08} and many subsequent works. This module was first used in \textsc{Phantom} by \citet{pricefederrath10} to compare the statistics of isothermal, supersonic turbulence between SPH and grid methods. Subsequent applications have been to the density variance-Mach number relation \citep{pfb11}, subsonic turbulence \citep{price12a}, supersonic MHD turbulence  \citep{tpf16}, and supersonic turbulence in a dust-gas mixture \citep{tpl17}. Adaptations of this module have also been incorporated into other SPH codes \citep{bauerspringel12,valdarnini16}. 

 The amplitude and phase of each Fourier mode is initialised by creating a set of six random numbers, $z_{n}$, drawn from a random Gaussian distribution with unit variance. These are generated by the usual Box-Muller transformation \citep[e.g.][]{pressetal92} by selecting two uniform random deviates $u_{1}, u_{2} \in [0,1]$ and constructing the amplitude according to
\begin{equation}
z = \sqrt{2 \log (1 /u_{1})} \cos (2 \pi u_{2}).
\end{equation}
The six Gaussian random numbers are set up according to
 \begin{equation}
{\bm x}_{n} = \sigma {\bm z}_{n},
\end{equation}
where the standard deviation, $\sigma$, is set to the square root of the energy per mode divided by the correlation time, $\sigma = \sqrt{E_{\rm m}/t_{\rm decay}}$, where both $E_{\rm m}$ and $t_{\rm decay}$ are user-specified parameters.
 
The `red noise' sequence \citep{uhlenbeckornstein30} is generated for each mode at each timestep according to \citep{bartosch01}
\begin{equation}
{\bm x}_{n+1}= f {\bm x}_{n} + \sigma \sqrt{(1 - f^{2})} {\bm z}_n,
\end{equation}
where $f = \exp(-\Delta t/t_{\rm decay})$ is the damping factor. The resulting sequence has zero mean with root-mean-square equal to the variance. The power spectrum in the time domain can vary from white noise ($P(f)$ constant) to ``brown noise'' ($P(f) = 1/f^{2}$).

The amplitudes and phases of each mode are constructed by splitting ${\bm x}_{n}$ into two vectors, ${\bm \Phi}_{a}$ and ${\bm \Phi}_{b}$ of length 3, employed to construct divergence- and curl-free fields according to
\begin{align}
{\bm A}_{m} & = w [{\bm \Phi}_{a} -  ( {\bm \Phi}_{a} \cdot\hat{\bm k} )\hat{\bm k}] + (1 - w)[({\bm \Phi}_{b} \cdot \hat{\bm k} )\hat{\bm k}] , \\
{\bm B}_{m} & = w [{\bm \Phi}_{b} -  ( {\bm \Phi}_{b} \cdot\hat{\bm k} )\hat{\bm k}] + (1 - w)[({\bm \Phi}_{a} \cdot \hat{\bm k} )\hat{\bm k}] ,
\end{align}
where ${\bm k} = [k_{x}, k_{y}, k_{z}]$ is the mode frequency. The parameter $w \in [0,1]$ is the `solenoidal weight', specifying whether the driving should be purely solenoidal ($w=1$) or purely compressive ($w=0$) (see \citealt{fks08,federrathetal10}).

The spectral form of the driving is defined in Fourier space, with options for three possible spectral forms 
\begin{equation}
C_{m} = 
\begin{cases}
1 & \textrm{uniform} \\
4 (a_{\rm min} - 1) \frac{(k - k_{c})^{2}}{(k_{\max} - k_{\min})^{2}} + 1 & \textrm{parabolic} \\
k/k_{\min}^{-5/3} & \textrm{Kolmogorov}
\end{cases}
\end{equation}
where $k = \sqrt{k_{x}^{2} + k_{y}^{2} + k_{z}^{2}}$ is the wavenumber, with non-zero amplitudes defined only for wavenumbers where $k_{\min} \leq k \leq k_{\max}$, and $a_{\rm min}$ is the amplitude of the modes at $k_{\rm min}$ and $k_{\rm max}$ in the parabolic case (we use $a_{\rm min} = 0$ in the code). The frequency grid is defined using frequencies from $k_{x}=n_x 2\pi/L_x$ in the $x$ direction, where $n_x \in [0,20]$ is an integer and $L_x$ is the box size in the $x-$direction, while $k_{y} = n_y 2\pi/L_y$ and $k_{z} = n_z 2\pi/L_z$ with $n_y \in [0,8]$ and $n_z \in [0,8]$. We then set up four modes for each combination of $n_x$, $n_y$ and $n_z$, corresponding to [$k_{x}$, $k_{y}$, $k_{z}$], $[k_{x}, -k_{y}, k_{z}]$, $[k_{x}, k_{y}, -k_{z}]$ and $[k_{x}, -k_{y}, -k_{z}]$. That is, we effectively sum from $[-(N-1)/2,(N -1)/2]$ in the $k_y$ and $k_z$ directions in the discrete Fourier transform, where $N = \max(n_x)$ is the number of frequencies. The default values for $k_{\min}$ and $k_{\max}$ are $2\pi$ and $6\pi$, respectively, corresponding to large scale driving of only the first few Fourier modes, so with default settings there are $112$ non-zero Fourier modes. The maximum number of modes, defining the array sizes needed to store the stirring information, is currently set to 1000.

We apply the forcing to the particles by computing the discrete Fourier transform over the stirring modes directly, i.e.
\begin{equation}
{\bm a}_{{\rm forcing}, a} = f_{\rm sol} \sum_{m=1}^{n_{\rm modes}} C_{m} \left[ {\bm A}_{m} \cos({\bm k}\cdot{\bm r}_{a}) - {\bm B}_{m}\sin({\bm k}\cdot{\bm r}_{a}) \right],
\end{equation}
where the factor $f_{\rm sol}$ is defined from the solenoidal weight, $w$, according to
\begin{equation}
f_{\rm sol} = \sqrt{\frac{3}{n_{\rm dim}}}  \sqrt{ \frac{3}{1 - 2w + n_{\rm dim} w^{2}}},
\end{equation}
such that the rms acceleration is the same irrespective of the value of $w$. We default to purely solenoidal forcing ($w=1$), with the factor $f_{\rm sol}$ thus equal to $\sqrt{3/2}$ by default. For individual timesteps, we update the driving force only when a particle is active.

 To aid reproducibility, it is often desirable to pre-generate the entire driving sequence prior to performing a calculation, which can then be written to a file and read back at runtime. This was the procedure used in \citet{pricefederrath10}, \citet{tpf16} and \citet{tpl17}.
 
\subsection{Accretion disc viscosity}
\label{sec:discviscosity}
 Accretion disc viscosity is implemented in \textsc{Phantom} via two different approaches, as described by \citet{lodatoprice10}. 
 
 \subsubsection{Disc viscosity using the shock viscosity term}
 \label{sec:discav}
 The default approach is to adapt the shock viscosity term to represent a \citet{shakurasunyaev73} $\alpha$-viscosity, as originally proposed by \citet{artymowiczlubow94} and \citet{murray96}. The key is to note that (\ref{eq:pishock}) and (\ref{eq:qvisc}) represent a Navier-Stokes viscosity term with a fixed ratio between the bulk and shear viscosity terms \citep[e.g.][]{murray96,jsd04,lodatoprice10,price12a,merubate12}. In particular, it can be shown \citep[e.g.][]{espanolrevenga03} that
\begin{equation}
\sum_{b} \frac{m_{b}}{\overline{\rho}_{ab}} ({\bm v}_{ab}\cdot \hat{\bm r}_{ab} )\frac{\nabla_{a} W_{ab}}{\vert r_{ab}\vert} \approx \frac15 \nabla \left( \nabla\cdot {\bm v}\right) + \frac{1}{10}\nabla^{2} {\bm v}, \label{eq:avns}
\end{equation}
where $\overline{\rho}_{ab}$ is some appropriate average of the density. This enables the artificial viscosity term, (\ref{eq:pishock}), to be translated into the equivalent Navier-Stokes terms. In order for the artificial viscosity to represent a disc viscosity, we make the following modifications \citep{lodatoprice10}:
\begin{enumerate}
\item the viscosity term is applied for both approaching and receding particles,
\item the speed in $v_{\rm sig}$ is set equal to $c_{\rm s}$,
\item a constant $\alpha^{\rm AV}$ is adopted, turning off shock detection switches (Section~\ref{sec:switches}), and
\item the viscosity term is multiplied by a factor $h/\vert r_{ab} \vert$.
\end{enumerate}
The net result is that (\ref{eq:qvisc}) becomes
\begin{equation}
q^{a}_{ab} = \begin{cases}- \frac{\rho_{a} h_{a}}{2 \vert r_{ab}\vert} \left(\alpha^{\rm AV} c_{{\rm s}, a}  + \beta_{\rm AV} \vert {\bm v}_{ab} \cdot \hat{\bm r}_{ab}\vert \right) {\bm v}_{ab} \cdot \hat{\bm r}_{ab}, &  {\bm v}_{ab} \cdot \hat{\bm r}_{ab} < 0 \\
-  \frac{\rho_{a} h_{a}}{2 \vert r_{ab}\vert} \alpha^{\rm AV} c_{{\rm s}, a} {\bm v}_{ab} \cdot \hat{\bm r}_{ab}. & \text{otherwise} \end{cases}\label{eq:qviscdisc}
\end{equation}

With the above modifications, the shear and bulk coefficients can be translated using (\ref{eq:avns}) to give \citep[e.g.][]{monaghan05,lodatoprice10,merubate12}
\begin{align}
\nu_{\rm AV} & \approx \frac{1}{10} \alpha^{\rm AV} c_{\rm s} h, \\
\zeta_{\rm AV} & = \frac53 \nu_{\rm AV} \approx  \frac{1}{6} \alpha^{\rm AV} c_{\rm s} h.
\end{align}
The Shakura-Sunyaev prescription is
\begin{equation}
\nu = \alpha_{\rm SS} c_{\rm s} H, \label{eq:ss}
\end{equation}
where $H$ is the scale height. This implies that $\alpha_{\rm SS}$ may be determined from $\alpha_{\rm AV}$ using
\begin{equation}
\alpha_{\rm SS} \approx \frac{\alpha^{\rm AV}}{10} \frac{\langle h \rangle}{H}, \label{eq:alphascale}
\end{equation}
where $\langle h \rangle$ is the mean smoothing length on particles in a cylindrical ring at a given radius.

In practice, this means that one must uniformly resolve the scale height in order to obtain a constant $\alpha_{\rm SS}$ in the disc. We have achieved this in simulations to date by choosing the initial surface density profile and the power-law index of the temperature profile (when using a locally isothermal equation of state) to ensure that this is the case \citep{lodatopringle07}. Confirmation that the scaling provided by (\ref{eq:alphascale}) is correct is shown in Figure 4 of \citet{lodatoprice10} and is checked automatically in the \textsc{Phantom} test suite.

 In the original implementation \citep{lodatoprice10} we also set the $\beta^{\rm AV}$ to zero, but this is dangerous if the disc dynamics are complex as there is nothing to prevent particle penetration (see Section~\ref{sec:av}). Hence in the current version of the code, $\beta^{\rm AV} = 2$ by default even if disc viscosity is set, but is only applied to approaching particles (c.f.~\ref{eq:qviscdisc}). Applying any component of viscosity to only approaching particles can affect the sign of the precession induced in warped discs \citep{lodatopringle07}, but in general retaining the $\beta^{\rm AV}$ term is safer with no noticeable effect on the overall dissipation due to the second order dependence of this term on resolution.

 Using $\alpha^{\rm AV}$ to set the disc viscosity has two useful consequences. First, it forces one to consider whether or not the scale height, $H$, is resolved. Second, knowing the value of $\alpha^{\rm AV}$ is helpful, as $\alpha^{\rm AV} \approx 0.1$ represents the lower bound below which a physical viscosity is not resolved in SPH (that is, viscosities smaller than this produce disc spreading independent of the value of $\alpha^{\rm AV}$; see \citealt{bate95,merubate12}), while $\alpha^{\rm AV} > 1$ constrains the timestep (Section~\ref{sec:timestep}).

\subsubsection{Disc viscosity using the Navier-Stokes viscosity terms}
An alternative approach is to compute viscous terms directly from the Navier-Stokes equation. Details of how the Navier-Stokes terms are represented are given below (Section~\ref{sec:viscosity}), but for disc viscosity a method for determining the kinematic viscosity is needed, which in turn requires specifying the scale height as a function of radius. We use
\begin{equation}
H_{a} \equiv \frac{c^{a}_{\rm s}}{\Omega(R_{a})},
\end{equation}
where we assume Keplerian rotation $\Omega = \sqrt{GM/R^{3}}$ and $c_{\rm s}$ is obtained for a given particle from the equation of state (which for consistency must be either isothermal or locally isothermal). It is important to note that this restricts the application of this approach only to discs where $R$ can be meaningfully defined, excluding, for example, discs around binary stars.  

The shear viscosity is then set using
\begin{equation}
\nu_{a} = \alpha_{\rm SS} c_{\rm s}^{a} H_{a},
\label{eq:nufromalpha}
\end{equation}
where $\alpha_{\rm SS}$ is a predefined/input parameter. The timestep is constrained using $C_{\rm visc} h^{2}/\nu$ as described in Section~\ref{sec:viscosity}. The advantage to this approach is that the shear viscosity is set directly and does not depend on the smoothing length. However, as found by \citet{lodatoprice10}, it remains necessary to apply some bulk viscosity to capture shocks and prevent particle penetration of the disc midplane, so one should apply the shock viscosity as usual. Using a shock-detection switch (Section~\ref{sec:switches}) means that this is usually not problematic. This formulation of viscosity was used in \citet{flp13}.

\subsection{Navier-Stokes viscosity}
\label{sec:viscosity}
 Physical viscosity is implemented as described in \citet{lodatoprice10}. Here, (\ref{eq:mom}) and (\ref{eq:dudt}) are replaced by the compressible Navier-Stokes equations, i.e.
\begin{align}
\frac{{\rm d}v^{i}}{{\rm d}t} = & -\frac{1}{\rho}\frac{\partial S_{\rm NS}^{ij}}{\partial x^{j}} + \Pi_{\rm shock} + {\bm a}_{\rm ext}({\bm r}, t) \nonumber \\
&  + {\bm a}_{\rm sink-gas} + {\bm a}_{\rm selfgrav}, \label{eq:ns} \\
\frac{{\rm d}u}{{\rm d}t} = & -\frac{P}{\rho} \left(\nabla\cdot{\bm v}\right) + \Lambda_{\rm visc} + \Lambda_{\rm shock} - \Lambda_{\rm cool}, \label{eq:dudtns}
\end{align}
with the stress tensor given by
\begin{equation}
S_{\rm NS}^{ij} = \left[ P - \left(\zeta - \frac23\eta \right) \frac{\partial v^{k}}{\partial x^{k}} \right] \delta^{ij} - \eta \left( \frac{\partial v^{i}}{\partial x^{j}} + \frac{\partial v^{j}}{\partial x^{i}}\right)
\label{eq:sijns}
\end{equation}
where $\delta^{ij}$ is the Kronecker delta, and $\zeta$ and $\eta$ are the bulk and shear viscosity coefficients, related to the volume and kinematic shear viscosity coefficients by $\zeta_{v} = \zeta/\rho$ and $\nu \equiv \eta/\rho$.

\subsubsection{Physical viscosity using two first derivatives}
\label{sec:twofirstderivs}
 As there is no clear consensus on the best way to implement physical viscosity in SPH, \textsc{Phantom} currently contains two implementations. The simplest is to use two first derivatives, which is essentially that proposed by \citet{flebbeetal94}, \citet{watkinsetal96} and \citet{sijackispringel06}. In this case, (\ref{eq:ns}) is discretised in the standard manner using
\begin{align}
\frac{{\rm d}v^{i}_{a}}{{\rm d}t}  =& -\sum_{b} m_{b}\left[ \frac{S^{ij}_{{\rm NS}, a}}{\Omega_{a}\rho_{a}^{2}} \frac{\partial W_{ab}(h_{a})}{\partial x^{j}_{a}} + \frac{S^{ij}_{{\rm NS}, b}}{\Omega_{b}\rho_{b}^{2}} \frac{\partial W_{ab}(h_{b})}{\partial x^{j}_{a}}\right] \nonumber\\
& + \Pi^{i}_{\rm shock} + a^{i}_{\rm ext}({\bm r}, t)   + a^{i}_{\rm sink-gas} + a^{i}_{\rm selfgrav}, \label{eq:nssph}
\end{align}
where the velocity gradients are computed during the density loop using
\begin{equation}
\frac{\partial v_{a}^{i}}{\partial x_{a}^{j}} = -\frac{1}{\Omega_{a}\rho_{a}} \sum_{b} m_{b} v_{ab}^{i} \nabla_a^{j} W_{ab} (h_{a}). 
\label{eq:strain}
\end{equation}
Importantly, the differenced SPH operator is used in (\ref{eq:strain}) whereas (\ref{eq:nssph}) uses the symmetric gradient operator. The use of conjugate operators\footnote{The SPH difference operator is discretely skew-adjoint to the symmetric operator \citep{cumminsrudman99}.} is a common requirement in SPH in order to guarantee energy conservation and a positive definite entropy increase from dissipative terms \citep[e.g.][]{price10,triccoprice12}. Total energy conservation means that
\begin{equation}
\frac{{\rm d}E}{{\rm d}t} = \sum_{a} m_{a} \left( \frac{{\rm d}u_{a}}{{\rm d}t} + v^{i}_{a} \frac{{\rm d}v^{i}_{a}}{{\rm d}t} \right) = 0.
\end{equation}
This implies a contribution to the thermal energy equation given by
\begin{equation}
\frac{{\rm d}u_{a}}{{\rm d}t} = \frac{S_{{\rm NS}, a}^{ij}}{\Omega_{a} \rho_{a}^{2}} \sum_{b} m_{b} v_{ab}^{i} \nabla_a^{j} W_{ab} (h_{a}),
\end{equation}
which can be seen to reduce to (\ref{eq:dudt}) in the inviscid case ($S_{\rm NS}^{ij} = P \delta^{ij}$), but in general is an SPH expression for
\begin{equation}
\frac{{\rm d}u_{a}}{{\rm d}t} = -\frac{S_{{\rm NS}, a}^{ij}}{\rho_{a}} \frac{\partial v_{a}^{i}}{\partial x_{a}^{j}}.
\end{equation}
Using $S_{\rm NS}^{ij} = S_{\rm NS}^{ji}$ we have
\begin{equation}
\frac{{\rm d}u_{a}}{{\rm d}t} = -\frac12 \frac{S_{{\rm NS}, a}^{ij}}{\rho_{a}} \left( \frac{\partial v_{a}^{i}}{\partial x_{a}^{j}} +  \frac{\partial v_{a}^{j}}{\partial x_{a}^{i}} \right),
\end{equation}
which, using (\ref{eq:sijns}), gives
\begin{align}
\Lambda_{\rm visc} = \left(\zeta_{v, a} - \frac23 \nu_{a} \right) (\nabla\cdot{\bm v})_{a}^{2} + \frac{\nu_{a}}{2} \left( \frac{\partial v_{a}^{i}}{\partial x_{a}^{j}}
 + \frac{\partial v_{a}^{j}}{\partial x_{a}^{i}} \right)^{2}.
 \label{eq:vischeating}
\end{align}
By the square in the last term we mean the tensor summation $\sum_{j} \sum_{i} T_{ij} T_{ij}$, where $T_{ij} \equiv {\partial v_{a}^{i}}/{\partial x_{a}^{j}}
 +{\partial v_{a}^{j}}/{\partial x_{a}^{i}} $.
The heating term is therefore positive definite provided that the velocity gradients and divergence are computed using the difference operator (\ref{eq:strain}), both in (\ref{eq:vischeating}) and when computing the stress tensor (\ref{eq:sijns}).

 The main disadvantage of the first derivatives approach is that it requires storing the strain tensor for each particle, i.e. six additional quantities when taking account of symmetries.

\subsubsection{Physical viscosity with direct second derivatives}
\label{sec:ns2nd}
 The second approach is to use SPH second derivative operators directly. Here we use modified versions of the identities given by \citet{espanolrevenga03} (see also \citealt{monaghan05,price12}), namely
\begin{align}
\nabla \left[A (\nabla\cdot{\bm v})\right] \approx & \nonumber \\
 -\sum_{b} m_{b} &\left[\frac{A_{a}}{\rho_{a}} G_{ab}(h_{a})+ \frac{A_{b}}{\rho_{b}} G_{ab}(h_{b})\right] ({\bm v}_{ab}\cdot \hat{\bm r}_{ab}) \hat{\bm r}_{ab} , \\
\nabla\cdot (C \nabla {\bm v}) \approx & \nonumber \\
-\sum_{b} m_{b} & \left[\frac{C_{a}}{\rho_{a}}  G_{ab}(h_{a}) + \frac{C_{b}}{\rho_{b}} G_{ab}(h_{b})\right]{\bm v}_{ab},
\end{align}
where $G_{ab} \equiv - 2 F_{ab} / \vert r_{ab} \vert$, i.e. the scalar part of the kernel gradient divided by the particle separation, which can be thought of as equivalent to defining a new ``second derivative kernel'' \citep{brookshaw85,brookshaw94,price12,pricelaibe15}.

 From the compressible Navier-Stokes equations, (\ref{eq:ns}) with (\ref{eq:sijns}), the coefficients in these two terms are
\begin{align}
A & \equiv \frac12 \left(\zeta + \frac{\eta}{3}\right), \\
C & \equiv \frac12 \eta,
\end{align}
so that we can simply use
\begin{align}
\left(\frac{{\rm d}v^{i}_{a}}{{\rm d}t}\right)_{\rm visc}  =& \sum_{b} \frac{m_{b}}{\overline{\rho}_{ab}} (\tau_{a} + \tau_{b}) ({\bm v}_{ab}\cdot \hat{\bm r}_{ab}) \hat{r}^{i}_{ab} G_{ab} \nonumber \\
 & + \sum_{b} \frac{m_{b}}{\overline{\rho}_{ab}}(\kappa_{a} + \kappa_{b}) {v}^{i}_{ab} G_{ab},
\end{align}
where
\begin{align}
\tau & = \frac52 A, \\
\kappa & = \left(C - \frac{A}{2}\right).
\end{align}

 The corresponding heating terms in the thermal energy equation are given by
\begin{align}
\Lambda_{\rm visc} =& \frac{\tau_{a}}{\rho_{a}} \sum_{b} m_{b} ({\bm v}_{ab}\cdot \hat{\bm r}_{ab})^{2} G_{ab}(h_{a}) \nonumber \\
& + \frac{\kappa_{a}}{\rho_{a}} \sum_{b} m_{b} ({\bm v}_{ab})^{2} G_{ab}(h_{a}).
\end{align}

This is the default formulation of Navier-Stokes viscosity in the code since it does not require additional storage. In practice we have found little difference between the two formulations of physical viscosity, but this would benefit from a detailed study. In general one might expect the two first derivatives formulation to offer a less noisy estimate at the cost of additional storage. However, direct second derivatives are the method used in `Smoothed Dissipative Particle Dynamics' \citep{espanolrevenga03}.
 
 \subsubsection{Timestep constraint}
 Both approaches to physical viscosity use explicit timestepping, and therefore imply a constraint on the timestep given by
\begin{equation}
\Delta t^{a}_{\rm visc} \equiv C_{\rm visc} \frac{h_{a}^{2}}{\nu_{a}},
\end{equation}
where $C_{\rm visc} = 0.25$ by default \citep{brookshaw94}. When physical viscosity is turned on, this constraint is included with other timestep constraints according to (\ref{eq:dtmin}).

\subsubsection{Physical viscosity and the tensile instability}
 Caution is required in the use of physical viscosity at high Mach number, since negative stress can lead to the tensile instability \citep{morris96,monaghan00,gms01}. For subsonic applications this is usually not a problem since the strain tensor and velocity divergence are small compared to the pressure. In the current code we simply emit a warning if physical viscosity leads to negative stresses during the calculation, but this would benefit from a detailed study.

\subsubsection{Physical viscosity and angular momentum conservation} 
 Neither method for physical viscosity exactly conserves angular momentum because the force is no longer directed along the line of sight joining the particles. However, the error is usually small (see discussion in \citealt{bonetlok99}, Section 5 of \citealt{pricemonaghan04a} or \citealt{huadams06a}).  Recently, \citet{mfg15} have proposed an algorithm for physical viscosity in SPH that explicitly conserves angular momentum by tracking particle spin, which may be worth investigating.

\subsection{Sink particles}
\label{sec:sinks}
Sink particles were introduced into SPH by \citet{bbp95} in order to follow star formation simulations beyond the point of fragmentation. In \textsc{Phantom}, these are treated separately to the SPH particles, and interact with other particles, including other sink particles, only via gravity. The differences with other point mass particles implemented in the code (e.g. dust, stars and dark matter) are that i) the gravitational interaction is computed using a direct $N^{2}$ summation which is \emph{not} softened by default (i.e., the $N-$body algorithm is collisional); ii) they are allowed to accrete gas; and iii) they store the accreted angular momentum and other extended properties, such as the accreted mass. Sink particles are evolved in time using the RESPA algorithm (Section~\ref{sec:respa}), which is second order accurate, symplectic, and allows sink particles to evolve on shorter timesteps compared to SPH particles.

\subsubsection{Sink particle accelerations}
\label{sec:sinkaccel}
The equations of motion for a given sink particle, $i$, are
\begin{align}
\frac{{\rm d}{\bm v}_{i}}{{\rm d}t} = & -\sum^{N_{\rm sink}}_{j=1} G M_{j} \phi'_{ij} (\epsilon) \hat{\bm r}_{ij} \nonumber \\
& - \sum^{N_{\rm part}}_{b=1} G m_{b} \phi'_{ib} (\epsilon_{ib}) \hat{\bm r}_{ib},
\end{align}
where $\phi'_{ab}$ is the usual softening kernel (Section~\ref{sec:ksoft}), $N_{\rm part}$ is the total number of gas particles, and $N_{\rm sink}$ is the total number of sink particles. The sink-gas softening length, $\epsilon_{ib}$, is defined as the maximum of the (fixed) softening length defined for the sink particles, $\epsilon$, and the softening length of the gas particle, $\epsilon_b$. That is, $\epsilon_{ib} \equiv\max(\epsilon, \epsilon_{b})$. SPH particles receive a corresponding acceleration
\begin{equation}
{\bm a}^{a}_{\rm sink-gas} = -\sum^{N_{\rm sink}}_{j=1} G M_{j} \phi'_{aj} (\epsilon_{aj}) \hat{\bm r}_{aj}.
\end{equation}
Softening of sink-gas interactions is not applied if the softening length for sink particles is set to zero, in which case the sink-gas accelerations reduce simply to
\begin{equation}
{\bm a}^{a}_{\rm sink-gas} = -\sum^{N_{\rm sink}}_{j=1} \frac{G M_{j}}{\vert {\bm r}_{a} - {\bm r}_{j}\vert^{3}} {\bm r}_{aj}.
\end{equation}
This is the default behaviour when sink particles are used in the code. Softening of sink-gas interactions is useful to model a point mass particle that does not accrete gas (e.g. by setting the accretion radius to zero). For example, we used a softened sink particle to simulate the core of the red giant in \cite{iaconietal17}. The sink-sink interaction is unsoftened by default ($\epsilon = 0$), giving the usual
\begin{equation}
{\bm a}^{i}_{\rm sink-sink} =  -\sum^{N_{\rm sink}}_{j=1} \frac{G M_{j}}{ \vert {\bm r}_{i} - {\bm r}_{j}\vert^{3}} {\bm r}_{ij}.
\end{equation}
Caution is required when integrating tight binary or multiple systems when $\epsilon = 0$ to ensure that the timestep conditions (Section~\ref{sec:timestep}) are strict enough.

\subsubsection{Accretion onto sink particles}
\label{sec:accrete}
Accretion of gas particles onto a sink particle occurs when a gas particle passes a series of accretion checks within the accretion radius $r_{\rm acc}$ of a sink particle (set in the initial conditions or when the sink particle is created; see Section~\ref{sec:sinkcreate}). First, a gas particle is indiscriminately accreted without undergoing any additional checks if it falls within $f_{\rm acc}r_{\rm acc}$, where $0 \le f_{\rm acc} \le 1$ (default $f_{\rm acc} = 0.8$). In the boundary region $f_{\rm acc}r_{\rm acc} < r < r_{\rm acc}$, a gas particle undergoes accretion if:
\begin{enumerate}
\item $\vert {\bm L}_{ai} \vert < \vert {\bm L}_{\rm acc} \vert$, that is, its specific angular momentum is less than that of a Keplerian orbit at $r_{\rm acc}$,
\item $e = \frac{v_{ai}^2}{2} - \frac{GM_i}{r_{ai}} < 0$, i.e., it is gravitationally bound to the sink particle, and
\item $e$ for this gas-sink pair is smaller than $e$ with any other sink particle, that is, out of all sink particles, the gas particle is most bound to this one.
\end{enumerate}
In the above conditions, ${\bm L}_{ai}$ is the relative specific angular momentum of the gas-sink pair, $a-i$, defined by
\begin{align}
\vert {\bm L}_{ai}^2\vert & \equiv \vert {\bm r}_{ai} \times {\bm v}_{ai}\vert^2 \nonumber \\
& = r_{ai}^2v_{ai}^2~-~\left({\bm r}_{ai}\cdot {\bm v}_{ai}\right)^2,
\end{align}
while $\vert {\bm L}_{\rm acc} \vert = r_{\rm acc}^2 \Omega_{\rm acc}$ is the angular momentum at $r_{\rm acc}$, where $\Omega_{\rm acc} = \sqrt{GM_i / r_{ai}^{3}}$ is the Keplerian angular speed at $r_{\rm acc}$, $v_{ai}$ and $r_{ai}$ are the relative velocity and position, respectively, and $M_i$ is the mass of the sink particle.
  
 When a particle, $a$, passes the accretion checks, then the mass, position, velocity, acceleration and spin angular momentum of the sink particle are updated according to
\begin{align}
{\bm r}_{i} & = \frac{({\bm r}_{a} m_{a} + {\bm r}_{i} M_{i})}{M_{i} + m_{a}}, \\
{\bm v}_{i} & = \frac{({\bm v}_{a} m_{a} + {\bm v}_{i} M_{i})}{M_{i} + m_{a}}, \\
{\bm a}_{i} & = \frac{({\bm a}_{a} m_{a} + {\bm a}_{i} M_{i})}{M_{i} + m_{a}}, \\
{\bm S}_{i} & = {\bm S}_{i} +\frac{m_{a} M_{i}}{M_{i} + m_{a}} \left[ \left({\bm r}_{a} - {\bm r}_{i}\right) \times  \left({\bm v}_{a} - {\bm v}_{i}\right)\right], \\
M_{i} & = M_{i} + m_{a}.
\end{align}
This ensures that mass, linear momentum and angular momentum (but not energy) are conserved by the accretion process. The accreted mass as well as the total mass for each sink particle is stored to avoid problems with round-off error in the case where the particle masses are much smaller than the sink mass. Accreted particles are tagged by setting their smoothing lengths negative. Those particles with $h \leq 0$ are subsequently excluded when the $k$d-tree is built.

\subsubsection{Sink particle boundary conditions}
No special sink particle boundary conditions are implemented in \textsc{Phantom} at present. More advanced boundary conditions to regularise the density, pressure and angular momentum near a sink have been proposed by \citet{bbp95} and used in \citet{batebonnell97}, and proposed again more recently by \citet{hww13}. While these conditions help to regularise the flow near the sink particle, they can also cause problems --- particularly the angular momentum boundary condition if the disc near the sink particle has complicated structure such as spiral density waves (Bate, private communication 2014).  Often it is more cost effective to simply reduce the accretion radius of the sink. This may change in future code versions.

\subsubsection{Dynamic sink particle creation}
\label{sec:sinkcreate}
As described in \citet{bbp95}, it is also possible to create sink particles on-the-fly provided certain physical conditions are met and self-gravity is turned on (Section~\ref{sec:gravity}). The primary conditions required for sink particle formation are that the density of a given particle exceeds some threshold physical density somewhere in the domain, and that this density peak occurs more than a critical distance $r_{\rm crit}$ from an existing sink. Once these conditions are met on a particular particle, $a$, the creation of a new sink particle occurs by passing the following conditions \citep{bbp95}:
\begin{enumerate}
\item the particle is a gas particle,
\item $\nabla\cdot{\bm v}_{a} \leq 0$, that is, gas surrounding the particle is at rest or collapsing,
\item $h_{a} < r_{\rm acc}/2$, i.e., the smoothing length of the particle is less than half of the accretion radius,
\item all neighbours within $r_{\rm acc}$ are currently active,
\item the ratio of thermal to gravitational energy of particles within $r_{\rm acc}$, $\alpha_{\rm J}$, satisfies $\alpha_{J}\leq 1/2$,
\item $\alpha_{\rm J} + \beta_{\rm rot} \leq 1$, where $\beta_{\rm rot} = \vert e_{\rm rot} \vert / \vert e_{\rm grav} \vert$ is the ratio of rotational energy to the magnitude of the gravitational energy for particles within $r_{\rm acc}$, and
\item  $e_{\rm tot} < 0$, that is, the total energy of particles within $r_{\rm acc}$ is negative (i.e.\ the clump is gravitationally bound).
\item the particle is at a local potential minimum, i.e. $\Phi$ is less than $\Phi$ computed on all other particles within $r_{\rm acc}$ \citep{federrathetal10a}
\end{enumerate}
A new sink particle is created at the position of particle $a$ if these checks are passed, and immediately the particles within $r_{\rm acc}$ are accreted by calling the routine described in Section~\ref{sec:accrete}. The checks above are the same as those in \citet{bbp95}, with the addition of the additional check from \citet{federrathetal10a} to ensure that sink particles are only created in a local minimum of the gravitational potential.

The various energies used to evaluate the criteria above are computed according to
\begin{align}
e_{\rm kin} = &\ \frac12 \sum^{N<r_{\rm acc}}_{b=1} m_{b} ({\bm v}_{b} - {\bm v}_{a})^{2}, \\
e_{\rm therm} = &\ \sum^{N<r_{\rm acc}}_{b=1} m_{b} u_{b}, \\
e_{\rm grav} = &\ -\frac{1}{2}\sum^{N<r_{\rm acc}}_{b=1}\sum^{N<r_{\rm acc}}_{c=b} G m_{b} m_{c} , \notag \\
                           & \times \left[\phi(\vert {\bm r}_{b} - {\bm r}_{c}\vert, h_b) + \phi(\vert {\bm r}_{b} - {\bm r}_{c}\vert, h_c)\right], \\
e_{\rm tot} = &\ e_{\rm kin} + e_{\rm therm} + e_{\rm grav}, \\
e_{\rm rot} \equiv &\ \sqrt{e^{2}_{{\rm rot}, x} + e^{2}_{{\rm rot}, y} + e^{2}_{{\rm rot}, z}}, \\
e_{{\rm rot}, x} \equiv &\ \frac12 \sum^{N<r_{\rm acc}}_{b=1} m_{b} \frac{L^{2}_{{ab},x}}{\sqrt{(y_{a} - y_{b})^{2} + (z_{a} - z_{b})^{2}}},\\
e_{{\rm rot}, y} \equiv &\ \frac12 \sum^{N<r_{\rm acc}}_{b=1} m_{b} \frac{L^{2}_{{ab},y}}{\sqrt{(x_{a} - x_{b})^{2} + (z_{a} - z_{b})^{2}}}, \\
e_{{\rm rot}, z} \equiv &\ \frac12 \sum^{N<r_{\rm acc}}_{b=1} m_{b} \frac{L^{2}_{{ab},z}}{\sqrt{(x_{a} - x_{b})^{2} + (y_{a} - y_{b})^{2}}},
\end{align}
where ${\bm L}_{ab} \equiv ({\bm r}_{a} - {\bm r}_{b}) \times ({\bm v}_{a} - {\bm v}_{b})$ is the specific angular momentum between a pair of particles, and $\phi$ is the gravitational softening kernel (defined in Section~\ref{sec:gravity}), which has units of inverse length. Adding the contribution from \emph{all} pairs, $b-c$, within the cloud is required to obtain the total potential of the cloud.

\subsubsection{Sink particle timesteps}
\label{sec:dtsinks}
 Sink particles are integrated together with a global, but adaptive, timestep, following the inner loop of the RESPA algorithm given in (\ref{eq:respa1})--(\ref{eq:respa4}) corresponding to a second-order Leapfrog integration. The timestep is controlled by the minimum of the sink-gas timestep, (\ref{eq:dtsinkgas}), and a sink-sink timestep \citep{dehnenread11}
\begin{equation}
\Delta t_{\rm sink-sink} \equiv \min_{i} \left( C_{\rm force} \eta_\Phi \sqrt{\frac{\vert\Phi^{\rm sink-sink}_{i}\vert}{\vert \nabla \Phi^{\rm sink-sink}_i \vert^{2}}} \right), \label{eq:dtsinksink}
\end{equation}
where the potential and gradient include other sink particles, plus any external potentials applied to sink particles except the sink-gas potential. We set $\eta_\Phi = 0.05$ by default, resulting in $\sim 300$--500 steps per orbit for a binary orbit with the default $C_{\rm force} = 0.25$ (see Section~\ref{sec:binaryorbit}). 

More accurate integrators such as the fourth-order Hermite scheme \citep{makinoaarseth92} or the fourth order symplectic schemes proposed by \citet{omf02a} or \citet{chinchen05} are not yet implemented in \textsc{Phantom}, but it would be a worthwhile effort to incorporate one of these in a future code version. See \citet{hubberetal13} for a recent implementation of a 4th order Hermite scheme for sink particles in SPH.

\subsection{Stellar physics}
\label{sec:stellar}

\label{sec:eosmesa}
 A tabulated equation of state (EOS) can be used to take account of the departure from an ideal gas, for example due to changes in ionization or molecular dissociation and recombination.  This tabulated EOS in {\sc Phantom} is adapted from the $\log{P_{\rm gas}}-T$ EOS tables provided with the open source package Modules for Experiments in Stellar Astrophysics {\sc mesa} \citep{paxtonetal11}.  Details of the data, originally compiled from blends of equations of state from \citet{scv95} (SCVH), \citet{timmesswesty00}, \citet[][also the 2005 update]{rogersnayfonov02}, \citet{potekhinchabrier10} and for an ideal gas, are outlined by \citet{paxtonetal11}.
 
  In our implementation (adapted from original routines for the {\sc Music} code; \citealt{goffreyetal16}), we compute the pressure and other required EOS variables for a particular composition by interpolation between sets of tables for different hydrogen abundance $X={0.0,0.2,0.4,0.6,0.8}$ and metallicity $Z={0.0,0.02,0.04}$.   Pressure is calculated with bicubic interpolation, and $\Gamma_1 \equiv \partial \ln P / \partial \ln \rho\vert_s$ with bilinear interpolation, in $\log{u}$ and $\log{V}\equiv \log{\rho}-0.7\log{u}+20$. The tables are currently valid in the ranges $10.5 \leq \log u \leq 17.5$ and $0.0 \leq \log V \leq 14.0$. Values requested outside the tables are currently computed by linear extrapolation. This triggers a warning to the user.
  
  We have not tested the thermodynamic consistency of our interpolation scheme from the tables, but this is an important consideration \citep{timmesarnett99}.

\subsection{Magnetohydrodynamics}
\label{sec:mhd}

\textsc{Phantom} implements the smoothed particle magnetohydrodynamics (SPMHD) algorithm described in \citet{price12} and \citet{triccoprice12,triccoprice13}, based on the original work by \citet{phillipsmonaghan85} and \citet{pricemonaghan04,pricemonaghan04a,pricemonaghan05}. \textsc{Phantom} was previously used to test a vector potential formulation \citep{price10}, but this algorithm has been subsequently removed from the code due to its poor performance (see \citealt{price10}). 

 The important difference between {\sc Phantom} and the \textsc{gadget} implementation of SPMHD \citep{dolagstasyszyn09,burzleetal11,burzleetal11a}, which also implements the core algorithms from \citet{pricemonaghan04,pricemonaghan04a,pricemonaghan05}, is our use of the divergence-cleaning algorithm from \citet{triccoprice12,triccoprice13} and \citet{tpb16}. This is vital for preserving the divergence-free (no monopoles) condition on the magnetic field. 
 
 For recent applications of \textsc{Phantom} to MHD problems, see e.g. \citet{tpf16}, \citet{dobbsetal16}, \citet{bonnerotetal17}, \citet{fpb17} and \citet{wpb16,wpb17}.

\subsubsection{Equations of magnetohydrodynamics}
\textsc{Phantom} solves the equations of magnetohydrodynamics in the form
\begin{align}
\frac{{\rm d}v^{i}}{{\rm d}t} & = -\frac{1}{\rho}\frac{\partial M^{ij}}{\partial x^{j}} + \Pi_{\rm shock} + f^{i}_{\rm divB} + a^{i}_{\rm ext} \nonumber \\
& \phantom{=} + a^{i}_{\rm sink-gas} + a^{i}_{\rm selfgrav}, \label{eq:mommhd} \\
\frac{{\rm d}u}{{\rm d}t} & = -\frac{P}{\rho} \left(\nabla\cdot{\bm v}\right) + \Lambda_{\rm shock} - \Lambda_{\rm cool}, \label{eq:dudtmhd} \\
\frac{{\rm d}}{{\rm d}t} \left(\frac{\bm B}{\rho} \right) & = \frac{1}{\rho} \left[ \left({\bm B}\cdot\nabla\right){\bm v} - \nabla\psi + \mathcal{\bm D}_{\rm diss} \right], \label{eq:dBdt} \\
\frac{{\rm d}}{{\rm d}t} \left( \frac{\psi}{c_{\rm h}} \right)& = -c_{\rm h} \left(\nabla\cdot{\bm B}\right) - \frac12 \frac{\psi}{c_{\rm h}} \left(\nabla\cdot{\bm v}\right) - \frac{\psi}{c_{\rm h}\tau_{\rm c}}, \label{eq:psi}
\end{align}
where ${\bm B}$ is the magnetic field, $\psi$ is a scalar used to control the divergence error in the magnetic field (see Section~\ref{sec:cleaning}, below), and $\mathcal{D}_{\rm diss}$ represents magnetic dissipation terms (Sections~\ref{sec:mhddiss} and \ref{sec:nonideal}, below). The Maxwell stress tensor, $M_{ij}$, is given by
\begin{equation}
M^{ij} = \left( P + \frac12 \frac{B^{2}}{\mu_{0}} \right) \delta^{ij} - \frac{B^{i} B^{j}}{\mu_{0}},
\label{eq:sij}
\end{equation}
where $\delta^{ij}$ is the Kronecker delta and $\mu_{0}$ is the permeability of free space. A source term related to the numerically-induced divergence of the magnetic field, given by
\begin{equation}
f^{i}_{\rm divB} \equiv - \frac{B^{i}}{\rho}\left(\nabla\cdot{\bm B}\right),
\end{equation}
is necessary to prevent the tensile instability in SPMHD \citep{phillipsmonaghan85,monaghan00,bot01,price12}. With this source term, the equation set for ideal MHD in the absence of the divergence cleaning field, $\psi$, is formally the same as in the \citet{powelletal99} 8-wave scheme \citep{price12}, meaning that the divergence errors in the magnetic field are advected by the flow, but not dissipated, unless cleaning is used.

\subsubsection{Discrete equations}
\label{sec:spmhd}
The discrete version of (\ref{eq:mommhd}) follows the same procedure as for physical viscosity (Section~\ref{sec:viscosity}), i.e.
\begin{align}
\frac{{\rm d}v^{i}_{a}}{{\rm d}t}  =& -\sum_{b} m_{b}\left[ \frac{M^{ij}_{a}}{\Omega_{a}\rho_{a}^{2}} \frac{\partial W_{ab}(h_{a})}{\partial x^{j}_{a}} + \frac{M^{ij}_{b}}{\Omega_{b}\rho_{b}^{2}} \frac{\partial W_{ab}(h_{b})}{\partial x^{j}_{a}}\right] \nonumber\\
&  + \Pi_{\rm shock}^{a} +  f^{i}_{{\rm divB}, a} + a^{i}_{{\rm ext}, a} + a^i_{\rm sink-gas} \nonumber \\
&  + a^i_{\rm selfgrav}, \label{eq:spmhdmom}
\end{align}
where $M_{a}^{ij}$ is defined according to (\ref{eq:sij}), $f_{{\rm divB}}$ is a correction term for stability (discussed below), and accelerations due to external forces are as described in Section~\ref{sec:extf}.

 Equations (\ref{eq:dBdt}) and (\ref{eq:psi}) are discretised according to \citep{pricemonaghan05,triccoprice12,tpb16}
\begin{align}
\frac{{\rm d}}{{\rm d}t} \left( \frac{{\bm B}}{\rho} \right)_a =& -\frac{1}{\Omega_{a} \rho_{a}^2} \sum_{b} m_{b}  {\bm v}_{ab} \left[ {\bm B}_{a} \cdot \nabla_a W_{ab}(h_{a})\right] \nonumber \\
& - \sum_{b} m_{b} \left[ \frac{\psi_{a}}{\Omega_{a}\rho_{a}^{2}} \nabla_a W_{ab}(h_{a})+ \frac{\psi_{b}}{\Omega_{b}\rho_{b}^{2}} \nabla_a W_{ab}(h_{b})\right]  \nonumber \\
& + \frac{1}{\rho_a} \mathcal{D}^{a}_{\rm diss}, \label{eq:dBrhodtsph} \\
\frac{{\rm d}}{{\rm d}t} \left(\frac{\psi}{c_{\rm h}} \right)_a = & \frac{c_{\rm h}^a}{\Omega_{a}\rho_{a}} \sum_{b} m_{b} {\bm B}_{ab} \cdot \nabla_a W_{ab}(h_{a}) \nonumber \\
& + \frac{\psi_{a}}{2c_{\rm h}^a\Omega_{a}\rho_{a}} \sum_{b} m_{b}  {\bm v}_{ab} \cdot \nabla_a W_{ab}(h_{a}) - \frac{\psi_{a}}{c_{\rm h}^a\tau^{a}_{\rm c}}. \label{eq:psisph}
\end{align}

 The first term in (\ref{eq:psisph}) uses the divergence of the magnetic field discretised according to
\begin{equation}
(\nabla\cdot{\bm B})_{a} = - \frac{1}{\Omega_{a}\rho_{a}} \sum_{b} m_{b} \left({\bm B}_{a} - {\bm B}_{b} \right) \cdot \nabla_a W_{ab}(h_{a}),
\label{eq:divB}
\end{equation}
which is therefore the operator we use when measuring the divergence error (c.f. \citealt{triccoprice12}).

\subsubsection{Code units}
\label{sec:magunits}
 An additional unit is required when magnetic fields are included to describe the unit of magnetic field. We adopt code units such that $\mu_{0} = 1$, as is common practice. The unit scalings for the magnetic field can be determined from the definition of the Alfv\'en speed,
\begin{equation}
v_{\rm A} \equiv \sqrt{\frac{B^{2}}{\mu_{0} \rho}}.
\end{equation}
Since the Alfv\'en speed has dimensions of length per unit time, this implies a unit for the magnetic field, $u_{\rm mag}$, given by
\begin{equation}
u_{\rm mag} = \left(\frac{\mu_{0} u_{\rm mass}}{u_{\rm dist} u_{\rm time}} \right)^{\frac12}.
\end{equation}
Converting the magnetic field in the code to physical units therefore only requires specifying $\mu_{0}$ in the relevant unit system. In particular, it avoids the differences between SI and cgs units in how the charge unit is defined, since $\mu_{0}$ is dimensionless and equal to $4\pi$ in cgs units but has dimensions that involve charge in SI units.

\subsubsection{Tensile instability correction}
 The correction term $f_{\rm divB}$ is necessary to avoid the tensile instability --- a numerical instability where particles attract each other along field lines --- in the regime where the magnetic pressure exceeds the gas pressure, that is, when plasma $\beta \equiv P / \tfrac{1}{2} B^2 < 1$ \citep{phillipsmonaghan85}. The correction term is computed using the symmetric divergence operator \citep{bot01,price12,triccoprice12}
\begin{align}
f^{i}_{{\rm divB}, a} = -\hat{B}_{a}^{i} \sum_{b} m_{b} & \left[ \frac{{\bm B}_{a} \cdot \nabla_{a} W_{ab} (h_{a})}{\Omega_{a} \rho_{a}^{2}} \right. \nonumber \\
 & \left. + \frac{{\bm B}_{b} \cdot \nabla_{a} W_{ab} (h_{b})}{\Omega_{b} \rho_{b}^{2}} \right]. \label{eq:fcorr}
\end{align}
Since this term violates momentum conservation to the extent that the $\nabla\cdot{\bm B}$ term is non-zero, several authors have proposed ways to minimise its effect. \citet{bot04} showed that stability could be achieved with $\hat{B}^i = \frac12 B^i$ and also proposed a scheme for scaling this term to zero for $\beta > 1$. \citet{bkw12} similarly advocated using a factor of $\frac12$ in this term. However, \citet{triccoprice12} showed that this could lead to problematic effects (their Figure~12). In {\sc Phantom}, we use
\begin{equation}
\hat{B}^i = \begin{cases} 
B^i & \beta < 2, \\
[(10 - \beta) B^i]/8 & 2 < \beta < 10, \\
0 & \textrm{otherwise,}
\end{cases}
\end{equation}
to provide numerical stability in the strong field regime while maintaining conservation of momentum when $\beta > 10$. This also helps to reduce errors in the MHD wave speed caused by the correction term \citep{iwasaki15}.

\subsubsection{Shock capturing}
\label{sec:mhddiss}
The shock capturing term in the momentum equation for MHD is identical to (\ref{eq:pishock}) and (\ref{eq:qvisc}) except that the signal speed becomes \citep{pricemonaghan04,pricemonaghan05,price12}
\begin{equation}
v_{{\rm sig}, a} = \alpha^{\rm AV}_{a} v_{a} + \beta^{\rm AV} \vert {\bm v}_{ab}\cdot\hat{\bm r}_{ab} \vert,
\end{equation}
where
\begin{equation}
v_{a} = \sqrt{c^{2}_{{\rm s}, a} + v_{{\rm A}, a}^{2}} \label{eq:vfast}
\end{equation}
is the fast magnetosonic speed. Apart from this, the major difference to the hydrodynamic case is the addition of an artificial resistivity term to capture shocks and discontinuities in the magnetic field (i.e.\ current sheets). This is given by
\begin{equation}
\mathcal{D}^{a}_{{\rm diss}} = \frac{\rho_{a}}{2} \sum_{b} m_{b} \left[ \frac{v^{B}_{{\rm sig}, a}}{\rho_{a}^{2}} \frac{F_{ab}(h_{a})}{\Omega_{a}} + \frac{v^{B}_{{\rm sig}, b}}{\rho_{b}^{2}} \frac{F_{ab}(h_{b})}{\Omega_{b}} \right] {\bm B}_{ab}, \label{eq:Bdiss}
\end{equation}
where $v^{B}_{{\rm sig}, a}$ is an appropriate signal speed (see below) multiplied by a dimensionless coefficient, $\alpha^{B}$. The corresponding contribution to the thermal energy from the resistivity term in (\ref{eq:shockheating}) is given by
\begin{align}
\Lambda_{\rm artres} = &  - \frac14 \sum_{b} m_{b} \left[ \frac{v^{B}_{{\rm sig}, a}}{\rho_{a}^{2}} \frac{F_{ab}(h_{a})}{\Omega_{a}} \right. \nonumber \\
& \left. \phantom{-\frac12\sum_{b} m_{b}}+ \frac{v^{B}_{{\rm sig}, b}}{\rho_{b}^{2}} \frac{F_{ab}(h_{b})}{\Omega_{b}} \right]
({\bm B}_{ab})^{2}.
\label{eq:resheating}
\end{align}
As with the artificial viscosity, (\ref{eq:Bdiss}) and (\ref{eq:resheating}) are the SPH representation of a physical resistivity term, $\eta \nabla^2 {\bm B}$, but with a coefficient that is proportional to resolution \citep{pricemonaghan04}. The resistive dissipation rate from the shock capturing term is given by
\begin{equation}
\eta \approx \frac12 \alpha^{B} v_{\rm sig}^{B} \vert r_{ab} \vert, \label{eq:eta}
\end{equation}
where $\vert r_{ab} \vert \propto h$.

\subsubsection{Switch to reduce resistivity}
\label{sec:artres}
  \textsc{Phantom} previously used the method proposed by \citet{triccoprice13} to reduce the dissipation in the magnetic field away from shocks and discontinuities. The signal velocity, $v_{\rm sig}^{B}$, was set equal to the magnetosonic speed (Equation~\ref{eq:vfast}) multiplied by the dimensionless coefficient $\alpha_B$, which was set according to
\begin{equation}
\alpha^{B}_{a} = \min \left( h_{a} \frac{\vert \nabla {\bm B}_{a} \vert}{\vert{\bm B}_{a}\vert}, \alpha^{B}_{\max}\right),
\end{equation}
where $\alpha^{B}_{\max} = 1.0$ by default and $\vert \nabla {\bm B}_{a} \vert$ is the 2-norm of the gradient tensor, i.e. the root mean square of all 9 components of this tensor. Unlike the viscous dissipation, this is set based on the instantaneous values of $h$ and ${\bm B}$ and there is no source/decay equation involved, as \citet{triccoprice13} found it to be unnecessary. Since $\alpha^{B}$ is proportional to resolution, from (\ref{eq:eta}) we see that this results in dissipation that is effectively second order ($\propto h^{2}$). When individual particle timesteps were used, inactive particles retained their value of $\alpha^{B}$ from the last timestep they were active.

 More recently we have found that a better approach, similar to that used for artificial conductivity, is to simply set $\alpha^{\rm B}=1$ for all particles and set the signal speed for artificial resistivity according to
\begin{equation}
v_{\rm sig}^{B} = \vert {\bm v}_{ab} \times \hat{\bm r}_{ab} \vert. 
\end{equation}
We find that this produces better results on all of our tests (Section~\ref{sec:mhdtests}), in particular, producing zero numerical dissipation on the current loop advection test (Section~\ref{sec:jadvect}). As with the \citet{triccoprice13} switch, it gives second-order dissipation in the magnetic field (demonstrated in Section~\ref{sec:alfven}; Figure~\ref{fig:alfven}). This is now the default treatment for artificial resistivity in \textsc{Phantom}.

\subsubsection{Conservation properties}
The total energy when using MHD is given by
\begin{equation}
E = \sum_{a} m_{a} \left( \frac12 v_{a}^{2} + u_{a} + \Phi_{a} + \frac12 \frac{B^{2}_{a}}{\mu_{0}\rho_{a}} \right).
\end{equation}
Hence total energy conservation, in the absence of divergence cleaning, corresponds to
\begin{align}
\frac{{\rm d}E}{{\rm d}t} =& \sum_{a} m_{a} \left[ {\bm v}_{a} \cdot \frac{{\rm d}{\bm v}_{a}}{{\rm d}t} +  \frac{{\rm d}u_{a}}{{\rm d}t} + \frac{{\rm d} \Phi_{a}}{{\rm d}t} \right. \nonumber\\
& + \left.  \frac{{B}^{2}_{a}}{2 \mu_{0}\rho_{a}^{2}} \frac{{\rm d}\rho_{a}}{{\rm d}t}  +  \frac{{\bm B}_{a}}{ \mu_{0}} \cdot \frac{{\rm d}}{{\rm d}t} \left( \frac{{\bm B}}{\rho} \right)_a \right] = 0. \label{eq:mhdenergyconservation}
\end{align}
Neglecting the $f_{\rm divB}$ correction term for the moment, substituting (\ref{eq:spmhdmom}), (\ref{eq:dudtsph}) and (\ref{eq:dBrhodtsph}) into (\ref{eq:mhdenergyconservation}) with the ideal MHD and shock capturing terms included demonstrates that the total energy is exactly conserved, using the same argument as the one given in Section~\ref{sec:energyconservation} (detailed workings can be found in \citealt{pricemonaghan04a}). The total linear momentum is also exactly conserved following a similar argument as in Section~\ref{sec:momconservation}. However, the presence of the $f_{\rm divB}$ correction term, though necessary for numerical stability, violates the conservation of both momentum and energy in the strong field regime (in the weak field regime, it is switched off and conservation is preserved). The severity of this non-conservation is related to the degree in which divergence errors are present in the magnetic field, hence inadequate divergence control (see below) usually manifests as a loss of momentum conservation in the code \citep[see][for details]{triccoprice12}.

\subsubsection{Divergence cleaning}
\label{sec:cleaning}
We adopt the `constrained' hyperbolic/parabolic divergence cleaning algorithm described by \citet{triccoprice12} and \citet{tpb16} to preserve the divergence-free condition on the magnetic field. This formulation addresses a number of issues with earlier formulations by \citet{dedneretal02} and \citet{pricemonaghan05}. 

 The main idea of the scheme is to propagate divergence errors according to a damped wave equation \citep{dedneretal02,pricemonaghan05}. This is facilitated by introducing a new scalar field, $\psi$, which is coupled to the magnetic field in (\ref{eq:dBdt}) and evolved according to (\ref{eq:psi}). 
  
 \citet{tpb16} generalised the method of \citet{dedneretal02} to include the case where the hyperbolic wave speed, $c_{\rm h}$, varies in time and space. This is the approach we use in {\sc Phantom}. The resulting `generalised wave equation' may be derived by combining the relevant term in (\ref{eq:dBdt}) with (\ref{eq:psi}) to give \citep{tpb16}
\begin{align}
\frac{{\rm d}}{{\rm d}t} \left[ \frac{1}{\sqrt{\rho} c_{\rm h}}\frac{{\rm d}}{{\rm d}t} \left( \frac{\psi}{\sqrt{\rho} c_{\rm h}} \right) \right] & - \frac{\nabla^{2} \psi}{\rho} \nonumber \\ 
& + \frac{{\rm d}}{{\rm d}t} \left[ \frac{1}{\sqrt{\rho} c_{\rm h}} \left(\frac{\psi}{\sqrt{\rho}c_{\rm h}\tau_{\rm c}}\right) \right] = 0. \label{eq:genwave}
\end{align}
When $c_{\rm h}$, $\rho$, $\tau_c$ and the fluid velocity are constant, this reduces to the usual damped wave equation in the form
\begin{equation}
\frac{\partial^{2} \psi}{\partial t^{2}} - c_{\rm h}^{2} \nabla^{2} \psi+ \frac{1}{\tau_{\rm c}} \frac{\partial \psi}{\partial t} = 0.
\end{equation}
The same equation holds for the evolution of ${\nabla\cdot{\bm B}}$ itself, i.e.,
\begin{equation}
\frac{\partial^{2} (\nabla\cdot{\bm B})}{\partial t^{2}} - c_{\rm h}^{2} \nabla^{2} (\nabla\cdot{\bm B}) + \frac{1}{\tau_{\rm c}} \frac{\partial (\nabla\cdot{\bm B})}{\partial t} = 0,
\end{equation}
from which it is clear that $c_{\rm h}$ represents the speed at which divergence errors are propagated and $\tau_{\rm c}$ is the decay timescale over which divergence errors are removed.

 \citet{triccoprice12} formulated a `constrained' SPMHD implementation of hyperbolic/parabolic cleaning which guarantees numerical stability of the cleaning. The constraint imposed by \cite{triccoprice12} is that, in the absence of damping, any energy removed from the magnetic field during cleaning must be conserved by the scalar field, $\psi$. This enforces particular choices of numerical operators for $\nabla \cdot {\bm B}$ and $\nabla \psi$ in (\ref{eq:dBrhodtsph}) and (\ref{eq:psisph}), respectively, in particular that they form a conjugate pair of difference and symmetric derivative operators. This guarantees that the change of magnetic energy is negative definite in the presence of the parabolic term (see below).
 
In {\sc Phantom}, we set the cleaning speed, $c_{\rm h}$, equal to the fast magnetosonic speed (Equation~\ref{eq:vfast}) so that its timestep constraint is satisfied already by (\ref{eq:dtcour}), as recommended by \citet{triccoprice13}. The decay timescale is set according to
\begin{equation}
\tau^a_{\rm c} \equiv \frac{h_a}{\sigma_{\rm c} c_{{\rm h},a}},
\end{equation}
where the dimensionless factor $\sigma_{\rm c}$ sets the ratio of parabolic to hyperbolic cleaning. This is set to $\sigma_{\rm c} = 1.0$ by default, which was empirically determined by \citet{triccoprice12} to provide optimal reduction of divergence errors in three dimensions.

The divergence cleaning dissipates energy from the magnetic field at a rate given by \citep{triccoprice12}
\begin{equation}
\left(\frac{{\rm d} E}{{\rm d}t}\right)_{\rm cleaning} = -\sum_{a} m_{a} \frac{\psi^{2}_{a}}{\mu_{0} \rho_{a} c_{{\rm h},a}^{2} \tau^{a}_{\rm c}}.
\end{equation}
In general, this is so small compared to other dissipation terms (e.g.\ resistivity for shock capturing) that it is not worth monitoring \citep{tpb16}. This energy is not added as heat, but simply removed from the calculation.

\subsubsection{Monitoring of divergence errors and over-cleaning}
\label{sec:divbcheck}
The divergence cleaning algorithm is guaranteed to either conserve or dissipate magnetic energy, and cleans divergence errors to a sufficient degree for most applications. However, the onus is on the user to ensure that divergence errors are not affecting simulation results. This may be monitored by the dimensionless quantity
\begin{equation}
\epsilon_{\rm divB} \equiv \frac{h\vert \nabla\cdot{\bm B}\vert}{\vert {\bm B} \vert}.
\end{equation}
The maximum and mean values of this quantity should be used to check the fidelity of simulations that include magnetic fields. A good rule-of-thumb is that the mean should remain $\lesssim 10^{-2}$ for the simulation to remain qualitatively unaffected by divergence errors.

The cleaning wave speed can be arbitrarily increased to improve the effectiveness of the divergence cleaning according to 
\begin{equation}
c_{{\rm h}, a} = f_{\rm clean} v_a ,
\end{equation}
where $f_{\rm clean}$ is an `over-cleaning' factor (by default, $f_{\rm clean} = 1$, i.e., no `over-cleaning'). \citet{tpb16} has shown that increasing $f_{\rm clean}$ leads to further reduction in divergence errors, without affecting the quality of obtained results, but with an accompanying computational expense associated with a reduction in the timestep size.

%

\subsection{Non-ideal magnetohydrodynamics}
\label{sec:nonideal}
\textsc{Phantom} implements non-ideal magnetohydrodynamics including terms for Ohmic resistivity, ambipolar (ion-neutral) diffusion and the Hall effect. Algorithms and tests are taken from~\citet{wpa14,wpb16}. See \citet{wpb16,wpb17,wbp18} for recent applications. Our formulation of non-ideal SPMHD in {\sc Phantom} is simpler than the earlier formulation proposed by \citet{hoskingwhitworth04} because we consider only one set of particles, representing a mixture of charged and uncharged species. Ours is similar to the implementation by \citet{tii13,tsukamotoetal15}.

\subsubsection{Equations of non-ideal MHD}
We assume the strong coupling or `heavy ion' approximation \citep[see e.g.][]{wardleng99,shuetal06,pandeywardle08}, which neglects ion pressure and assumes $\rho_i \ll \rho_n$ where the subscripts $i$ and $n$ refer to the ionised and neutral fluids, respectively. In this case, (\ref{eq:dBdt}) contains three additional terms in the form
\begin{equation}
\frac{{\rm d}}{{\rm d}t} \left( \frac{{\bm B}}{\rho} \right)_{\rm nimhd} = \frac{1}{\rho} \nabla \times \left[ \frac{\bm J}{\sigma_e} + \frac{{\bm J} \times {\bm B}}{e n_e} +  \frac{({\bm J} \times {\bm B}) \times {\bm B}}{\gamma_{\rm AD} \rho_i}\right],
\end{equation}
where $\sigma_e$ is the electrical conductivity, $n_e$ is the number density of electrons, $e$ is the charge on an electron and $\gamma_{\rm AD}$ is the collisional coupling constant between ions and neutrals \citep{pandeywardle08}. We write this in the form
\begin{align}
\frac{{\rm d}}{{\rm d}t} \left( \frac{{\bm B}}{\rho} \right)_{\rm nimhd} = \frac{1}{\rho} \nabla \times \bigg[ & \eta_{\rm O} {\bm J} +  \eta_{\rm H} {\bm J} \times \hat{\bm B} \nonumber \\
& +  \left. \eta_{\rm AD} ({\bm J} \times \hat{\bm B}) \times \hat{\bm B} \right],
\end{align}
where $\hat{\bm B}$ is the unit vector in the direction of ${\bm B}$ such that $\eta_{\rm O}$, $\eta_{\rm AD}$ and $\eta_{\rm Hall}$ are the coefficients for resistive and ambipolar diffusion and the Hall effect, respectively, each with dimensions of area per unit time.

 To conserve energy, we require the corresponding resistive and ambipolar heating terms in the thermal energy equation in the form
\begin{equation}
\left( \frac{{\rm d}u}{{\rm d}t}\right)_{\rm nimhd} = \frac{ \eta_{\rm O}}{\rho} J^2  + \frac{\eta_{\rm AD}}{\rho} \left[ J^2 - ({\bm J}\cdot{\hat{\bm B}})^2\right].
\end{equation}
The Hall term is non-dissipative, being dispersive rather than diffusive, so does not enter the energy equation.

 We currently neglect the `Biermann battery' term \citep{biermann50} proportional to $\nabla P_e/(e n_e)$ in our non-ideal MHD implementation, both because it is thought to be negligible in the interstellar medium \citep{pandeywardle08} and because numerical implementations can produce incorrect results \citep{grazianietal15}. This term is mainly important in generating seed magnetic fields for subsequent dynamo processes (e.g. \citealt{khomenkoetal17}).

\subsubsection{Discrete equations}
 Our main constraint is that the numerical implementation of the non-ideal MHD terms should exactly conserve energy, which is achieved by discretising in the form \citep{wpa14}
\begin{align}
\frac{{\rm d}}{{\rm d}t} \left( \frac{{\bm B}}{\rho} \right)_{{\rm nimhd},a} = - \sum_b \Bigg[ & \frac{{\bm D}_a}{\Omega_a \rho_a^2} \times \nabla_a W_{ab} (h_a) \nonumber \\
 + &  \frac{{\bm D}_b}{\Omega_b \rho_b^2} \times \nabla_a W_{ab} (h_b) \Bigg],
\end{align}
where
\begin{equation}
{\bm D} \equiv \eta_{\rm O} {\bm J} + \eta_{\rm H} ({\bm J} \times \hat{\bm B}) + \eta_{\rm AD} [({\bm J} \times \hat{\bm B}) \times \hat{\bm B}].
\end{equation}

The corresponding term in the energy equation is given by
\begin{equation}
\left( \frac{{\rm d}u_a}{{\rm d}t}\right)_{\rm nimhd} = - \frac{{\bm D}_a \cdot {\bm J}_a}{\rho_a},
\end{equation}
where the magnetic current density is computed alongside the density evaluation according to
\begin{equation}
{\bm J} = \frac{1}{\Omega_a\rho_a} \sum_b m_b ({\bm B}_a - {\bm B}_b) \times \nabla_a W_{ab} (h_a).
\end{equation}
Non-ideal MHD therefore utilises a `two first derivatives' approach, similar to the formulation of physical viscosity described in Section~\ref{sec:twofirstderivs}. This differs from the `direct second derivatives' approach used for our artificial resistivity term, and in previous formulations of physical resistivity \citep{bonafedeetal11}. In practice the differences are usually minor. Our main reason for using two first derivatives for non-ideal MHD is that it is easier to incorporate the Hall effect and ambipolar diffusion terms.

\subsubsection{Computing the non-ideal MHD coefficients}
 To self-consistently compute the coefficients $\eta_{\rm O}$, $\eta_{\rm H}$ and $\eta_{\rm AD}$ from the local gas properties, we use the {\sc nicil} library \citep{wurster16} for cosmic ray ionisation chemistry and thermal ionisation.  We refer the reader to \citet{wurster16} and \citet{wpb16} for details, since this is maintained and versioned as a separate package.

\subsubsection{Timestepping}
 With explicit timesteps, the timestep is constrained in a similar manner to other terms, using
\begin{equation}
\label{eq:nimhd:dt}
\Delta t = \frac{C_{\rm nimhd} h^2}{\max(\eta_{\rm O},\eta_{\rm AD}, \vert \eta_{\rm H}\vert )}
\end{equation}
where $C_{\rm nimhd} = 1/(2\pi)$ by default. This can prove prohibitive, so we employ the so-called `super-timestepping' algorithm from \citet{aag96} to relax the stability criterion for the Ohmic and ambipolar terms (only). The implementation is described in detail in \citet{wpb16}. Currently the Hall effect is always timestepped explicitly in the code.


\subsection{Self-gravity}
\label{sec:gravity}
 {\sc Phantom} includes self-gravity between particles. By self-gravity we mean a solution to Poisson's equation,
\begin{equation}
\nabla^{2} \Phi = 4\pi G \rho({\bm r}), \label{eq:poisson}
\end{equation}
where $\Phi$ is the gravitational potential and $\rho$ represents a continuous fluid density. The corresponding acceleration term in the equation of motion is
\begin{equation}
{\bm a}_{\rm selfgrav} = -\nabla\Phi.
\end{equation}
Since (\ref{eq:poisson}) is an elliptic equation, implying instant action, it requires a global solution. This solution is obtained in {\sc Phantom} by splitting the acceleration into `short-range' and `long-range' contributions,
\begin{equation}
{\bm a}^{a}_{\rm selfgrav} = {\bm a}^{a}_{\rm short} + {\bm a}^{a}_{\rm long} ,
\end{equation}
where the `short-range' interaction is computed by direct summation over nearby particles, and the `long-range' interaction is computed by hierarchical grouping of particles using the $k$d-tree.

 The distance at which the gravitational acceleration is treated as `short-' or `long-range' is determined for each node-node pair, $n$-$m$, either by the tree opening criterion,
\begin{equation}
\theta^{2} < \left(\frac{s_{m}}{r_{nm}}\right)^{2},
\label{eq:treeacc}
\end{equation}
where $0 \leq \theta \leq 1$ is the tree opening parameter, or by nodes whose smoothing spheres intersect,
\begin{equation}
r_{nm}^{2} < \left[s_{n} + s_{m} + \max(R_{\rm kern} h_{\max}^{n}, R_{\rm kern} h_{\max}^{m})\right]^{2}.
\label{eq:isneighb}
\end{equation}
Here, $s$ is the node size, which is the minimum radius about the centre of mass that contains all the particles in the node, and $h_{\max}$ is the maximum smoothing length of the particles within the node. Node pairs satisfying either of these criteria have the particles contained within them added to a trial neighbour list, used for computing the short-range gravitational acceleration. Setting $\theta = 0$ therefore leads to the gravitational acceleration computed entirely as `short-range', that is, only via direction summation, while $\theta = 1$ gives the fastest possible, but least accurate, gravitational force evaluation. The code default is $\theta = 0.5$.

\subsubsection{Short-range interaction}
\label{sec:softening}
How to solve (\ref{eq:poisson}) on particles is one of the most widely misunderstood problems in astrophysics. In SPH or collisionless N-body simulations (i.e. stars, dark matter), the particles do not represent physical point mass particles, but rather interpolation points in a density field that is assumed to be continuous. Hence one needs to first reconstruct the density field on the particles, then solve (\ref{eq:poisson}) in a manner which is consistent with this \citep[e.g.][]{hernquistbarnes90}.

 How to do this consistently using a spatially adaptive softening length was demonstrated by \citet{pricemonaghan07}, since an obvious choice is to use the iterative kernel summation in (\ref{eq:rhosum}) to both reconstruct the density field and set the softening length, i.e.\footnote{Strictly one should use the number density instead of the mass density when computing the softening length via (\ref{eq:geteps}), but as we enforce equal masses for each particle type in \textsc{Phantom}, the two methods are equivalent.}
\begin{equation}
\rho_{a} = \sum_{b} m_{b} W_{ab} (\epsilon_{a}); \hspace{5mm} \epsilon_{a} = h_{\rm fac} (m_{a}/\rho_{a})^{1/3}.
\label{eq:geteps}
\end{equation}
  It can then be shown that the gravitational potential consistent with this choice is given by
\begin{equation}
\Phi_{a} = -G \sum_{b} m_{b} \phi(\vert {\bm r}_{a} - {\bm r}_{b} \vert, \epsilon_{a}),
\end{equation}
where $\phi$ is the softening kernel derived from the density kernel via Poisson's equation (Section~\ref{sec:ksoft}, below). For a variable smoothing length, energy conservation necessitates taking an average of the kernels, i.e.
\begin{equation}
\Phi_{a} = -G \sum_{b} m_{b} \left[ \frac{\phi_{ab}(\epsilon_{a}) + \phi_{ab}(\epsilon_{b})}{2} \right].
\end{equation}
\citet{pricemonaghan07} showed how the equations of motion could then be derived from a Lagrangian in order to take account of the softening length gradient terms, giving equations of motion in the form
\begin{align}
{\bm a}^{a}_{\rm selfgrav} = & -\nabla\Phi_{a}, \nonumber \\
= & -G \sum_{b} m_{b} \left[ \frac{\phi'_{ab}(\epsilon_{a}) + \phi'_{ab}(\epsilon_{b})}{2}\right] \hat{\bm r}_{ab} \nonumber \\
& -\frac{G}{2} \sum_{b} m_{b} \left[ \frac{\zeta_{a}}{\Omega^{\epsilon}_{a}} \nabla_{a} W_{ab} (\epsilon_{a}) + \frac{\zeta_{b}}{\Omega^{\epsilon}_{b}} \nabla_{a} W_{ab} (\epsilon_{b})\right],
\label{eq:agrav}
\end{align}
where $\Omega^{\epsilon}$ and $\zeta$ are correction terms necessary for energy conservation, with $\Omega$ as in (\ref{eq:omega}) but with $h$ replaced by $\epsilon$, and $\zeta$ defined as
\begin{equation}
\zeta_{a} \equiv \frac{\partial \epsilon_{a}}{\partial \rho_{a}} \sum_{b} m_{b} \frac{\partial \phi_{ab}(\epsilon_{a})}{\partial \epsilon_{a}}.
\label{eq:zeta}
\end{equation}
The above formulation satisfies all of the conservation properties, namely conservation of linear momentum, angular momentum, and energy. 

The short range acceleration is evaluated for each particle in the leaf node $n$ by summing (\ref{eq:agrav}) over the trial neighbour list obtained by node-node pairs that satisfy either of the criteria in (\ref{eq:treeacc}) or (\ref{eq:isneighb}). For particle pairs separated outside the softening radius of either particle, the short range interaction reduces to
\begin{equation}
{\bm a}^{a}_{\rm short, r > R_{\rm kern}\epsilon} = -G \sum_{b} m_{b} \frac{ {\bm r}_{a} - {\bm r}_{b}}{\vert {\bm r}_{a} - {\bm r}_{b}\vert^{3}}.
\end{equation}
We use this directly for such pairs to avoid unnecessary evaluations of the softening kernel.

It is natural in SPH to set the gravitational softening length equal to the smoothing length $\epsilon = h$, since both derive from the same density estimate. Indeed, \citet{bateburkert97} showed that this is a crucial requirement to resolve gas fragmentation correctly. For pure $N$-body calculations, \citet{pricemonaghan07} also showed that setting the (variable) softening length in the same manner as the SPH smoothing length (Sections~\ref{sec:hset}--\ref{sec:hrho}) results in a softening that is always close to the `optimal' fixed softening length \citep{merritt96,athanassoulaetal00,dehnen01}. In collisionless $N$-body simulations, this has been found to increase the resolving power, giving results at lower resolution comparable to those obtained at higher resolution with a fixed softening length \citep{baglakhandai09,iannuzzidolag11}. It also avoids the problem of how to `choose' the softening length, negating the need for `rules of thumb' such as the one given by \citet{springel05} where the softening length is chosen to be 1/40 of the average particle spacing in the initial conditions.

\subsubsection{Functional form of the softening kernel}
\label{sec:ksoft}
The density kernel and softening potential kernel are related using Poisson's equation (\ref{eq:poisson}), i.e.
\begin{equation}
W(r, \epsilon) = \frac{1}{4\pi r^{2}} \frac{\partial}{\partial r} \left( r^{2}  \frac{\partial \phi}{\partial r} \right),
\end{equation}
where $r \equiv \vert{\bm r} - {\bm r}'\vert$. Integrating this equation gives the softening kernels used for the force and gravitational potential. As with the standard kernel functions (Section~\ref{sec:kfunc}), we define the potential and force kernels in terms of dimensionless functions of the scaled interparticle separation, $q\equiv r/h$, according to
\begin{align}
\phi(r, \epsilon) & \equiv \frac{1}{\epsilon} \tilde{\phi} (q), \label{eq:phisoft}\\
\phi'(r, \epsilon) & \equiv \frac{1}{\epsilon^{2}} \tilde{\phi}' (q), \label{eq:fsoft}
\end{align}
where the dimensionless force kernel is obtained from the density kernel $f(q)$ (Section~\ref{sec:kfunc}--\ref{sec:kdefs}) using
\begin{equation}
\tilde{\phi}' (q) = \frac{4\pi}{q^{2}} C_{\rm norm} \int f(q') q'^{2} {\rm d}q',
\label{eq:fq}
\end{equation}
with the integration constant set such that $\tilde{\phi}'(q) = 1/q^{2}$ at $q = R_{\rm kern}$. The potential function is
\begin{equation}
\tilde{\phi}(q) = \int \tilde{\phi}'(q') dq'.
\label{eq:phiq}
\end{equation}
The derivative of $\phi$ with respect to $\epsilon$ required in (\ref{eq:zeta}) is also written in terms of a dimensionless function, i.e.
\begin{equation}
\frac{\partial \phi(r,\epsilon)}{\partial \epsilon} \equiv \frac{1}{\epsilon^{2}} h(q),
\end{equation}
where from differentiating (\ref{eq:phisoft}) we have
\begin{equation}
h(q) = -\tilde{\phi}(q) - q\tilde{\phi}'(q).
\label{eq:hq}
\end{equation}
Included in the \textsc{Phantom} source code is a \textsc{Python} script using the {\sc sympy} library to solve (\ref{eq:fq}), (\ref{eq:phiq}) and (\ref{eq:hq}) using symbolic integration to obtain the softening kernel from the density kernel. This makes it straightforward to obtain the otherwise laborious expressions needed for each kernel (expressions for the cubic spline kernel are in Appendix~A of \citealt{pricemonaghan07} and for the quintic spline in Appendix~A of \citealt{hubberetal11}). Figure~\ref{fig:kernels-softening} shows the functional form of the softening kernels for each of the kernels available in \textsc{Phantom}. The kernel function $f(q)$ is shown for comparison (top row in each case).

\begin{figure}
   \centering
   \includegraphics[width=\columnwidth]{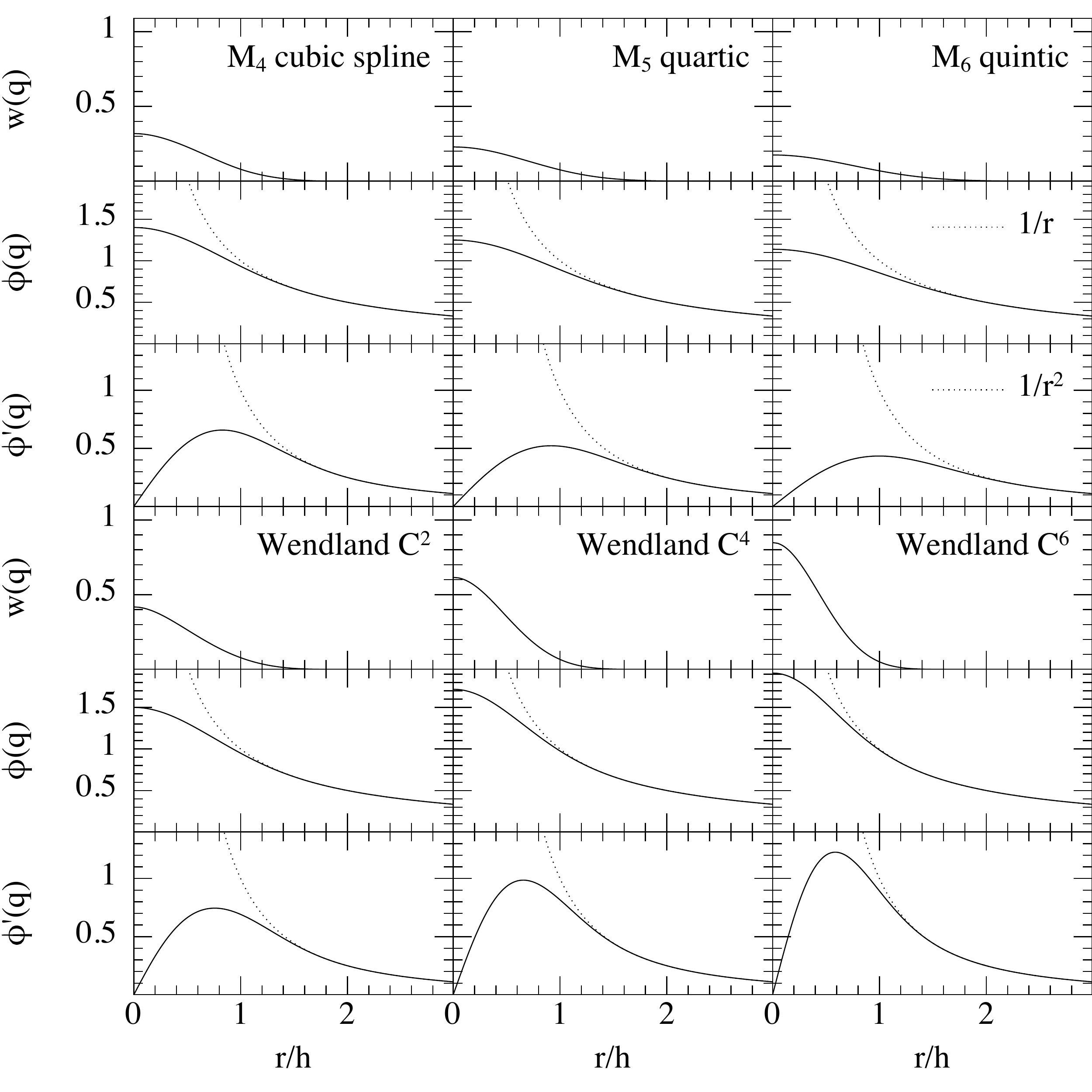} 
   \caption{Functional form of the softening kernel functions $-\phi(r,h)$ and $\phi'(r,h)$ used to compute the gravitational force in \textsc{Phantom}, shown for each of the available kernel functions $w(r,h)$ (see Figure~\ref{fig:kernels}). Dotted lines show the functional form of the unsoftened potential ($-1/r$) and force ($1/r^{2}$) for comparison.}
   \label{fig:kernels-softening}
\end{figure}

\subsubsection{Softening of gravitational potential due to stars, dark matter and dust}
\label{sec:softeningdm}
In the presence of collisionless components (e.g. stars, dark matter and dust) we require estimates of the density in order to set the softening lengths for each component. We follow the generalisation of the SPH density estimate to multi-fluid systems described by \citet{laibeprice12} where the density (and hence the softening length and $\Omega^{\epsilon}$) for each component is computed in the usual iterative manner (Section~\ref{sec:hrho}), but using only neighbours of the same type (c.f. Section~\ref{sec:twofluid}). That is, the softening length for stars is determined based on the local density of stars, the softening length for dark matter is based on the local density of dark matter particles, and so on. The gravitational interaction both within each type and between different types is then computed using (\ref{eq:agrav}). This naturally symmetrises the softening between different types, ensuring that momentum, angular momentum and energy are conserved.


\subsubsection{Long-range interaction}
\label{sec:longrangegravity}
At long range, that is $r > R_{\rm kern} \epsilon_{a}$ and $r > R_{\rm kern} \epsilon_{b}$, the second term in (\ref{eq:agrav}) tends to zero since $\zeta = 0$ for $q \geq R_{\rm kern}$, while the first term simply becomes $1/r^{2}$. Computing this via direct summation would have an associated $\mathcal{O}(N^{2})$ computational cost, thus we adopt the usual practice of using the $k$d-tree to reduce this cost to $\mathcal{O}(N\log N)$.

The main optimisation in \textsc{Phantom} compared to a standard tree code (e.g. \citealt{hernquistkatz89,benzetal90}) is that we compute the long-range gravitational interaction once \emph{per leaf-node} rather than once \emph{per-particle} and that we use Cartesian rather than spherical multipole expansions to enable this \citep{dehnen00,gaftonrosswog11}.

 The long range acceleration on a given leaf node $n$ consists of a sum over distant nodes $m$ that satisfy neither (\ref{eq:treeacc}) nor (\ref{eq:isneighb}),
\begin{equation}
{\bm a}_{n} = \sum_{m} {\bm a}_{nm}.
\end{equation}
The acceleration along the node centres, between a given pair $n$ and $m$, is given (using index notation) by
\begin{equation}
a^{i}_{nm} = -\frac{GM_{m}}{r^{3}} r^{i} + \frac{1}{r^{4}} \left(\hat{r}^{k} Q^{m}_{ik} - \frac52 \hat{r}^{i} \mathcal{Q}^{m} \right),
\label{eq:anm}
\end{equation} 
where $r^i \equiv x^{i}_{n} - x^{i}_{m}$ is the relative position vector, $\hat{r}$ is the corresponding unit vector, $M_{m}$ is the total mass in node $m$, $Q^{m}_{ij}$ is the quadrupole moment of node $m$, and repeated indices imply summation. We define the following scalar and vector quantities for convenience:
\begin{align}
\mathcal{Q} & \equiv \hat{r}^{i} \hat{r}^{j} Q_{ij}, \\
\mathcal{Q}_{i} & \equiv \hat{r}^{j} Q_{ij}.
\end{align}

Alongside the acceleration, we compute the six independent components of the first derivative matrix,
\begin{equation}
\frac{\partial a^{i}_{n}}{\partial r^{j}} = \sum_{m} \frac{\partial a^{i}_{nm}}{\partial r^{j}},
\end{equation}
where
\begin{align}
\frac{\partial a_{nm}^{i}}{\partial r^{j}} = & \frac{GM_{m}}{r^{3}} \bigg [ 3\hat{r}^{i} \hat{r}^{j} - \delta^{ij}  \bigg] \nonumber \\
& + \frac{1}{r^{5}} \left[ Q_{ij}^{m} + \left(\frac{35}{2} \hat{r}^{i} \hat{r}^{j} - \frac52 \delta_{ij}\right) \mathcal{Q}^{m} \right. \nonumber \\
& \phantom{+\frac{1}{r^{5}}\bigg[} - 5 \hat{r}^{i} \mathcal{Q}_{j}^{m} - 5 \hat{r}^{j} \mathcal{Q}_{i}^{m}  \bigg],
\end{align}
and the ten independent components of the second derivatives, given by
\begin{align}
\frac{\partial^{2} a^{i}_{nm}}{\partial r^{j} \partial r^{k}} = & -\frac{3GM_{m}}{r^{4}} \bigg[ 5 \hat{r}^{i} \hat{r}^{j} \hat{r}^{k} - \delta_{jk} \hat{r}^{i} - \delta_{ik} \hat{r}^{j} - \delta_{ij} \hat{r}^{k} \bigg] \nonumber \\
& + \frac{1}{r^{6}} \bigg[ -5(\hat{r}^{k} Q^{m}_{ij} + \hat{r}^{i} Q^{m}_{jk} + \hat{r}^{j} Q^{m}_{ik}) \nonumber \\
& \phantom{+ \frac{1}{r^{6}} \bigg[} - \frac{315}{2}\hat{r}^{i}\hat{r}^{j}\hat{r}^{k} \mathcal{Q}^{m}  \nonumber \\
& \phantom{+ \frac{1}{r^{6}} \bigg[} + \frac{35}{2} \left(\delta_{ij} \hat{r}^{k} + \delta_{ik} \hat{r}^{j} + \delta_{jk} \hat{r}^{i} \right) \mathcal{Q}^{m} \nonumber \\
& \phantom{+ \frac{1}{r^{6}} \bigg[} + 35 \left( \hat{r}^{j} \hat{r}^{k} \mathcal{Q}^{m}_{i} + \hat{r}^{i} \hat{r}^{k} \mathcal{Q}^{m}_{j} + \hat{r}^{i} \hat{r}^{j} \mathcal{Q}^{m}_{k} \right) \nonumber \\
& \phantom{+ \frac{1}{r^{6}} \bigg[} - 5 (\delta_{ij} \mathcal{Q}^{m}_{k} + \delta_{ik} \mathcal{Q}^{m}_{j} + \delta_{jk} \mathcal{Q}^{m}_{i}) \bigg].
\end{align}
The acceleration on each individual particle inside the leaf node $n$ is then computed using a second-order Taylor series expansion of ${\bm a}_{\rm node}^{n}$ about the node centre, i.e.
\begin{equation}
a^{i}_{{\rm long}, a} = a^{i}_{n} + \Delta x^{j} \frac{\partial a_{n}^{i}}{\partial r^{j}} + \frac12 \Delta x^{j} \Delta x^{k} \frac{\partial^{2} a_{n}^{i}}{\partial r^{j}\partial r^{k}},
\end{equation}
where $\Delta x_{a}^{i} \equiv x^{i}_{a} - x^{i}_{0}$ is the relative distance of each particle from the node centre of mass. Pseudo-code for the resulting force evaluation is shown in Figure~\ref{fig:forcecalc}.

 The quadrupole moments are computed during the tree build using
\begin{equation}
Q_{ij} = \sum_{a} m_{a} \left[ 3 \Delta x^{i} \Delta x^{j} -  (\Delta x)^{2} \delta^{ij}\right],
\end{equation}
where the sum is over all particles in the node. Since $Q$ is a symmetric tensor, only six independent quantities need to be stored ($Q_{xx}$, $Q_{xy}$, $Q_{xz}$, $Q_{yy}$, $Q_{yz}$ and $Q_{zz}$). 
 
 The current implementation in \textsc{Phantom} is $\mathcal{O}(N_{\textrm{leafnodes}} \log N_{\rm part})$ rather than the $\mathcal{O}(N)$ treecode implementation proposed by \citet{dehnen00} since we do not currently implement the symmetric node-node interactions required for $\mathcal{O}(N)$ scaling. Neither does our treecode conserve linear momentum to machine precision, except when $\theta = 0$. Implementing these additional features would be desirable.

\subsection{Dust-gas mixtures}
\label{sec:dust}
 Modelling dust-gas mixtures is the first step in the `grand challenge' of protoplanetary disc modelling \citep{haworthetal16}. The public version of \textsc{Phantom} implements dust-gas mixtures using two approaches. One models the dust and gas as two separate types of particles (two-fluid), as presented in \citet{laibeprice12,laibeprice12a}, and the other, for small grains, using a single type of particle that represents the combination of dust and gas together (one-fluid), as described in \citet{pricelaibe15}. Various combinations of these algorithms have been used in our recent papers using \textsc{Phantom}, including \citet{dipierroetal15,dipierroetal16, ragusaetal17} and \citet{tpl17} (see also \citealt{hutchisonetal16}).
 
 In the two-fluid implementation, the dust and gas are treated as two separate fluids coupled by a drag term with only explicit timestepping. In the one-fluid implementation, the dust is treated as part of the mixture, with an evolution equation for the dust fraction.
  
 
\subsubsection{Continuum equations}
The two-fluid formulation is based on the continuum equations in the form
\begin{align}
\frac{\partial \rho_{\rm g}}{\partial t} +  (\vg\cdot \nabla) \rho_{\rm g} & = -\rho_{\rm g} (\nabla\cdot{\bm v}_{\rm g}), \label{eq:rhog} \\
\frac{\partial \rho_{\rm d}}{\partial t} +  (\vd\cdot\nabla) \rho_{\rm d} & = -\rho_{\rm d} (\nabla\cdot{\bm v}_{\rm d}), \label{eq:rhod}\\
\frac{\partial {\bm v}_{\rm g}}{\partial t} + (\vg\cdot\nabla) \vg & = -\frac{\nabla P}{\rho_{\rm g}} + \frac{K}{\rho_{\rm g}} (\vd - \vg), \label{eq:vg} \\
\frac{\partial \vd}{\partial t} + (\vd\cdot\nabla) \vd & = - \frac{K}{\rho_{\rm d}} (\vd - \vg), \label{eq:vd} \\
\frac{\partial u}{\partial t} + (\vg\cdot\nabla) u& = -\frac{P}{\rho_{\rm g}} (\nabla\cdot \vg) + \Lambda_{\rm drag} , \label{eq:dudttwof}
\end{align}
where the subscripts ${\rm g}$ and ${\rm d}$ refer to gas and dust properties, $K$ is a drag coefficient and the drag heating is
\begin{equation}
\Lambda_{\rm drag} \equiv K (\vd - \vg)^{2}.
\end{equation}
The implementation in \textsc{Phantom} currently neglects any thermal coupling between the dust and the gas (see \citealt{laibeprice12}), aside from the drag heating. Thermal effects are important for smaller grains since they control the heating and cooling of the gas (e.g. \citealt{dopckeetal11}). Currently the internal energy ($u$) of dust particles is simply set to zero.

\subsubsection{Stopping time}
The stopping time
\begin{equation}
t_{\rm s} = \frac{\rho_{\rm g} \rho_{\rm d}}{K(\rho_{\rm g} + \rho_{\rm d})},
\label{eq:ts}
\end{equation}
is the characteristic timescale for the exponential decay of the differential velocity between the two phases caused by the drag. In the code, $t_{\rm s}$ is by default specified in physical units, which means that code units need to be defined appropriately when simulating dust-gas mixtures. 

\subsubsection{Two-fluid dust-gas in SPH}
\label{sec:twofluid}
In the two-fluid method, the mixture is discretised into two distinct sets of `dust' and `gas' particles. In the following, we adopt the convention from \citet{monaghankocharyan95} that subscripts $a$, $b$ and $c$ refer to gas particles while $i$, $j$ and $k$ refer to dust particles. Hence, (\ref{eq:rhog})--(\ref{eq:rhod}) are discretised with a density summation \emph{over neighbours of the same type} (c.f. Section~\ref{sec:softeningdm}), giving
\begin{equation}
\rho_{a} = \sum_{b} m_{b} W_{ab} (h_{a}); \hspace{5mm} h_{a} = h_{\rm fact} \left( \frac{m_{a}}{\rho_{a}} \right ) ^{1/3},
\end{equation}
for a gas particle, and
\begin{equation}
\rho_{i} = \sum_{j} m_{j} W_{ij} (h_{i}); \hspace{5mm} h_{i} = h_{\rm fact} \left( \frac{m_{i}}{\rho_{i}} \right ) ^{1/3},
\end{equation}
for a dust particle. The kernel used for density is the same as usual (Section~\ref{sec:kdefs}). We discretise the equations of motion for the gas particles, (\ref{eq:vg}), using
\begin{equation}
\left(\frac{{\rm d}{\bm v}_{a}}{{\rm d}t}\right)_{\rm drag} = -3 \sum_{j} m_{j} \frac{{\bm v}_{aj} \cdot \hat{\bm r}_{aj}}{(\rho_{a} + \rho_{j}) t^{\rm s}_{aj}}  \hat{\bm r}_{aj} D_{aj} (h_{a}), \label{eq:vgsph}
\end{equation}
and for dust, (\ref{eq:vd}), using
\begin{equation}
\left(\frac{{\rm d}{\bm v}_{i}}{{\rm d}t}\right)_{\rm drag} = -3 \sum_{b} m_{b} \frac{{\bm v}_{ib} \cdot \hat{\bm r}_{ib}}{(\rho_{i} + \rho_{b}) t^{\rm s}_{ib}}  \hat{\bm r}_{ib} D_{ib} (h_{b}), \label{eq:vdsph}
\end{equation}
where $D$ is a `double hump' kernel, defined in Section~\ref{sec:dragkernel}, below. The drag heating term in the energy equation, (\ref{eq:dudttwof}), is discretised using
\begin{equation}
\Lambda_{\rm drag} = 3 \sum_{j} m_{j} \frac{({\bm v}_{aj} \cdot \hat{\bm r}_{aj})^{2}}{(\rho_{a} + \rho_{j}) t^{\rm s}_{aj}}  D_{aj} (h_{a}). \label{eq:lambdasph}
\end{equation}
Notice that gas properties are only defined on gas particles and dust properties are defined only on dust particles, greatly simplifying the algorithm. Buoyancy terms caused by dust particles occupying a finite volume \citep{monaghankocharyan95,laibeprice12} are negligible in astrophysics because typical grain sizes ($\mu$m) are negligible compared to simulation scales of $\sim$au or larger.

\subsubsection{Drag kernels}
\label{sec:dragkernel}

\begin{figure}
   \centering
   \includegraphics[width=\columnwidth]{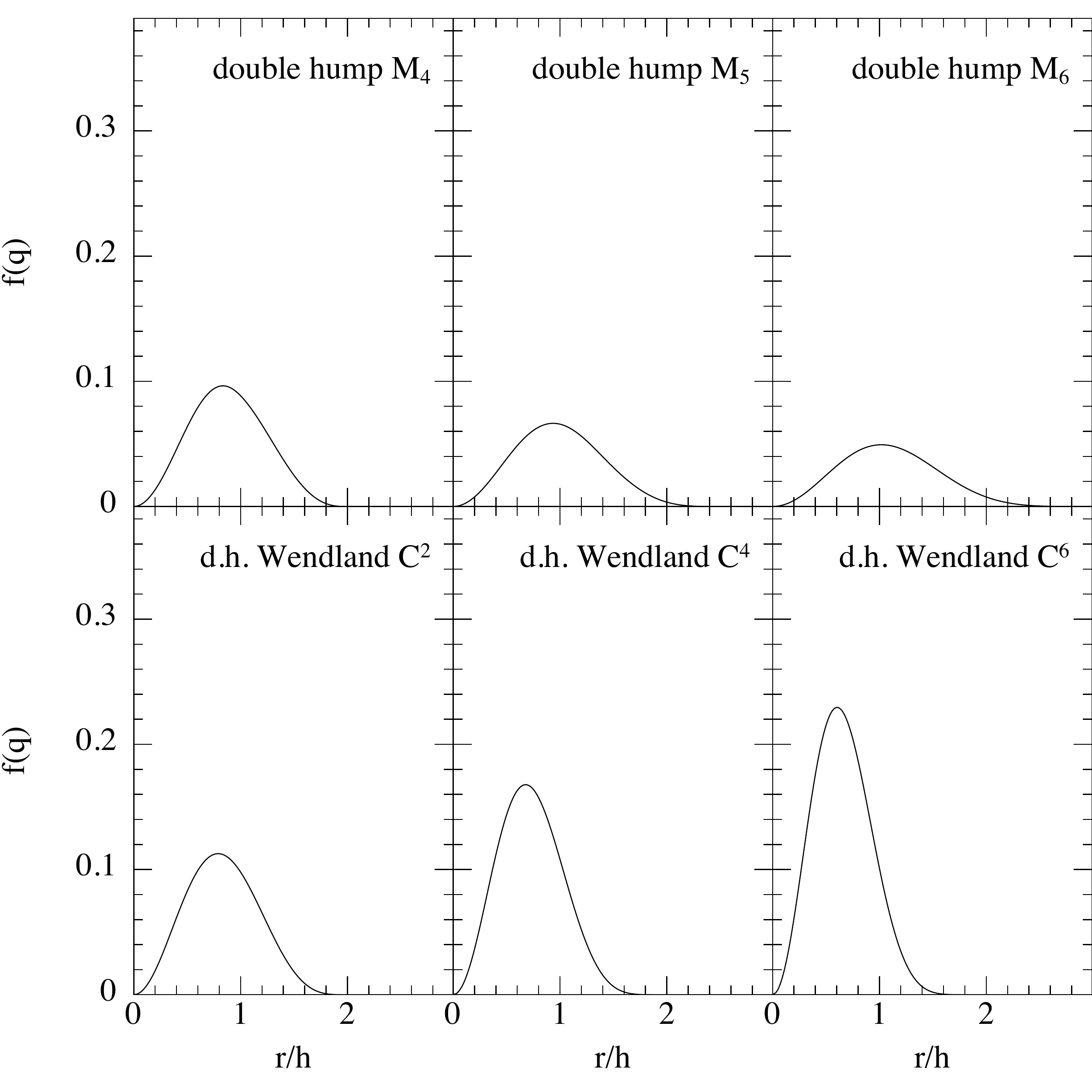} 
   \caption{Double hump smoothing kernels $D(r,h)$ available in \textsc{Phantom}, used in the computation of the dust-gas drag force.}
   \label{fig:doublehump}
\end{figure}

Importantly, we use a `double-hump' shaped kernel function $D$ \citep{fq96} instead of the usual bell-shaped kernel $W$ when computing the drag terms. Defining $D$ in terms of a dimensionless kernel function as previously (c.f.~Section~\ref{sec:kfunc}),
\begin{equation}
D(r,h) = \frac{\sigma}{h^{3}} g(q),
\end{equation}
then the double hump kernels are defined from the usual kernels according to
\begin{equation}
g(q) = q^{2} f(q),
\end{equation}
where the normalisation constant $\sigma$ is computed by enforcing the usual normalisation condition
\begin{equation}
\int D (r, h) {\rm d}V = 1.
\end{equation}
 Figure~\ref{fig:doublehump} shows the functional forms of the double hump kernels used in {\sc Phantom}. Using double hump kernels for the drag terms was found by \citet{laibeprice12} to give a factor of 10 better accuracy at no additional cost. The key feature is that these kernels are zero at the origin putting more weight in the outer parts of the kernel where the velocity difference is large. This also solves the problem of how to define the unit vector in the drag terms (\ref{eq:vgsph}), (\ref{eq:vdsph}) and (\ref{eq:lambdasph}) --- it does not matter since $D$ is also zero at the origin.
 
\subsubsection{Stopping time in SPH}
The stopping time is defined between a pair of particles, using the properties of gas and dust defined on the particle of the respective type, i.e.
\begin{equation}
t^{\rm s}_{aj} = \frac{\rho_{a} \rho_{j}}{K_{aj} (\rho_{a} + \rho_{j})}.
\end{equation}
The default prescription for the stopping time in \textsc{Phantom} automatically selects a physical drag regime appropriate to the grain size, as described below and in \citet{laibeprice12a}. Options to use either a constant $K$ or a constant $t_{\rm s}$ between pairs are also implemented, useful for testing and debugging (c.f. Section~\ref{sec:dusttest}).

\subsubsection{Epstein drag}
\label{sec:epstein}
\begin{figure}
   \centering
   \includegraphics[width=\columnwidth]{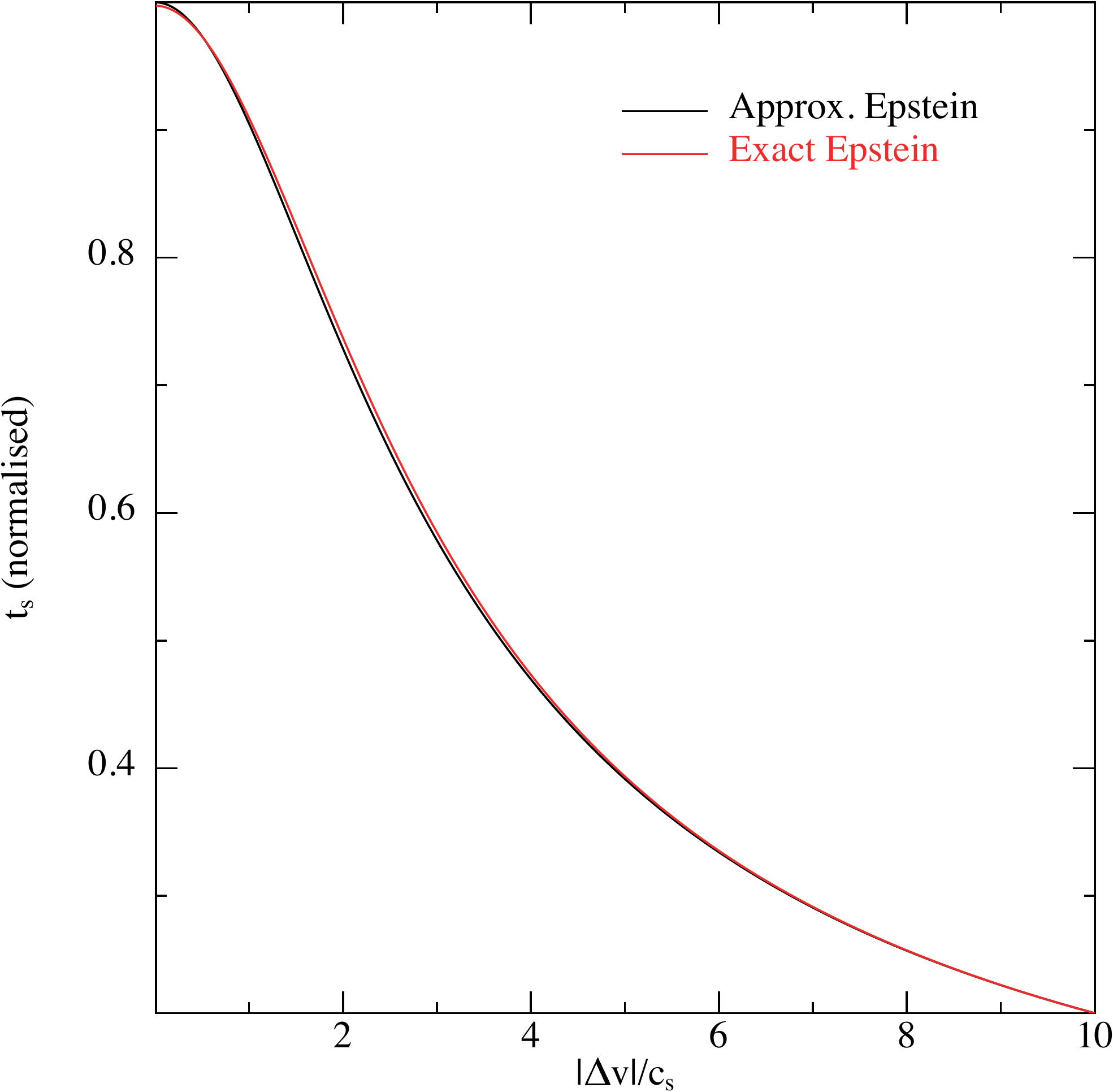} 
   \caption{Dependence of the drag stopping time $t_{\rm s}$ on differential Mach number, showing the increased drag (decrease in stopping time) as the velocity difference between dust and gas increases. The black line shows the analytic approximation we employ (Equation~\ref{eq:tseps}) which may be compared to the red line showing the exact expression from \citet{epstein24}. The difference is less than 1 per cent everywhere.}
   \label{fig:ts-deltav}
\end{figure}

 To determine the appropriate physical drag regime we use the procedure suggested by \citet{stepinskivalageas96} where we first evaluate the Knudsen number
\begin{equation}
K_{\rm n} = \frac{9\lambda_{\rm g}}{4 s_{\rm grain}},
\end{equation}
where $s_{\rm grain}$ is the grain size and $\lambda_{\rm g}$ is the gas mean free path (see Section~\ref{sec:mfp}, below, for how this is evaluated). For $K_{n} \geq 1$, the drag between a particle pair is computed using the generalised formula for Epstein drag from \citet{kwok75}, as described in \citet{paardekoopermellema06} and \citet{laibeprice12a}, giving
\begin{equation}
K_{aj} = \rho^{a}_{\rm g} \rho^{j}_{\rm d} \frac43 \sqrt\frac{8\pi}{\gamma} \frac{s_{\rm grain}^{2}}{m_{\rm grain}} c^{a}_{\rm s} f,
\end{equation}
where
\begin{equation}
m_{\rm grain} \equiv \frac43 \pi \rho_{\rm grain} s_{\rm grain}^{3},
\end{equation}
and $\rho_{\rm grain}$ is the intrinsic grain density, which is 3~g/cm$^{3}$ by default. The variable $f$ is a correction for supersonic drift velocities given by \citep{kwok75}
\begin{equation}
f \equiv \sqrt{1 + \frac{9\pi}{128} \frac{\Delta v^{2}}{c_{\rm s}^{2}}},
\end{equation}
where $\Delta v \equiv \vert \vd - \vg \vert = v_{\rm d}^{j} - v_{\rm g}^{a}$. The stopping time is therefore
\begin{equation}
t_{\rm s} = \frac{ \rho_{\rm grain} s_{\rm grain}}{\rho c_{\rm s}f}  \sqrt{\frac{\pi\gamma}{8}},
\label{eq:tseps}
\end{equation}
where $\rho \equiv \rho_{\rm d} + \rho_{\rm g}$. This formula, (\ref{eq:tseps}), reduces to the standard expression for the linear Epstein regime in the limit where the drift velocity is small compared to the sound speed (i.e. $f\to 1$). Figure~\ref{fig:ts-deltav} shows the difference between the above simplified prescription and the exact expression for Epstein drag (\citealt{epstein24}; c.f. equations~11 and 38 in \citealt{laibeprice12a}) as a function of $\Delta v/c_{\rm s}$, which is less than 1 per cent everywhere.

\subsubsection{Stokes drag}
For $K_{n} < 1$, we adopt a Stokes drag prescription, describing the drag on a sphere with size larger than the mean free path of the gas \citep{fanzhu98}. Here we use \citep{laibeprice12a}
\begin{equation}
K_{aj} = \rho^{a}_{\rm g} \rho^{j}_{\rm d} \frac12 C_{\rm D} \frac{\pi s_{\rm grain}^{2}}{m_{\rm grain}} \vert \Delta v\vert,
\end{equation}
where the coefficient $C_{\rm D}$ is given by \citep{fassioprobstein70} (see \citealt{whipple72,weidenschilling77})
\begin{equation}
C_{\rm D} = 
\begin{cases}
24 R_{e}^{-1}, & R_{\rm e} < 1, \\
24 R_{e}^{-0.6}, & 1 < R_{\rm e} < 800, \\
0.44, & R_{\rm e} > 800,
\end{cases}
\label{eq:cd}
\end{equation}
where $R_{\rm e}$ is the local Reynolds number around the grain
\begin{equation}
R_{\rm e} \equiv \frac{2 s_{\rm grain} \vert \Delta v \vert}{\nu},
\end{equation}
and $\nu$ is the microscopic viscosity of the gas (see below; not to be confused with the disc viscosity). Similar formulations of Stokes drag can be found elsewhere (see e.g. discussion in \citealt{woitkehelling03} and references therein). The stopping time in the Stokes regime is therefore given by
\begin{equation}
t_{\rm s} = \frac{8  \rho_{\rm grain} s_{\rm grain}}{3\rho \vert \Delta v\vert C_{\rm D}},
\label{eq:tsstokes}
\end{equation}
where it remains to evaluate $\nu$ and $\lambda_{\rm g}$.

\subsubsection{Kinematic viscosity and mean free path}
\label{sec:mfp}
We evaluate the microscopic kinematic viscosity, $\nu$, assuming gas molecules interact as hard spheres, following \citet{chapmancowling70}. The viscosity is computed from the mean free path and sound speed according to
\begin{equation}
\nu = \sqrt{\frac{2}{\pi \gamma}} c_{\rm s} \lambda_{\rm g},
\end{equation}
with the mean free path defined by relating this expression to the expression for the dynamic viscosity of the gas \citep{chapmancowling70} given by
\begin{equation}
\mu_{\nu} = \frac{5m}{64\sigma_{\rm s}} \sqrt{\frac{\pi}{\gamma}} c_{\rm s}, \label{eq:munu}
\end{equation}
with $\mu_{\nu} = \rho_{\rm g} \nu$, giving
\begin{equation}
\lambda_{\rm g} = \frac{5\pi}{64 \sqrt{2}} \frac{1}{n_{\rm g} \sigma_{s}},
\end{equation}
where $n_{\rm g}= \rho_{\rm g}/m$ is the number density of molecules and $\sigma_{s}$ is the collisional cross section. To compute this, \textsc{Phantom} currently assumes the gas is molecular Hydrogen, such that the mass of each molecule and the collisional cross section are given by
\begin{align}
m & = 2 m_{\rm H}, \\
\sigma_{\rm s} & = 2.367 \times 10^{-15} {\rm cm}^{2}.
\end{align}

\begin{figure}
   \centering
   \includegraphics[width=\columnwidth]{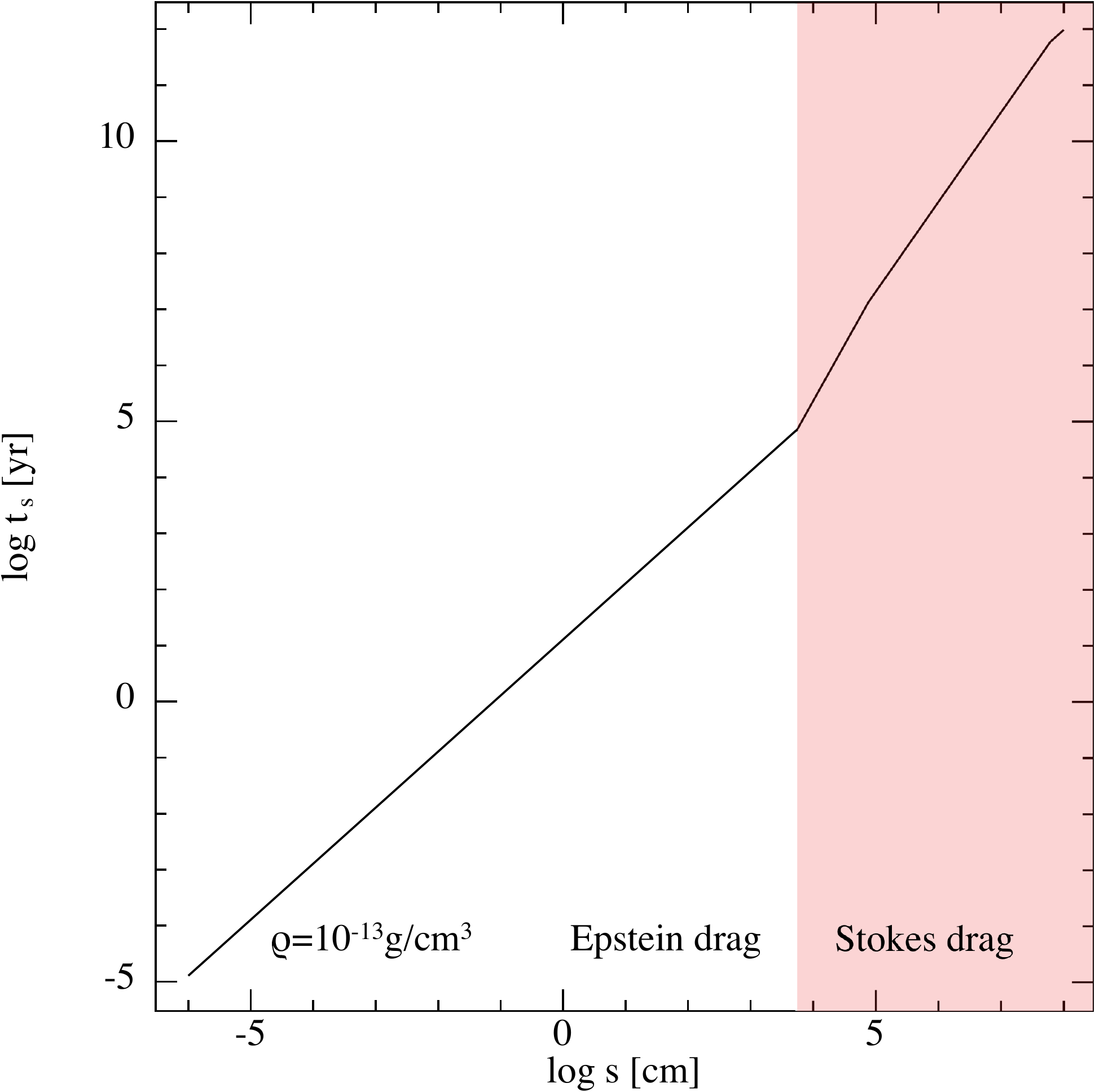} 
   \caption{Drag stopping time $t_{\rm s}$ (in years) as a function of grain size, showing the continuous transition between the Epstein and Stokes drag regimes. The example shown assumes fixed density $\rho = 10^{-13}$g/cm$^{3}$ and sound speed $c_{\rm s} = 6 \times 10^{4}$ cm/s with subsonic drag $\Delta v = 0.01 c_{\rm s}$ and material density $\rho_{\rm grain} = 3$g/cm$^{3}$.}
   \label{fig:ts}
\end{figure}

\subsubsection{Stokes/Epstein transition}
 At the transition between the two regimes, assuming $R_{\rm e} < 1$, (\ref{eq:tsstokes}) reduces to
\begin{equation}
t_{\rm s} = \frac{2 \rho_{\rm grain} s_{\rm grain}^{2}}{9 \rho c_{\rm s} \lambda_{\rm g}} \sqrt{\frac{\pi \gamma}{2}},
\end{equation}
which is the same as the Epstein drag in the subsonic regime when $\lambda_{\rm g} = 4 s_{\rm grain}/9$, i.e. $K_{\rm n} = 1$.  That this transition is indeed continuous in the code is demonstrated in Figure~\ref{fig:ts}, which shows the transition from Epstein to Stokes drag and also through each of the Stokes regimes in (\ref{eq:cd}) by varying the grain size while keeping the other parameters fixed. For simplicity, we assumed a fixed $\Delta v = 0.01 c_{\rm s}$ in this plot, even though in general one would expect $\Delta v$ to increase with stopping time (see Equation~\ref{eq:deltava}).

\subsubsection{Self gravity of dust}
With self-gravity turned on, dust particles interact in the same way as stars or dark matter (Section~\ref{sec:softeningdm}), with a softening length equal to the smoothing length determined from the density of neighbouring dust particles. Dust particles can be accreted by sink particles (Section~\ref{sec:accrete}), but a sink cannot currently be created from dust particles (Section~\ref{sec:sinkcreate}). There is currently no mechanism in the code to handle the collapse of dust to form a self-gravitating object independent of the gas.

\subsubsection{Timestep constraint}
\label{sec:dtdrag}
For the two-fluid method, the timestep is constrained by the stopping time according to
\begin{equation}
\Delta t^{a}_{\rm drag} = \min_{j} ( t_{\rm s}^{aj}). \label{eq:dtdrag}
\end{equation}
This requirement, alongside the spatial resolution requirement $h \lesssim c_{\rm s} t_{\rm s}$ \citep{laibeprice12}, means the two-fluid method becomes both prohibitively slow and increasingly inaccurate for small grains. In this regime one should switch to the one-fluid method, as described below.

\subsubsection{Continuum equations: One-fluid}
\label{sec:onefluid}
In \citet{laibeprice14}, we showed that the two-fluid equations, (\ref{eq:rhog})--(\ref{eq:dudttwof}), can be rewritten as a single fluid mixture using a change of variables given by
\begin{align}
\rho & \equiv \rho_{\rm g} + \rho_{\rm d}, \\
\epsilon & \equiv \rho_{\rm d}/\rho, \\
{\bm v} & \equiv \frac{\rho_{\rm g} \vg + \rho_{\rm d} \vd}{ \rho_{\rm g} + \rho_{\rm d} }, \\
\Delta {\bm v} & \equiv \vd - \vg,
\end{align}
where $\rho$ is the combined density, $\epsilon$ is the dust fraction, ${\bm v}$ is the barycentric velocity, and $\Delta {\bm v}$ is the differential velocity between the dust and gas.

In \citet{laibeprice14} we derived the full set of evolution equations in these variables, and in \citet{laibeprice14b}, implemented and tested an algorithm to solve these equations in SPH. However, using a fluid approximation cannot properly capture the velocity dispersion of large grains, as occurs for example when large planetesimals stream simultaneously in both directions through the midplane of a protoplanetary disc. For this reason, the one-fluid equations are better suited to treating small grains, where the stopping time is shorter than the computational timestep. In this limit we employ the `terminal velocity approximation' \citep[e.g.][]{youdingoodman05} and the evolution equations reduce to \citep{laibeprice14,pricelaibe15}
\begin{align}
\frac{{\rm d}\rho}{{\rm d}t} & = -\rho (\nabla\cdot{\bm v}), \\
\frac{{\rm d}{\bm v}}{{\rm d}t} & = -\frac{\nabla P}{\rho} + {\bm a}_{\rm ext}, \\
\frac{{\rm d}\epsilon}{{\rm d}t} & = -\frac{1}{\rho} \nabla\cdot \left[{\epsilon (1 - \epsilon) \rho \Delta{\bm v}}\right], \label{eq:depsdt} \\
\frac{{\rm d}u}{{\rm d}t} & = -\frac{P}{\rho} (\nabla\cdot{\bm v}) + \epsilon (\Delta {\bm v} \cdot \nabla) u,  \label{eq:dudteps}
\end{align}
where 
\begin{equation}
\Delta{\bm v} \equiv t_{\rm s} \left({\bm a}_{\rm d} - {\bm a}_{\rm g}\right), \label{eq:deltava}
\end{equation}
where ${\bm a}_{\rm d}$ and ${\bm a}_{\rm g}$ refers to any acceleration acting only on the dust or gas phase, respectively. For the simple case of pure hydrodynamics, the only difference is the pressure gradient, giving
\begin{equation}
\Delta{\bm v} \equiv t_{\rm s} \frac{\nabla P}{\rho_{\rm g}} = \frac{t_{\rm s}}{(1 - \epsilon)} \frac{\nabla P}{\rho}, \label{eq:deltav}
\end{equation}
such that (\ref{eq:depsdt}) becomes
\begin{equation}
\frac{{\rm d}\epsilon}{{\rm d}t} = -\frac{1}{\rho} \nabla\cdot \left({\epsilon t_{\rm s}\nabla P}\right). \label{eq:depsdth}
\end{equation}

Importantly, the one-fluid dust algorithm does not result in any heating term in ${\rm d}u/{\rm d}t$ due to drag, because this term is $\mathcal{O}(\Delta {\bm v}^{2})$ and thus negligible \citep{laibeprice14}.

\subsubsection{Visualisation of one-fluid results}
\label{sec:onefluidviz}
Finally, when visualising results of one-fluid calculations, one must reconstruct the properties of the dust and gas in post-processing. We use
\begin{align}
\rho_{\rm g} & = (1 - \epsilon) \rho, \\
\rho_{\rm d} & = \epsilon \rho, \\
{\bm v}_{\rm g} & = {\bm v} - \epsilon \Delta {\bm v}, \\
{\bm v}_{\rm d} & = {\bm v} + (1 - \epsilon) \Delta {\bm v}.
\end{align}
To visualise the one-fluid results in a similar manner to those from the two-fluid method we reconstruct a set of `dust' and `gas' particles with the same positions but with the respective properties of each type of fluid particle copied onto them. We compute $\Delta {\bm v}$ from (\ref{eq:deltav}) using the pressure gradient computed using (\ref{eq:sphmom}), multiplied by the stopping time and the gas fraction. See \citet{pricelaibe15} for more details.

\subsubsection{One-fluid dust-gas implementation}
\label{sec:onefluidsph}
 Our implementation of the one-fluid method in \textsc{Phantom} follows \citet{pricelaibe15} with a few minor changes and corrections\footnote{One should be aware that we derived several of the above equations incorrectly in Appendix B of \citet{pricelaibe15} (see \citealt{pricelaibe15erratum}). The above equations are the correct versions and reflect what is implemented in \textsc{Phantom}.}. In particular, we use the variable $s = \sqrt{\epsilon \rho}$ described in Appendix~B of \citet{pricelaibe15} to avoid problems with negative dust fractions.  The evolution equation (\ref{eq:depsdt}) expressed in terms of $s$ is given by
\begin{equation}
\frac{{\rm d}s}{{\rm d}t} = -\frac{1}{2s} \nabla\cdot \left( \frac{\rho_{\rm g} \rho_{\rm d}}{\rho} \Delta {\bm v} \right) - \frac{s}{2} (\nabla\cdot{\bm v}),
\end{equation}
which for the case of hydrodynamics becomes
\begin{align}
\frac{{\rm d}s}{{\rm d}t} & = -\frac{1}{2s} \nabla\cdot \left( \frac{s^{2}}{\rho} t_{\rm s} \nabla P \right) - \frac{s}{2} (\nabla\cdot{\bm v}), \nonumber \\
& = -\frac12 \nabla\cdot \left( \frac{s}{\rho} t_{\rm s} \nabla P \right) - \frac{t_{\rm s}}{2\rho} \nabla P \cdot \nabla s - \frac{s}{2} (\nabla\cdot{\bm v}).
\end{align}
 The SPH discretisation of this equation is implemented in the form
\begin{align}
\frac{{\rm d}s_{a}}{{\rm d}t} = & - \frac{1}{2} \sum_{b} \frac{m_{b} s_{b}}{\rho_{b}} \left (\frac{t_{{\rm s}, a}}{\rho_{a}} + \frac{t_{{\rm s}, b}}{\rho_b} \right) (P_{a} - P_{b}) \frac{\overline{F}_{ab}}{\vert r_{ab} \vert}  \nonumber \\
& + \frac{s_{a}}{2\rho_{a} \Omega_{a}} \sum_{b} m_{b} {\bm v}_{ab} \cdot \nabla_{a} W_{ab} (h_{a}), \label{eq:dsdtsph}
\end{align}
where $\overline{F}_{ab} \equiv \frac12 [ F_{ab} (h_{a}) + F_{ab}(h_{b}) ]$. The thermal energy equation, (\ref{eq:dudteps}), takes the form
\begin{equation}
\frac{{\rm d}u}{{\rm d}t} = - \frac{P}{\rho} (\nabla\cdot {\bm v}) + \frac{s^{2} t_{\rm s}}{\rho \rho_{\rm g}}\nabla P \cdot \nabla u,
\end{equation}
the first term of which is identical to (\ref{eq:dudtsph}) and the second term of which is discretised in \textsc{Phantom} according to
\begin{equation}
-\frac{\rho_{a}}{2\rho^{\rm g}_{a}}  \sum_{b} m_{b}\frac{s_{a} s_{b}}{\rho_{a}\rho_{b}} \left (\frac{t_{{\rm s}, a}}{\rho_{a}} + \frac{t_{{\rm s}, b}}{\rho_b} \right) (P_{a} - P_{b}) (u_{a} - u_{b})\frac{\overline{F}_{ab}}{\vert r_{ab} \vert} .
\end{equation}

\subsubsection{Conservation of dust mass}
Conservation of dust mass with the one-fluid scheme is in principle exact because \citep{pricelaibe15}
\begin{equation}
\frac{{\rm d}}{{\rm d}t} \left( \sum_{a} \frac{m_{a}}{\rho_{a}} s_{a}^{2}\right) = \sum_{a} m_{a} \left( \frac{2s_{a}}{\rho_{a}} \frac{{\rm d}s_{a}}{{\rm d}t} - \frac{s_{a}^{2}}{\rho_{a}^{2}} \frac{{\rm d}\rho_{a}}{{\rm d}t}\right) = 0.
\end{equation}
In practice, some violation of this can occur because although the above algorithm guarantees positivity of the dust fraction, it does not guarantee that $\epsilon$ remains less than unity. Under this circumstance, which occurs only rarely, we set $\epsilon = \max(\epsilon,1)$ in the code. However, this violates the conservation of dust mass. This specific issue has been recently addressed in detail in the study by \citet{ballabioetal18}. Therefore, in the latest code there are two main changes:
\begin{itemize}
\item Rather than evolve $s = \sqrt{\epsilon \rho}$, in the most recent code we instead evolve a new variable $s' = \sqrt{\rho_{\rm d}/\rho_{\rm g}}$. This prevents the possibility of $\epsilon > 1$.
\item We limit the stopping time such that the timestep from the one fluid algorithm does not severely restrict the computational performance
\end{itemize}
For details of these changes we refer the reader to \citet{ballabioetal18}.

\subsubsection{Conservation of energy and momentum}
 Total energy with the one-fluid scheme can be expressed via
\begin{equation}
E =\sum_{a} m_{a}  \left[ \frac12 v^{2}_{a} + (1 - \epsilon_{a}) u_{a} \right],
\end{equation}
which is conserved exactly by the discretisation since
\begin{align}
\sum_{a} m_{a} & \left[ {\bm v}_{a}\cdot \frac{{\rm d} {\bm v}_{a}}{{\rm d}t} + (1 - \epsilon_{a}) \frac{{\rm d}u_{a}}{{\rm d}t} \right. \nonumber \\
&\left. - u_{a} \left(\frac{2s_{a}}{\rho_{a}}\frac{{\rm d}s_{a}}{{\rm d}t} - \frac{s^{2}_{a}}{\rho_{a}^{2}} \frac{{\rm d}\rho_{a}}{{\rm d}t}\right)\right] = 0.
\end{align}
Conservation of linear and angular momentum also hold since the discretisation of the momentum equation is identical to the hydrodynamics algorithm.

\subsubsection{Timestep constraint}
\label{sec:dtdragonef}
For the one-fluid method, the timestep is constrained by the \emph{inverse} of the stopping time according to
\begin{equation}
\Delta t^{a}_{\rm drag} = C_{\rm force}  \frac{h^{2}}{ \epsilon t_{\rm s} c_{\rm s}^{2}}.
\end{equation}
This becomes prohibitive precisely in the regime where the one-fluid method is no longer applicable ($t_{\rm s} > t_{\rm Cour}$; see \citealt{laibeprice14}), in which case one should switch to the two-fluid method instead. There is currently no mechanism to automatically switch regimes, though this is possible in principle and may be implemented in a future code version.
 


\begin{table*}
\begin{center}
\begin{tabular}{lcc }
\hline
\hline
Process & Description & Reference\\
\hline
\mion{H}{I} (atomic) cooling  & Electron collisional excitation/ & \citet{1993ApJS...88..253S} \\ 
   & resonance line emission &  \\ 
H$_2$ (molecular) cooling  & Vibrational/rotational excitation cooling& \citet{1999MNRAS.305..802L} \\ 
  & by collisions with H, He and H$_2$ & \\ 
Fine structure cooling  & \mion{C}{II}, \mion{Si}{II} and \mion{O}{I} collisions with  & \citet{2007ApJ...666....1G} \\ 
 & H, H$_2$, free e$^{-}$ and H$^{+}$ & \\
CO rotational cooling & CO collisions with H$_{2}$, H and free e$^{-}$ & \citet{1993ApJ...418..263N}, \\
& & \citet{1995ApJS..100..132N} \\
Recombination cooling & Free e$^{-}$ recombining with ionised gas & \citet{2003ApJ...587..278W} \\ 
 & on PAH and dust grain surfaces & \\
Gas-grain cooling & Dust-gas collisional heat transfer & \citet{1989ApJ...342..306H} \\ 
\hline
Cosmic-ray heating  & Energy deposition associated with  & \citet{1978ApJ...222..881G} \\ 
 & cosmic ray ionization & \\
Photo-electric heating  & UV e$^-$ excitation from dust and PAH & \citet{2003ApJ...587..278W} \\ 
\hline
\hline
\end{tabular}
\caption{Heating and cooling processes in the \textsc{Phantom} cooling module.}
\label{CoolTable}
\end{center}
\end{table*}

\begin{figure}
\begin{center}
   \includegraphics[width=\columnwidth]{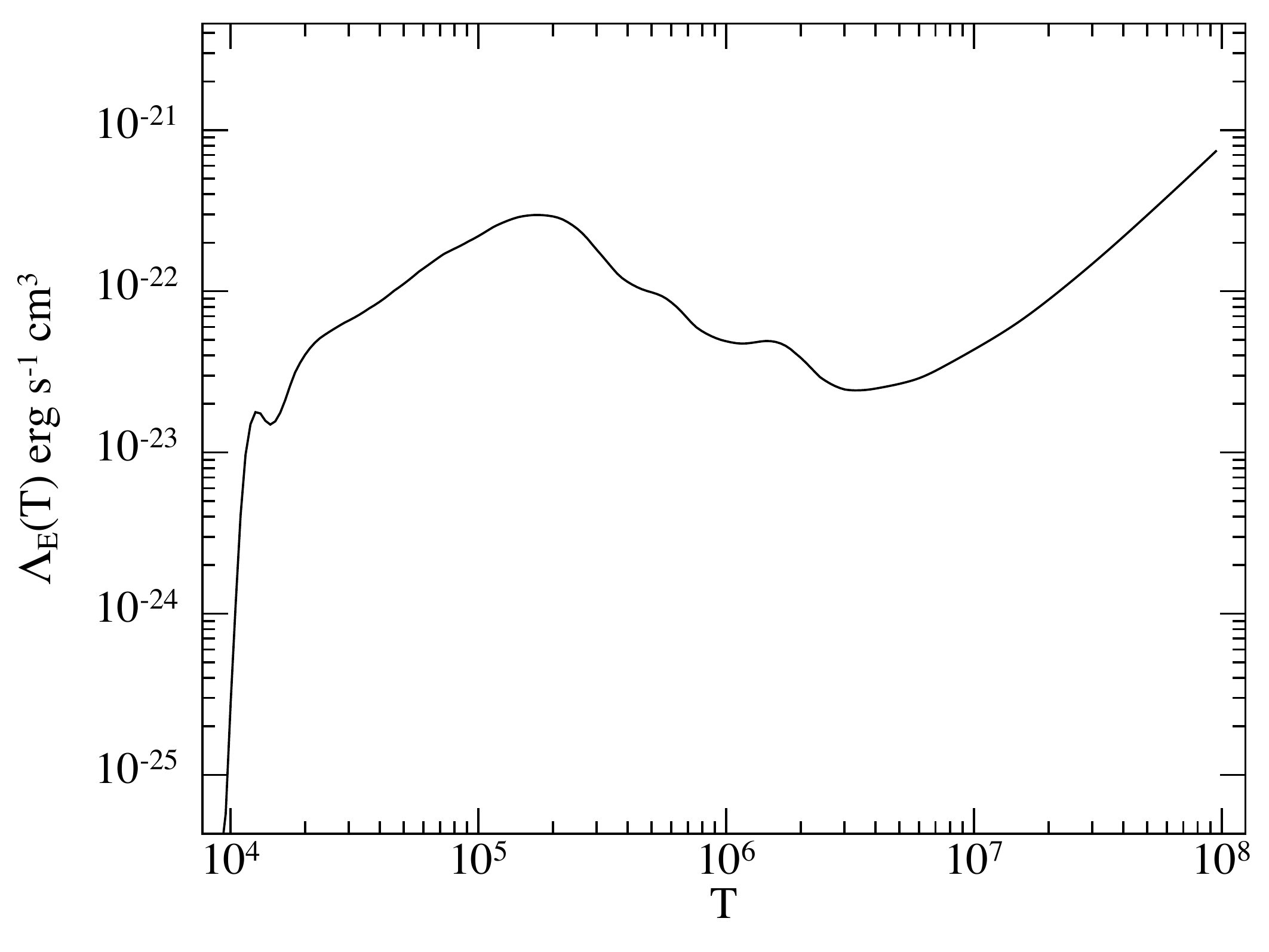} 
   \caption{Emissivity, $\Lambda_{\rm E}$ [erg s$^{-1}$ cm$^3$] as a function of temperature for the ISM cooling assuming default abundances appropriate for the Warm Neutral Medium (WNM). Note that as we treat cooling from atomic hydrogen using a full non-equilibrium treatment, the behaviour of $\Lambda_{\rm E}$ close to $10^{4}$~K is highly sensitive to the electron fraction, which in the case shown here is much smaller than it would be in collisional ionization equilibrium. Values of $\Lambda_{\rm E}$ below $10^{4}$~K depend strongly on the current chemical state of the gas and are not shown in this plot.}
   \label{fig:emissivity}
\end{center}
\end{figure}

\subsection{Interstellar medium (ISM) physics}
\label{sec:ism}

\subsubsection{Cooling function}
\label{sec:ismcooling}
The cooling function in \textsc{Phantom} is based on a set of {\sc Fortran} modules written by \citet{2007ApJS..169..239G}, updated further by \citet{2010MNRAS.404....2G}. It includes cooling from atomic lines (\mion{H}{I}), molecular excitation (H$_2$), fine structure metal lines (\mion{Si}{I}, \mion{Si}{II}, \mion{O}{I}, \mion{C}{I}, \mion{C}{II}),  gas-dust collisions and polycyclic aromatic hydrocarbon (PAH) recombination (see \citealt{2007ApJS..169..239G} for references on each process). Heating is provided by cosmic rays and the photoelectric effect. The strength of the cooling is a function of the temperature, density, and various chemical abundances. Table \ref{CoolTable} summarises these various heating and cooling processes. Figure~\ref{fig:emissivity} shows the resulting emissivity $\Lambda_{\rm E}(T)$ for temperatures between $10^4$ and $10^8$K. The cooling rate per unit volume is related to the emissivity according to
\begin{equation}
\Lambda_{\rm cool} \equiv n^2 \Lambda_{\rm E} (T) \,  {\rm erg} \phantom{s}{\rm s}^{-1} {\rm cm}^{-3},
\end{equation}
where $n$ is the number density. The cooling in the energy equation corresponds to
\begin{equation}
\left(\frac{{\rm d}u}{{\rm d}t}\right)_{\rm cool} = -\frac{\Lambda_{\rm cool}}{\rho}.
\label{eq:dudtcool}
\end{equation}
These routines were originally adapted for use in \textsc{sphng} \citep{2008MNRAS.389.1097D} and result in an approximate two-phase ISM with temperatures of 100~K and 10~000~K. Note that the cooling depends on a range of parameters (dust-to-gas ratio, cosmic ray ionisation rate, etc.), many of which can be specified at runtime. Table~\ref{tab:abund} lists the default abundances, which are set to values appropriate for the Warm Neutral Medium taken from \citet{sembachetal00}. The abundances listed in Table~\ref{tab:abund}, along with the dust-to-gas ratio, are the only metallicity-dependent parameters that can be changed at runtime. An option for cooling appropriate to the zero-metallicity early universe is also available.

\begin{table}
\begin{center}
\begin{tabular}{cc}
\hline
\hline
Element & Abundance \\
\hline
C & $1.4\times 10^{-4}$ \\
O & $3.2\times 10^{-4}$ \\
Si & $1.5\times 10^{-5}$ \\
e$^{-}$ & $2 \times 10^{-4}$ \\
\hline
\hline
\end{tabular}
\end{center}
\caption[ISM chemistry]{Default fractional abundances for C, O, Si and $e^{-}$ in the ISM cooling and chemistry modules. Abundances are taken from \citet{sembachetal00} appropriate for the Warm Neutral Medium (WNM). These are lower than solar because it is assumed some fraction of the metals are locked up in dust rather than being available in the gas phase.}
\label{tab:abund}
\end{table}


\subsubsection{Timestep constraint from cooling}
\label{sec:dtcool}
When cooling is used, we apply an additional timestep constraint in the form
\begin{equation}
\Delta t _{\rm cool}^a = C_{\rm cool} \left| \frac{u}{({\rm d}u/{\rm d}t)_{\rm cool}} \right|,
\label{eq:dtcool}
\end{equation}
where $C_{\rm cool} = 0.3$ following \citet{2007ApJS..169..239G}. The motivation for this additional constraint is to not allow the cooling to completely decouple from the other equations in the code \citep{suttneretal97}, and to avoid cooling instabilities that may be generated by numerical errors \citep{zyk96}.

Cooling is currently implemented only with explicit timestepping of (\ref{eq:dudtcool}), where $u$ is evolved alongside velocity in our leapfrog timestepping scheme. However, the substepping scheme described below for the chemical network (Section~\ref{sec:h2chemdt}) is also used to update the cooling on the chemical timestep, meaning that the cooling can evolve on a much shorter timestep than the hydrodynamics when it is used in combination with chemistry, which it is by default. Implementation of an implicit cooling method, namely the one proposed by \citet{townsend09}, is under development.

\subsubsection{Chemical network}
A basic chemical network is included for ISM gas that evolves the abundances of H, H$^{+}$, e$^-$, and the molecules H$_2$ and CO.
The number density of each species, $n_X$, is evolved using simple rate equations of the form
\begin{equation}
\frac{dn_X}{dt} = C_X - D_X n_X,
\end{equation}
where $C_X$ and $D_X$ are creation and destruction coefficients for each species. In general, $C_X$ and $D_X$ are functions of density, temperature, and abundances of other species. The number density of each species, $X$, is time integrated according to
\begin{equation}
n_X(t+\Delta t) = n_X(t) + \frac{dn_X}{dt} \Delta t. \label{eq:h2chem}
\end{equation}
There are in effect only three species to evolve (H, H$_2$ and CO), as the H$^{+}$ and e$^-$ abundances are directly related to the H abundance. 

The chemistry of atomic hydrogen is effectively the same as in \citet{2007ApJS..169..239G}. H is created by recombination in the gas phase and on grain surfaces, and destroyed by cosmic ray ionisation and free electron collisional ionisation. H$_2$ chemistry is based on the model of \citet{2004ApJ...612..921B}, with H$_2$ created on grain surfaces and destroyed by photo-dissociation and cosmic-rays (see \citealt{2008MNRAS.389.1097D} for computational details). 

The underlying processes behind CO chemistry are more complicated, and involve many intermediate species in creating CO from C and O by interactions with H species. Instead of following every intermediate step we use the model of \citet{1997ApJ...482..796N} (see \citealt{pettittetal14} for computational details). CO is formed by a gas phase reaction from an intermediate CH$_Z$ step after an initial reaction of C$^{+}$ and H$_2$ (where Z encompasses many similar type species). CO and CH$_Z$ are subject to independent photodestruction, which far outweighs the destruction by cosmic-rays. Abundances of C$^{+}$ and O are used in the CO chemistry, and their abundance is simply the initial value excluding what has been used up in CO formation. \citet{2012MNRAS.421..116G} test this and a range of simpler and more complicated models, and show that the model adopted here is sufficient for large-scale simulations of CO formation, although it tends to over-produce CO compared to more sophisticated models.

The details for each reaction in the H, H$_2$ and CO chemistry are given in Table \ref{ReactionTable}, with relevant references for each process.

\begin{table*}
\begin{center}
\begin{tabular}{lcc }
\hline
\hline
Reaction & Description & Reference\\
\hline
${\rm H^{+} +e^- +grain\rightarrow H +grain }$ & Grain surface recombination & \citet{2001ApJ...563..842W}\\
$\rm H^{+} +e^- \rightarrow H +\gamma$ &  Gas-phase recombination & \citet{1992ApJ...387...95F}\\
${\rm H+e^- \rightarrow H^{+} + 2e^-}$ & $\rm e^-$ collisional ionisation &\citet{1997NewA....2..181A} \\
${\rm H+c.r. \rightarrow H^{+} + e^-}$ & Cosmic ray ionisation & \citet{2007ApJS..169..239G} \\
\hline
${\rm H + H + grain \rightarrow H_2 + grain}$ & Grain surface formation & \citet{2004ApJ...612..921B}\\
$\rm H_2 +\gamma \rightarrow 2H$ & UV photodissociation &  \citet{1996ApJ...468..269D}\\
$\rm H_2 +c.r. \rightarrow H_2^+ + e^-$ & Cosmic ray ionisation$^\ast$ &  \citet{2004ApJ...612..921B}\\
\hline
$\rm C^{+} + H_2 \rightarrow CH_2^+ +\gamma$ & Radiative association & \citet{1997ApJ...482..796N}\\ 
$\rm CH_2^+ + various \rightarrow CH_X + various$ & Rapid neutralisation$^\dag$& - \\ 
$\rm CH_Z + O \rightarrow CO + H $ & Gas phase formation & - \\ 
$\rm CH_Z + \gamma \rightarrow C + H $ & UV photodissociation & - \\ 
$\rm CO + \gamma \rightarrow C^+ + O + e^-$ & UV photodissociation$^{\dag\dag}$ & - \\ 
\hline
\hline
\end{tabular}
\caption[{\sc Phantom} ISM chemistry]{Processes and references for the \text{Phantom} ISM chemistry module tracing the evolution of H, $\rm H_2$ and CO.
{\footnotesize
\newline $^\ast$ H$_2^{+}$ ions produced by cosmic ray ionisation of H$_2$ are assumed to dissociatively recombine to $\rm H + H$, so that the effective reaction in the code is actually $\rm H_2 + c.r. \rightarrow H + H$.}
\newline $^\dag$ Process is intermediate and is assumed rather than fully represented.
\newline $^{\dag\dag}$ C is not present in our chemistry, but is assumed to rapidly photoionise to C$^{+}$.}
\label{ReactionTable}
\end{center}
\end{table*}

\subsubsection{Timestep constraint from chemistry}
\label{sec:h2chemdt}
H chemistry is evolved on the cooling timestep, since the timescale on which the H$^+$ abundance changes significantly is generally comparable to or longer than the cooling time. This is not true for H$_2$ and CO. Instead, these species are evolved using a chemical time-stepping criterion, where (\ref{eq:h2chem}) is subcycled during the main time step at the interval $\Delta t_{\rm chem}$. If the abundance is decreasing then the chemical timestep is 
\begin{equation}
\Delta t_{\rm chem} = -\frac{1}{10} \frac{n_X}{(C_X - D_X n_X)},
\end{equation}
i.e., 10 per cent of the time needed to completely deplete the species. If the abundance is increasing,
\begin{equation}
\Delta t_{\rm chem} = \frac{\Delta t_{\rm hydro}}{200} ,
\end{equation}
where $\Delta t_{\rm hydro}$ is the timestep size for the hydrodynamics, and was found to be an appropriate value by test simulations. These updated abundances feed directly into the relevant cooling functions. Although the cooling function includes \mion{Si}{I} and \mion{C}{I}, the abundances of these elements are set to zero in the current chemical model.

\subsection{Particle injection}
\label{sec:inject}
 We implement several algorithms for modelling inflow boundary conditions (see \citealt{toupinetal15,toupinetal15a} for recent applications). This includes injecting SPH particles in spherical winds from sink particles (both steady and time dependent), in a steady Cartesian flow and for injection at the $L_{1}$ point between a pair of sink particles to simulate the formation of accretion discs by Roche Lobe overflow in binary systems. 
 
%
%

\section{Initial conditions} 
\label{sec:initial}

\subsection{Uniform distributions}
\label{sec:unifdis}
 The simplest method for initialising the particles is to set them on a uniform Cartesian distribution. The lattice arrangement can be cubic (equal particle spacing in each direction, $\Delta x = \Delta y = \Delta z$), close-packed ($\Delta y = \sqrt{3/4}\Delta x$, $\Delta z = \sqrt{6}/3 \Delta x$, repeated every 3 layers in $z$), hexagonal close-packed (as for close-packed but repeated every two layers in $z$), or uniform random. The close-packed arrangements are the closest to a `relaxed' particle distribution, but require care with periodic boundary conditions due to the aspect ratio of the lattice. The cubic lattice is not a relaxed arrangement for the particles, but is convenient and sufficient for problems where the initial conditions are quickly erased (e.g. driven turbulence). For problems where initial conditions matter, it is usually best to relax the particle distribution by evolving the simulation for a period of time with a damping term (Section~\ref{sec:damping}). This is the approach used, for example, in setting up stars in hydrostatic equilibrium (Section~\ref{sec:starsetup}).
 
 \subsection{Stretch mapping}
\label{sec:stretchmap}
General non-uniform density profiles may be set up using `stretch mapping' \citep{herant94}. The procedure for spherical distributions is the most commonly used \citep[e.g.][]{fhr07,rosswogprice07,rrh09}, but we have generalised the method for any density profile that is along one coordinate direction \citep[e.g.][]{pricemonaghan04a}. Starting with particles placed in a uniform distribution, the key is that a particle should keep the same relative position in the mass distribution. For each particle with initial coordinate $x_{0}$ in the relevant coordinate system, we solve the equation
\begin{equation}
f(x) = \frac{M(x)}{M(x_{\rm max})} - \frac{x_{0} - x_{\rm min}}{(x_{\rm max} - x_{\rm min})} = 0, \label{eq:stretchmap}
\end{equation}
where $M(x)$ is the desired density profile integrated along the relevant coordinate direction, i.e.
\begin{equation}
M(x) \equiv \int_{x_{\rm min}}^{x} \rho(x') dS(x') dx', \label{eq:mx}
\end{equation}
where the area element ${\rm d}S(x')$ depends on the geometry and the chosen direction, given by
\begin{equation}
dS(x) = \left\{ \begin{array}{ll}
1 & \textrm{Cartesian or cyl./sph. along $\phi$, $\theta$ or $z$}, \\
2\pi x & \textrm{cylindrical along r}, \\
4\pi x^{2} & \textrm{spherical along r}.
\end{array}\right.
\end{equation}
We solve (\ref{eq:stretchmap}) for each particle using Newton-Raphson iterations
\begin{equation}
x = x - \frac{f(x)}{f'(x)},
\end{equation}
where
\begin{equation}
f'(x) = \frac{\rho(x)dS(x)}{M(x_{\rm max})},
\end{equation}
iterating until $\vert f(x) \vert < 10^{-9}$. The Newton-Raphson iterations have second order convergence, but may fail in extreme cases. Therefore, if the root-finding has failed to converge after 30 iterations, we revert to a bisection method, which is only first order but guaranteed to converge.

Stretch mapping is implemented in such a way that only the desired density function need be specified, either as through an analytic expression (implemented as a function call) or as a tabulated dataset. Since the mass integral in (\ref{eq:mx}) may not be known analytically, we compute this numerically by integrating the density function using the trapezoidal rule.

The disadvantage of stretch mapping is that in spherical or cylindrical geometry it produces defects in the particle distribution arising from the initial Cartesian distribution of the particles. In this case, the particle distribution should be relaxed into a more isotropic equilibrium state before starting the simulation. For stars, for example, this may be done by simulating the star in isolation with artificial damping added (Section~\ref{sec:damping}). Alternative approaches are to relax the simulation using an external potential chosen to produce the desired density profile in equilibrium (e.g. \citealt{zurekbenz86,nnm88}) or to iteratively `cool' the particle distribution to generate `optimal' initial conditions \citep{diehletal15}.

\subsection{Accretion discs}

\subsubsection{Density field}
\label{sec:icdisc}
The accretion disc setup module uses a Monte-Carlo particle placement (details in Section~\ref{sec:montecarlo}) in cylindrical geometry to construct density profiles of the form
\begin{equation}
\rho(x,y,z) = \Sigma_{0} f_{\rm s} \left(\frac{R}{R_{\rm in}}\right)^{-p} \exp{\left(\frac{-z^{2}}{2H^{2}}\right)} ,
\end{equation}
where $\Sigma_{0}$ is the surface density at $R=R_{\rm in}$ (if $f_{\rm s} = 1$), $H \equiv c_{\rm s}/\Omega$ is the scale height (with $\Omega \equiv \sqrt{GM/R^{3}}$), $p$ is the power-law index (where $p=3/2$ by default following \citealt{lodatopringle07}) and $f_{\rm s} \equiv ( 1 - \sqrt{R_{\rm in}/R} )$ is an optional factor to smooth the surface density near the inner disc edge.

 Several authors have argued that a more uniform particle placement is preferable for setting up discs in SPH \citep{csw09,vanaverbekeetal09}. This may be important if one is interested in transient phenomena at the start of a simulation, but otherwise the particle distribution settles to a smooth density distribution within a few orbits (c.f. \citealt{lodatoprice10}).
 
\subsubsection{Velocity field}
 The orbital velocities of particles in the disc are set by solving the equation for centrifugal equilibrium, i.e.
\begin{equation}
v_{\phi}^{2} = \frac{GM}{R} - f_{p} - 2v_{\phi} f_{\rm BH}, \label{eq:vphi}
\end{equation}
where the correction from radial pressure gradients is given by
\begin{equation}
f_{p} = -c_{\rm s}^{2}(R) \left(\frac32 + p + q + \frac{1}{2f_{\rm s}}\right),
\end{equation}
where $q$ is the index of the sound speed profile such that $c_{\rm s}(R) = c_{{\rm s}, in} (R/R_{\rm in})^{-q}$ and $f_{\rm BH}$ is a correction used for discs around a spinning black hole \citep{npn15}
\begin{equation}
f_{\rm BH} = - \frac{2a}{c^{3}} \left(\frac{GM}{R}\right)^{2},
\end{equation}
where $a$ is the black hole spin parameter. The latter assumes Lense-Thirring precession is implemented as in Section~\ref{sec:lt}. Where self-gravity is used, $M$ is the enclosed mass at a given radius $M(< R)$, otherwise it is simply the mass of the central object. Using the enclosed mass for the self-gravitating case is an approximation since the disc is not spherically symmetric, but the difference is small and the disc relaxes quickly into the true equilibrium. Equation~(\ref{eq:vphi}) is a quadratic for $v_{\phi}$ which we solve analytically.
 
\subsubsection{Warps}
Warps are applied to the disc \citep[e.g.][]{lodatoprice10,npn15} by rotating the particles about the $y$-axis by the inclination angle $i$ [in general a function of radius $i\equiv i(R)$], according to
\begin{align}
x' & =  x\cos(i) + z\sin(i), \\
y' & = y, \\
z' & = -x\sin(i) + z\cos(i),
\end{align}
with the velocities similarly adjusted using
\begin{align}
v'_{x} & = v_{x}\cos(i) + v_{z}\sin(i), \\
v'_{y} & = v_{y}, \\
v'_{z} & = -v_{x}\sin(i) + v_{z} \cos(i).
\end{align}

\subsubsection{Setting an $\alpha$-disc viscosity}
The simplest approach to mimicking an $\alpha$-disc viscosity in SPH is to employ a modified shock viscosity term, setting the desired $\alpha_{\rm SS}$ according to (\ref{eq:alphascale}) as described in more detail in Section~\ref{sec:discav}. Since the factor $\langle h \rangle / H$ is dependent both on resolution and temperature profile (i.e. the $q$-index), it is computed in the initial setup by taking the desired $\alpha_{\rm SS}$ as input in order to give the required $\alpha_{\rm AV}$. Although this does not guarantee that $\alpha_{\rm SS}$ is constant with radius and time (this is only true with specific choices of $p$ and $q$ and if the disc is approximately steady), it provides a simple way to prescribe the disc viscosity.

\subsection{Stars and binary stars}
\label{sec:starsetup}
We implement a general method for setting up `realistic' stellar density profiles, based on either analytic functions (e.g.\ polytropes) or tabulated data files output from stellar evolution codes (see \citealt{iaconietal17} for a recent application of this to common envelope evolution). 

The basic idea is to set up a uniform density sphere of particles and set the density profile by stretch mapping (see below). The thermal energy of the particles is set so that the pressure gradient is in hydrostatic equilibrium with the self-gravity of the star for the chosen equation of state. We then relax the star into equilibrium for several dynamical times using a damping parameter (Section~\ref{sec:damping}), before re-launching the simulation with the star on an appropriate orbit.
 
  For simulating red giants it is preferable to replace the core of the star by a sink particle \citep[see][]{passyetal12,iaconietal17}. When doing so one should set the accretion radius of the sink to zero and set a softening length for the sink particle consistent with the original core radius (see Section~\ref{sec:sinkaccel}).

\subsection{Galactic initial conditions }

 In addition to simulating ISM gas in galactic discs with analytic stellar potentials, one may represent bulge-halo-disc components by collisionless $N$-body particles (see Section \ref{sec:softeningdm}). To replace a potential with a resolved system requires care with the initial conditions (i.e. position, velocity, mass). If setup incorrectly the system will experience radial oscillations and undulations in the rotation curve, which will have strong adverse effects on the gas embedded within. We include algorithms for initialising the static-halo models of \citet{2015MNRAS.449.3911P} (which used the \textsc{sphng} SPH code). These initial conditions require the NFW profile to be active and care must be taken to ensure the mass and scale lengths correspond to the rotation curve used to generate the initial conditions. Other codes may alternatively be used to seed multi-component $N$-body disc galaxies (e.g. \citealt{1995MNRAS.277.1341K,2001NewA....6...27B,2007MNRAS.378..541M,2014MNRAS.444...62Y}), including \textsc{magalie} \citep{2001NewA....6...27B} and \textsc{galic} \citep{2014MNRAS.444...62Y} for which we have implemented format readers.

The gas in galactic scale simulations can be setup either in a uniform surface density disc, or according to the Milky Way's specific surface density. The latter is based on the radial functions given in \citet{1995ApJ...443..152W}. As of yet we have not implemented a routine for enforcing hydrostatic equilibrium \citep{2005MNRAS.361..776S,2010MNRAS.407..705W}; this may be included in a future update.

\subsection{Damping}
\label{sec:damping}
 To relax a particle distribution into equilibrium, we adopt the standard approach \citep[e.g.][]{gingoldmonaghan77} of adding an external acceleration in the form  
\begin{equation}
{\bm a}_{\rm ext, damp}^{a} = - f_{\rm d} {\bm v},
\end{equation}
such that a percentage of the kinetic energy is removed each timestep. The damping parameter, $f_{\rm d}$, is specified by the user. A sensible value for $f_{\rm d}$ is of order a few percent (e.g.\ $f_{\rm d} = 0.03$) such that a small fraction of the kinetic energy is removed over a Courant timescale.  

\section{Software Engineering}
\label{sec:software}
No code is completely bug-free (experience is the name everyone gives to their mistakes; \citealt{wilde1892}). However, we have endeavoured to apply the principles of good software engineering to \textsc{Phantom}. These include:
\begin{enumerate}
\item a modular structure,
\item unit tests of important modules,
\item nightly builds,
\item automated nightly tests,
\item automated code maintenance scripts,
\item version control with {\sc git},
\item wiki documentation, and
\item a bug database and issue tracker.
\end{enumerate}

Together these simplify the maintenance, stability and usability of the code, meaning that \textsc{Phantom} can be used direct from the development repository without fear of regression, build failures or major bugs.

Specific details of how the algorithms described in Section~\ref{sec:methods} are implemented are given in the Appendix. Details of the test suite are given in Appendix~\ref{sec:testsuiteapp}.


\section{Numerical tests}
\label{sec:tests}
 Unless otherwise stated, we use the M$_{6}$ quintic spline kernel with $h_{\rm fac}=1.0$ by default, giving a mean neighbour number of 113 in 3D. Almost all of the test results are similar when adopting the cubic spline kernel with $h_{\rm fac} = 1.2$ (requiring $\approx 58$ neighbours), apart from the tests with the one-fluid dust method where the quintic is required. Since most of the algorithms used in {\sc Phantom} have been extensively tested elsewhere, our aim is merely to demonstrate that the implementation in the code is correct, and to illustrate the typical results that should be achieved on these tests when run by the user. The input files used to run the entire test suite shown in the paper are available on the website, so it should be straightforward for even a novice user to reproduce our results.
  
Unless otherwise indicated, we refer to dimensionless ${\cal L}_{1}$ and ${\cal L}_{2}$ norms when referencing errors, computed according to
\begin{align}
{\cal L}_{1} & \equiv \frac{1}{N C_0}\sum_{i=1}^{N} \vert y_{i} - y_{\rm exact} \vert, \\
{\cal L}_{2} & \equiv \sqrt{\frac{1}{N C_0}\sum_{i=1}^{N} \vert y_{i} - y_{\rm exact} \vert^{2}},
\end{align}
where $y_{\rm exact}$ is the exact or comparison solution interpolated or computed at the location of each particle $i$ and N is the number of points. The norms are the standard error norms divided by a constant, which we set to the maximum value of the exact solution within the domain, $C_0 = \max{(y_{\rm exact})}$, in order to give a dimensionless quantity. Otherwise quoting the absolute value of ${\cal L}_{1}$ or ${\cal L}_{2}$ is meaningless. Dividing by a constant has no effect when considering convergence properties. These are the default error norms computed by {\sc splash} \citep{price07}.

 Achieving formal convergence in SPH is more complicated than in mesh-based codes where linear consistency is guaranteed \citep[see][]{price12}. The best that can be achieved with regular (positive, symmetric) kernels is second order accuracy away from shocks provided the particles remain well ordered \citep{monaghan92}. The degree to which this remains true depends on the smoothing kernel and the number of neighbours. Our tests demonstrate that formal second order convergence can be achieved with {\sc Phantom} on certain problems (e.g. Section~\ref{sec:alfven}). More typically one obtains something between first and second order convergence in smooth flow depending on the degree of spatial disordering of the particles. The other important difference compared to mesh-based codes is that there is no intrinsic numerical dissipation in SPH due to its Hamiltonian nature --- numerical dissipation terms must be explicitly added. We perform all tests with these terms included.
 
 We use timestep factors of $C_{\rm Cour} = 0.3$ and $C_{\rm force} = 0.25$ by default for all tests (Section~\ref{sec:timestep}). 

\subsection{Hydrodynamics}
\label{sec:hydrotest}

\begin{figure*}
   \centering
   \includegraphics[width=0.75\textwidth]{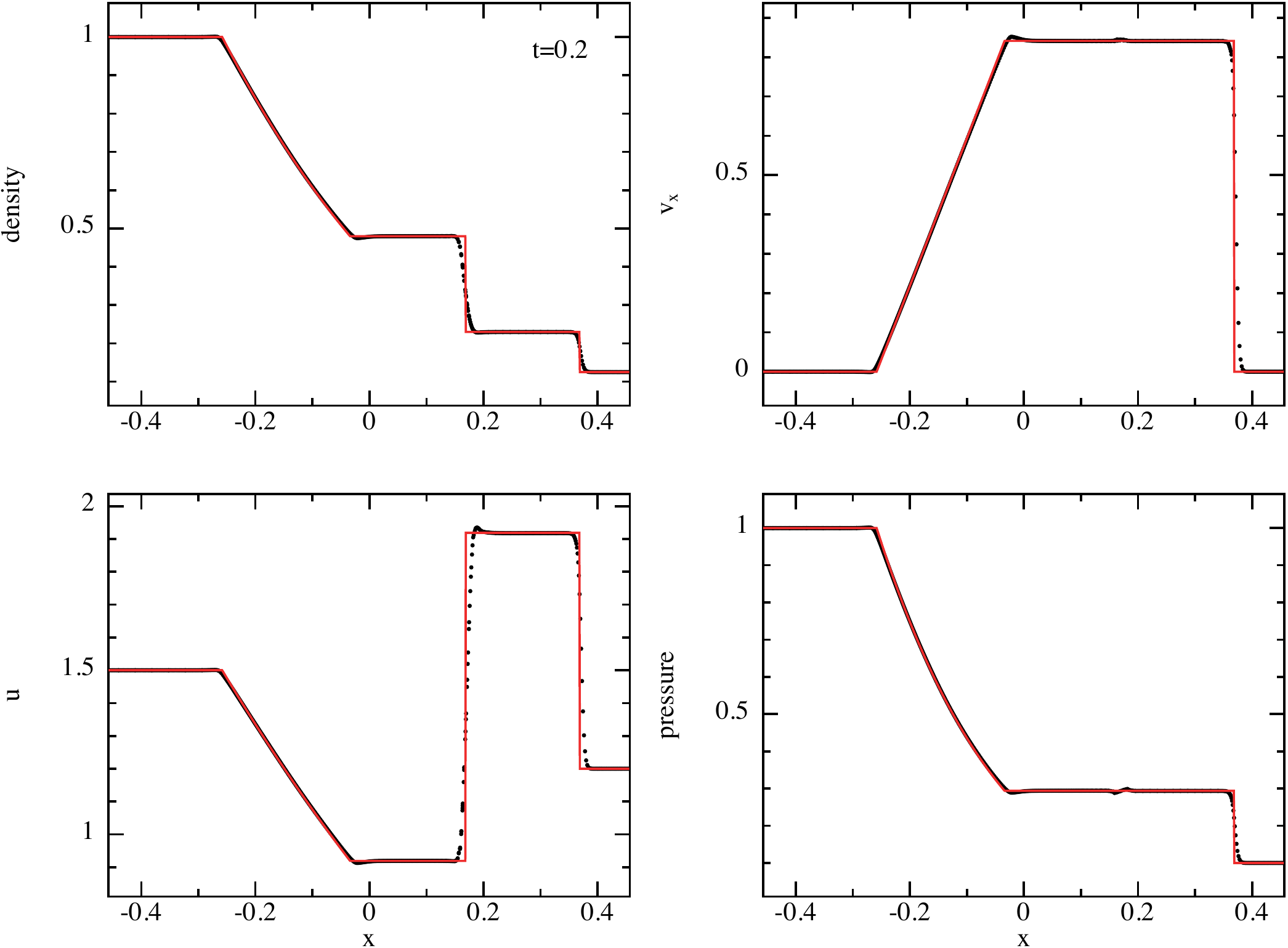} 
   \caption{Results of the Sod shock tube test in 3D, showing projection of all particles (black dots) compared to the analytic solution (red line). The problem is set up with $[\rho, P] = [1,1]$ for $x \leq 0$ and $[\rho, P] = [0.125,0.1]$ for $x > 0$ with $\gamma = 5/3$, with zero velocities and no magnetic field. The density contrast is initialised using equal mass particles placed on a close packed lattice with $256 \times 24 \times 24$ particles initially in $x \in [-0.5,0]$ and $128 \times 12 \times 12$ particles initially in $x \in [0,0.5]$. The results are shown with constant $\alpha^{\rm AV} = 1$.}
\label{fig:sod}
\end{figure*}

\begin{figure*}
   \centering
   \includegraphics[width=0.75\textwidth]{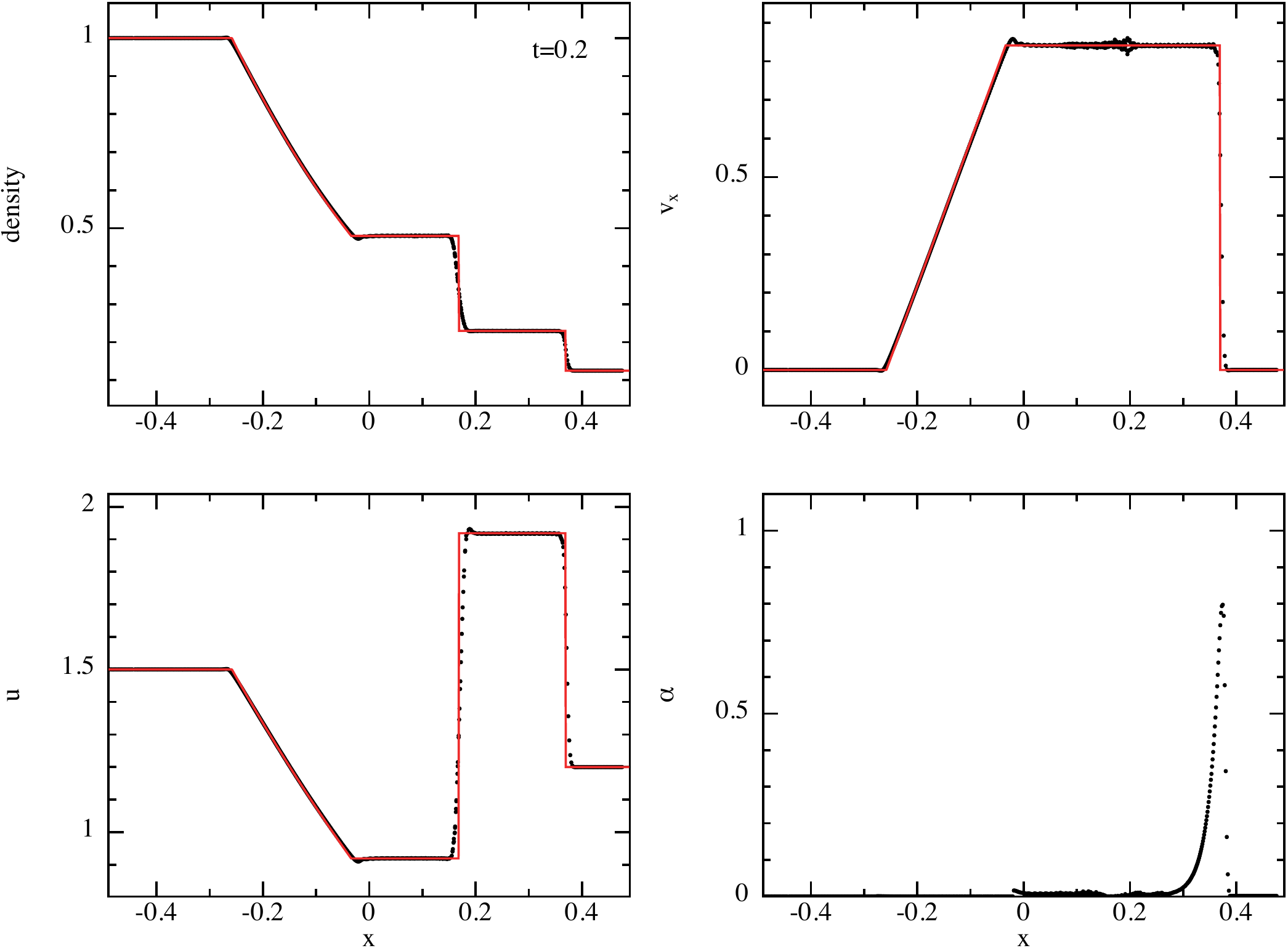} 
   \caption{As in Figure~\ref{fig:sod}, but with code defaults for all dissipation terms ($\alpha^{\rm AV} \in [0,1]$; $\alpha_u = 1$). These leave more noise in the velocity field behind the shock but provide second order convergence in smooth flow. The lower right panel in this case shows the resultant values for the viscosity parameter $\alpha^{\rm AV}$.}
\label{fig:sodcd}
\end{figure*}

\begin{figure*}
   \centering
   \includegraphics[width=0.75\textwidth]{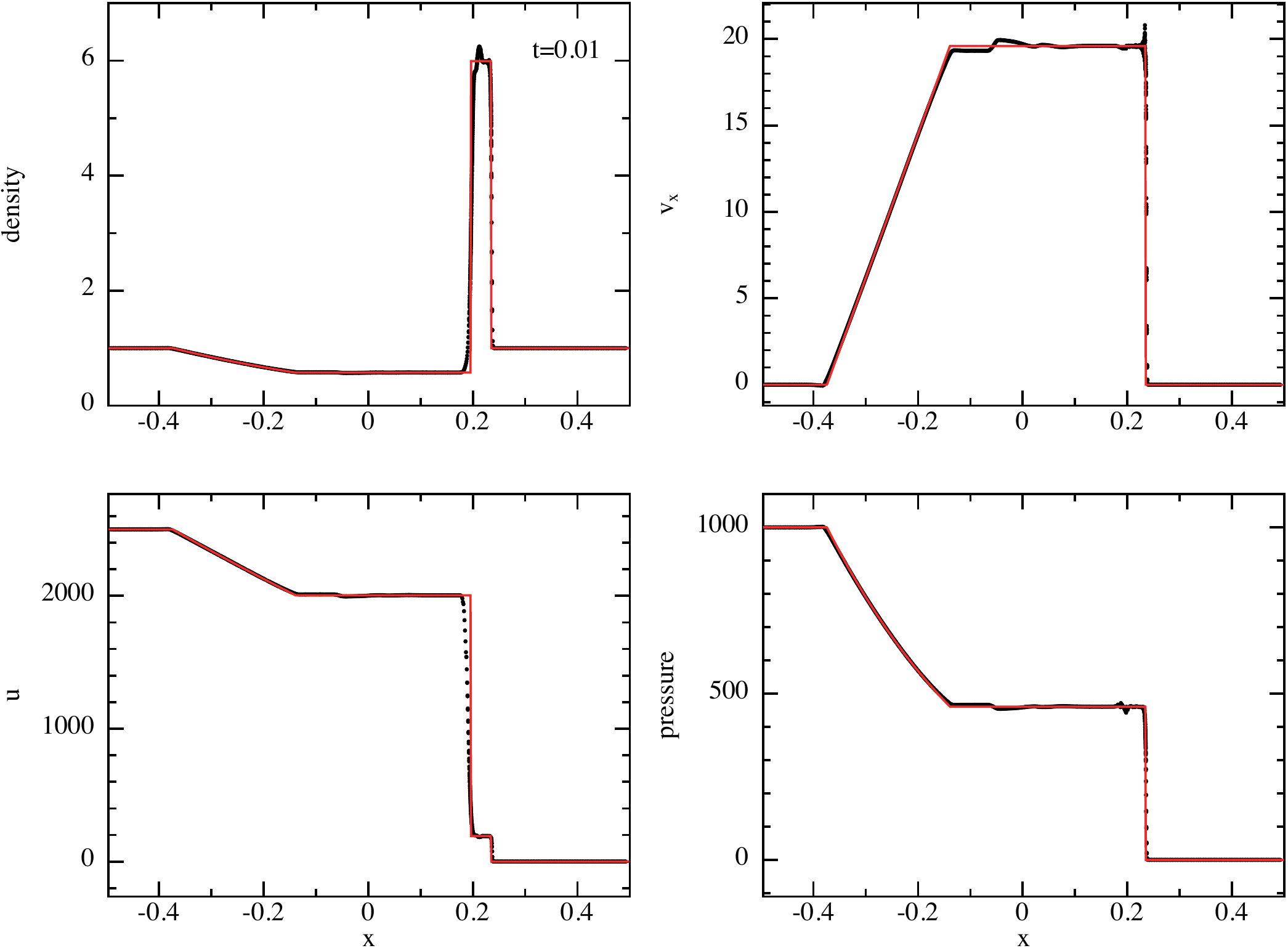} 
   \caption{Results of the 3D blast wave test, showing projection of all particles (black dots) compared to the analytic solution (red line). The problem is set up with $[\rho, P] = [1,1000]$ for $x \leq 0$ and $[\rho, P] = [1,0.1]$ for $x > 0$ with $\gamma = 7/5$, with zero velocities and no magnetic field. We use equal mass particles placed on a close packed lattice with $800 \times 12 \times 12$ particles initially in $x \in [-0.5,0.5]$. Results are shown with constant $\alpha^{\rm AV} = 1$}
\label{fig:blast}
\end{figure*}

\begin{figure*}
   \centering
   \includegraphics[width=0.75\textwidth]{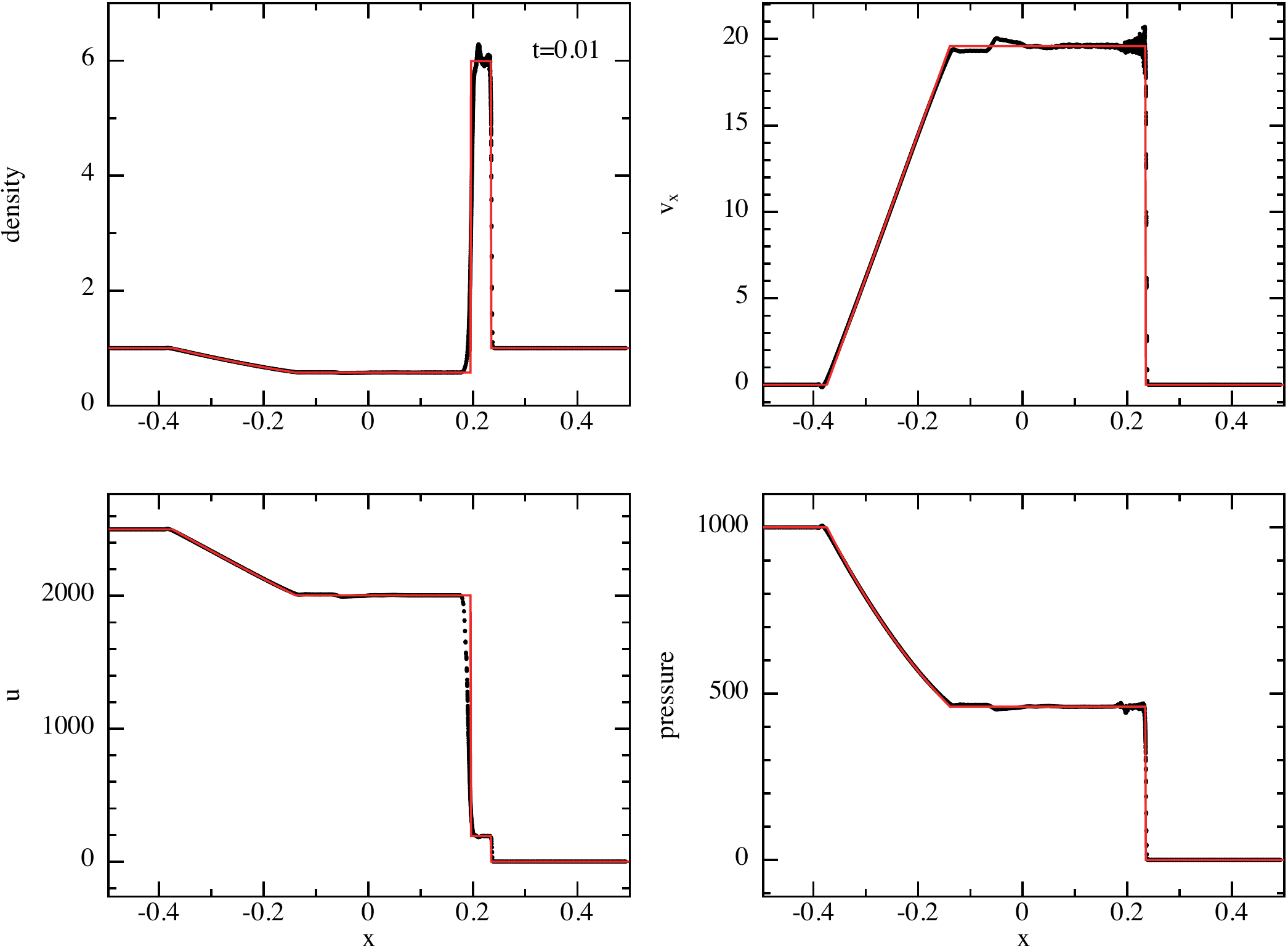} 
   \caption{As in Figure~\ref{fig:blast} but with code defaults for dissipation switches ($\alpha^{\rm AV} \in [0,1]$; $\alpha_u = 1$). As in Figure~\ref{fig:sodcd} the velocity field behind the shock is more noisy with switches applied, but the switches reduce the numerical dissipation away from shocks.}
\label{fig:blastalpha1}
\end{figure*}

\begin{figure}
   \centering
   \includegraphics[width=\columnwidth]{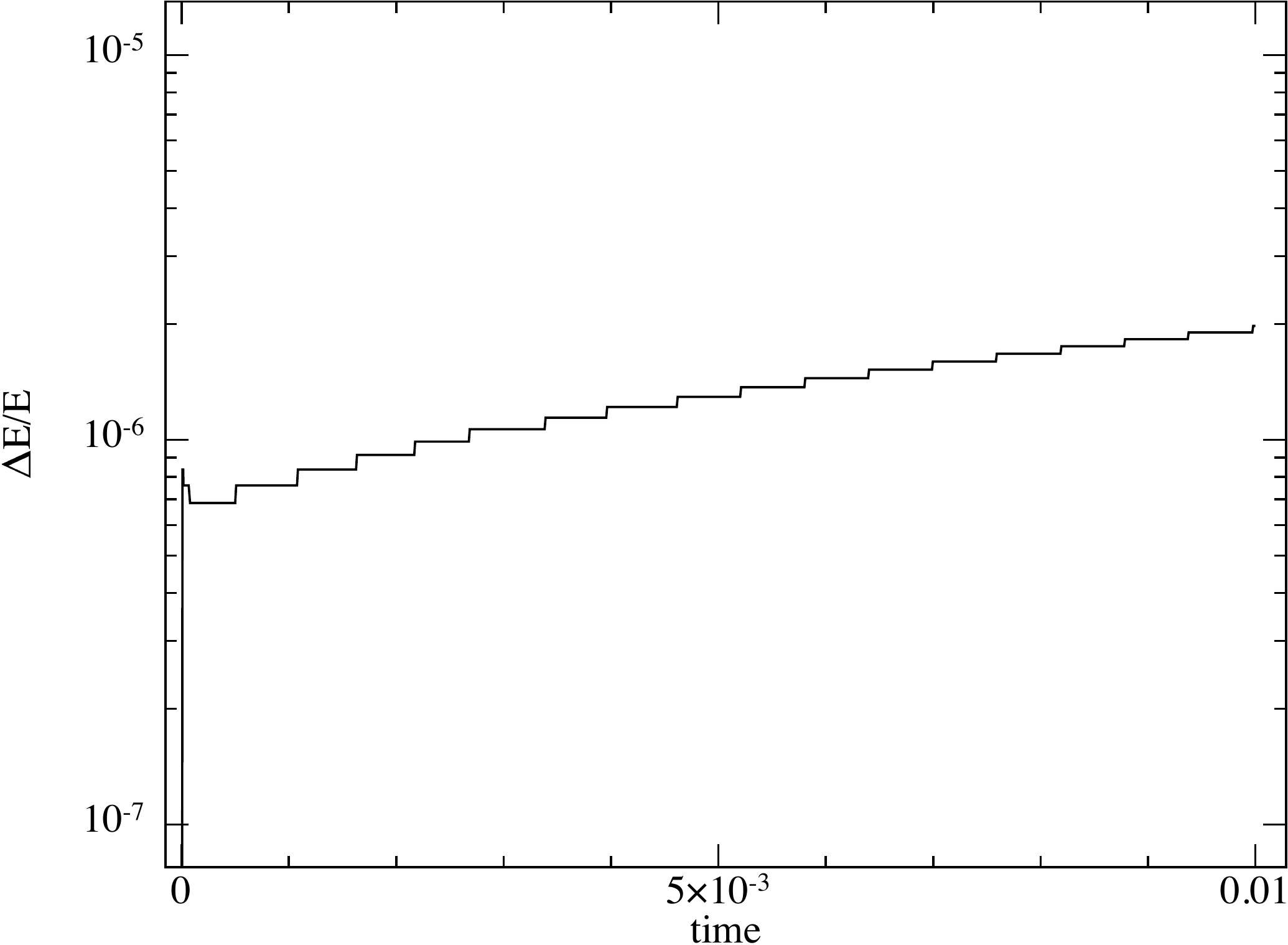} 
   \caption{Relative error in energy conservation for the 3D blast wave test. Energy is conserved to an error of $10^{-6}$ in this test.}
\label{fig:blasten}
\end{figure}

\subsubsection{Sod shock tube}
\label{sec:sod}
 Figure~\ref{fig:sod} shows the results of the standard \citet{sod78} shock tube test, performed in 3D using $[\rho, P] = [1,1]$ in the `left state' ($x \leq 0$) and $[\rho, P] = [0.125,0.1]$ for the `right state' ($x > 0$) with a discontinuity initially at $x=0$ and zero initial velocity and magnetic field. We perform the test using an adiabatic equation of state with $\gamma = 5/3$ and periodic boundaries in $y$ and $z$. While many 1D solutions appear in the literature, only a handful of results on this test have been published for SPH in 3D (e.g. \citealt{dolagetal05,hubberetal11,becketal16}; a 2D version is shown in \citealt{price12}). The tricky part in a 3D SPH code is how to set up the density contrast. Setting particles on a cubic lattice is a poor choice of initial condition since this is not a stable arrangement for the particles \citep{morris96,morrisphd,lombardietal99,bot04}. The approach taken in \citet{springel05} (where only the density was shown, being the easiest to get right) was to relax the two halves of the box into a stable arrangement using the gravitational force run with a minus sign, but this is time consuming. 
 
 Here we take a simpler approach which is to set the particles initially on a close-packed lattice (Section~\ref{sec:unifdis}), since this is close to the relaxed arrangement \citep[e.g.][]{lombardietal99}. To ensure continuity of the lattice across periodic boundaries we fix the number of particles in the $y$ ($z$) direction to the nearest multiple of 2 (3) and adjust the spacing in the $x-$direction accordingly to give the correct density in each subdomain. We implement the boundary condition in the x-direction by tagging the first and last few rows of particles in the $x$ direction as boundary particles, meaning that their particle properties are held constant. The results shown in Figure~\ref{fig:sod} use $256 \times 24 \times 24$ particles initially in $x \in [-0.5,0]$ and $128 \times 12 \times 12$ particles in $x \in [0,0.5]$ with code defaults for the artificial conductivity ($\alpha_{u} = 1$ with $v_{\rm sig}^{u}$ given by equation~\ref{eq:vsigupr}) and artificial viscosity ($\alpha^{\rm AV} = 1, \beta^{\rm AV} = 2$). The results are identical whether global or individual particle timesteps (Section~\ref{sec:indtimesteps}) are used. Figure~\ref{fig:sodcd} shows the results when code defaults for viscosity are also employed, resulting in a time-dependent $\alpha^{\rm AV}$ (see lower right panel). There is more noise in the velocity field around the contact discontinuity in this case, but the results are otherwise identical.
 
  The ${\cal L}_2$ errors for the solutions shown in Figure~\ref{fig:sod} are 0.0090, 0.0022, 0.0018 and 0.0045 for the density, velocity, thermal energy and pressure, respectively. With dissipation switches turned on (Figure~\ref{fig:sodcd}) the corresponding errors are 0.009, 0.0021, 0.0019 and 0.0044 respectively. That is, our solutions are within 1 per cent of the analytic solution for all four quantities in both cases, with the density profile showing the strongest difference (mainly due to smoothing of the contact discontinuity).
 
  \citet{puriramachandran14} compared our shock capturing algorithm with other SPH variants, including Godunov SPH. Our scheme was found to be the most robust of those tested.


\subsubsection{Blast wave}
 As a more extreme version of the shock test, Figures~\ref{fig:blast} and \ref{fig:blastalpha1} shows the results of the blast wave problem from \citet{monaghan97}, set up initially with $[\rho, P] = [1,1000]$ for $x \leq 0$ and $[\rho, P] = [1.0,0.1]$ for $x > 0$ and with $\gamma = 1.4$ (appropriate to air). As previously we set the particles on a close-packed lattice with a discontinuous initial pressure profile. We employ $800 \times 12 \times 12$ particles in the domain $x \in [-0.5,0.5]$. Results are shown at $t=0.01$. This is a more challenging problem than the Sod test due to the higher Mach number. As previously, we show results with both $\alpha^{\rm AV} = 1$ (Figure~\ref{fig:blast}) and with the viscosity switch $\alpha \in [0,1]$. Both calculations use $\alpha_{u} = 1$ with (\ref{eq:vsigupr}) for the signal speed in the artificial conductivity. For the solution shown in Figure~\ref{fig:blast}, we find normalised ${\cal L}_2$ errors of 0.057, 0.063, 0.051 and 0.018 in the density, velocity, thermal energy and pressure, respectively, compared to the analytic solution. Employing switches (Figure~\ref{fig:blastalpha1}), we find corresponding ${\cal L}_2$ errors of 0.056, 0.059, 0.052 and 0.017. That is, our solutions are within 6 per cent of the analytic solution at this resolution.
 
 The main source of error is that the contact discontinuity is over-smoothed due to the artificial conductivity, while the velocity profile shows a spurious jump at the location of the contact discontinuity. This glitch is a startup error caused by our use of purely discontinuous initial conditions --- it could be removed by adopting smoothed initial conditions but we prefer to perform the more difficult version of this test. There is also noise in the post-shock velocity profile because the viscosity switch does not apply enough dissipation here. As in the previous test, this noise can be reduced by increasing the numerical viscosity, e.g. by adopting a constant $\alpha^{\rm AV}$ (compare Figures~\ref{fig:blast} and \ref{fig:blastalpha1}).
 
  Figure~\ref{fig:blasten} quantifies the error in energy conservation, showing the error in the total energy as a function of time, i.e $\vert E - E_0 \vert / \vert E_0 \vert$. Energy is conserved to a relative accuracy of better than $2 \times 10^{-6}$.

\begin{figure}
   \centering
   \includegraphics[width=\columnwidth]{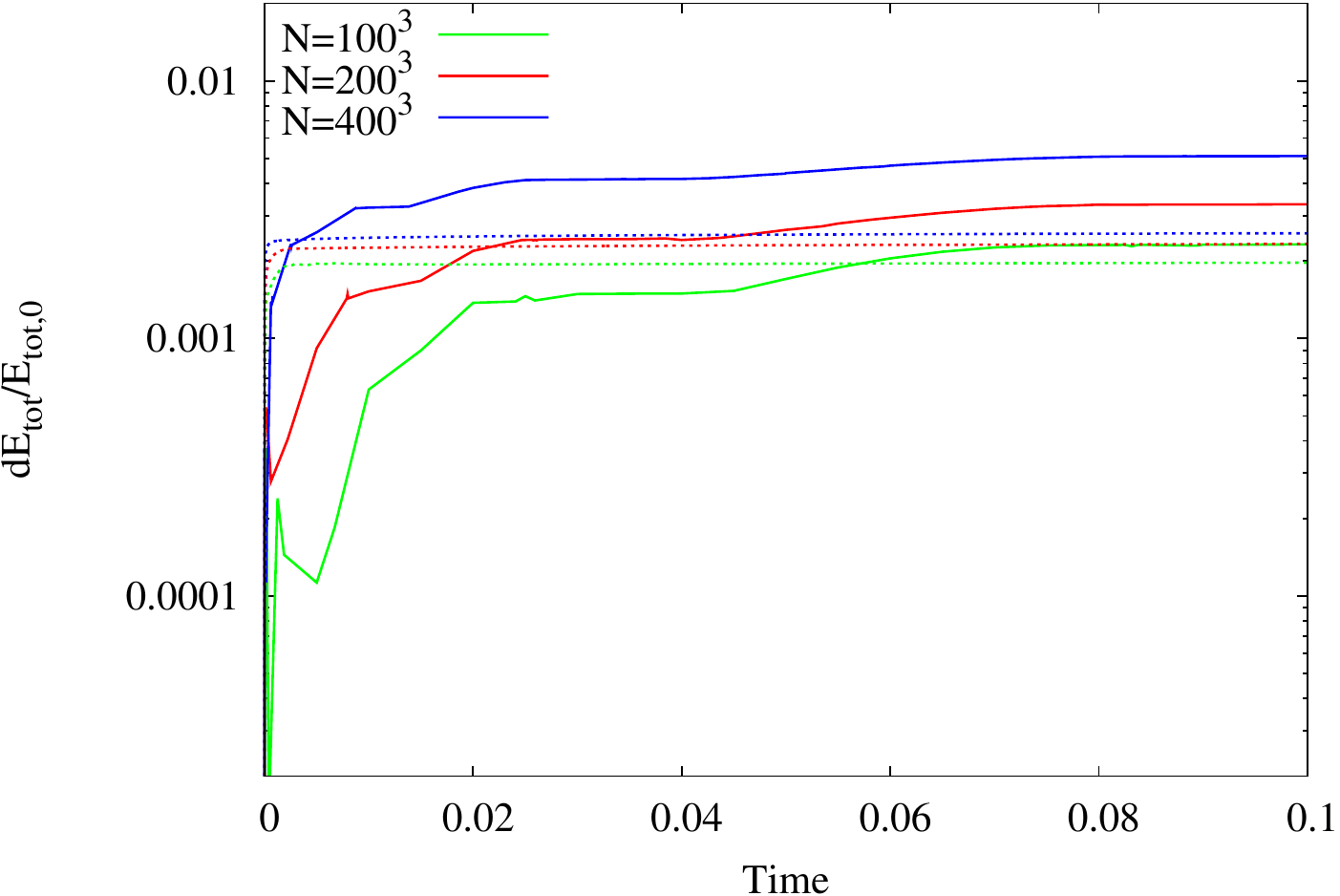} 
   \caption{Evolution of the relative error in total energy for the Sedov blast wave problem.  Resolutions are given in the legend; solid lines use individual timestepping while dotted lines show global timestepping.}
\label{fig:sedovEtot}
\end{figure}

\subsubsection{Sedov blast wave}
\label{sec:sedov}

 The Sedov-Taylor blast wave \citep{taylor50,taylor50a,sedov59} is a similar test to the previous but with a spherical geometry. This test is an excellent test for the individual timestepping algorithm, since it involves propagating a blast wave into an ambient medium of `inactive' or `asleep' particles, which can cause severe loss of energy conservation if they are not carefully awoken \citep{saitohmakino09}. For previous uses of this test with SPH, see e.g. \citet{springelhernquist02,rosswogprice07} and for a comparison between SPH and mesh codes on this test see \citet{taskeretal08}.
 
  We set up the problem in a uniform periodic box $x, y, z \in [-0.5,0.5]$, setting the thermal energy on the particles to be non-zero in a barely-resolved sphere around the origin. We assume an adiabatic equation of state with $\gamma = 5/3$. The total energy is normalised such that the total thermal energy in the blast is $E_{0} = \sum_{a} m_{a} u_{a} = 1$, distributed on the particles within $r < R_{\rm kern} h_{0}$ using the smoothing kernel, i.e.
\begin{equation}
u_{a} =  \begin{cases}
E_{0}W(r, h_{0}), & r/h_{0} \leq R_{\rm kern}\\
0 & r/h_{0} > R_{\rm kern}
\end{cases}
\end{equation}
where $r = \sqrt{x^{2} + y^{2} + z^{2}}$ is the radius of the particle and we set $h_{0}$ to be twice the particle smoothing length.

\begin{figure*}
   \centering
   \includegraphics[width=0.8\textwidth]{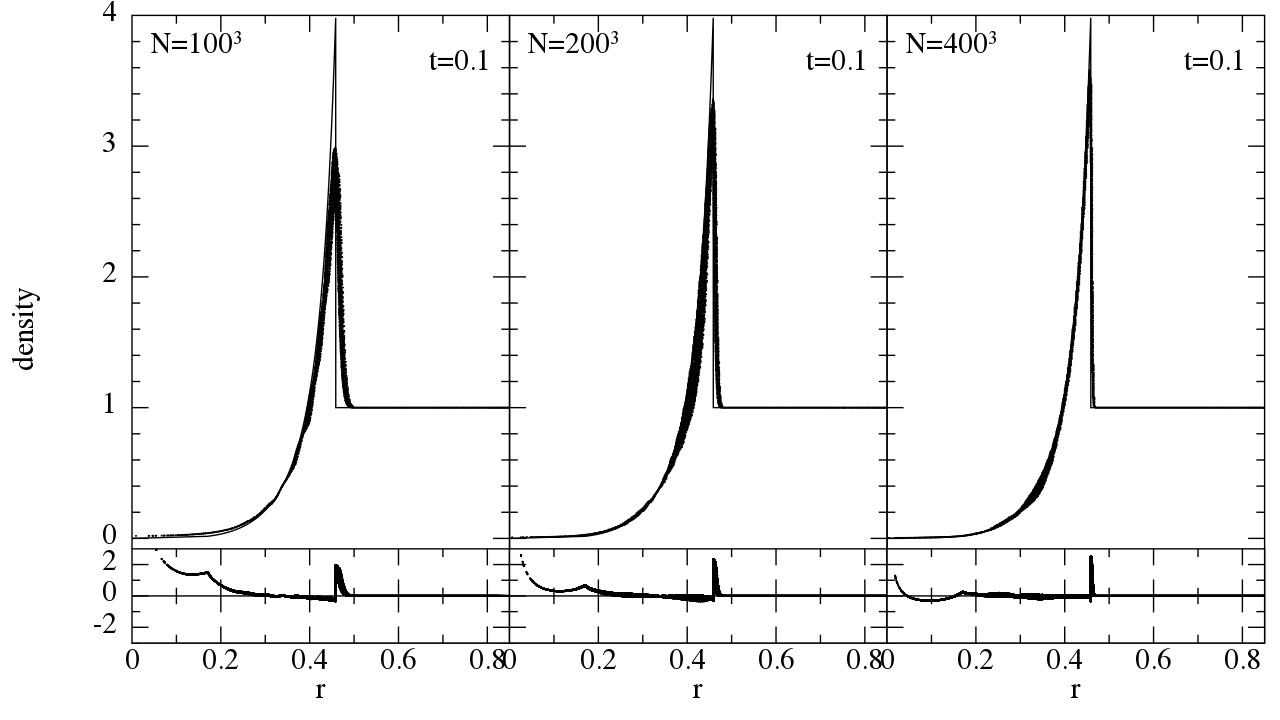}
   \caption{Density as a function of radius in the Sedov blast wave problem at three resolutions.  All particles are placed initially on a closepacked lattice, are evolved using individual timesteps and we use the quintic kernel.  The analytic solution is given by the solid line, and the bottom panels show the residuals compared to the analytical solution.}
\label{fig:sedov}
\end{figure*}

We simulate the Sedov blast wave using both global and individual timesteps at several different resolutions.  Figure~\ref{fig:sedovEtot} shows the evolution of the relative error in total energy for our suite, while Figure~\ref{fig:sedov} shows the density at $t=0.1$ compared to the analytical solution given by the solid line.
Energy is conserved to better than 1 per cent in all cases.  Using higher spatial resolution results in a better match of the post-shock density with the analytic solution. The scatter at the leading edge of the shock is a result of the default artificial conductivity algorithm.
Given the initial strong gradient in thermal energy, artificial conductivity is also important for reducing the noise on this problem, as first noted by \citet{rosswogprice07}.

%
%

\begin{figure*}
   \centering
   \includegraphics[width=\textwidth]{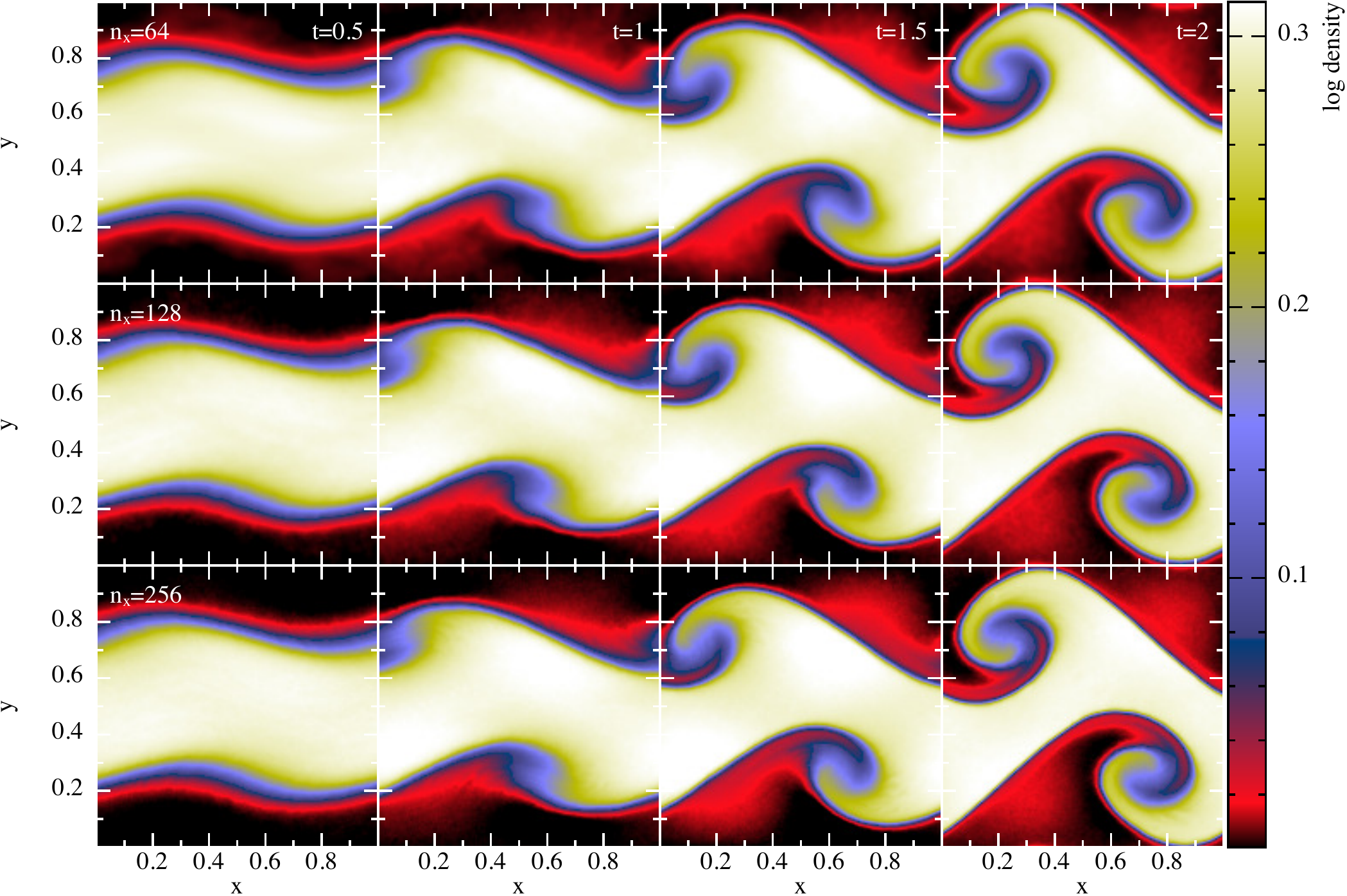} 
   \caption{Results of the well-posed Kelvin-Helmholtz instability test from \citet{robertsonetal10}, shown at a resolution of (from top to bottom) $64\times74\times12$, $128\times148\times12$ and $256\times296\times12$ equal mass SPH particles. We use stretch mapping (Section~\ref{sec:stretchmap}) to achieve the initial density profile, consisting of a 2:1 density jump with a smoothed transition.}
\label{fig:kh}
\end{figure*}

\subsubsection{Kelvin-Helmholtz instability}
\label{sec:kh}
 Much has been written about Kelvin-Helmholtz instabilities with SPH \citep[e.g.][]{agertzetal07,price08,abel11,valdarnini12,readhayfield12,hfg13}. For the present purpose it suffices to say that the test problems considered by \citet{agertzetal07} and \citet{price08} are not well posed. That is, the number of small-scale secondary instabilities will always increase with numerical resolution because high wavenumber modes grow fastest in the absence of physical dissipation or other regularising forces such as magnetic fields or surface tension. The ill-posed nature of the test problem has been pointed out by several authors \citep{robertsonetal10,mlp12,lecoanetetal16}, who have each proposed well-posed alternatives.
 
  We adopt the setup from \citet{robertsonetal10}, similar to the approach by \citet{mlp12}, where the initial density contrast is smoothed using a ramp function. This should suppress the formation of secondary instabilities long enough to allow a single large scale mode to grow. The density and shear velocity in the $y$ direction are given by
\begin{equation}
\rho(y) = \rho_{1} + R(y)[\rho_{2} - \rho_{1}], \label{eq:khrhoy}
\end{equation}
and
\begin{equation}
v_{x}(y) = v_{1} + R(y)[v_{2} - v_{1}],
\end{equation}
where $\rho_{1} = 1$, $\rho_{2} = 2$, $v_{1} = -0.5$ and $v_{2} = 0.5$ with constant pressure $P = 2.5$, $\gamma = 5/3$. The ramp function is given by
\begin{equation}
R(y) \equiv \left[1 - f(y)\right]\left[1 -g(y)\right],
\end{equation}
where 
\begin{align}
f & \equiv \frac{1}{1 + \exp\left[2(y-0.25)/\Delta\right]}, \nonumber \\
g & \equiv \frac{1}{1 + \exp\left[2(0.75-y)/\Delta\right]},
\end{align}
and we set $\Delta = 0.25$. Finally, a perturbation is added in the velocity in the $y$ direction, given by
\begin{equation}
v_{y} = 0.1 \sin (2\pi x). \label{eq:khperturb}
\end{equation}

The setup in \citet{robertsonetal10} is 2D, but since \textsc{Phantom} is a 3D code, we instead set up the problem using a thin three dimensional box. We first set up a uniform close packed lattice in a periodic box with dimensions $1 \times 1 \times \sqrt{24}/n_{x}$, where $n_{x}$ is the initial resolution in the $x$ direction such that the box thickness is set to be exactly 12 particle spacings in the $z$ direction independent of the resolution used in the $x$ and $y$ direction. The box is set between $[0,1]$ in $x$ and $y$, consistent with the ramp function. We then set up the density profile by stretch-mapping in the $y$ direction using (\ref{eq:khrhoy}) as the input function (c.f.~Section~\ref{sec:stretchmap}).
 
\begin{figure}
\includegraphics[width=\linewidth]{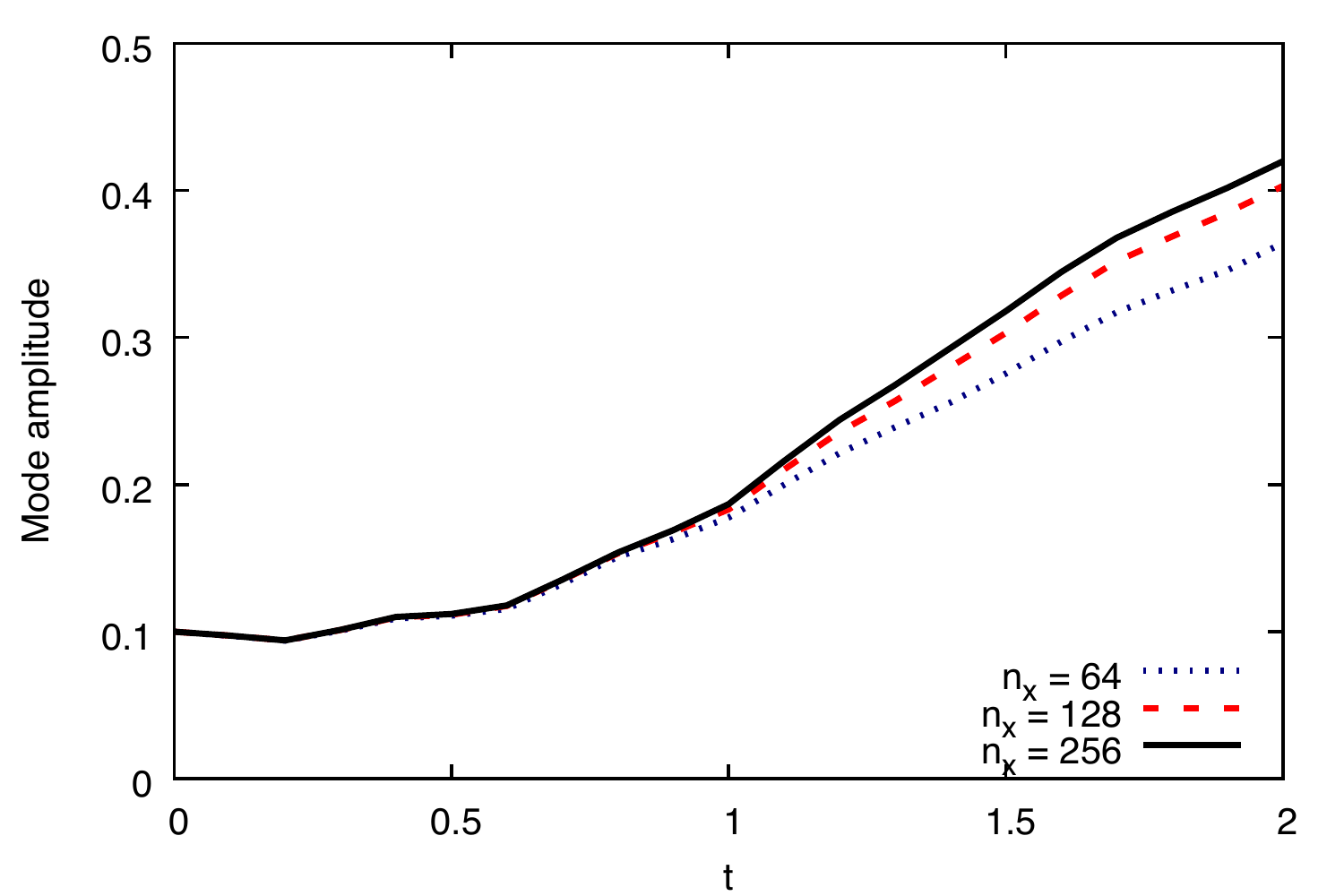}
\caption{Growth of the amplitude of the seeded mode for the Kevin-Helmholtz instability test. The amplitude at $t=2$ between the $n_x=128$ and $n_x=256$ calculations is within 4 per cent.}
\label{fig:khmodeamp}
\end{figure}

  Figure~\ref{fig:kh} shows the results of this test, showing a cross section of density at $z=0$ for three different resolutions (top to bottom) and at the times corresponding to those shown in \citet{robertsonetal10}. We compute the quantitative difference between the calculations by taking the root mean square difference of the cross section slices shown above interpolated to a 1024 $\times$ 1024 pixel map. The error between the $n_x = 64$ calculation and the $n_x = 256$ calculation is $1.3 \times 10^{-3}$, while this reduces to $4.9 \times 10^{-4}$ for the $n_x = 128$ calculation. Figure~\ref{fig:khmodeamp} shows the growth of the amplitude of the mode seeded in (\ref{eq:khperturb}). We follow the procedure described in \citet{mlp12} to calculate the mode amplitude. At $t=2$, the amplitude between the $n_x = 128$ and $n_x = 256$ calculations is within 4 per cent. The artificial viscosity and conductivity tend to slow convergence on this problem, so it is a good test of the dissipation switches (we use the default code viscosity switch as discussed in Section~\ref{sec:switches}). 

\subsection{External forces}
\label{sec:extftest}
%

\begin{figure}
   \centering
   \includegraphics[width=\columnwidth]{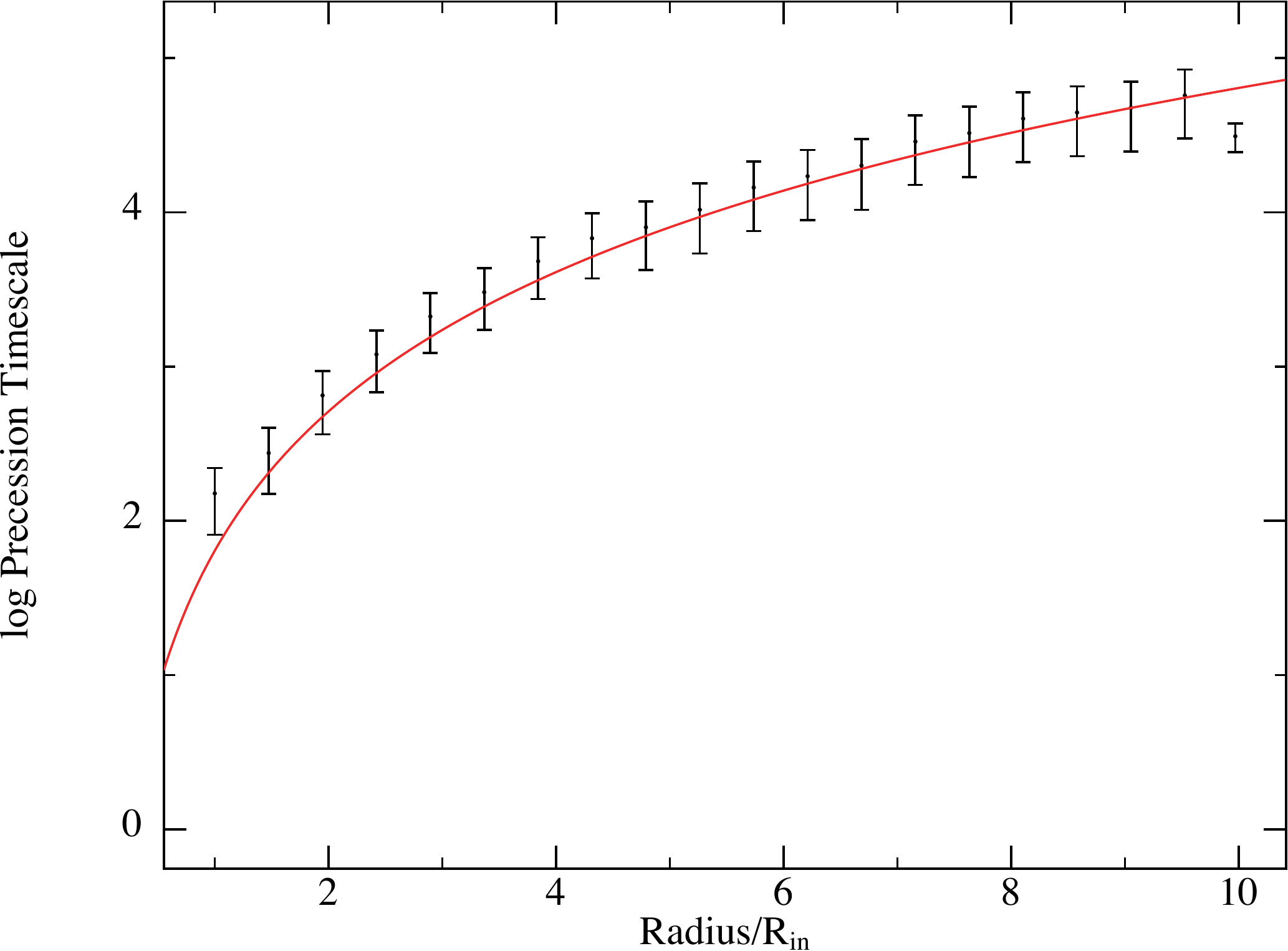} 
   \caption{Lense-Thirring precession test from a disc inclined by $30^{\circ}$. Here the precession time-scale is measured from the cumulative twist in the disc and the exact solution, $t_p = R^3/2a$ is represented by the red line.}
\label{fig:lt_precession}
\end{figure}

\subsubsection{Lense-Thirring precession}
We test the implementation of the Lense-Thirring precession by computing the precession induced on a pressure-less disc of particles, as outlined in \citet{npn15}. This disc is simulated for one orbit at the outer edge such that the inner part of the disc has precessed multiple times but the outer region has not yet completed a full precession. The precession timescale is estimated by measuring the twist as a function of time for each radial bin; in the inner region this is the time taken for the twist to go from a minimum (zero twist) to a maximum (fully twisted) and in the outer region the gradient of the twist against time is used to calculate the equivalent time. Figure~\ref{fig:lt_precession} shows the precession timescale measured from the simulation as a function of the radius compared to the analytically derived precession timescale, with uncertainties derived from the calculation of the gradient.

\begin{figure}
\includegraphics{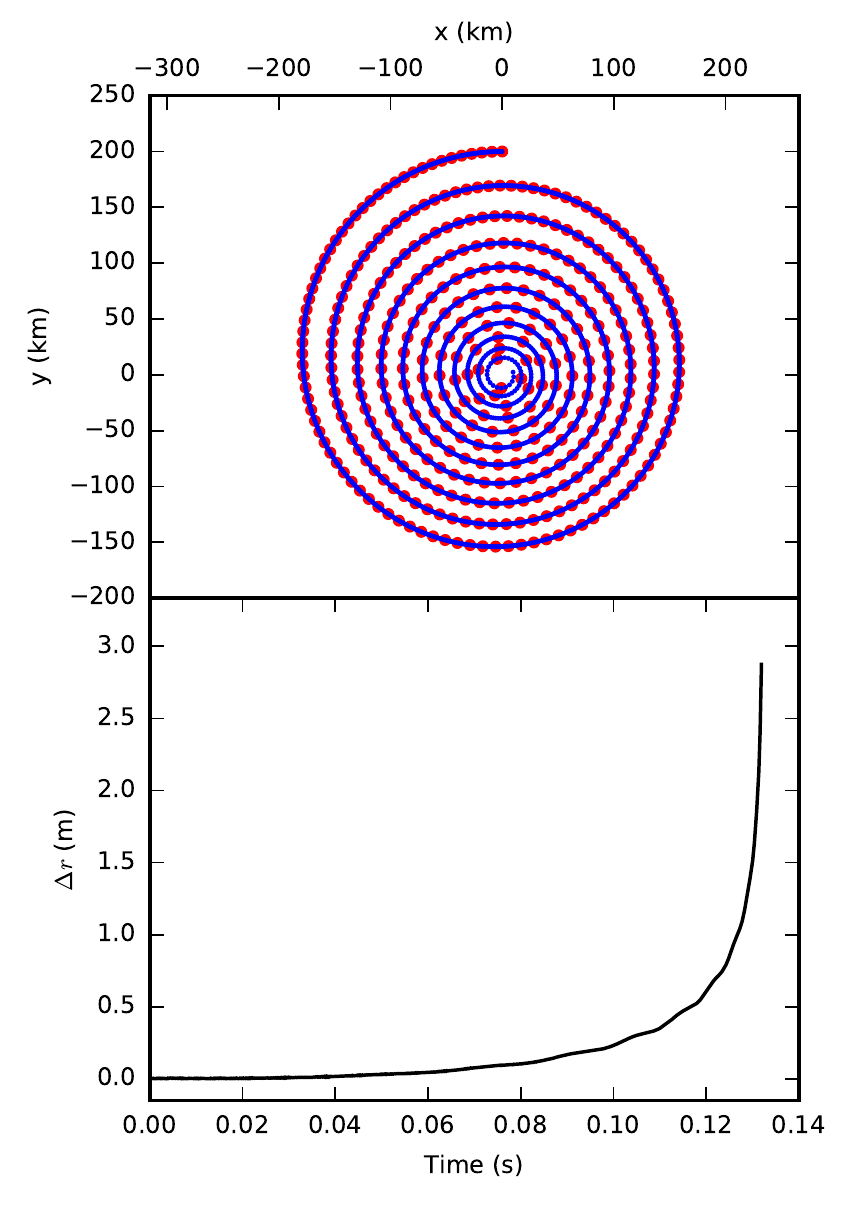}
\caption{Inspiral of a test particle subjected to Poynting-Robertson drag using \textsc{Phantom} (blue curve) compared to the expected solution (red curve). For clarity, only some of the points are plotted. The results are indistinguishable on this scale. Bottom panel shows the distance ($\Delta r$) between the PHANTOM particle and the test code particle. This demonstrates that the implementation of Poynting-Robertson drag in \textsc{Phantom} is consistent with a 4th order numerical solution to (\ref{eq:prdrag}).}
\label{fig:prdrag}
\end{figure}

\subsubsection{Poynting-Robertson drag}
Figure~\ref{fig:prdrag} shows the trajectory of a spherical assembly of 89 pressureless SPH particles subject to Poynting-Robertson drag with a fixed value of $\beta_{\rm PR} = 0.1$, assuming a central neutron star of mass 1.4M$_{\odot}$ and 10\,km radius and a particle initially orbiting at $R=200$~km with initial $v_\phi$ of 0.9 times the Keplerian orbital speed. We compare this to the trajectory of a test particle produced by direct numerical integration of the equations of motion, (\ref{eq:prdrag}), with a 4th order Runge Kutta scheme. As shown in Figure \ref{fig:prdrag}, there is no significant difference between the codes. We therefore expect that the behaviour of SPH gas or dust particles under the influence of any given $\beta$ will be correct.

 \begin{figure*}
   \centering
   \includegraphics[width=\textwidth]{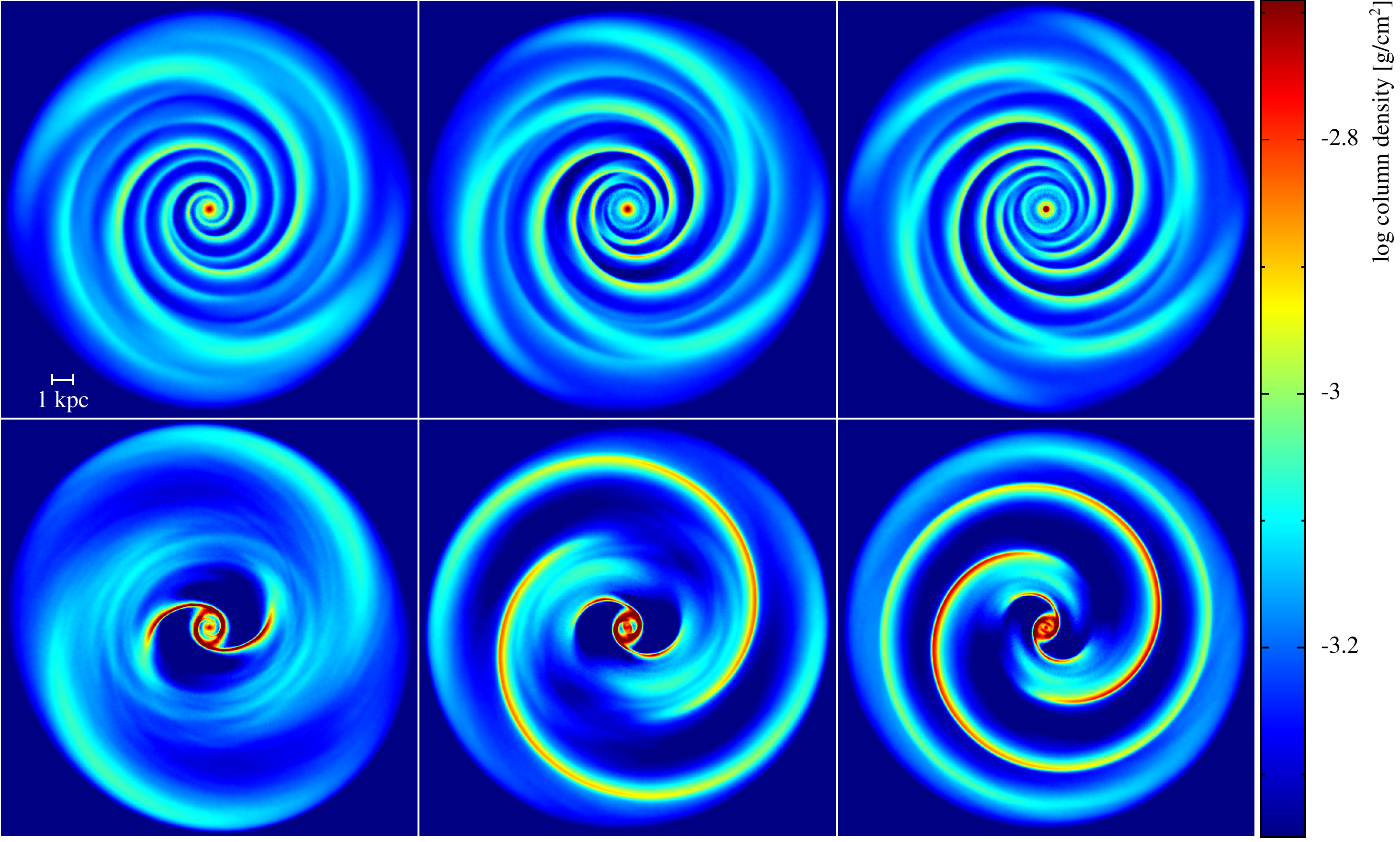} 
   \caption{Gas in a galactic disc under the effect of different galactic potentials. Top row shows models with a 2, 3, and 4 armed spiral (left to right) with a pitch angle of $15^\circ$ and pattern speed of 20 km s$^{-1}$ kpc$^{-1}$. Bottom row shows a bar potential with pattern speeds of 40, 60 and 80 km s$^{-1}$ kpc$^{-1}$ (left to right).}
\label{fig:spirals}
\end{figure*}

\subsubsection{Galactic potentials}
 Figure~\ref{fig:spirals} shows 6 calculations with gas embedded within different galactic potentials (Section~\ref{sec:galdisc}). We set up an isothermal gas disc with $T=10~000$K, with a total gas mass of $1\times10^9$ M$_\odot$ set up in a uniform surface density disc from 0--10 kpc in radius. A three-part potential model for the Milky Way provides the disc with an axisymmetric rotation curve (bulge plus disc plus halo, the same as \citealt{pettittetal14}). The top row shows gas exposed to spiral potentials of \citet{2002ApJS..142..261C} with three different arm numbers (2, 3, 4), while the bottom row shows simulations within the bar potential of \citet{2001PASJ...53.1163W} at three different pattern speeds (40, 60 and 80 km s$^{-1}$ kpc$^{-1}$). All models are shown after approximately one full disc rotation (240 Myr). Gas can be seen to trace the different spiral arm features, with the two armed model in particular showing branches characteristic of such density wave potentials (e.g. \citealt{2004MNRAS.350L..47M}). The bars drive arm features in the gas, the radial extent of which is a function of the pattern speed. Also note the inner elliptical orbits of the bar at the location of the Lindblad resonance which is an effect of the peaked inner rotation curve resulting from the central bulge.

\subsection{Accretion discs}
\label{sec:disctest}
 SPH has been widely used for studies of accretion discs, ever since the first studies by \citet{artymowiczlubow94,artymowiczlubow96,murray96} and \citet{mmm96} showed how to use the SPH artificial viscosity term to mimic a \citet{shakurasunyaev73} disc viscosity.

\subsubsection{Measuring the disc viscosity}
\begin{figure}
   \centering
   \includegraphics[width=\columnwidth]{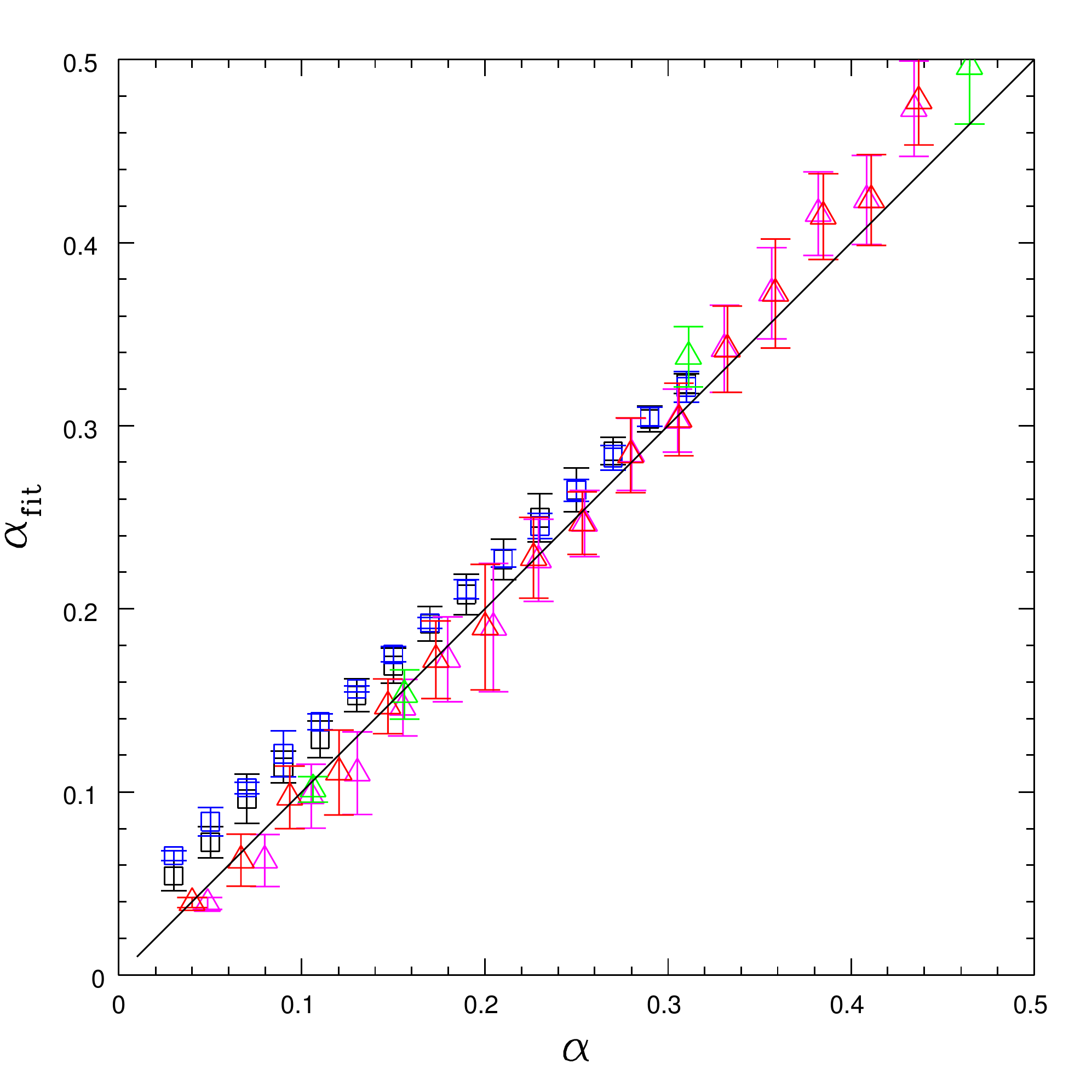} 
   \caption{Calibration of the disc viscosity in {\sc Phantom}, comparing the input value of the Shakura-Sunyaev $\alpha$ from (\ref{eq:alphascale}) ($x-$axis) with the measured diffusion rate of the surface density by fitting to a 1D code ($y$-axis). Triangles indicate simulations with the disc viscosity computed using the artificial viscosity (Section~\ref{sec:discav}), while squares represent simulations using physical viscosity (Section~\ref{sec:twofirstderivs}). All simulations use 2 million particles except for the green, cyan and red triangles which use 20 million particles. Figure taken from \citet{lodatoprice10}.}
\label{fig:alpha}
\end{figure}

 The simplest test is to measure the disc viscosity from the diffusion rate of the disc surface density. Figure~\ref{fig:alpha} shows the results of an extensive study of this with {\sc Phantom} perfomed by \citet{lodatoprice10}. For this study we set up a disc from $R_{\rm in} = 0.5$ to $R_{\rm out} = 10$ with surface density profile
 \begin{equation}
 \Sigma = \Sigma_0 R^{-p} \left( 1 - \sqrt{\frac{R_{\rm in}}{R}} \right),
 \end{equation}
and a locally isothermal equation of state $c_{\rm s} = c_{{\rm s}, 0} R^{-q}$. We set $p=3/2$ and $q=3/4$ such that the disc is uniformly resolved, i.e. $h/H \sim$ constant \citep{lodatopringle07}, giving a constant value of the \citet{shakurasunyaev73} $\alpha$ parameter according to (\ref{eq:alphascale}). We set $c_{{\rm s}, 0}$ such that the aspect ratio is $H/R = 0.02$ at $R=1$. We used 2 million particles by default, with several additional calculations perfomed using 20 million particles.  The simulation is performed to $t=1000$ in code units.
 
The diffusion rate is measured by fitting the surface density evolution obtained from {\sc Phantom} with the results of a `ring code' solving the standard 1D diffusion equation for accretion discs \citep{lynden-bellpringle74,pringle81,pringle92}. Details of the fitting procedure are given in \citet{lodatoprice10}. In short, we use Newton-Raphson iterations to find the minimum error between the 1D code and the surface density profile from {\sc Phantom} at the end of the simulation, which provides the best fit ($\alpha_{\rm fit}$) and error bars. Figure~\ref{fig:alpha} shows that the measured diffusion rates agree with the expected values to within the error bars. The exception is for low viscosity discs with physical viscosity, where contribution from artificial viscosity becomes significant. Triangles in the figure show the results with disc viscosity computed from the artificial viscosity (Section~\ref{sec:discav}), while squares represent simulations with physical viscosity set according to (\ref{eq:nufromalpha}).
 
  This test demonstrates that the implementation of disc viscosity matches the analytic theory to within measurement errors. This also demonstrates that the translation of the artificial viscosity term according to (\ref{eq:alphascale}) is correct.

\subsubsection{Warp diffusion}
 A more demanding test of disc physics involves the dynamics of warped discs. Extensive analytic theory exists, starting with the linear theory of \citet{papaloizoupringle83}, subsequent work by \citet{pringle92}, and culminating in the work by \citet{ogilvie99} which provides the analytic expressions for the diffusion rate of warps in discs for non-linear values of both disc viscosity and warp amplitude. Importantly, this theory applies in the `diffusive' regime where the disc viscosity exceeds the aspect ratio, $\alpha > H/R$. For $\alpha \lesssim H/R$ the warp propagation is wave-like and no equivalent non-linear theory exists \citep[see][]{lubowogilvie00,lop02}.
 
  \begin{figure}
   \centering
   \includegraphics[width=\columnwidth]{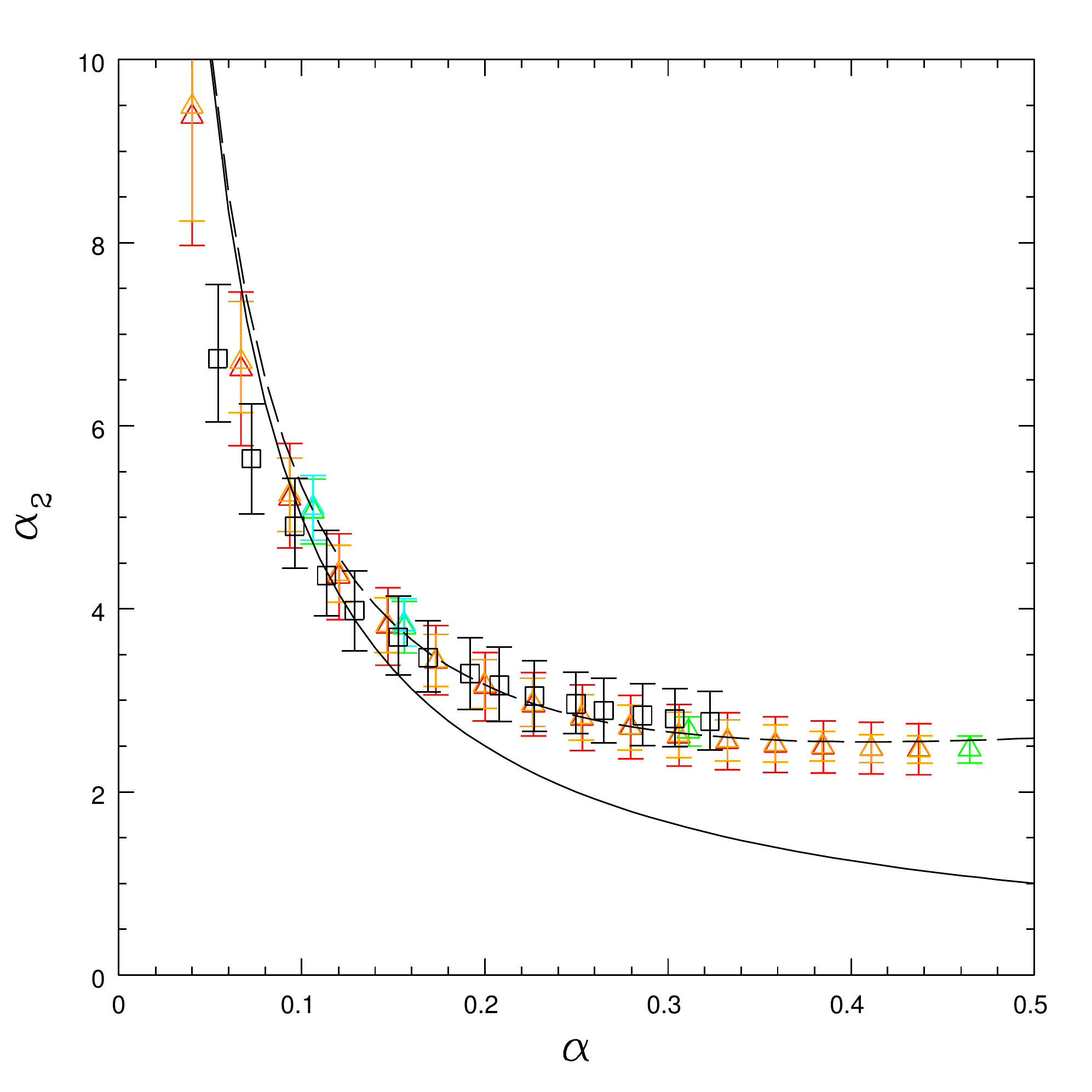} 
   \caption{Warp diffusion rate as a function of disc viscosity, showing the {\sc Phantom} results compared to the non-linear theory of \citet{ogilvie99} (dashed line). The linear prediction, $\alpha_2 = 1/(2\alpha)$, is shown by the solid line, highlighting the agreement of {\sc Phantom} with the non-linear theory. Colouring of points is as in Figure~\ref{fig:alpha}. Figure taken from \citet{lodatoprice10}.}
\label{fig:alpha2}
\end{figure}

  {\sc Phantom} was originally written to simulate warped discs --- with our first study in \citet{lodatoprice10} designed to test the \citet{ogilvie99} theory in 3D simulations. Figure~\ref{fig:alpha2} shows the results of this study, showing the measured warp diffusion rate as a function of disc viscosity. The setup of the simulations is as in the previous test but with a small warp added to the disc, as outlined in Section 3.3 of \citet{lodatoprice10}. Details of the fitting procedure used to measure the warp diffusion rate are also given in Section 4.2 of that paper. The dashed line shows the non-linear prediction of \citet{ogilvie99}, namely
\begin{equation}
\alpha_2 = \frac{1}{2\alpha} \frac{4(1 + 7\alpha^2)}{4 + \alpha^2}.
\end{equation}
Significantly, the {\sc Phantom} results show a measurable difference between the predictions of the non-linear theory and the prediction from linear theory \citep{papaloizoupringle83} of $\alpha_2 = 1/(2\alpha)$, shown by the solid black line. 

 In addition to the results shown in Figure~\ref{fig:alpha2}, {\sc Phantom} also showed a close match to both the predicted self-induced precession of the warp and to the evolution of non-linear warps (see Figures~13 and 14 in \citealt{lodatoprice10}, respectively). From the success of this initial study we have used {\sc Phantom} to study many aspects of disc warping, either with isolated warps or breaks \citep{lodatoprice10,nkp12}, warps induced by spinning black holes \citep{nixonetal12,npn15,nealonetal16} and warps in circumbinary \citep{nkp13,flp13} or circumprimary \citep{doganetal15,martinetal14,martinetal14a} discs. In particular, {\sc Phantom} was used to discover the phenomenon of `disc tearing' where sections of the disc are `torn' from the disc plane and precess effectively independently \citep{nixonetal12,nkp13,npn15}.


\subsubsection{Disc-planet interaction}
\begin{figure}
\centering
\includegraphics[width=\columnwidth]{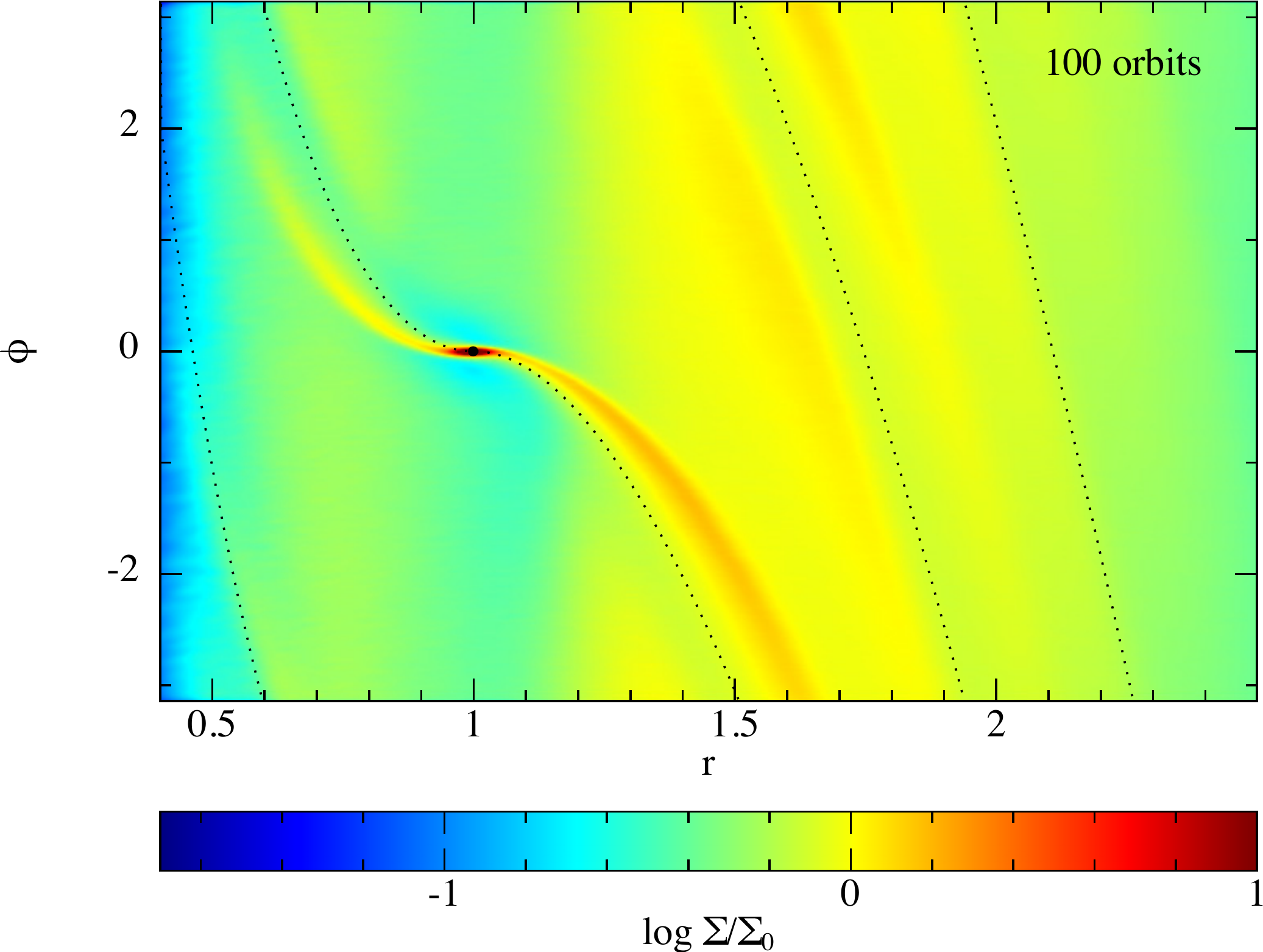}
\caption{Planet-disc interaction in 3D, showing the `viscous Jupiter' calculation comparable to the 2D results shown with various grid and SPH codes in Figure~10 of \citet{de-val-borroetal06}. The dotted lines shows the estimated position of the planetary shocks from \citealt{ogilvielubow02}. The offset between this solution and the numerical shock position is due to the approximate nature of the analytic solution (see \citealt{de-val-borroetal06}).}
\label{fig:planetdisc}
\end{figure}

 Although there is no `exact' solution for planet-disc interaction, an extensive code comparison was performed by \citet{de-val-borroetal06}. Figure~\ref{fig:planetdisc} shows the column density of a 3D \textsc{Phantom} calculation, plotted in r-$\phi$ with the density integrated through the $z$ direction, comparable to the `viscous Jupiter' setup in \citet{de-val-borroetal06}.

Two caveats apply when comparing our results with those in Figure 10 of \citet{de-val-borroetal06}. The first is that the original comparison project was performed in 2D and mainly with grid-based codes with specific `wave damping' boundary conditions prescribed. We chose simply to ignore the prescribed boundary conditions and two dimensionality and instead modelled the disc in 3D with a central accretion boundary at $r=0.25$ with a free outer boundary, with the initial disc set up from $r=0.4$ to $r=2.5$. We used $10^6$ SPH particles. Second, the planetary orbit was prescribed on a fixed circular orbit with no accretion onto either the planet or the star.  Although we usually use sink particles in \textsc{Phantom} to model planet-disc interaction \citep[e.g.][]{dipierroetal15}, for this test we thus employed the fixed binary potential (Section~\ref{sec:extbinary}) to enable a direct comparison. We thus used $M = 10^{-3}$ in the binary potential, corresponding to the `Jupiter' simulation in \citet{de-val-borroetal06} with the planet on a fixed circular orbit at $r = 1$.
 
 As per the original comparison project, we implemented Plummer softening of the gravitational force from the planet,
\begin{equation}
\phi_{\rm planet} = - \frac{-m_{\rm planet}}{\sqrt{r^{2} + \epsilon^{2}}},
\end{equation}
where $\epsilon = 0.6H$. We also implemented the prescribed increase of planet mass with time for the first 5 orbits according to
\begin{equation}
m_{\rm planet} = \sin^{2} \left( \frac{\pi t}{10 P_{\rm orbit}} \right),
\end{equation}
where $P_{\rm orbit}$ is the orbital period ($2\pi$ in code units), though we found this made little or no difference to the results in practice. We assumed a locally isothermal equation of state, $c_{\rm s} \propto R^{-0.5}$, such that $H/R \approx 0.05 \approx$ constant. We assumed an initially constant surface density, $\Sigma_{0} = 0.002 M_{\rm *}/(\pi a^{2})$. We also employed a Navier-Stokes viscosity with $\nu= 10^{-5}$ \emph{on top of} the usual settings for shock dissipation with the viscosity switch, namely $\alpha_{\rm AV} \in [0,1]$ and $\beta_{\rm AV} = 2$. We use the Navier-Stokes viscosity implementation described in Section~\ref{sec:ns2nd}, which is the code default. We also employed the cubic spline kernel (with $h_{\rm fac} = 1.2$) rather than the quintic to reduce the computational expense (using the quintic made no difference to the results).
 
 Despite the different assumptions, the results in Figure~\ref{fig:planetdisc} are strikingly similar to those obtained with most of the grid-based codes in \citet{de-val-borroetal06}. The main difference is that our gap is shallower, which is not surprising since this is where resolution is lowest in SPH. There is also some difference in the evolution of the surface density, particularly at the inner boundary, due to the difference in assumed boundary conditions. However, the dense flow around the planet and in the shocks appear well resolved compared to the other codes. What is interesting is that the SPH codes used in the original comparison performed poorly on this test. This may be simply due to the low resolution employed, as the two SPH calculations used 250~000 and 300~000 particles, respectively, (though performed only in 2D rather than 3D), but given the extent of other differences between adopted setup and SPH algorithms, it is hard to draw firm conclusions.
 
As per the original comparison, Figure~\ref{fig:planetdisc} shows the estimated position of the planetary shocks from \citet{ogilvielubow02} plotted as dotted lines, namely
\begin{equation}
\phi(r,t) = \begin{cases}
 t - \frac{2}{3 \varepsilon} \left ( r^{3/2} - \frac32 \ln r - 1\right); & r > r_{\rm planet}; \\
 t + \frac{2}{3 \varepsilon} \left ( r^{3/2} - \frac32 \ln r - 1\right); & r < r_{\rm planet},
\end{cases}
\end{equation}
where $\varepsilon = 0.05$ is the disc aspect ratio. The disagreement between the shock position and the \citet{ogilvielubow02} solution seen in Figure~\ref{fig:planetdisc} was also found in every simulation shown in \citet{de-val-borroetal06}, so more likely reflects the approximate nature of the analytic solution rather than numerical error.

 We found this to be a particularly good test of the viscosity limiter, since there is both a shock and a shear flow present. Without the limiter in (\ref{eq:avsource}), we found the shock viscosity switch would simply trigger to $\alpha_{\rm AV} \approx 1$. The original \citet{morrismonaghan97} switch (Section~\ref{sec:switches}) also performs well on this test, suggesting that the velocity divergence is better able to pick out shocks in differentially rotating discs compared to its time derivative.

\subsection{Physical viscosity}
\label{sec:visctest}

\subsubsection{Taylor-Green vortex}
The Taylor-Green vortex \citep{taylorgreen37} consists of a series of counter-rotating vortices. We perform this test using four vortices set in a thin 3D slab. The initial velocity fields are given by: $v_x = v_0 \sin(2 \pi x) \cos(2 \pi y)$, $v_y = -v_0 \cos(2 \pi x) \sin(2 \pi y)$ with $v_0 = 0.1$. The initial density is uniform $\rho = 1$, and an isothermal equation of state is used ($P = c_{\rm s}^2 \rho$) with speed of sound $c_{\rm s}=1$. Viscosity will cause each component of the velocity field to decay at a rate $\propto \exp(-16 \pi^2 \nu t)$, where $\nu$ is the kinematic shear viscosity.

\begin{figure}
\centering
\includegraphics[width=\columnwidth]{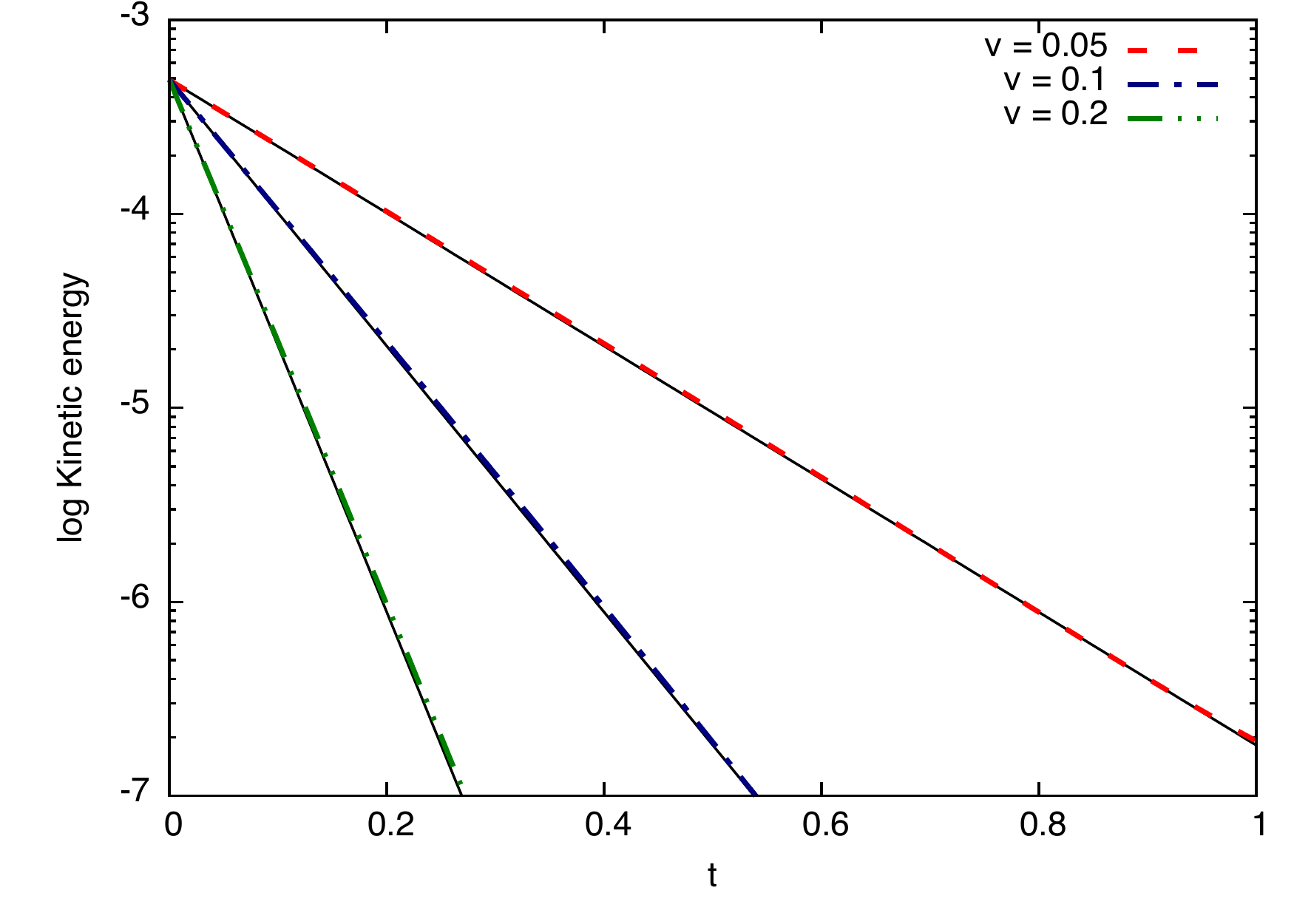}
\caption{Test of physical Navier-Stokes viscosity in the Taylor-Green vortex using kinematic shear viscosity $\nu = 0.05$, $0.1$, $0.2$. The exponential decay rate of kinetic energy may be compared to the analytic solution in each case (solid black lines), demonstrating that the calibration of physical viscosity in \textsc{Phantom} is correct.}
\label{fig:tgvortex}
\end{figure}

Figure~\ref{fig:tgvortex} shows the kinetic energy for a series of calculations using $\nu = 0.05$, $0.1$, $0.2$. In each case, the kinetic energy exponentially decays by several orders of magnitude. The corresponding analytic solutions are shown by the solid black lines for comparison, demonstrating that the implementation of physical viscosity in \textsc{Phantom} is correct.

\subsection{Sink particles}
\label{sec:sinktest}

\subsubsection{Binary orbit}
\label{sec:binaryorbit}
\begin{figure}
\centering
\includegraphics[width=\columnwidth]{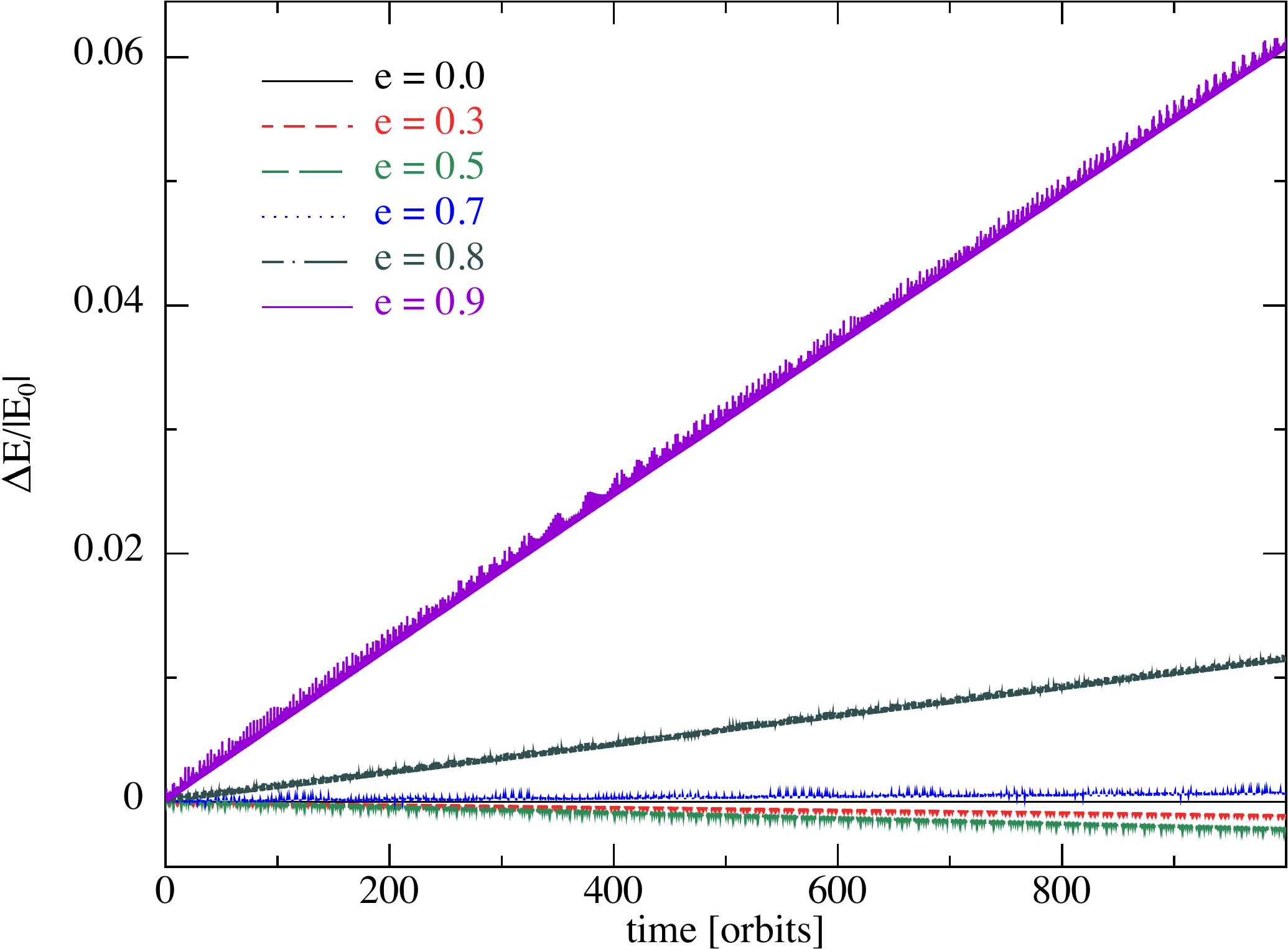}
\caption{Errors in energy conservation in a sink particle binary integration with code default parameters for the timestep control, showing the energy drift caused by the adaptive timestepping. Angular momentum is conserved to machine precision.}
\label{fig:sinkbinary}
\end{figure}

 Figure~\ref{fig:sinkbinary} shows the error in total energy conservation $\Delta E/\vert E_0 \vert$ for a set of simulations consisting of two sink particles set up in a binary orbit, a common test of $N$-body integrators \citep[e.g.][]{hmm95,quinnetal97,farrbertschinger07,dehnenread11}. We fix the initial semi-major axis $a=1$ with masses $m_1 = m_2 = 0.5$ and with the two sink particles initially at periastron, corresponding to ${\bm x}_1 = [- m_2/(m_1 + m_2)\Delta,0,0]$ and ${\bm x}_2 = [ m_1/(m_1 + m_2)\Delta,0,0]$ where $\Delta = a(1-e)$ is the initial separation. The corresponding initial velocities are ${\bm v}_1 = [0,- m_2/(m_1 + m_2)\vert v \vert,0]$ and ${\bm v}_2 = [ 0,m_1/(m_1 + m_2)\vert v \vert,0]$, where $\vert v \vert = \sqrt{a(1-e^2)(m_1 + m_2)}/\Delta$. The orbital period is thus $P = \sqrt{4\pi^2 a^3 / (G (m_1 + m_2))} = 2\pi$ in code units for our chosen parameters. Importantly, we use an adaptive timestep which is not time-symmetric so there remains some drift in the energy error which is absent if the timestep is constant (see e.g. \citealt{hmm95,quinnetal97} and \citealt{dehnenread11} for discussion of this issue).
 
 Figure~\ref{fig:sinkbinary} shows the error in total energy as a function of time for the first 1000 orbits for calculations with initial eccentricities of $e=0.0$ (a circular orbit), $0.3$, $0.5$, $0.7$ and $0.9$. Energy conservation is worse for more eccentric orbits, as expected, with $\Delta E/ \vert E_0 \vert \sim 6\%$ after 1000 orbits for our most extreme case ($e=0.9$). The energy error can be reduced arbitrarily by decreasing the timestep, so this is mainly a test of the default settings for the sink particle timestep control. For this problem the timestep is controlled entirely by (\ref{eq:dtphi}), where by default we use $\eta_\Phi = 0.05$, giving 474 steps per orbit for $e=0.9$. For simulations with more eccentric orbits we recommend decreasing $C_{\rm force}$ from the default setting of $0.25$ to obtain more accurate orbital dynamics.

In addition to calibrating the timestep constraint, Figure~\ref{fig:sinkbinary} also validates the sink particle substepping via the RESPA algorithm (Section~\ref{sec:respa}) since for this problem the ``gas" timestep is set only by the desired interval between output files (to ensure sufficient output for the figure we choose $\Delta t_{\rm max} = 1$, but we also confirmed that this choice is unimportant for the resultant energy conservation). This means that increasing the accuracy of sink particle interactions adds little or no cost to calculations involving gas particles.
 
 The corresponding plot for angular momentum conservation (not shown) merely demonstrates that angular momentum is conserved to machine precision ($\Delta L / \vert L \vert \sim 10^{-15}$), as expected. Importantly, angular momentum remains conserved to machine precision even with adaptive timestepping.
 
\subsubsection{Restricted three-body problem}
 \citet{chinchen05} proposed a more demanding test of $N$-body integrators, consisting of a test particle orbiting in the potential of a binary on a fixed circular orbit. We set up this problem with a single sink particle with ${\bm x} = [0,0.0580752367,0]$ and ${\bm v} = [0.489765446,0,0]$, using the time-dependent binary potential as described in Section~\ref{sec:extbinary} with $M = 0.5$. This is therefore a good test of the interaction between a sink particle and external potentials in the code, as well as the sink particle timestepping algorithm. For convenience we set the sink mass $m=1$ and accretion radius $r_{\rm acc} = 0.1$, although both are irrelevant to the problem.
 
\begin{figure}
\centering
\includegraphics[width=\columnwidth]{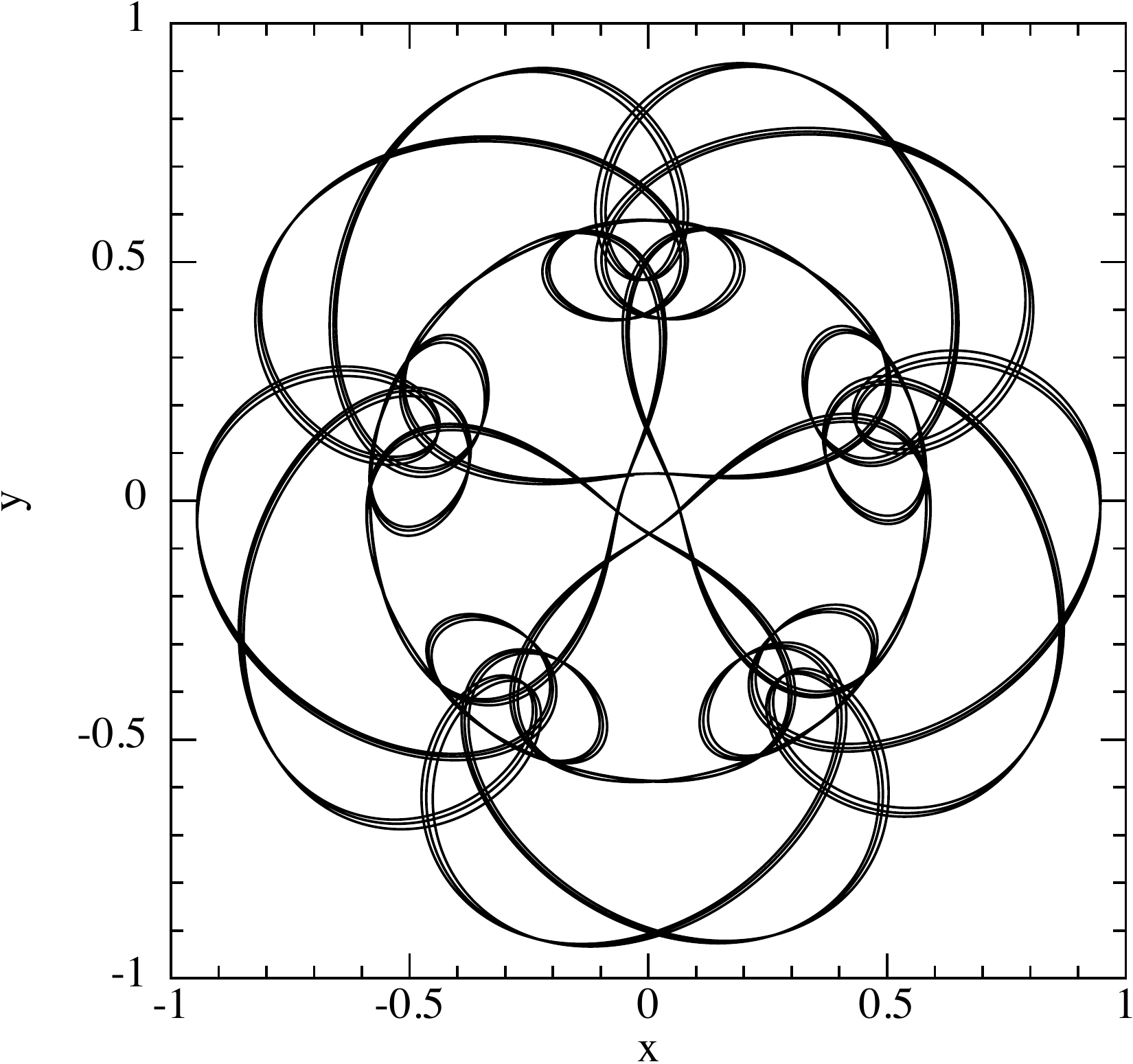}
\caption{Orbit of a sink particle in the restricted 3-body problem from \citet{chinchen05} using default code parameters. This tests both the time-integrator for sink particles and the interaction with a time-dependent binary potential (Section~\ref{sec:extbinary}). The trajectory of the sink is plotted every timestep for 3 periods ($t=27\pi$).}
\label{fig:chinesecoin}
\end{figure}

 Figure~\ref{fig:chinesecoin} shows the resulting orbit using the default code parameters, where we plot the trajectory of the sink particle up to $t=27\pi$, as in Figures 1 and 2 of \citet{chinchen05}. Considering that we use only a second-order integrator, the orbital trajectory is remarkably accurate, showing no chaotic behaviour and only a slight precession consistent or better than the results with some of the fourth order schemes shown in their paper (albeit computed with a larger timestep). We are thus satisfied that our time integration scheme and the associated timestep settings can capture complex orbital dynamics with sufficient accuracy.
 
\subsection{Magnetohydrodynamics}
\label{sec:mhdtests}


\subsubsection{3D circularly polarised Alfv\'en wave}
\label{sec:alfven}
 \citet{toth00} introduced the circularly polarised Alfv\'en wave test, an exact non-linear solution to the MHD equations which can therefore be performed using a wave of arbitrarily large amplitude. Most results of this test are shown in 2D \citep[e.g.][]{pricemonaghan05,rosswogprice07,price12,triccoprice13}. Here we follow the 3D setup outlined by \citet{gardinerstone08}. We use a periodic domain of size $L \times L/2 \times L/2$ where $L=3$, with the wave propagation direction defined using angles $a$ and $b$ where $\sin a = 2/3$ and $\sin b = 2/\sqrt{5}$, with the unit vector along the direction of propagation given by ${\bm r} = [\cos a \cos b, \cos a \sin b, \sin a]$. We use an initial density $\rho = 1$, an adiabatic equation of state with $\gamma = 5/3$ and $P = 0.1$. We perform the `travelling wave test' from \citet{gardinerstone08}, where the wavelength $\lambda = 1$ and the vectors $[v_{1}, v_{2}, v_{3}] = [0, 0.1 \sin(2\pi x_{1}/\lambda), 0.1 \cos(2\pi x_{1}/\lambda)]$ and $[B_{1}, B_{2}, B_{3}] = [1, 0.1 \sin(2\pi x_{1}/\lambda), 0.1 \cos(2\pi x_{1}/\lambda)]$ are projected back into the $x$, $y$ and $z$ components using the transformations given by \citep{gardinerstone08}
\begin{align}
x & = x_{1} \cos a \cos b - x_{2} \sin b - x_{3} \sin a \cos b, \nonumber \\
y & = x_{1} \cos a \sin b + x_{2} \cos b - x_{3} \sin a \sin b, \nonumber \\
z & = x_{1} \sin a + x_{3} \cos a.
\end{align}

\begin{figure}
   \centering
   \includegraphics[width=\columnwidth]{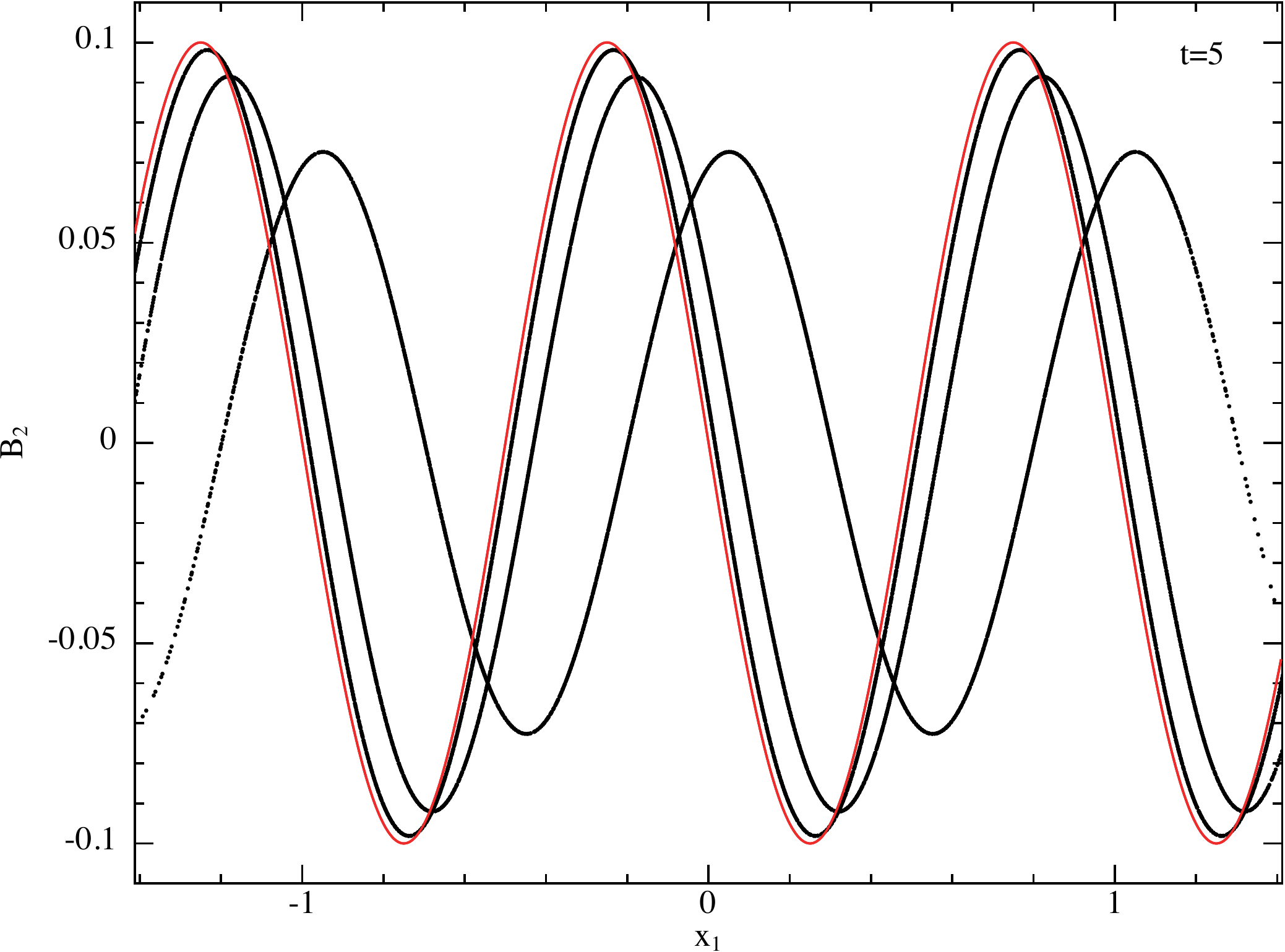} 
   \caption{Results of the 3D circularly polarised Alfven wave test after 5 periods, showing perpendicular component of the magnetic field on all particles (black dots) as a function of distance along the axis parallel to the wave vector. Results are shown using $32 \times 18 \times 18$, $64 \times 36 \times 39$ and $128 \times 74 \times 78$ particles (most to least damped, respectively), compared to the exact solution given by the solid red line. Convergence is shown in Figure~\ref{fig:alfvenconv}.}
\label{fig:alfven}
\end{figure}

  Figure~\ref{fig:alfven} shows the results of this test using $32 \times 18 \times 18$, $64 \times 36 \times 39$ and $128 \times 74 \times 78$ particles initially set on a close packed lattice, compared to the exact solution given by the red line (the same as the initial conditions for the wave). We plot the transverse component of the magnetic field $B_{2}$ as a function of $x_{1}$, where $B_{2}~\equiv~(B_{y}-2B_{x})/\sqrt{5}$ and $x_{1}\equiv(x + 2y + 2z)/3$ for our chosen values of $a$ and $b$. There is both a dispersive and dissipative error, with the result converging in both phase and amplitude towards the undamped exact solution as the resolution is increased.
  
\begin{figure}
   \centering
   \includegraphics[width=0.8\columnwidth]{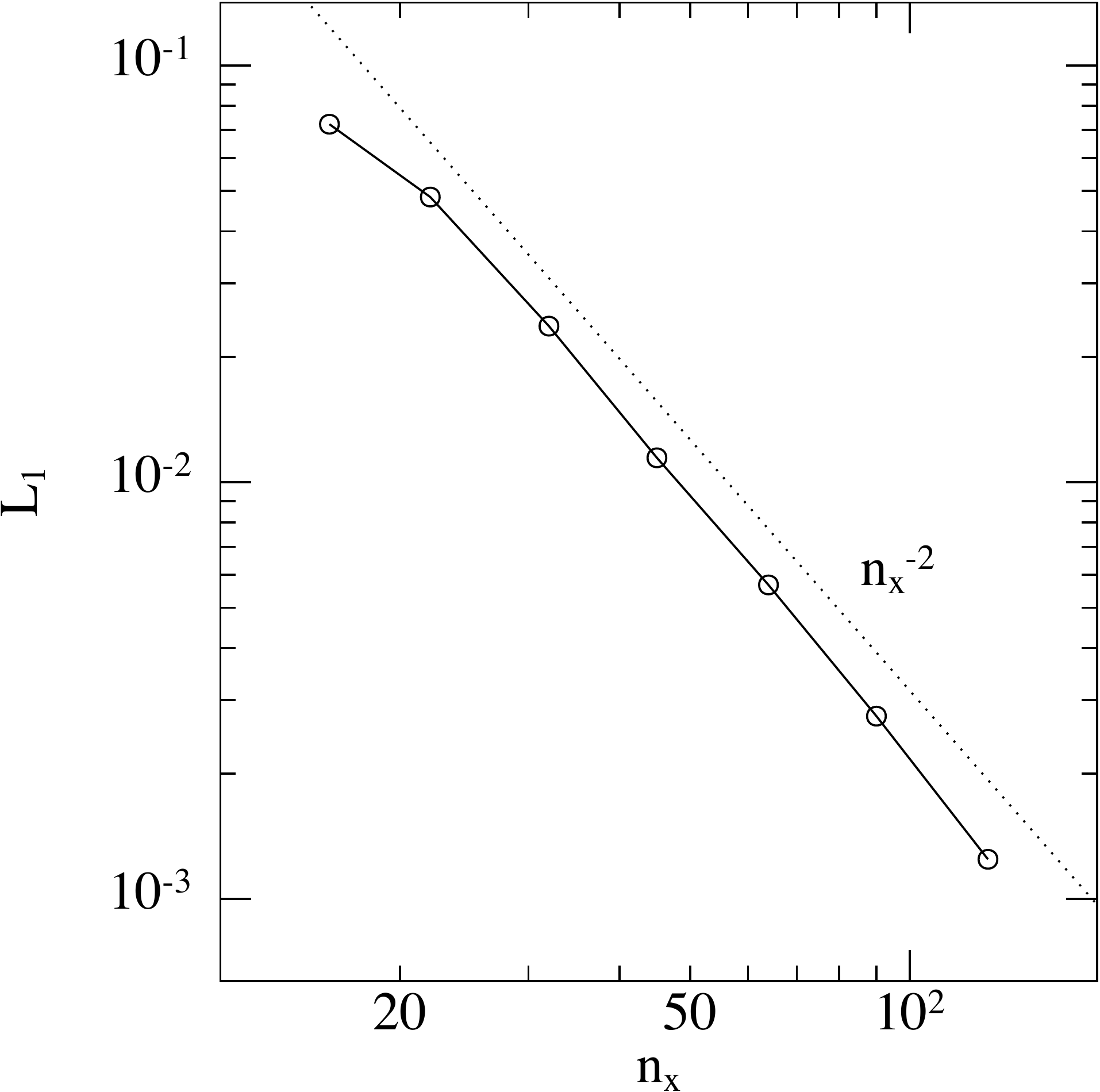} 
   \caption{Convergence in the 3D circularly polarised Alfven wave test, showing the $L_{1}$ error as a function of the number of particles along the x-axis, $n_{x}$ alongside the expected slope for second order convergence (dotted line). Significantly, this demonstrates second order convergence \emph{with all dissipation switched on}.}
\label{fig:alfvenconv}
\end{figure}

 Figure~\ref{fig:alfvenconv} shows a convergence study on this problem, showing, as in \citet{gardinerstone08}, the $L_{1}$ error as a function of the number of particles in the $x-$direction. The convergence is almost exactly second order. This is significant because we have performed the test with code defaults for all dissipation and divergence cleaning terms. This plot therefore demonstrates the second order convergence of both the viscous and resistive dissipation in \textsc{Phantom} (see Sections~\ref{sec:switches} and \ref{sec:artres}). By comparison, the solution shown by \citet{pricemonaghan05} (Figure 6 in their paper) was severely damped when artificial resistivity was applied.

 As noted by \citet{pricemonaghan05} and illustrated in Figure~12 of \citet{price12}, this problem is unstable to the SPMHD tensile instability \citep[e.g.][]{phillipsmonaghan85} in the absence of force correction terms since the plasma $\beta \equiv P/ \frac12 B^{2} \approx 0.2$. Our results demonstrate that the correction term (Section~\ref{sec:spmhd}) effectively stabilises the numerical scheme without affecting the convergence properties.

\begin{figure*}
   \centering
   \includegraphics[width=0.75\textwidth]{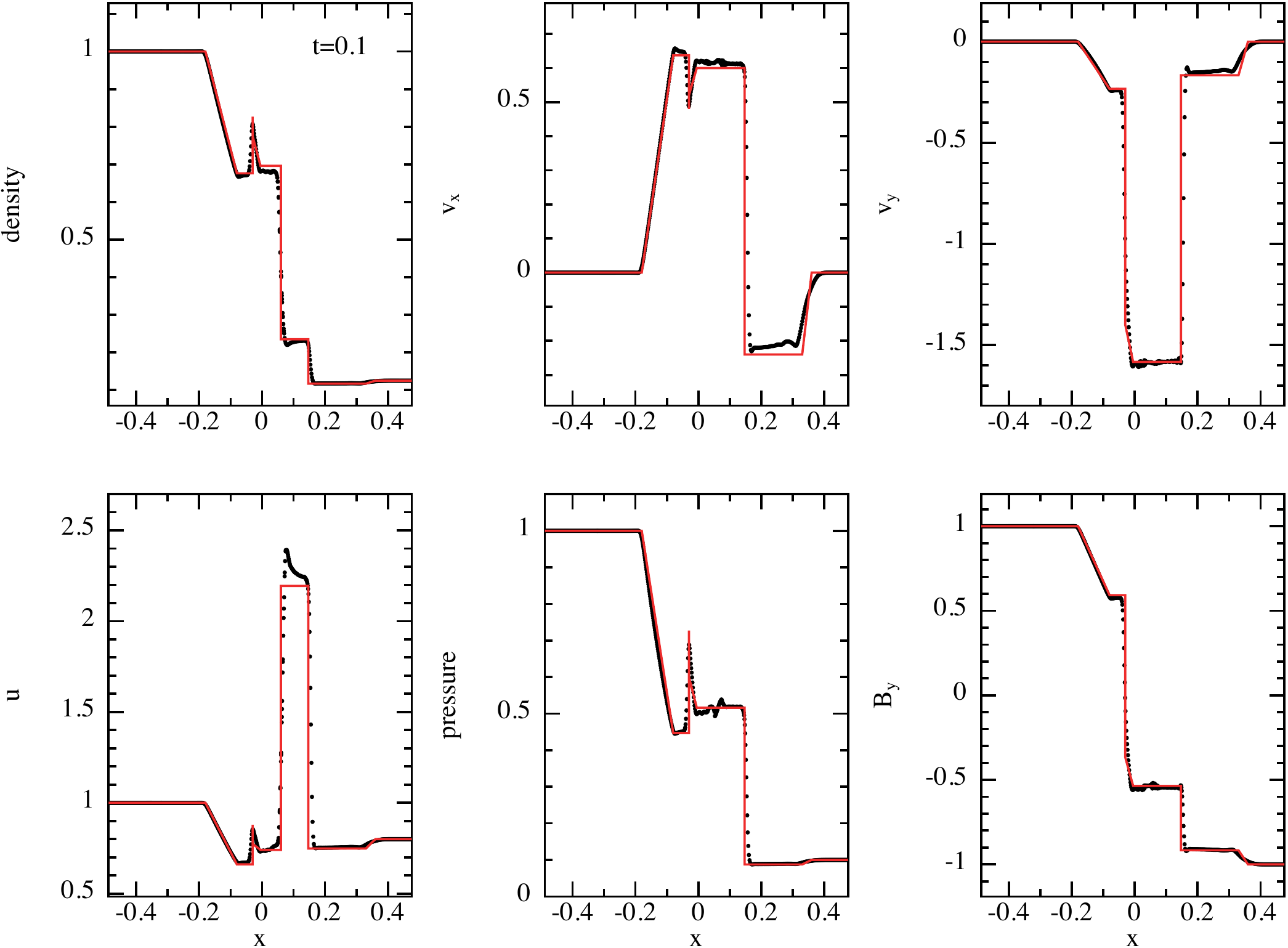} 
   \caption{Results of the \citet{briowu88} shock tube test in 3D, showing projection of all particles (black dots) compared to the reference solution (red line). The problem is set up with $[\rho, P, B_{y}, B_{z}] = [1,1,1,0]$ for $x \leq 0$ and $[\rho, P, B_{y}, B_{z}] = [0.125,0.1,-1,0]$ for $x > 0$ with zero initial velocities, $B_{x} = 0.75$ and $\gamma = 2$. The density contrast is initialised using equal mass particles placed on a close packed lattice with $256 \times 24 \times 24$ particles initially in $x \in [-0.5,0]$ and $128 \times 12 \times 12$ particles initially in $x \in [0,0.5]$.}
\label{fig:briowu}
\end{figure*}

\begin{figure*}
   \centering
   \includegraphics[width=0.75\textwidth]{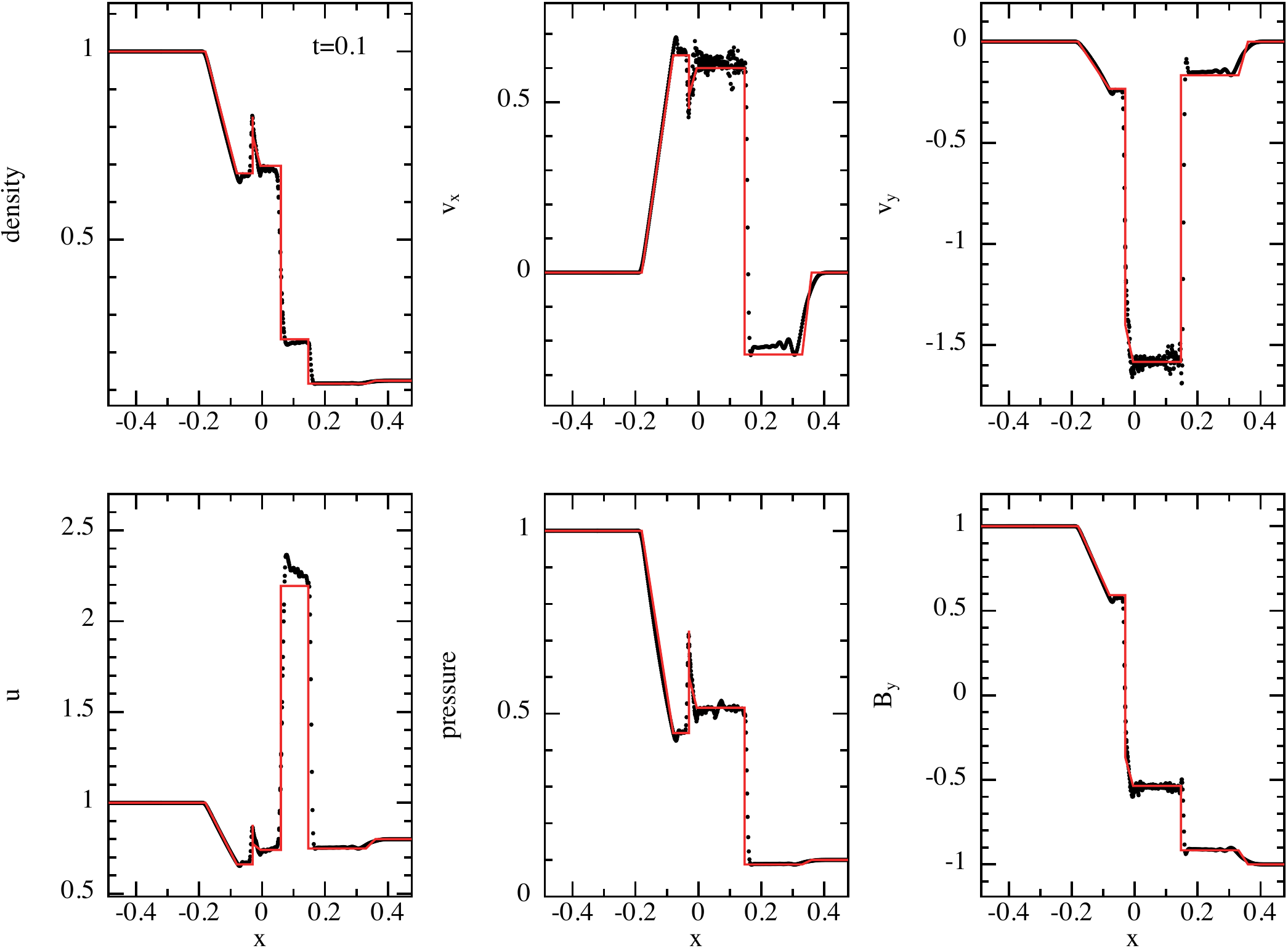} 
   \caption{As in Figure~\ref{fig:briowu} but using code defaults which give second order convergence away from shocks. Some additional noise in the velocity field is visible, while otherwise the solutions are similar.}
\label{fig:briowusw}
\end{figure*}

\begin{figure*}
   \centering
   \includegraphics[width=0.9\textwidth]{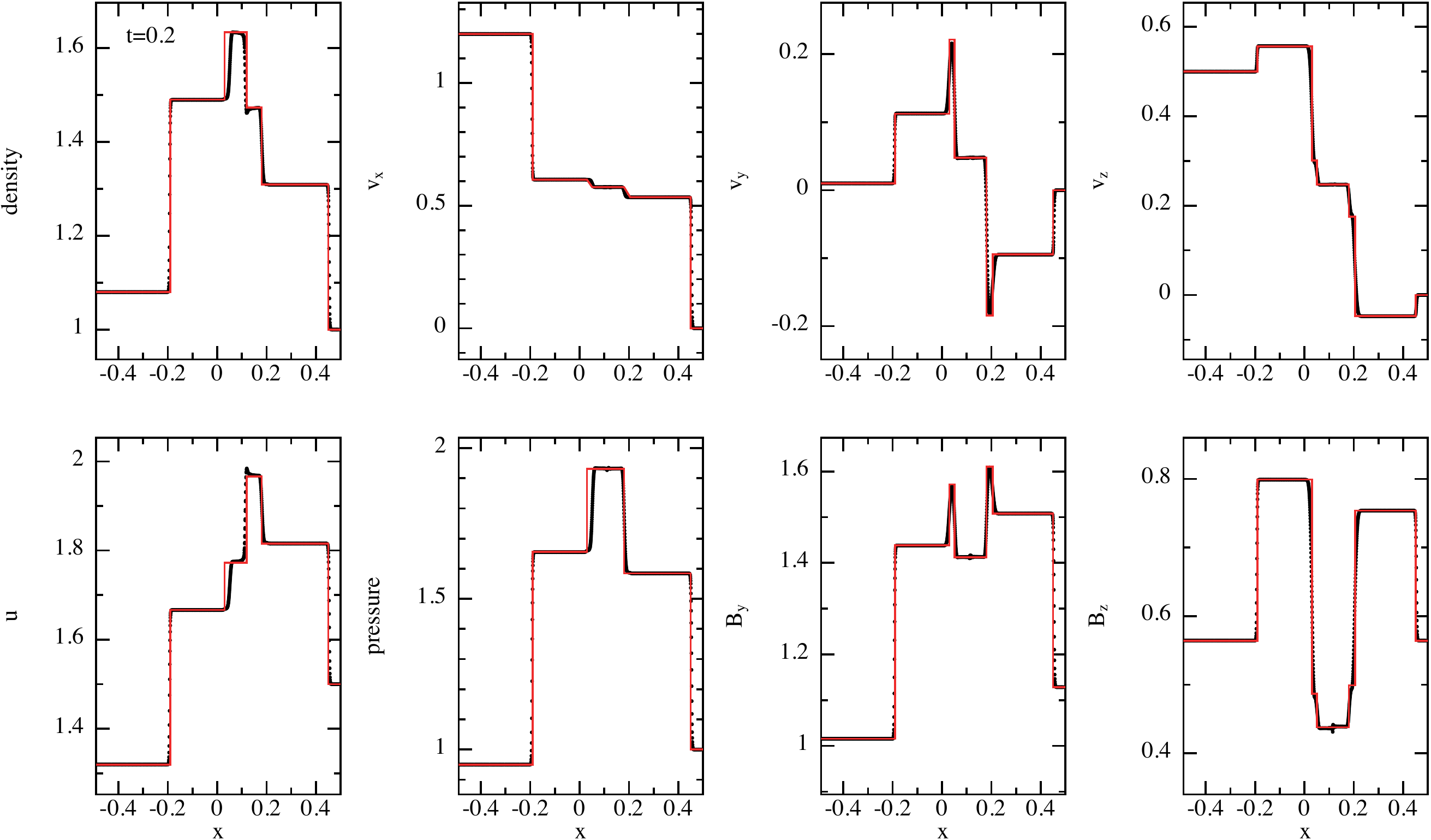} 
   \caption{Results of the 7-discontinuity MHD shock tube test 2a from \citet{ryujones95} in 3D, showing projection of all particles (black dots) compared to the reference solution (red line). The problem is set up with $[\rho, P, v_{x}, v_{y}, v_{z}, B_{x}, B_{y}, B_{z}] = [1.08,0.95,1.2,0.01,0.5,2/\sqrt{4\pi},3.6/\sqrt{4\pi},2/\sqrt{4\pi}]$ for $x \leq 0$ and $[\rho, P, v_{x}, v_{y}, v_{z}, B_{x}, B_{y}, B_{z}] = [1,1,0,0,0,2/\sqrt{4\pi},4/\sqrt{4\pi},2/\sqrt{4\pi}]$ for $x > 0$ with $\gamma = 5/3$. The density contrast is initialised using equal mass particles placed on a close packed lattice with $379 \times 24 \times 24$ particles initially in $x \in [-0.5,0]$ and $238 \times 12 \times 12$ particles initially in $x \in [0,0.5]$.}
\label{fig:mshk3}
\end{figure*}

\begin{figure*}
   \centering
   \includegraphics[width=0.9\textwidth]{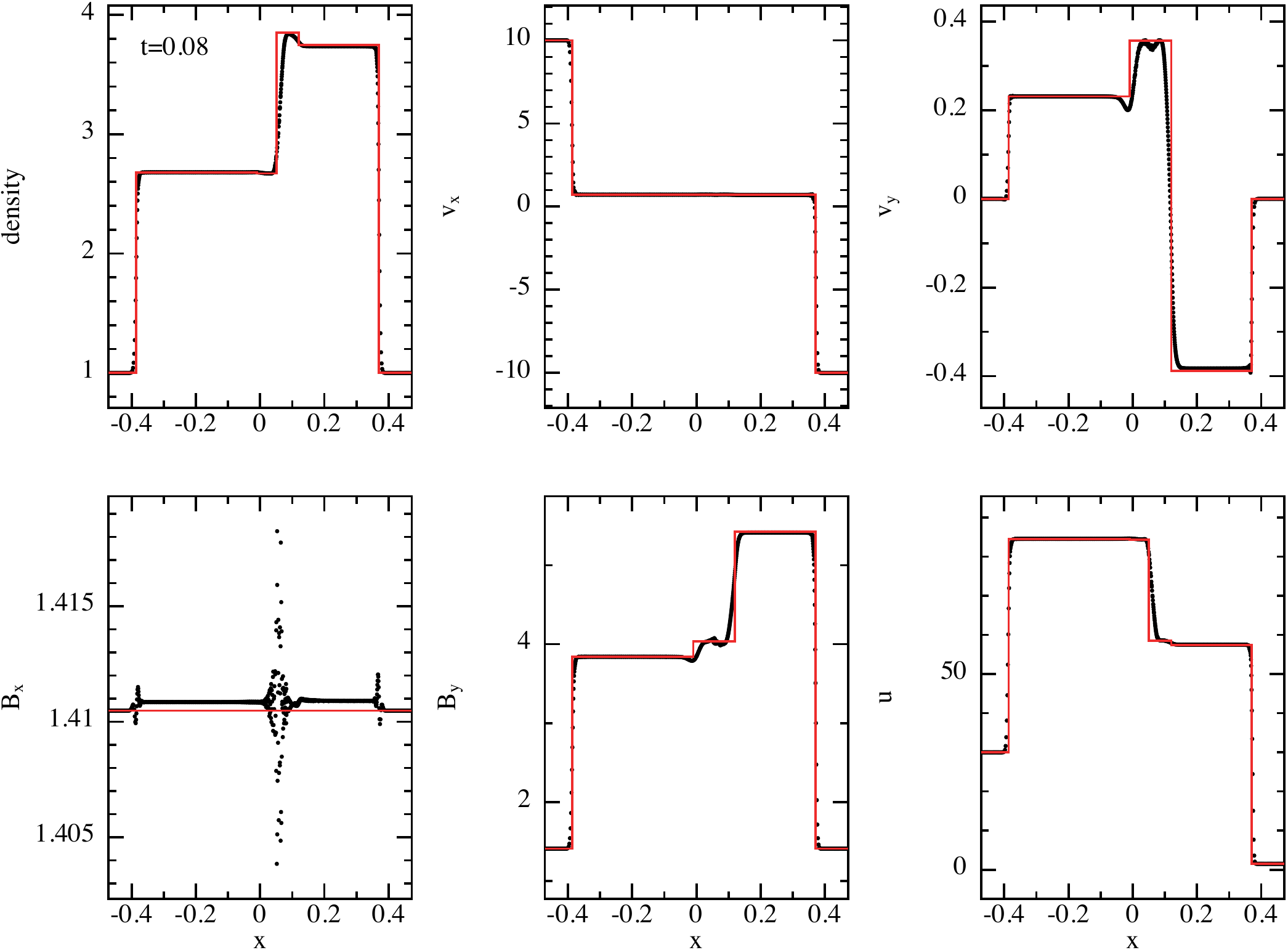} 
   \caption{Results of the MHD shock tube test 1a from \citet{ryujones95} in 3D, showing projection of all particles (black dots) compared to the reference solution (red line). The problem is set up with $[\rho, P, v_{x}, v_{y}, v_{z}, B_{x}, B_{y}, B_{z}] = [1.0,20.0,10,0,0,5/\sqrt{4\pi},5/\sqrt{4\pi},0]$ for $x \leq 0$ and $[\rho, P, v_{x}, v_{y}, v_{z}, B_{x}, B_{y}, B_{z}] = [1,1,-10,0,0,5/\sqrt{4\pi},5/\sqrt{4\pi},0]$ for $x > 0$ with $\gamma = 5/3$. We show results using $652 \times 12 \times 12$ particles. This has historically proven difficult for SPMHD codes. We find ${\cal L}_1$ within 1\% of the reference solution except in $v_y$ (5\%).}
\label{fig:mshk7}
\end{figure*}

\subsubsection{MHD shock tubes}
\label{sec:mhdshocks}
 The classic \citet{briowu88} shock tube test generalises the Sod shock tube (Section~\ref{sec:sod}) to MHD. It has provoked debate over the years (e.g.~\citealt{wu88a,daiwoodward94a,fallekomissarov01,tyy13}) because of the presence of a compound slow shock and rarefaction in the solution, which is stable only when the magnetic field is coplanar and there is no perturbation to the tangential ($B_{z}$) magnetic field \citep{bkp96}. Whether or not such solutions can exist in nature remains controversial (e.g.~\citealt{fengetal07}). Nevertheless it has become a standard benchmark for numerical MHD \citep[e.g.][]{stoneetal92,daiwoodward94a,balsara98,ryujones95}. It was first used to benchmark SPMHD by \citet{pricemonaghan04,pricemonaghan04a} and 1.5D results on this test with SPMHD, for comparison, can be found in e.g. \citet{pricemonaghan05,rosswogprice07,dolagstasyszyn09,price10} and \citet{vkp14}, with 2D versions shown in \citet{price12}, \citet{triccoprice13} and \citet{tpb16}. We handle the boundary conditions by setting the first and last few planes of particles to be `boundary particles' (Section~\ref{sec:types}), meaning that the gas properties on these particles are fixed.
 
 Figure~\ref{fig:briowu} shows the results of the \citet{briowu88} problem using \textsc{Phantom}, performed in 3D with $256 \times 24 \times 24$ particles initially in $x \in [-0.5,0]$ and $128 \times 12 \times 12$ particles initially in $x \in [0,0.5]$ set on close packed lattices with purely discontinuous initial conditions in the other variables (see caption). The projection of all particles onto the $x$-axis are shown as black dots, while the red lines shows a reference solution taken from \citet{balsara98}. Figure~\ref{fig:briowu} shows the results when a constant $\alpha^{\rm AV} = 1$ is employed, while Figure~\ref{fig:briowusw} shows the results with default code parameters, giving second order dissipation away from shocks. For constant $\alpha^{\rm AV}$ (Figure~\ref{fig:briowu}), we find the strongest deviation from the reference solution is in $v_x$, with ${\cal L}_1 = 0.015$ and ${\cal L}_2 = 0.065$ at this resolution. The remaining ${\cal L}_2$ errors are within 5 per cent of the reference solution, while the ${\cal L}_1$ norms are all smaller than 1.5 per cent in the other variables. Similar errors are found with code defaults (Figure~\ref{fig:briowusw}), with ${\cal L}_2 = 0.074$ in $v_x$ and ${\cal L}_1$ norms smaller than 1.6 per cent in all variables. That is, our solutions are within 1.6 per cent of the reference solution. Using the default Courant factor of 0.3, total energy is conserved to better than 0.5\%, with maximum $|\Delta E|/|E_0| = 4.2 \times 10^{-3}$ up to $t=0.1$.
  
 Figure~\ref{fig:mshk3} shows the result of the ``7 discontinuity'' test from \citet{ryujones95}. This test is particularly sensitive to over-dissipation by resistivity given the sharp jumps in the transverse magnetic and velocity fields. A reference solution with intermediate states taken from the corresponding table in \citet{ryujones95} is shown by the red lines for comparison. Here the boundary particles are moved with a fixed velocity in the $x$-direction. The largest deviation from the reference solution is in the $v_y$ component, mainly due to the over-dissipation of the small spikes, with ${\cal L}_1 = 0.02$ and ${\cal L}_2 = 0.07$ at this resolution. The remaining ${\cal L}_2$ errors are within 3\% of the reference solution while the ${\cal L}_1$ norms are smaller than 0.9\% in all other variables.

 Finally, Figure~\ref{fig:mshk7} shows the results of shock tube 1a from \citet{ryujones95}. This test is interesting because it has historically proven to be a difficult test for SPMHD codes in more than 1D. In particular, to obtain reasonable results on this problem \citet{pricemonaghan05} had to employ both an explicit shear viscosity term and a large neighbour number. Even then the jumps were found to show significant deviation from the analytic solution (see figure~10 in \citealt{pricemonaghan05}). We did find that using $h_{\rm fac} = 1.2$ significantly improved the results of this test compared to our default $h_{\rm fac} = 1.0$ for the quintic kernel. Likewise we found the results with the cubic spline kernel could be noisy. However, Figure~\ref{fig:mshk7} demonstrates that with only this change to the default code parameters we can obtain results with ${\cal L}_1$ errors of better than 4.4\% in $v_y$ and less than 0.9\% in all other variables at this resolution. The bottom left panel shows the errors induced in $B_x$, with the largest error (${\cal L}_\infty$) only 0.6\%. This is a substantial improvement over the 2D results shown in \citet{pricemonaghan05}.

\begin{figure*}
   \centering
   \includegraphics[width=\textwidth]{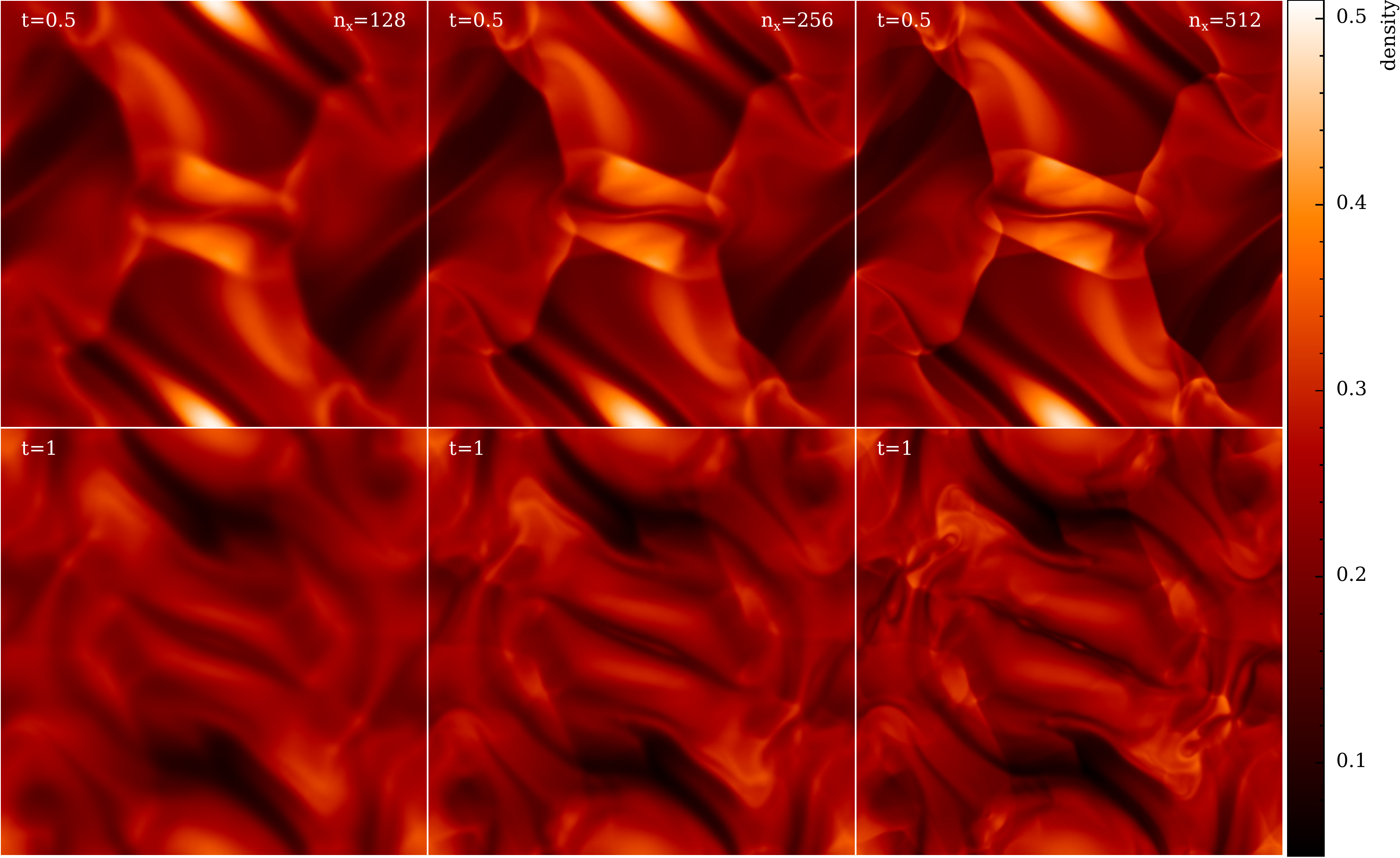} 
   \caption{Density in a $z=0$ cross section of the Orszag-Tang vortex test performed in 3D. Results are shown at $t=0.5$ (top) and $t=1$ (bottom) at a resolution of $128 \times 148 \times 12$, $256 \times 296 \times 12$ and $512 \times 590 \times 12$ particles (left to right). Compare e.g. to Figure~4 in \citet{daiwoodward94} or Figure~22 in \citet{stoneetal08}, while improvements in the SPMHD method over the last decade can be seen by comparing to Figure~14 in \citet{pricemonaghan05}.}
\label{fig:orstang}
\end{figure*}

\begin{figure}
   \centering
   \includegraphics[width=\columnwidth]{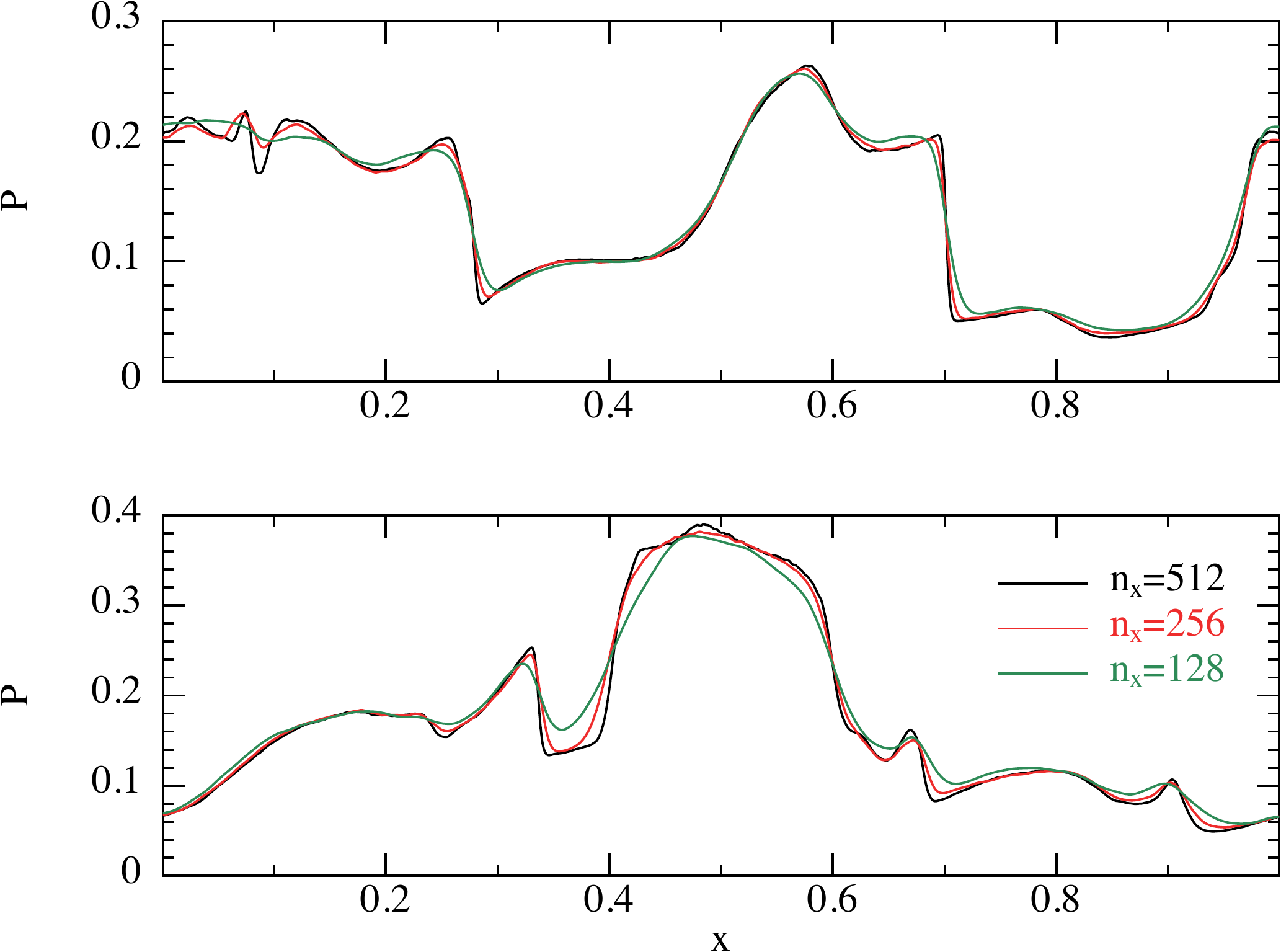} 
   \caption{Horizontal slices of pressure shown at t=0.5 in the Orszag-Tang vortex test. We show cuts along $y=0.3125$ (top) and $y=0.4277$ (bottom) in the $z=0$ plane for three different numerical resolutions (see legend).}
\label{fig:orstangcut}
\end{figure}

\subsubsection{Orszag-Tang vortex}
 The Orszag-Tang vortex \citep{orszagtang79,dahlburgpicone89,piconedahlburg91} has been used widely to test astrophysical MHD codes \citep[e.g.][]{stoneetal92,rjf95,daiwoodward98,toth00,londrillodel-zanna00}. Similar to our hydrodynamic tests, we perform a 3D version of the original 2D test problem, similar to the `thin box' setup used by \citet{dolagstasyszyn09}. Earlier results on this test with 2D SPMHD can be found in \citet{pricemonaghan05}, \citet{rosswogprice07}, \citet{triccoprice12, triccoprice13}, \citet{tpb16} and \citet{hopkinsraives16} and in 3D by \citet{dolagstasyszyn09} and \citet{price10}.
 
  The setup is a uniform density, periodic box $x,y \in [-0.5,0.5]$ with boundary in the $z$ direction set to $\pm 2\sqrt{6}/n_x$, where $n_x$ is the initial number of particles in $x$, in order to setup the 2D problem in 3D (c.f. Section~\ref{sec:kh}). We use an initial plasma $\beta_{0} = 10/3$, initial Mach number $\mathcal{M}_{0} = v_{0}/c_{\rm s, 0} = 1$, initial velocity field $[v_{x}, v_{y}, v_{z}] = [-v_{0} \sin (2\pi y'), v_{0} \sin(2\pi x'), 0]$ and magnetic field $[B_{x}, B_{y}, B_{z}] = [-B_{0} \sin (2\pi y'), B_{0} \sin(4\pi x'), 0]$, where $v_{0} = 1$, $B_{0} = 1/\sqrt{4\pi}$, $x' \equiv x - x_{\min}$ and $y' \equiv y - y_{\min}$; giving $P_{0} = \frac12 B_{0}^{2} \beta_{0} \approx 0.133$ and $\rho_{0} = \gamma P_{0} \mathcal{M}_{0} \approx 0.221$. We use an adiabatic equation of state with $\gamma = 5/3$.

   Figure~\ref{fig:orstang} shows the results at $t=0.5$ (top row) and at $t=1$ (bottom) at resolutions of $n_x$ = 128, 256, and 512 particles (left to right). At $t=0.5$, the main noticeable change as the resolution is increased is that the shocks become more well defined, as does the dense filament consisting of material trapped in the reconnecting layer of magnetic field in the centre of the domain. This current sheet eventually becomes unstable to the tearing mode instability \citep[e.g.][]{fkr63, syrovatskii81, priest85}, seen by the development of small magnetic islands or `beads' at $t=1$ at high resolution (bottom right panel; c.f. \citealt{pps89}). The appearance of these islands occurs only at high resolution and when the numerical dissipation is small (compare to the results using Euler potentials in 2D shown in Figure~13 of \citealt{triccoprice12}), indicating that our implementation of artificial resistivity (Section~\ref{sec:artres}) and divergence cleaning (Section~\ref{sec:spmhd}) are effective in limiting the numerical dissipation. 
   
   One other feature worth noting is that the slight `ringing' behind the shock fronts visible in the results of \citet{pricemonaghan05} is absent from the low resolution calculation. This is because the \citet{cullendehnen10} viscosity switch does a better job of detecting and responding to the shock compared to the previous \citet{morrismonaghan97}-style switch used in that paper. It is also worth noting that the results on this test, in particular the coherence of the shocks, are noticeably worse \emph{without} artificial resistivity, indicating that a small amount of dissipation in the magnetic field \emph{is} necessary to capture MHD shocks correctly in SPMHD \citep[c.f.][]{pricemonaghan04,pricemonaghan05,triccoprice12}.
   
   Figure~\ref{fig:orstangcut} shows horizontal slices of the pressure at $t=0.5$, showing cuts along $y=0.3125$ (top) and $y=0.4277$ (bottom) following (e.g.) \citet{londrillodel-zanna00} and \citet{stoneetal08}. The main difference at higher resolution is that the shocks become sharper and more well defined. Most of the smooth flow regions are converged with $n_x=256$ (i.e. the red and black lines are indistinguishable), but the parts of the flow where dissipation is important are can be seen to converge more slowly. This is expected.
 
\begin{figure*}
   \centering
   \includegraphics[width=0.75\textwidth]{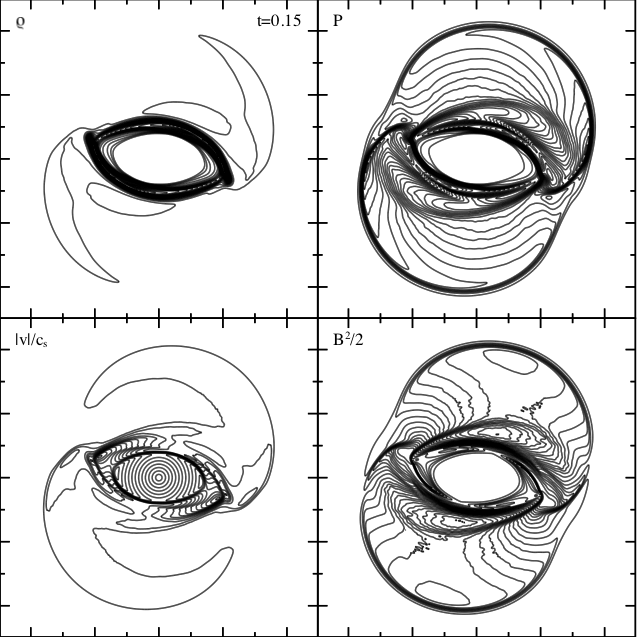} 
   \caption{Density, pressure, Mach number and magnetic pressure shown in a $z=0$ cross section at $t=0.15$ in the 3D MHD rotor problem, using $n_{x} = 256$, equivalent to $\sim 300^{2}$ resolution elements in 2D. The plots show 30 contours with limits identical to those given by \citet{toth00}; $0.483 < \rho < 12.95$, $0.0202 < P < 2.008$, $0 < \vert v \vert / c_{\rm s} < 1.09$ and $0 < \frac12 B^{2} < 2.642$.}
\label{fig:rotor}
\end{figure*}

\subsubsection{MHD rotor problem}
\label{sec:rotor}
 \citet{balsaraspicer99} introduced the `MHD rotor problem' to test the propagation of rotational discontinuities. Our setup follows \citet{toth00}'s `first rotor problem' as used by \citet{pricemonaghan05}, except that we perform the test in 3D. A rotating dense disc of material with $\rho = 10$ is set up with cylindrical radius $R = 0.1$, surrounded by a uniform periodic box $[x,y] \in [-0.5,0.5]$ with the $z$ boundary set to $[-\sqrt{6}/(2n_{x}),\sqrt{6}/(2n_{x})]$, or 12 particle spacings on a close packed lattice. The surrounding medium has density $\rho = 1$. Initial velocities are $v_{x, 0} = -v_{0} (y - y_{0})/r$ and $v_{y, 0} = v_{0} (x - x_{0})/r$ for $r < R$, where $v_{0} = 2$ and $r = \sqrt{x^{2} + y^{2}}$. The initial pressure $P=1$ everywhere while the initial magnetic field is given by $[B_{x}, B_{y}, B_{z}]~=~[5/\sqrt{4\pi}, 0, 0]$ with $\gamma = 1.4$. We set up the initial density contrast unsmoothed, as in \citet{pricemonaghan05}, by setting up two uniform lattices of particles masked to the initial cylinder, with the particle spacing adjusted inside the cylinder by the inverse cube root of the density contrast. At a resolution of $n_{x} = 256$ particles for the closepacked lattice, this procedure uses 1~145~392 particles, equivalent to a 2D resolution of $\sim 300^{2}$, inbetween the $200^{2}$ results shown in \citet{toth00} and \citet{pricemonaghan05} and the $400^{2}$ used in \citet{stoneetal08}.
 
  Figure~\ref{fig:rotor} presents the results of this test, showing 30 contours in density, pressure, Mach number and magnetic pressure using limits identical to those given in \citet{toth00}. The symmetry of the solution is preserved by the numerical scheme and the discontinuities are sharp, as discussed in \citet{stoneetal08}. The contours we obtain with \textsc{Phantom} are noticeably less noisy than the earlier SPMHD results given in \citet{pricemonaghan05}, a result of the improvement in the treatment of dissipation and divergence errors in SPMHD since then (c.f. Section~\ref{sec:mhd}; see also recent results in \citealt{tpb16})

\subsubsection{Current loop advection}
\label{sec:jadvect}
 The current loop advection test was introduced by \citet{gardinerstone05,gardinerstone08}, and regarded by \citet{stoneetal08} as the most discerning of their code tests. We perform this test in 3D, as in the `first 3D test' from \citet{stoneetal08} by using a thin 3D box with non-zero $v_{z}$. The field setup is with a vector potential $A_{z} = A_{0} (R - r)$ for $r < R$, giving $B_{x} = -A_{0} y/r$, $B_{y} = A_{0} x/r$ and $B_{z} = 0$ where $r = \sqrt{x^{2} + y^{2}}$, $R=0.3$ and we use $A_{0} = 10^{-3}$, $\rho_{0} = 1$, $P_{0} = 1$ and an adiabatic equation of state with $\gamma = 5/3$. We use a domain $[x, y, z] \in [-1:1,-0.5:0.5,-\sqrt{6}/(2n_{x}):\sqrt{6}/(2n_{x})]$ with $[v_{x}, v_{y}, v_{z}] = [2, 1, 0.1 /\sqrt{5}]$. The test is difficult mainly because of the cusp in the vector potential gradient at $r=R$ leading to a cylindrical current sheet at this radius. The challenge is to advect this infinite current without change (in numerical codes the current is finite but with a magnitude that increases with resolution). We choose the resolution to be comparable to \citet{stoneetal08}.
 
\begin{figure}
   \centering
   \includegraphics[width=\columnwidth]{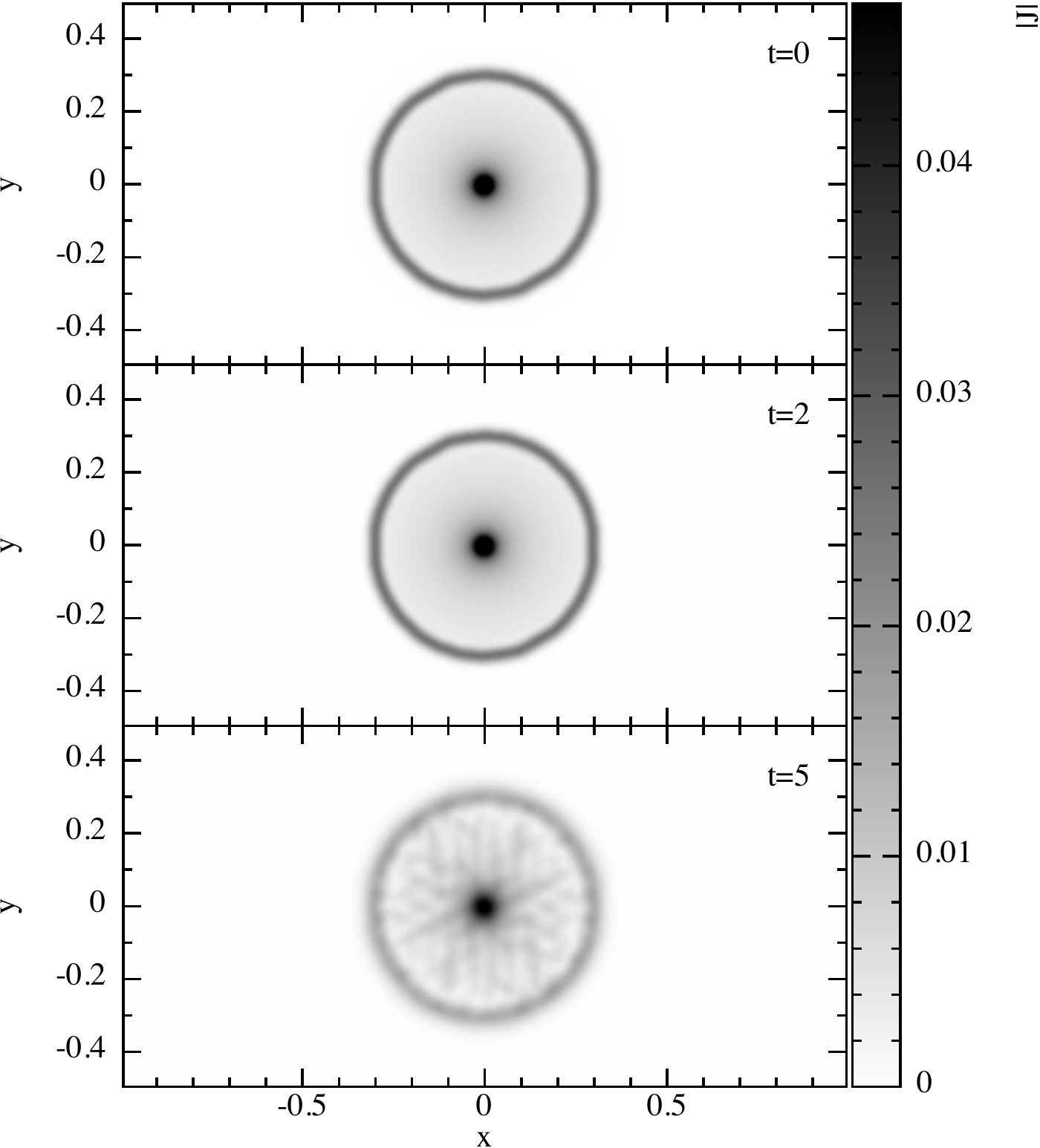} 
   \caption{Magnitude of the current density $\vert \nabla \times {\bm B} \vert$ in the current loop advection test performed in 3D, showing comparing the initial conditions (top) to the result after two (middle) and five (bottom) crossings of the box, using $128 \times 74 \times 12$ particles. \emph{Full dissipation, shock capturing and divergence cleaning terms were applied for this test}, without which the advection is exact to machine precision. The advection is affected mainly by the divergence cleaning acting on the outer (infinite) current sheet.}
   \label{fig:jadvect}
\end{figure}

 For SPMHD, this is mainly a test of the shock dissipation and divergence cleaning terms, since in the absence of these terms the advection can be computed to machine precision (c.f. 2D results shown in \citealt{rosswogprice07} and Figure 11 of \citealt{price12}, shown after one thousand crossings of the computational domain). Figure~\ref{fig:jadvect} shows the results of this test in \textsc{Phantom} with $128 \times 74 \times 12$ particles after two box crossings, computed \emph{with all dissipation and divergence cleaning terms switched on}, precisely as in the previous tests including the shock tubes (Sections~\ref{sec:mhdshocks}--\ref{sec:rotor}). Importantly, our implementation of artificial resistivity (Section~\ref{sec:artres}) guarantees that the dissipation is identically zero when there is no relative velocity between the particles, meaning that simple advection of the current loop is not affected by numerical resistivity. However, the problem remains sensitive to the divergence cleaning (Section~\ref{sec:spmhd}), in particular to any spurious divergence of ${\bm B}$ that is measured by the SPMHD divergence operator, (\ref{eq:divB}). For this reason the results using the quintic kernel, (\ref{eq:quinticspline}), are substantially better than those using the cubic spline, because the initial measurement of $\nabla\cdot{\bm B}$ is smaller and so the evolution is less affected by the divergence cleaning.

\subsubsection{MHD blast wave}
The MHD blast wave problem consists of an over-pressurised central region that expands preferentially along the strong magnetic field lines. Our setup uses the 3D initial conditions of \citet{stoneetal08}, which follows from the work of \citet{londrillodel-zanna00} and \citet{balsaraspicer99}. For a recent application of SPMHD to this problem, see \citet{triccoprice12} and \citet{tpb16}. Set in a periodic box $[x, y, z] \in [-0.5, 0.5]$, the fluid has uniform $\rho = 1$ and ${\bm B} = [10 / \sqrt{2}, 0, 10 / \sqrt{2}]$. The pressure is set to $P=1$, using $\gamma = 1.4$, except for a region in the centre of radius $R=0.125$ which has its pressure increased to $P=100$. This yields initial plasma beta $\beta = 2$ inside the blast and $\beta=0.02$ outside. The particles are arranged on a close-packed triangular lattice using $n_x = 256$.

%
\begin{figure*}
\centering
\includegraphics[width=0.75\textwidth]{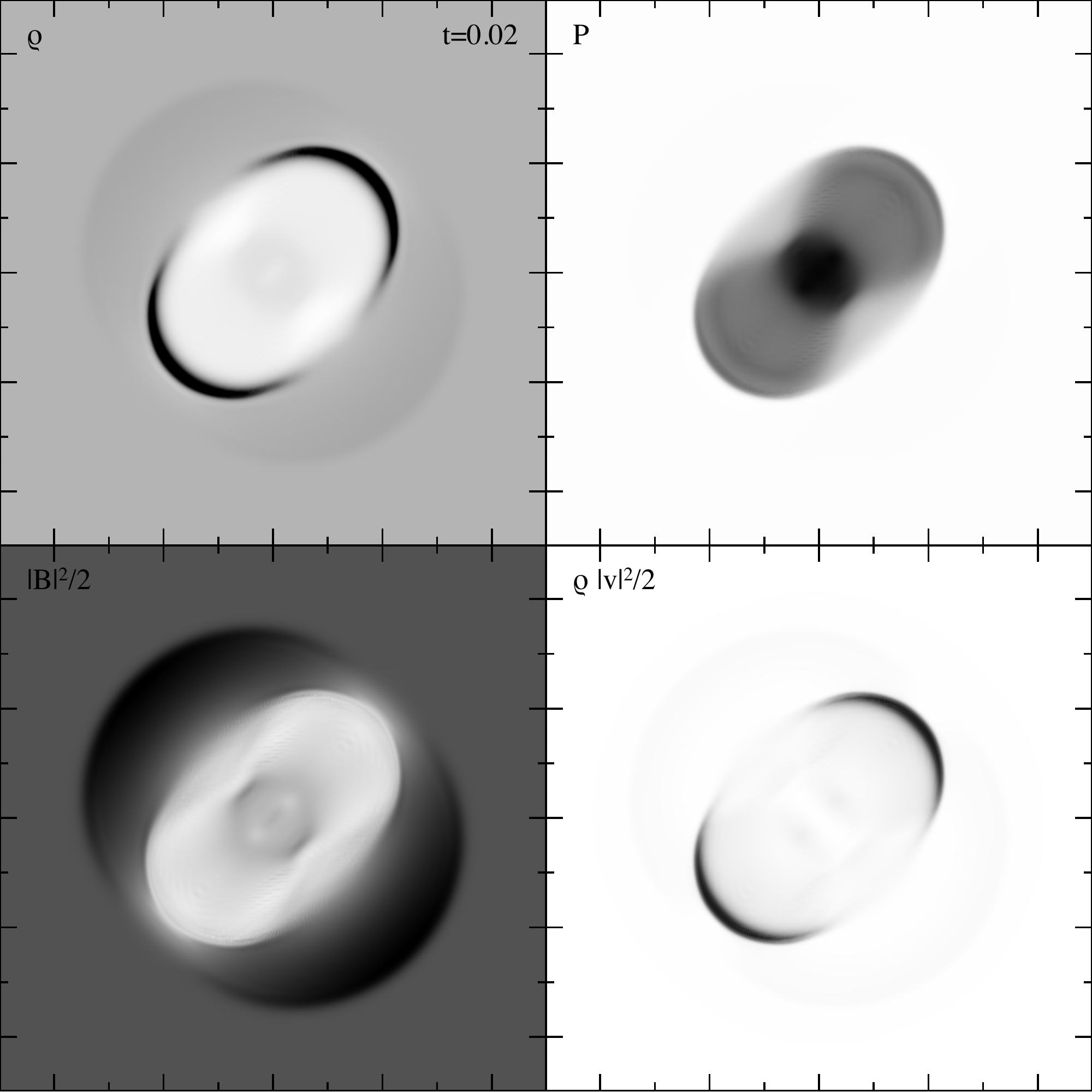}
\caption{Slices through $y=0$ for the MHD blast wave problem showing density (top left), gas pressure (top right), magnetic energy density (bottom left) and kinetic energy density (bottom right). The plot limits are $\rho \in [0.19, 2.98]$, $P \in [1, 42.4]$, $[25.2, 64.9]$ for the magnetic energy density and $[0, 33.1]$ for the kinetic energy density. These are directly comparable to Figure~8 in \citet{gardinerstone08}.}
\label{fig:mhdblast}
\end{figure*}

Figure~\ref{fig:mhdblast} shows slices through $y=0$ of density, pressure, magnetic energy density and kinetic energy density, which may be directly compared to results in \citet{gardinerstone08}. The shock positions and overall structure of the blast wave in all four variables show excellent agreement with the results shown in their paper. The main difference is that their solution appears less smoothed, suggesting that the overall numerical dissipation is lower in {\sc Athena}.

\subsubsection{Balsara-Kim supernova-driven turbulence}
We reproduce the `test problem' of \citet{balsarakim04} (hereafter \citetalias{balsarakim04}) modelling supernova-driven turbulence in the interstellar medium. \citetalias{balsarakim04} used this test to argue strongly against the use of divergence cleaning for problems involving strong shocks. Specifically, they compared three different divergence cleaning schemes against a constrained transport method, finding that divergence cleaning was unusable for such problems, with all three divergence cleaning schemes producing strong temporal fluctuations in magnetic energy during the growth phase of the supernova-driven dynamo. The problems were attributed to issues with the non-locality of divergence cleaning.
 
 We follow the setup in \citetalias{balsarakim04} as closely as possible, but several issues make a direct comparison difficult. Chief among these is their use of a physical ISM cooling prescription. We implement a similar algorithm (Section~\ref{sec:ism}), but our cooling prescription is not identical (e.g. our implementation includes live chemistry and thus the possibility for a cold phase of the ISM, which theirs does not). Secondly, they give the setup parameters for the problem in computational units, but the use of ISM cooling requires the physical units of the problem to be specified. Since these are not stated in their paper, one must guess the units by reading descriptions of the same problem in physical units given in \citet{kbm01}, \citet{balsaraetal04} and \citet{mac-lowetal05}.
  
  We set up the problem as follows. Particles are initialised on a close-packed lattice in a periodic box with $x, y, z \in [-0.1,0.1]$, with $\rho = 1$ and the initial thermal pressure set to $P = 0.3$ (all in code units; as specified in \citetalias{balsarakim04}). We infer a length unit of kpc since this is described as a `200 pc$^3$ box' in their other papers. We employ an adiabatic equation of state with $\gamma = 5/3$ and turn on interstellar cooling and chemistry in the code with default initial abundances (i.e. atomic Hydrogen everywhere). We choose the mass unit such that a density of $2.3 \times 10^{-23}$ g/cm$^3$, as described in \citet{kbm01}, corresponds to $\rho = 1$ in code units as described in \citetalias{balsarakim04}. Finally, we set the time unit such that $G = 1$ in code units (a common choice, even though gravity is not involved in the problem). We compute the problem using resolutions of $64 \times 74 \times 78$ particles and $128 \times 148 \times 156$ particles.
 
  Supernovae are injected into the simulation every 0.00125 in code units at the positions listed in Table~1 of \citetalias{balsarakim04}. We infer this to correspond to the `$12 \times$ Galactic' rate described in \citet{balsaraetal04}, i.e. 12 times faster than 1 per 1.26 Myr. We inject supernovae following the description in \citetalias{balsarakim04} by setting the pressure to $P=13649.6$ on particles within a distance of 0.005 code units from the injection site (corresponding to a radius of 5 pc in physical units). Our choice of units means that this corresponds to an energy injection within a few percent of $10^{51}$ ergs in physical units, corresponding to the description given in their other papers. However, this is only true if the density equals the initial value, since \citetalias{balsarakim04} specify pressure rather than the energy. We follow the description in \citetalias{balsarakim04} (i.e., we set the pressure), even though this gives an energy not equal to $10^{51}$ ergs if injected in a low or high density part of the computational domain.
  
  The initial magnetic field is uniform in the $x$-direction with $B_x = 0.056117$, as stated in \citetalias{balsarakim04}. We could not reconcile this with the magnetic energy plotted in their paper, which show an initial magnetic energy of $10^{-6}$. Nor could we reconcile this with the statement in \citet{balsaraetal04} that ``the magnetic energy is $10^{-6}$ times smaller than the thermal energy''. So any comparison is approximate. Nevertheless, \citetalias{balsarakim04} state that the problems caused by divergence cleaning are not dependent on specific details of the implementation. 
  
\begin{figure*}
\centering
\includegraphics[width=\textwidth]{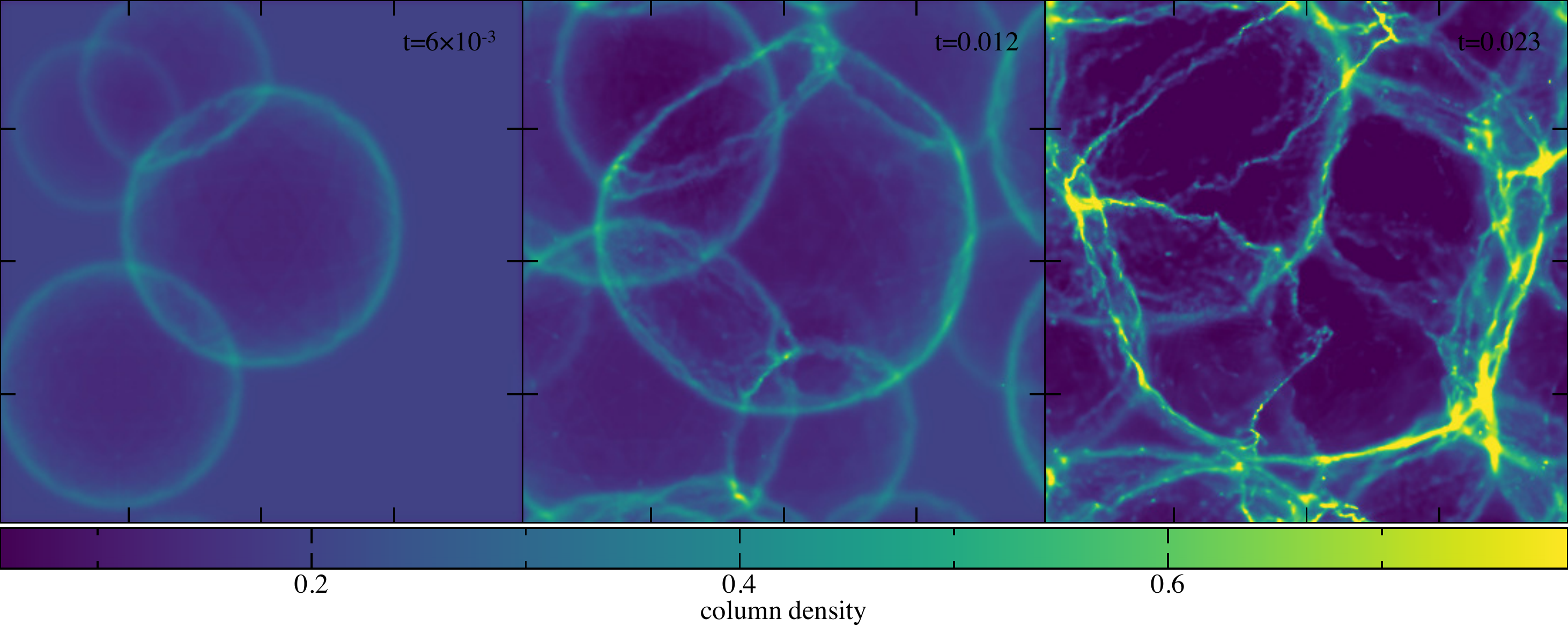}
\caption{Balsara-Kim supernova-driven turbulence, showing column density at three different times at a resolution of $64 \times 74 \times 78$ particles. Supernovae are injected every 0.00125 in code units, leading to a series of interacting blast waves. Interstellar chemistry and cooling is turned on, producing a dense filaments in a turbulent interstellar medium.}
\label{fig:balsarakimdens}
\end{figure*}

   Figure~\ref{fig:balsarakimdens} shows the evolution of the column density in the lowest resolution calculation using $64 \times 74 \times 78$ particles. The combination of supernovae injection and cooling drives turbulence and significant structure in the density field. 
   
\begin{figure}
\centering
\includegraphics[width=\columnwidth]{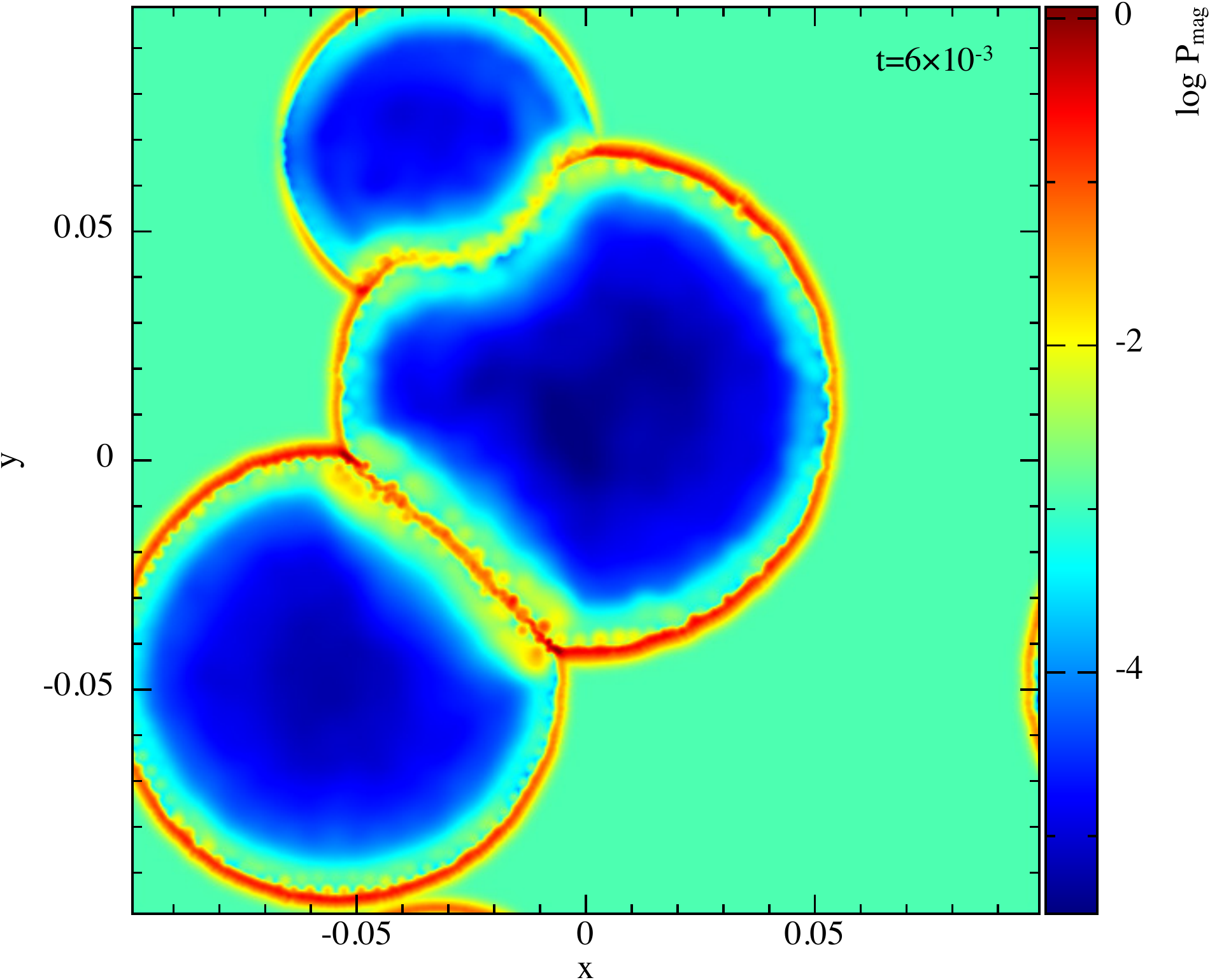}
\caption{Cross-section slice of magnetic pressure at $t = 0.006$ in the Balsara-Kim supernova-driven turbulence test using $128 \times 148 \times 156$ particles. No large scale artefacts in magnetic energy are visible, indicating that the simulation is not corrupted by divergence cleaning.}
\label{fig:balsarakimpmag}
\end{figure}

 To specifically address the issues found by \citetalias{balsarakim04}, Figure~\ref{fig:balsarakimpmag} show a cross section slice of the magnetic pressure at a time similar to the one shown in Figure 3 of \citetalias{balsarakim04} (they do not indicate which slice they plotted; we chose $z=0.0936$). The magnetic pressure in the interior of the supernovae shells is smooth, and does not display any of the large scale artefacts of the type found in their paper. 
  
\begin{figure}
\centering
\includegraphics[width=\columnwidth]{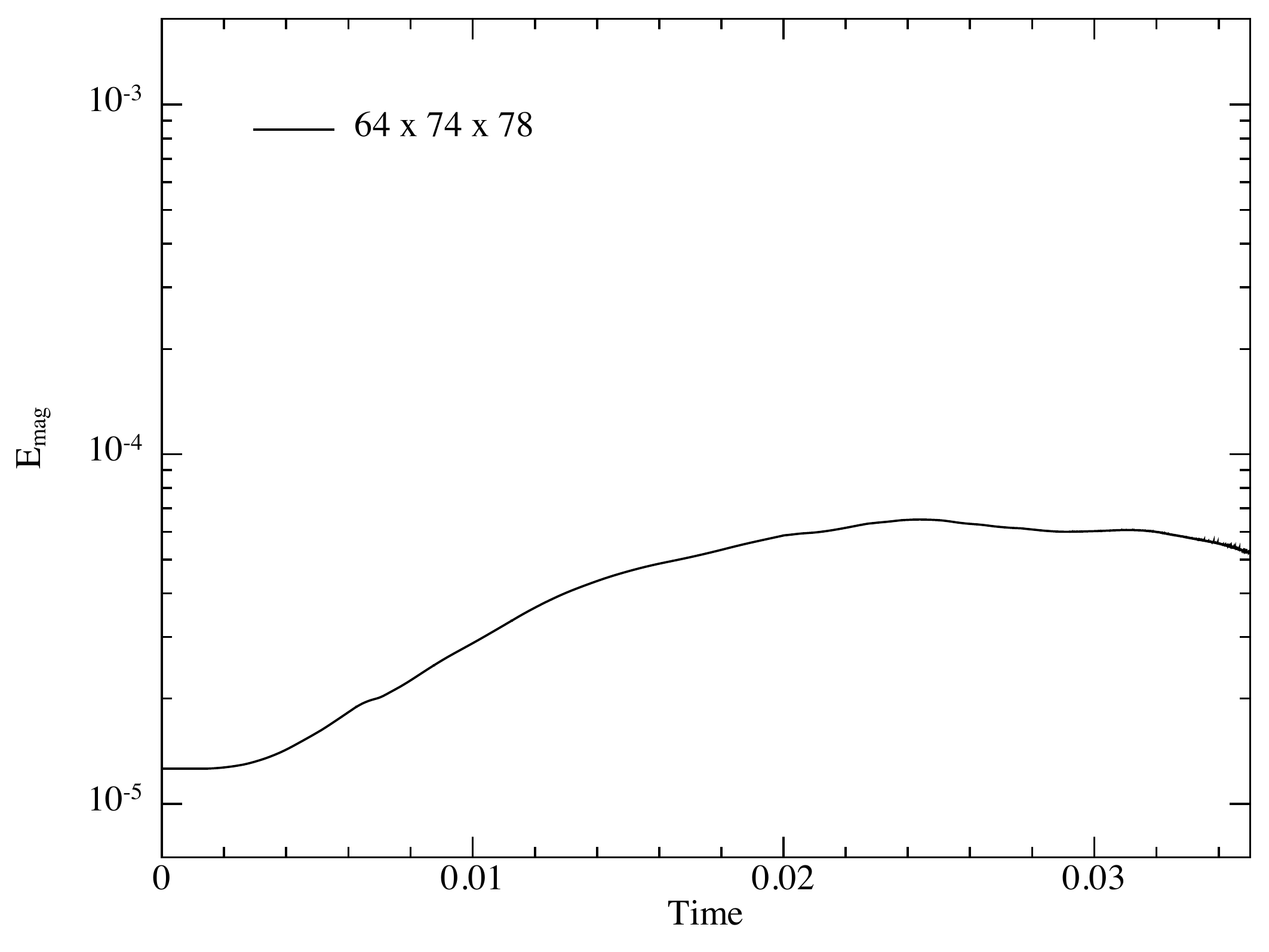}
\caption{Magnetic energy as a function of time in the Balsara-Kim supernova-driven turbulence problem. The magnetic energy increases monotonically by approximately an order of magnitude before reaching its saturation value at $t \approx 0.02$. There are no spurious spikes in magnetic energy caused by divergence cleaning, in contrast to what was found by \citet{balsarakim04}.}
\label{fig:balsarakimemag}
\end{figure}

 Figure~\ref{fig:balsarakimemag} shows the time evolution of the magnetic energy in the low resolution calculation. The magnetic energy rises monotonically up to $t \approx 0.02$ before the magnetic energy saturates, similar to what was found by \citetalias{balsarakim04} for their staggered mesh / constrained transport scheme (compare with Figure 2 in their paper). There are no large temporal fluctuations in the magnetic energy of the kind they report for their divergence cleaning methods.
 
 In summary, the results we obtain for the magnetic field energy and structure within supernova-driven turbulence in the interstellar medium matchs closest the constrained transport result of \citetalias{balsarakim04}. There is no evidence that our simulations experience the numerical issues encountered by \citetalias{balsarakim04} with their divergence cleaning schemes, suggesting that the problems they reported are primarily code dependent rather than being fundamental. The most probable reason our results are of the same quality as \citetalias{balsarakim04}'s constrained transport result is our use of constrained divergence cleaning, which guarantees that energy removed from the magnetic field is negative definite. A follow up study to examine this in more depth would be worthwhile. For our present purposes, we can conclude that our divergence cleaning algorithm is sufficiently robust --- even in `real world' problems --- to be useful in practice (see also Section~\ref{sec:jet}).
  
\subsection{Non-ideal MHD}
\label{sec:nimhdtest}
The following tests demonstrate the non-ideal MHD algorithms (Section~\ref{sec:nonideal}). We adopt periodic boundary conditions for both tests, initialising the particles on a close-packed lattice with an isothermal equation of state, $P = c_\text{s}^2 \rho$, and using the $C^4$ Wendland kernel.

\subsubsection{Wave damping test}
\label{sec:ni:tests:wave}
To test ambipolar diffusion in the strong coupling approximation, we follow the evolution of decaying Alfv{\'e}n waves, as done in (e.g.) \citet{ckw09} and \citet{wpa14}. 

In arbitrary units, the initial conditions are a box of size $L_\text{x} \times \frac{\sqrt{3}}{2}L_\text{x} \times \frac{\sqrt{6}}{2}L_\text{x}$ with $L_\text{x}=1$, a density of $\rho = 1$, magnetic field of $\bm{B} = B_0\hat{\bm{x}}$ with $B_0 = 1$, sound speed of $c_\text{s} = 1$, and velocity of $\bm{v} = v_0\sin(kx)\hat{\bm{z}}$ where $k = 2\pi/L_\text{x}$ is the wave number and $v_0 = 0.01v_\text{A}$ where $v_\text{A}$ is the Alfv{\'e}n velocity. We adopt an ambipolar diffusion coefficient of $\eta_{\rm AD} = 0.01v_\text{A}^2$.  All artificial dissipation terms are turned off.  We use $n_x = 128$ particles.

The solution to the dispersion relation for Alfv{\'e}n waves \citep{Balsara1996} is
\begin{equation}
\omega^2 + \eta_{\rm AD}k^2 \omega i- v_\text{A}^2k^2 = 0,
\end{equation}
where $\omega = \omega_\text{R} + \omega_\text{I}i$ is the complex angular frequency of the wave, giving a damped oscillation in the form
\begin{equation}
h(t) = h_0 \left| \sin\left(\omega_\text{R}t\right)\right|\text{e}^{\omega_\text{I}t}.
\end{equation}
In our test, $h(t)$ corresponds the the root-mean-square of the magnetic field in the $z$-direction, $\left< B_\text{z}^2 \right>^{1/2}$, and  $h_0 = v_0 B_0 / (v_\text{A} \sqrt{2})$.

\begin{figure}
\begin{center}
\includegraphics[width=\columnwidth]{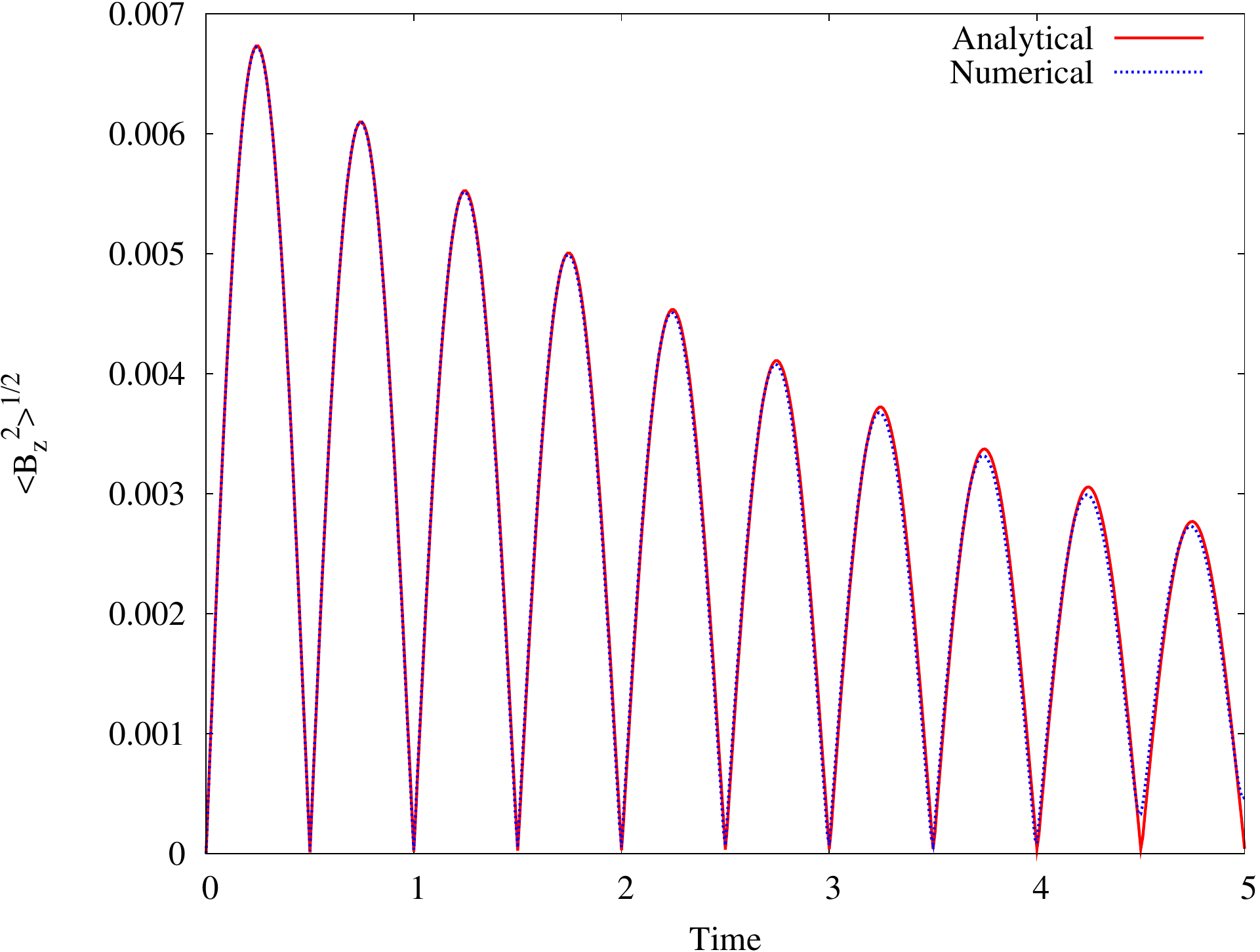}
\caption{Wave damping test showing the decay of Alfv{\'e}n waves in the presence of ambipolar diffusion, using the coefficient $\eta_\text{AD} = 0.01v_\text{A}^2$.  The ${\cal L}_2$ error between the analytic and numerical solution is $7.5\times 10^{-5}$.}
\label{fig:nitests:wavedamp}
\end{center}
\end{figure}

Figure~\ref{fig:nitests:wavedamp} shows the time evolution of $\left< B_\text{z}^2 \right>^{1/2}$ to $t = 5$ for both the numerical results (blue line) and the analytic solution (red line).  At the end of the test, the  ${\cal L}_2$ error is $7.5\times 10^{-5}$ (evaluated at intervals of d$t = 0.01$), demonstrating close agreement between the numerical and analytical results.

Given that, by design, there is motion of the particles and that we have excluded artificial dissipation, the particles tend to `break' from the initial lattice.  For both the $M_6$ quintic kernel and the $C^4$ Wendland kernel on a cubic lattice, the particles fall off the lattice at $t \approx 0.75$.  Prior to this, however, the ${\cal L}_2$ error is smaller than that calculated using the $C^4$ Wendland kernel and a close-packed lattice, thus there is a trade off between accuracy and long-term stability.

\subsubsection{Standing shock}
\label{sec:ni:tests:shock}
To test the Hall effect, we compare our solutions against the 1D isothermal steady-state equations for the the strong Hall effect regime. The numerical solution to this problem is given in \citet{falle03} and \citet{osullivandownes06}, and also summarised in Appendix C1.2 of \citet{wpb16}.

The left- and right-hand side of the shock are initialised with $\left(\rho_0, v_\text{x,0},v_\text{y,0},v_\text{z,0},B_\text{x,0},B_\text{y,0},B_\text{z,0}\right) = 
\left(1.7942,-0.9759,-0.6561,0.0,1.0,1.74885,0.0\right)$ and $\left(1.0,-1.751, 0.0,0.0,1.0,0.6,0.0\right)$, respectively, with the discontinuity at $x=0$.  We use boundary particles at the $x$-boundary, superseding the periodicity in this direction. To replicate inflowing boundary conditions in the $x$-direction, when required, the initial domain of interest $x_l < x < x_r$ is automatically adjusted to $x'_l < x < x'_r$ where $x'_r = x_r - v_0 t_{\rm max}$, where $t_{\rm max}$ is the end time of the simulation. Note that for inflowing conditions, $x_r$ and $v_0$ will have opposite signs.  Thus, at the end of the simulation, the entire range of interest will still be populated with particles.

The non-ideal MHD coefficients are  $\eta_\text{OR} = 1.12\times 10^{-12}$, $\eta_\text{HE} = -3.53\times 10^{-2}B$, and $\eta_\text{AD} = 7.83\times 10^{-3}v_\text{A}^2$.  We include all artificial dissipation terms using their default settings. We initialise $512 \times 14 \times 15$ particles in $x < 0$ and $781 \times 12 \times 12$ particles in $x \geq 0$, with the domain extending from $x_l = x'_l = -2$ to $x_r = 2$ with $x'_r = 3.75$.

\begin{figure}
\begin{center}
\includegraphics[width=\columnwidth]{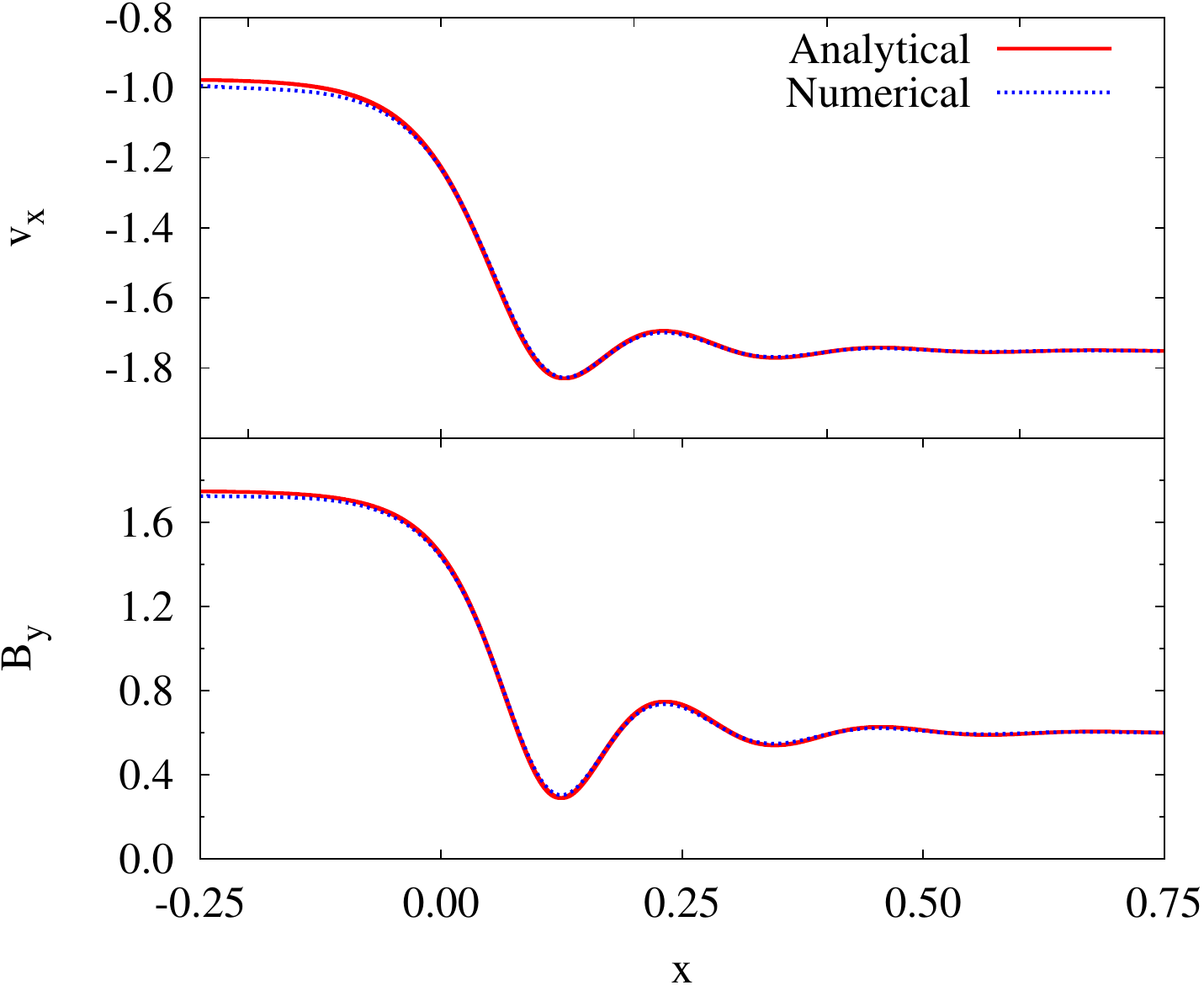}
\caption{Hall-dominated standing shock using $\eta_\text{OR} = 1.12\times 10^{-12}$, $\eta_\text{HE} = -3.53\times 10^{-2}B$, and $\eta_\text{AD} = 7.83\times 10^{-3}v_\text{A}^2$.  The numerical and analytical results agree to within 3 per cent everywhere.}
\label{fig:nitests:standing}
\end{center}
\end{figure}

Figure~\ref{fig:nitests:standing} shows $v_\text{x}$ and $B_\text{y}$ for both the numerical and analytical results, which agree to within 3 per cent at any given position.
On the left-hand side of the shock interface, the numerical results are lower than the analytical solution because of the artificial dissipation terms, which are required to properly model a shock.  These results do not depend on either the kernel choice or the initial particle lattice configuration.

\subsection{Self-gravity}
\label{sec:sg}

\subsubsection{Polytrope}
\label{sec:polytrope}
The simplest test of self-gravity is to model a spherical polytrope in hydrostatic equilibrum. Similar tests have been shown for SPH codes dating back to the original papers of \citet{gingoldmonaghan77,gingoldmonaghan78,gingoldmonaghan80}.  Modern calculations have used these simple models in more complex applications, ranging from common envelope evolution \citep{iaconietal17} to tidal disruption events \citep{coughlinnixon15,coughlinetal16,coughlinetal16a,bonnerotetal16,bonnerotetal17}.

The equation of state is $P=K\rho^\gamma$, with $\gamma = 1 +1/n$ where $n$ is the polytropic index. The exact hydrostatic solution is given by
\begin{equation}
\label{eq:polytrope}
\frac{\gamma K}{4\pi G\left(\gamma - 1\right) }\frac{\text{d}^2}{\text{d}r^2}\left(r\rho^{\gamma-1}\right) + r\rho = 0.
\end{equation}
Our initial setup uses a solution scaled to a radius $R = 1$, for a polytropic index of $n = 3/2$, corresponding to $\gamma = 5/3$, with $K=0.4244$. We solve (\ref{eq:polytrope}) numerically. We place the particles initially on a hexagonal close-packed lattice, truncated to a radius of $R=1$, which we then stretch map (see Section~\ref{sec:stretchmap}) such that the initial radial density profile matches the exact solution. 

\begin{figure}
   \centering
   \includegraphics[width=0.45\textwidth]{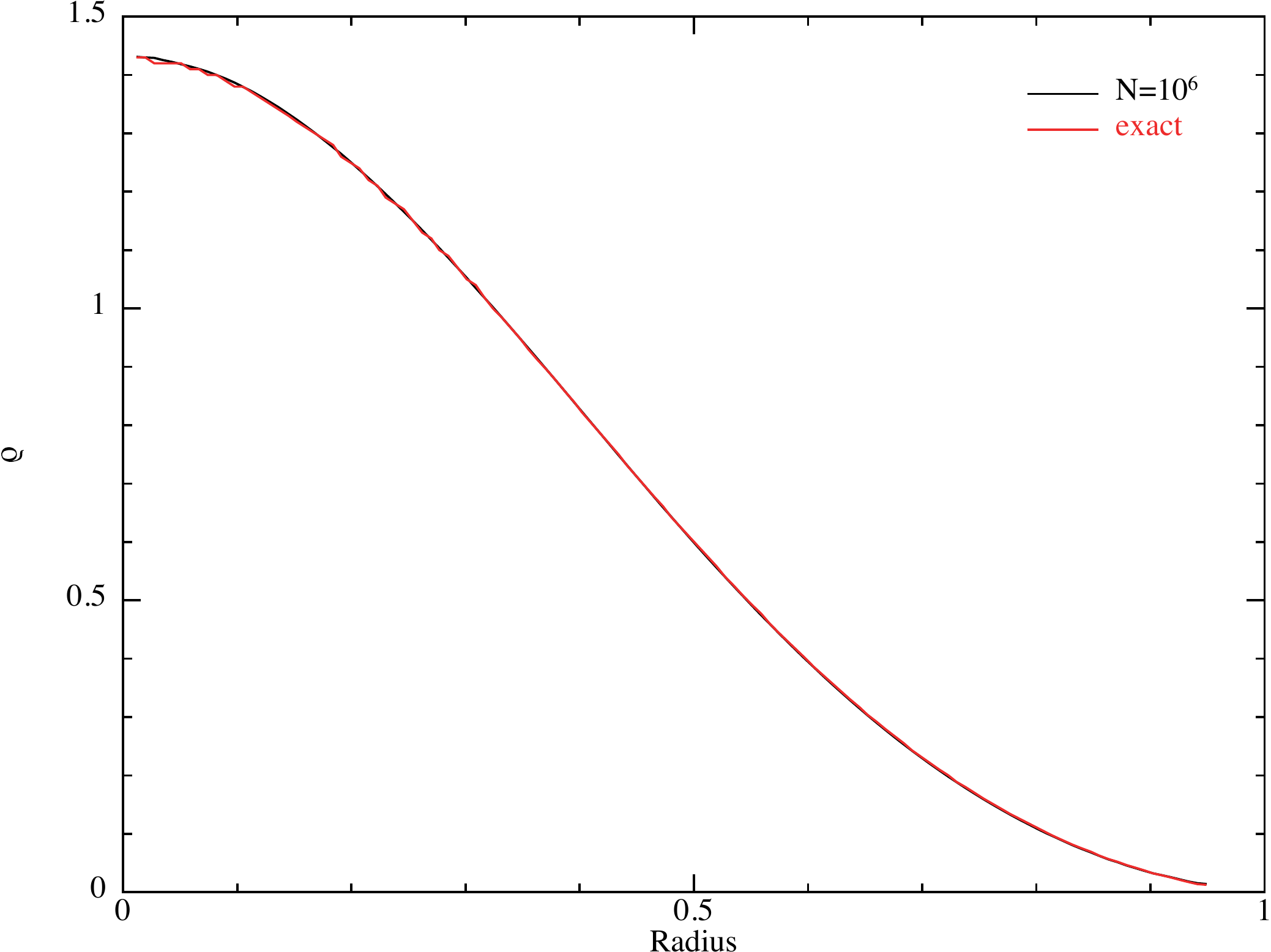} 
   \caption{Polytrope static structure using $10^6$ particles (black), compared to the exact solution (red), shown at $t = 100$.}
\label{fig:polyres}
\end{figure}

The relaxation time depends on the initial density profile and on how far the initial particle configuration is from equilibrium.  Figure~\ref{fig:polyres} shows the solution at $t=100$ in code units.  The polytrope relaxes within a few dynamical times, with only a slight rearrangement of the particles from the stretched lattice. The density profile at all times is equal to within 3 per cent of the exact solution for $r \leq 0.7$ and the polytrope remains in hydrostatic equilibrium.

Once the static solution is obtained, we tested the energy conservation by giving the star a radial perturbation. That is, we applied a velocity perturbation of the form $v_r = 0.2r$ to the $N \approx 10^5$ model, and evolved the polytrope for 100 time units. We turned off the artificial viscosity for this test.  The total energy --- including contributions from thermal, kinetic and gravitational energy --- remained conserved to within 3 per cent.

\subsubsection{Binary polytrope}
\label{sec:binarypolytrope}
\begin{figure}
   \centering
   \includegraphics[width=0.45\textwidth]{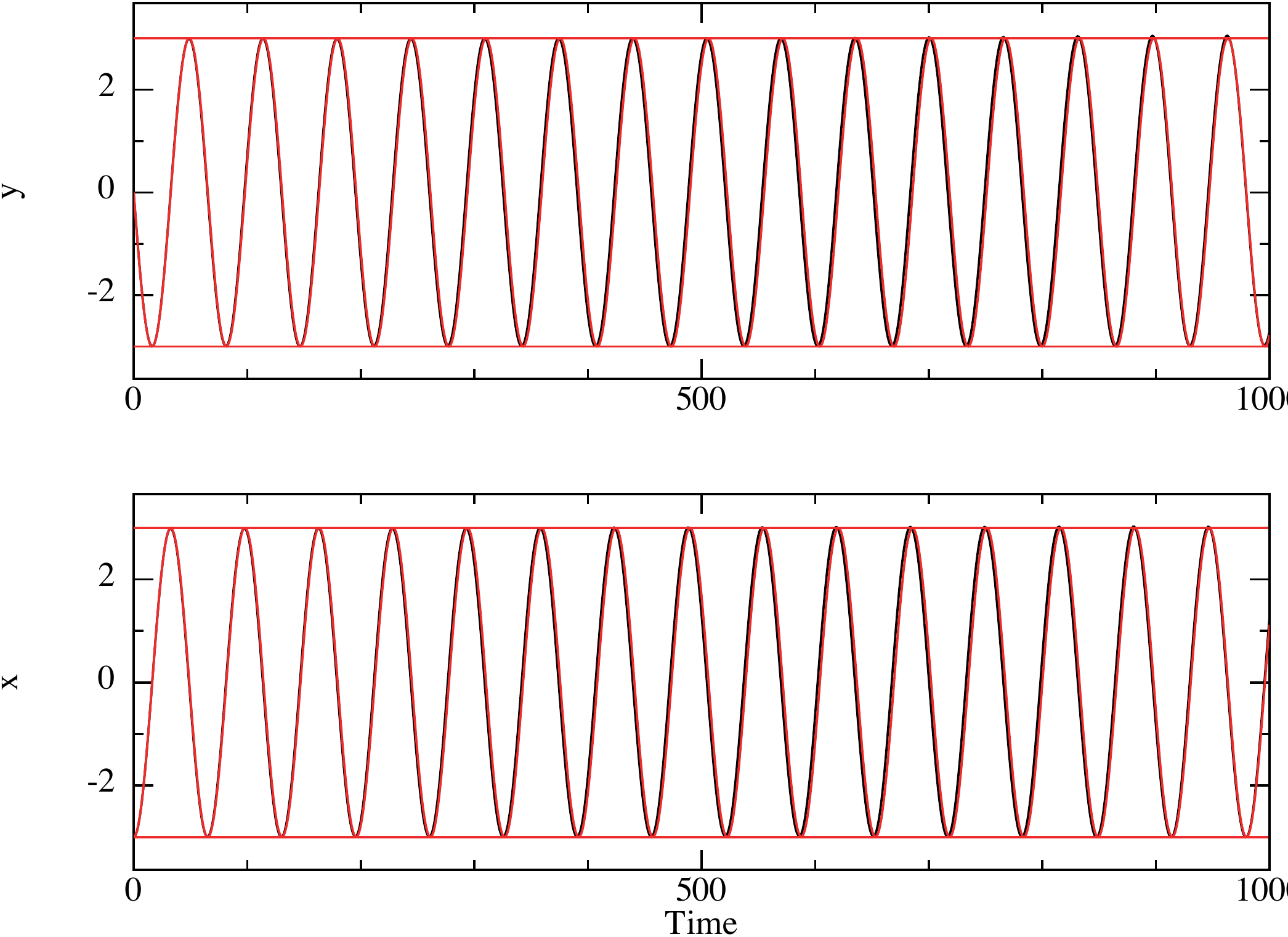} 
   \caption{$y$- and $x$-positions of the centre of mass of one star of a binary system (top and bottom, respectively).  Each star is a polytrope with mass and radius of unity, and $N = 10^4$ particles; the initial separation is 6.0 in code units.  After $\sim15$ orbits, the separation remains within 1 percent, and the period remains constant within the given time resolution.  The red line represents the analytical position with respect to time, and the black line represents the numerical position.}
\label{fig:polyperiod}
\end{figure}

Next, we placed two initially unrelaxed, identical polytropes, each with $N = 10^4$ particles in a circular orbit around each other with a separation of $6R$ and evolved for $\sim 15$ orbits (1000 code units). Figure~\ref{fig:polyperiod} demonstrates that the separation remains within 1 per cent of the initial separation over 15 orbits, and that the orbital period remains constant. After the initial relaxation, total energy is conserved to within 0.06 per cent.

The stars are far enough apart that any tidal deformation as they orbit is insignificant.  The final density profile of each star agrees with the expected profile within 3 per cent for $r \leq 0.7$. 

\subsubsection{Evrard collapse}
A more complex test, relevant especially for star formation, is the so-called `Evrard collapse' \citep{evrard88} modelling the adiabatic collapse of a cold gas sphere. It has been used many times to test SPH codes with self-gravity, e.g.\ \citet{hernquistkatz89}, \citet{steinmetzmueller93}, \citet{ThackerEtAl2000}, \citet{escalaetal04} and many others. 

Following the initial conditions of \citet{evrard88}, we setup the particles initially in a sphere of radius $R = 1$ and mass $M = 1$, with density profile
\begin{equation}
\rho(r) = \frac{M(R)}{2\pi R^2}\frac{1}{r}.
\end{equation}
The density profile is created using the same stretch mapping method as for the polytrope (see Section \ref{sec:polytrope}). The sphere is initially isothermal, with the specific internal energy set to $u = 0.05GM/R$ with an adiabatic index of $\gamma = 5/3$. The sphere initially undergoes gravitational collapse.

In the literature, the results of the Evrard collapse are typically normalised to a characteristic value.  Here, we simply show the results in code units (Section~\ref{sec:units}), since these units already represent a normalised state. A distance unit of $R = 1$ and mass unit $M(R) = 1$ is adopted, with the time unit set such that $G \equiv 1$, where $G$ is the gravitational constant.

\begin{figure}
   \centering
   \includegraphics[width=0.45\textwidth]{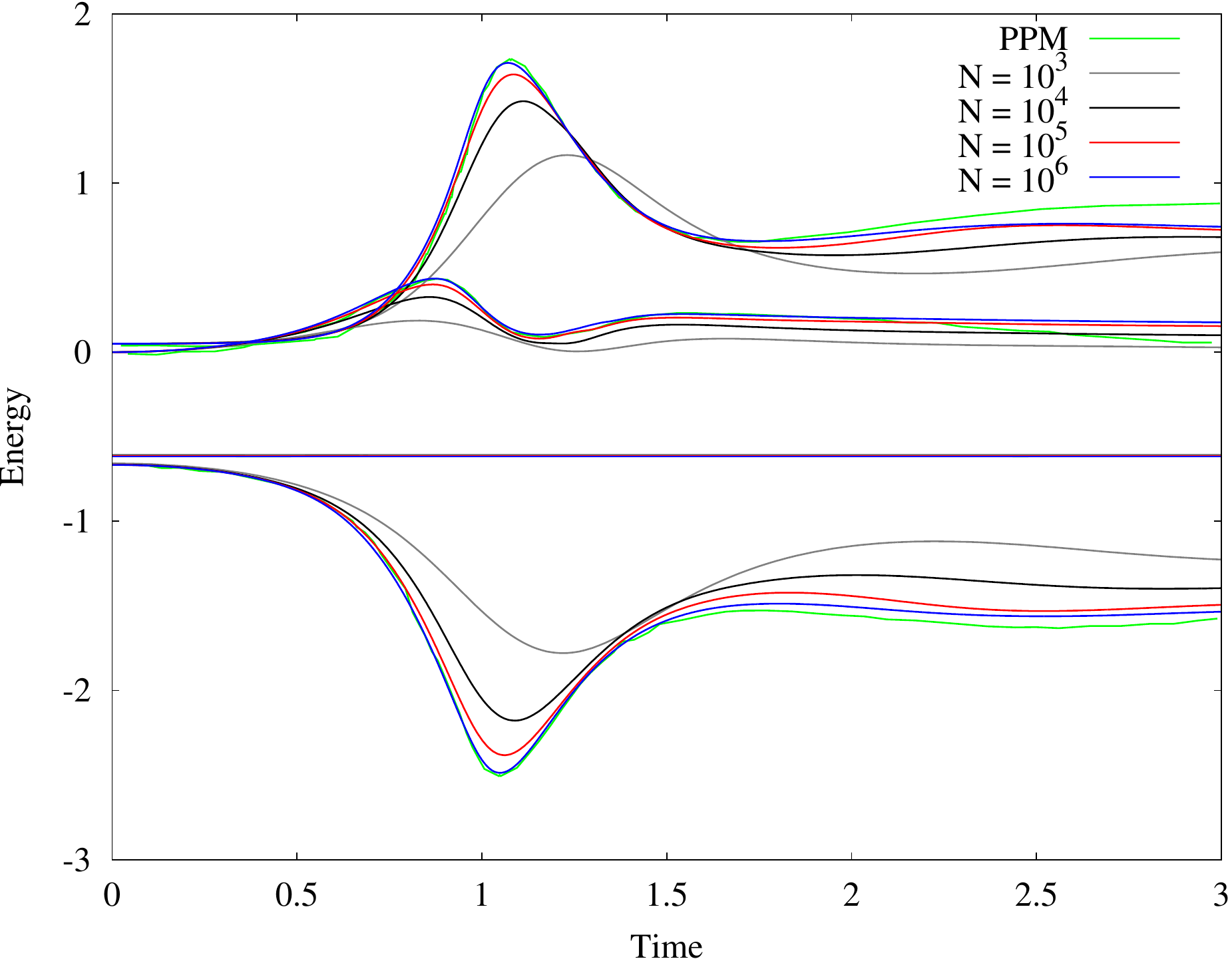} 
   \caption{Thermal, kinetic, total and potential energies as a function of time during the Evrard collapse \citep{evrard88}. Green lines are calculated using a 1D PPM code, taken from Figure 6 of \citet{steinmetzmueller93}, while remaining colours show SPH simulations of different resolutions.  Both energy and time are given in code units, where $R=M=G=1$.}
\label{fig:sgEvrard:energy}
\end{figure}

Figure~\ref{fig:sgEvrard:energy} shows the kinetic, thermal, total and potential energies as a function of time, at four different numerical resolutions. The green line shows the reference solution, computed using a 1D piecewise parabolic method (PPM) code using 350 zones, which we transcribed from Figure 6 of \citet{steinmetzmueller93}.
As the number of particles increases, the energies for $t \lesssim 1.5$ converge to the results obtained from the PPM code.  At $t \gtrsim 2$, the SPH results appear to converge to energies that differ slightly from the PPM code. Given that we are not able to perform a comparable convergence study with the PPM code, we are unable to assess whether or not this discrepancy is significant.

\begin{figure}
   \centering
   \includegraphics[width=0.45\textwidth]{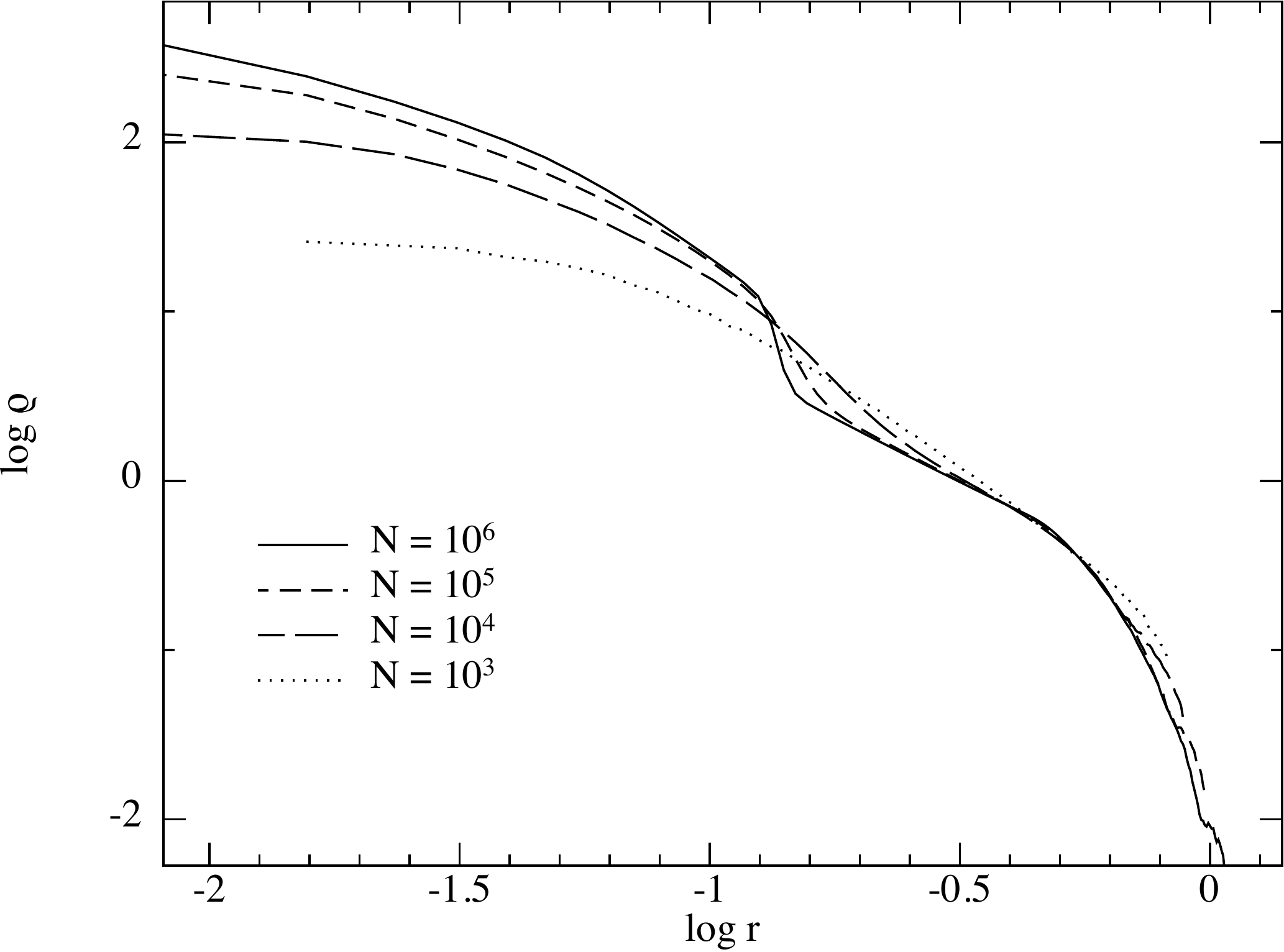}
   \includegraphics[width=0.45\textwidth]{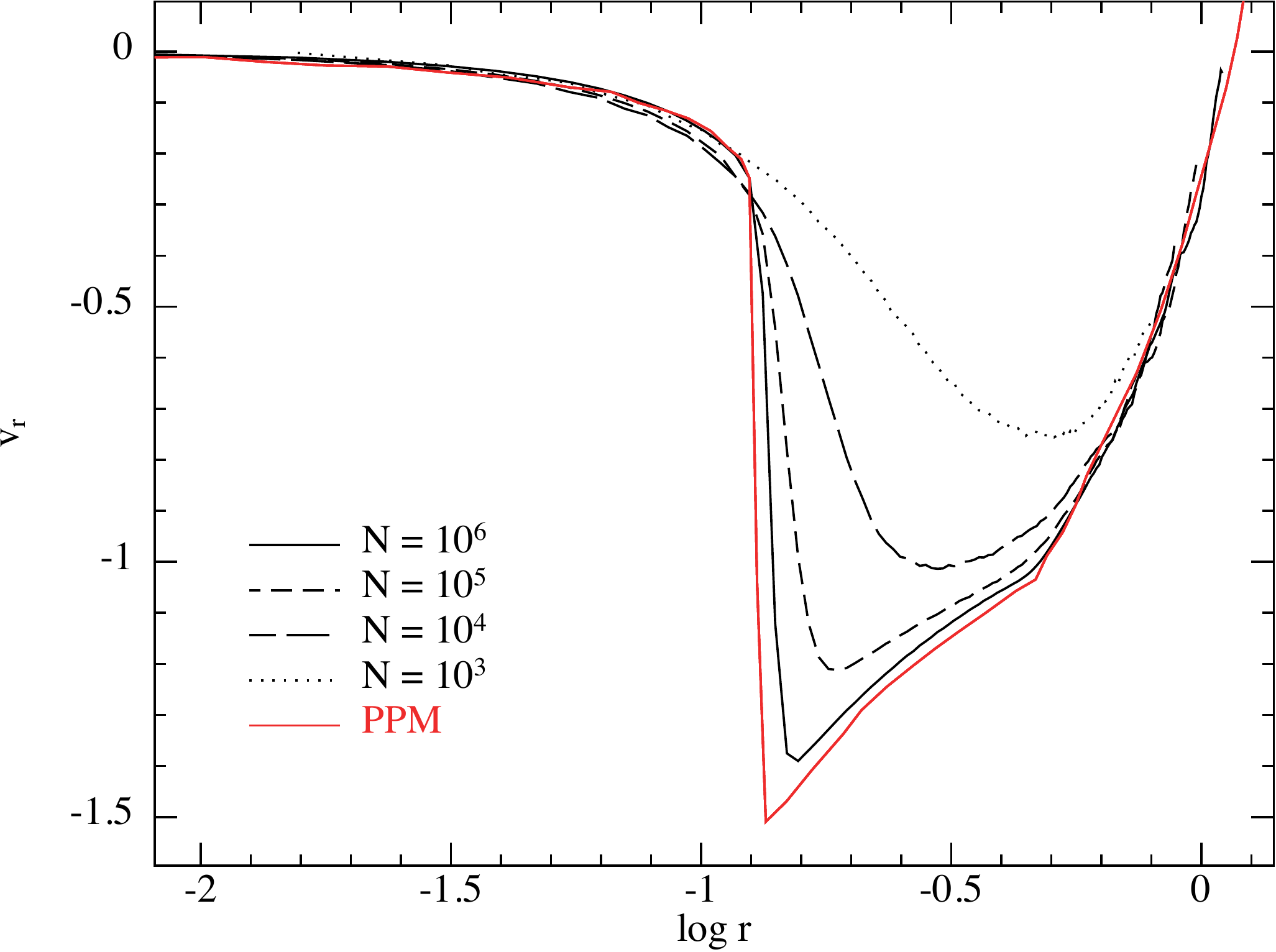}    
   \caption{Radial profile of the Evrard collapse \citep{evrard88} at $t = 0.77$; red lines are taken from Figure 7 of \citet{steinmetzmueller93}, with panels showing density (top) and radial velocity (bottom) as a function of (log) radius.  All values are given in code units, where $R=M=G=1$.  The outward propagating shock at $r \approx 0.1$ is sharper at high resolution.}
\label{fig:sgEvrard:radial}
\end{figure}

Figure~\ref{fig:sgEvrard:radial} shows enclosed mass, density, thermal energy and radial velocity as a function of radius at $t = 0.77$, where the SPH results may be compared to the PPM results presented by \citet{steinmetzmueller93} shown with the red line in the Figure. At this time, the outward propagating shock is at $r \approx 0.1$, with the shock profile in agreement with the reference solution at high resolution.   

\subsection{Dust-gas mixtures}
\label{sec:dusttest}
 The SPH algorithms used in \textsc{Phantom} for dust-gas mixtures have been extensively benchmarked in \citet{laibeprice12,laibeprice12a} (for the two-fluid method) and in \citet{pricelaibe15} (for the one-fluid method; hereafter \citetalias{pricelaibe15}). Here we merely demonstrate that the implementation of these algorithms in \textsc{Phantom} gives satisfactory results on these tests. For recent applications of \textsc{Phantom} to more realistic problems involving dust/gas mixtures see \citet{dipierroetal15,dipierroetal16}, \citet{ragusaetal17} and \citet{tpl17}.

\subsubsection{\sc Dustybox}
\label{sec:dustybox}
\begin{figure}
   \centering
   \includegraphics[width=\columnwidth]{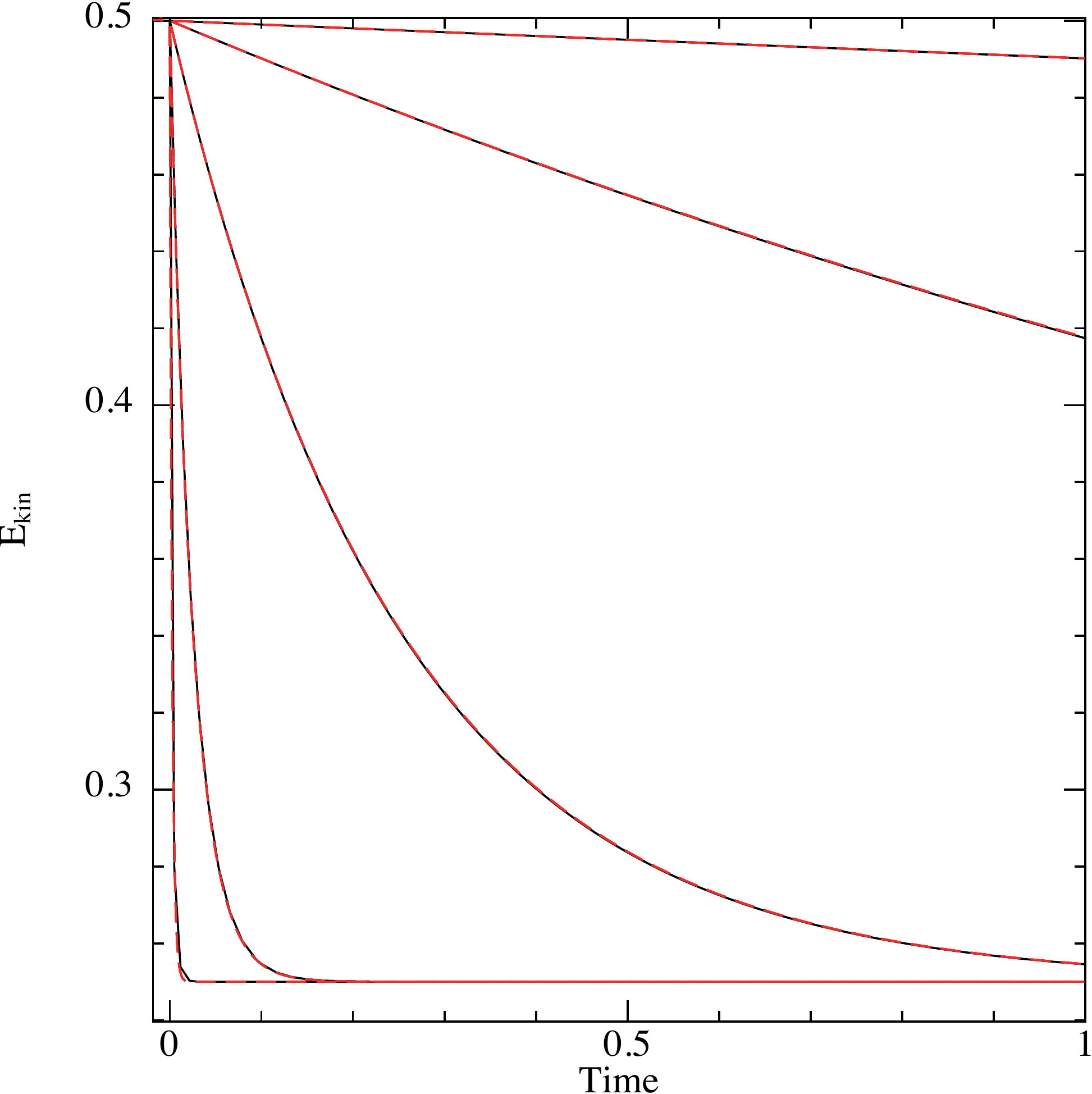} 
   \caption{Decay of kinetic energy as a function of time in the {\sc dustybox} test, involving a uniformly translating mixture of gas and dust coupled by drag. Solid lines show the \textsc{Phantom} results for drag coefficients $K=0.01$, 0.1, 1.0, 10 and 100 (top to bottom), which may be compared to the corresponding analytic solutions given by the dashed red lines.}
\label{fig:dustybox}
\end{figure}

Figure~\ref{fig:dustybox} shows the results of the \textsc{dustybox} problem \citep{monaghankocharyan95,paardekoopermellema06,laibeprice11}. We setup a uniform, periodic box $x,y,z \in [-0.5,0.5]$ with $32 \times 36 \times 39$ gas particles set on a close-packed lattice and $32 \times 36 \times 39$ dust particles also set on a close-packed lattice. The gas particles are initially at rest while the dust is given a uniform velocity $v_{x} = 1$ in the $x$-direction. We employ an isothermal equation of state with $c_{\rm s} = 1$, uniform gas and dust densities of $\rho_{\rm g} = \rho_{\rm d} = 1$, using the cubic spline kernel for the SPH terms and the double-hump cubic spline kernel (Section~\ref{sec:dragkernel}) for the drag terms, following \citet{laibeprice12}.

 The red dashed lines in Figure~\ref{fig:dustybox} show the exact solution for kinetic energy as a function of time. For our chosen parameters the barycentric velocity is $v_{x} = 0.5$, giving $v_{\rm g}(t) = 0.5[1 + \Delta v_{x}(t)]$, $v_{\rm d}(t)  = 0.5[1 - \Delta v_{x}(t)]$, where $\Delta v_{x}(t) = \exp{(-2Kt)}$ \citep{laibeprice11} and the red lines show $E_{\rm kin}(t) = \frac12[v_{\rm g}(t)^{2} + v_{\rm d}(t)^{2}]$. The close match between the numerical and analytic solutions (${\cal L}_{2} \sim 3 \times 10^{-4}$) demonstrates that the drag terms are implemented correctly.

The {\sc dustybox} test is irrelevant for the one-fluid method (Section~\ref{sec:onefluidsph}) since this method implicitly assumes that the drag is strong enough so that the terminal velocity approximation holds --- implying that the relative velocites are simply the barycentric values at the end of the {\sc dustybox} test.

\begin{figure*}
   \centering
   \includegraphics[width=0.45\textwidth]{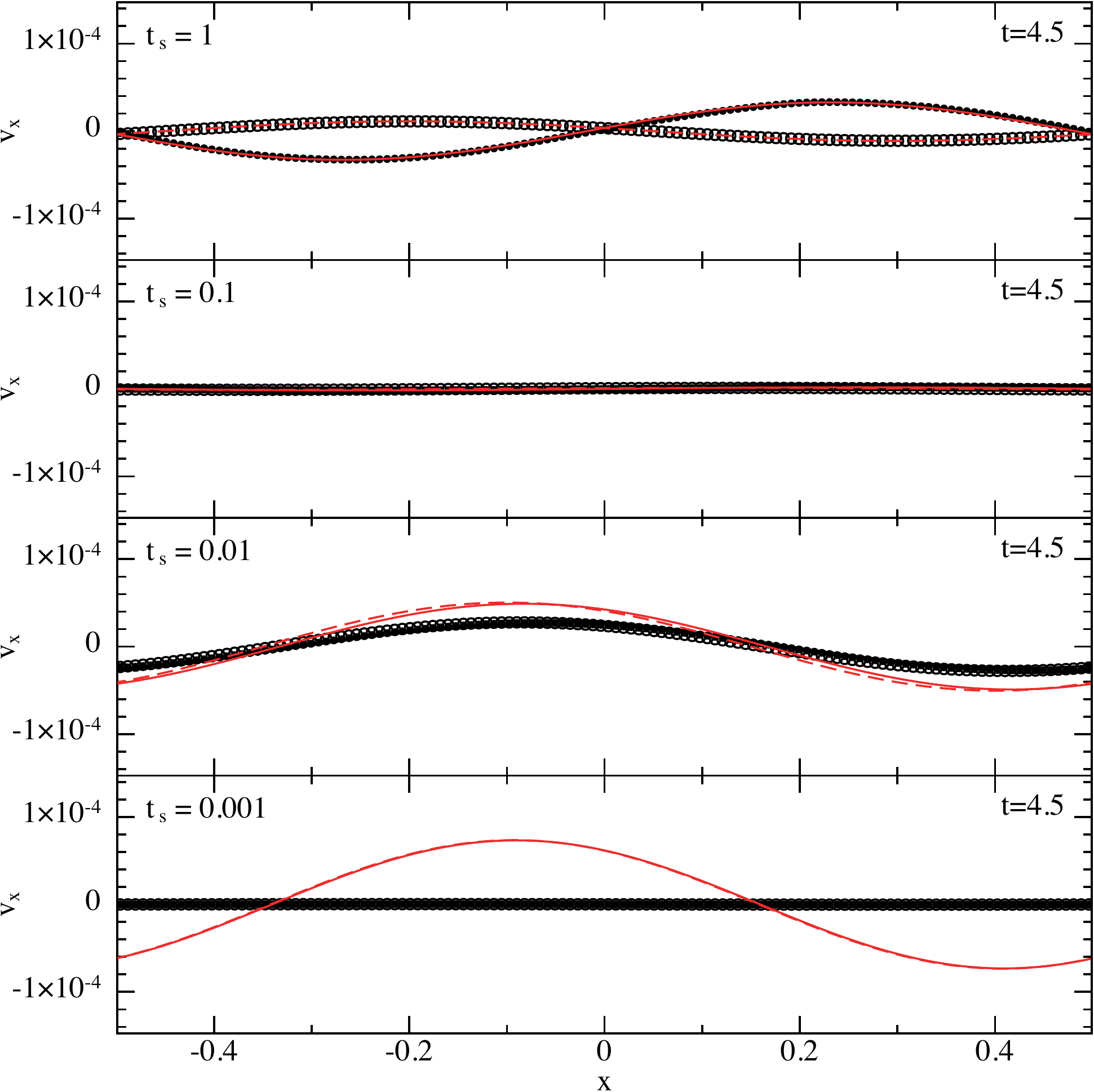} \hspace{0.5cm}
   \includegraphics[width=0.45\textwidth]{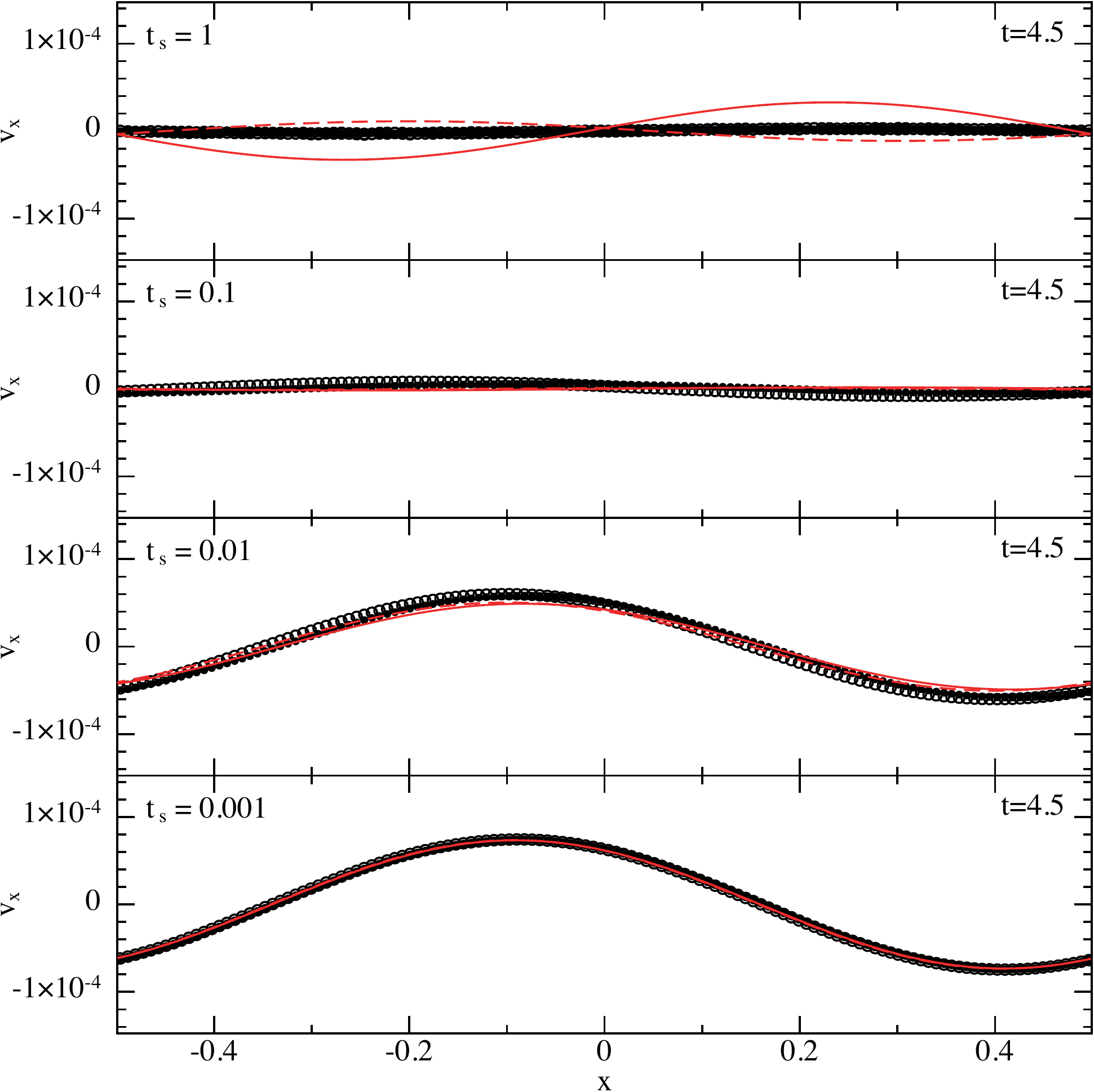}
   \caption{Velocity of gas (solid) and dust (circles) at $t=4.5$ in the {\sc dustywave} test, using the two-fluid method (left; with $64 \times 12 \times 12$ gas particles, $64 \times 12 \times 12$ dust particles) and the one-fluid method (right; with $64 \times 12 \times 12$ mixture particles) and a dust-to-gas ratio of unity. Panels show results with $K=0.5,5,50$ and $500$ (top to bottom), corresponding to the stopping times indicated. The two-fluid method is accurate when the stopping time is long (top three panels in left figure) but requires $h \lesssim c_{\rm s} t_{\rm s}$ to avoid overdamping (bottom two panels on left; \citealt{laibeprice12}). The one-fluid method should be used when the stopping time is short (right figure).}
\label{fig:dustywave} 
\label{fig:dustywave-onef}
\end{figure*}

\subsubsection{\sc Dustywave}
\citet{maddison98} and \citet{laibeprice11} derived the analytic solution for linear waves in a dust-gas mixture: the `{\sc dustywave}' test. The corresponding dispersion relation is given by \citep{maddison98,laibeprice11,laibeprice12}
\begin{equation}
\omega^3 + iK \left(\frac{1}{\rho_{\rm g}} + \frac{1}{\rho_{\rm d}} \right) \omega^2-  c_{\rm s}^2 k^2 \omega - iK \frac{k^2 c_{\rm s}^2}{\rho_{\rm d}} = 0,
\end{equation}
which can be more clearly expressed as
\begin{equation}
(\omega^2 - c_{\rm s}^2 k^2) + \frac{i}{\omega t_{\rm s}} (\omega^2 - \tilde{c}_{\rm s}^2 k^2) = 0,
\end{equation}
where $\tilde{c}_{\rm s} = c_{\rm s} (1 + \rho_{\rm d}/\rho_{\rm g})^{-1/2}$ is the modified sound speed \citep[e.g.][]{miuraglass82}. This demonstrates the two important limits i) $t_{\rm s} \to \infty$, giving undamped sound waves in the gas and ii) $t_{\rm s} \to 0$, giving undamped sound waves in the mixture propagating at the modified sound speed. In between these limits, the mixture is dissipative and waves are damped by the imaginary term. This is seen in the analytic solutions shown in Figure~\ref{fig:dustywave}.

\paragraph{Two-fluid.}
We perform this test first with the two-fluid algorithm, using $64 \times 12 \times 12$ gas particles and $64 \times 12 \times 12$ dust particles set up on a uniform, close-packed lattice in a periodic box with $x \in [-0.5,0.5]$ and the $y$ and $z$ boundaries set to correspond to 12 particle spacings on the chosen lattice. The wave is set to propagate along the $x-$axis with $v_{\rm g} = v_{\rm d} =A \sin(2\pi x)$, $\rho = \rho_{0}[1 + A\sin(2\pi x)]$ with $\rho_{0} = 1$ and $A=10^{-4}$. The density perturbation is initialised using stretch mapping (Section~\ref{sec:stretchmap}; see also Appendix B in \citealt{pricemonaghan04a}). We perform this test using an adiabatic equation of state with $c_{{\rm s}, 0} = 1$. We adopt a simple, constant $K$ drag prescription, choosing $K = 0.5, 5, 50$ and $500$ such that the stopping time given by (\ref{eq:ts}) is a multiple of the wave period ($t_{\rm s} = 1, 0.1, 0.01$ and $0.001$, respectively).
 
  The left panels in Figure~\ref{fig:dustywave} show the results of this test using the two-fluid method, showing velocity in each phase compared to the analytic solution after 4.5 wave periods (the time is chosen to give a phase offset between the phases). For stopping times $t_{\rm s} \gtrsim 0.1$ the numerical solution matches the analytic solution to within 4 per cent. For short stopping times, \citet{laibeprice12} showed that the resolution criterion $h \lesssim c_{\rm s} t_{\rm s}$ needs to be satisfied to avoid overdamping of the mixture. For the chosen number of particles, the smoothing length is $h = 0.016$, implying in this case that the criterion is violated when $t_{\rm s} \lesssim 0.016$. This is evident from the lower two panels in Figure~\ref{fig:dustywave}, where the numerical solution is overdamped compared to the analytic solution.
  
  This problem is not unique to SPH codes, but represents a fundamental limitation of two-fluid algorithms in the limit of short stopping times due to the need to resolve the physical separation between the phases (which becomes ever smaller as $t_{\rm s}$ decreases) when they are modelled with separate sets of particles (or with a grid and a physically separate set of dust particles). The need to resolve a physical length scale results in first-order convergence of the algorithm in the limit of short stopping times, as already noticed by \citet{miniati10} in the context of grid-based codes. The problem is less severe when the dust fraction is small \citep{lorenbate14}, but is difficult to ameliorate fully.
  

\paragraph{One-fluid.}  
  The limit of short stopping time (small grains) is the limit in which the mixture is well described by the one-fluid formulation in the terminal velocity approximation (Section~\ref{sec:onefluid}). To compare and contrast the two methods for simulating dust in \textsc{Phantom}, the right half of Figure~\ref{fig:dustywave} shows the results of the {\sc dustywave} test computed with the one-fluid method. To perform this test we set up a single set of $64 \times 12 \times 12$ `mixture' particles placed on a uniform closepacked lattice, with an initially uniform dust fraction $\epsilon = 0.5$. The particles are given a mass corresponding to the combined mass of the gas \emph{and} dust, with the density perturbation set as previously.
  
  The one-fluid solution is accurate precisely where the two-fluid method is inaccurate, and vice-versa. For short stopping times ($t_{\rm s} = 0.001$; bottom row) the numerical solution is within 1.5 per cent of the analytic solution, compared to errors greater than 60 per cent for the two-fluid method (left figure). For long stopping times ($t_{\rm s} \lesssim 1$; top two rows) the one-fluid method is both inaccurate and slow, but this is precisely the regime in which the two-fluid method (left figure) is explicit and therefore cheap. Thus, the two methods are complementary.

\subsubsection{Dust diffusion}
\label{sec:dustdiffuse}
\begin{figure}
   \centering
   \includegraphics[width=\columnwidth]{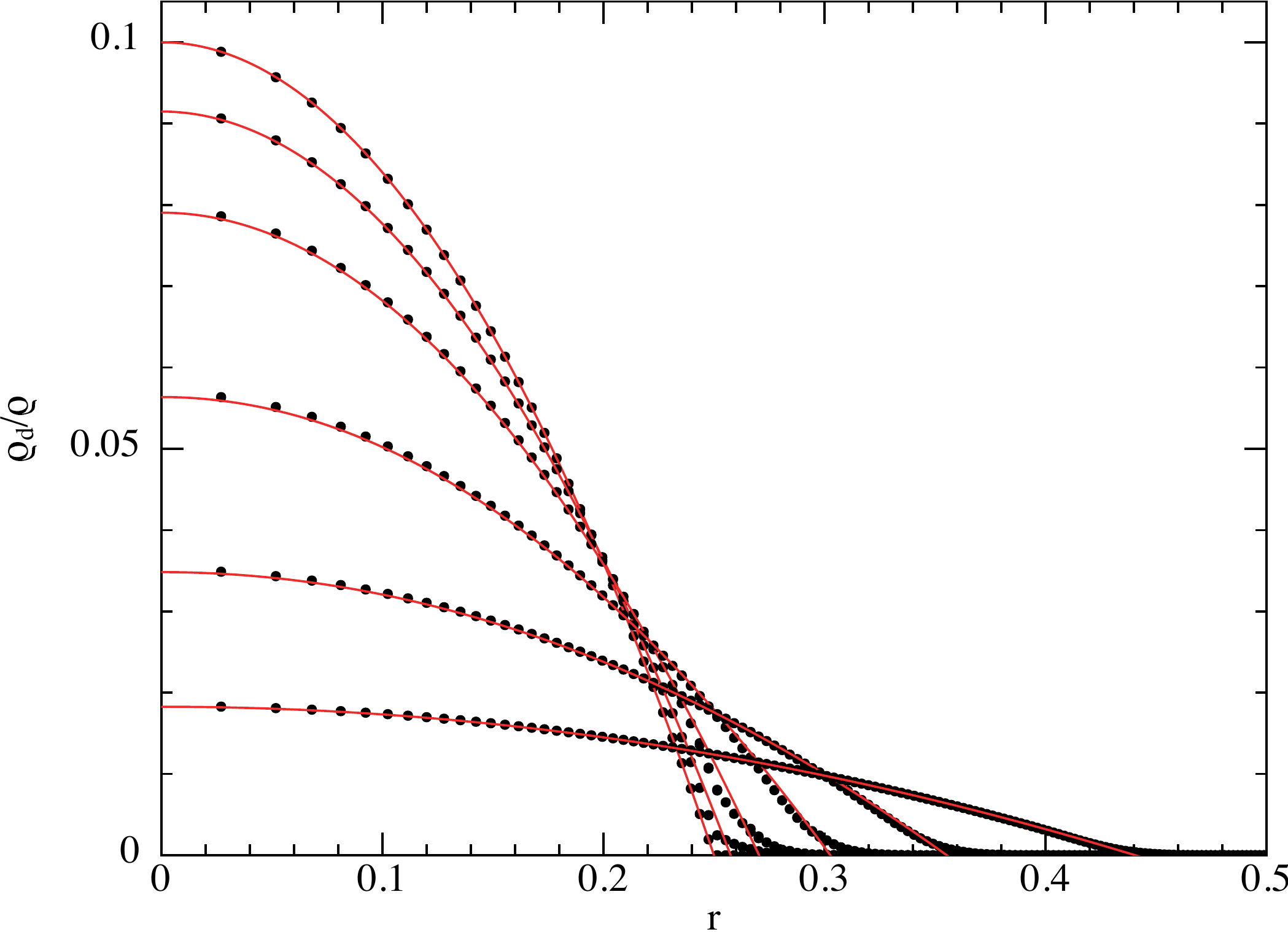}
   \caption{Dust diffusion test from \citet{pricelaibe15}, showing the evolution of the dust fraction on the particles (black dots) as a function of radius at 6 different times (top to bottom), which may be compared to the analytic solution given by the red lines.}
\label{fig:dustydiffuse}
\end{figure}

 A simple test of the one-fluid dust diffusion algorithm is given by \citetalias{pricelaibe15}. For this test we set up the particles on a uniform cubic lattice in a 3D periodic box $x,y,z \in [-0.5,0.5]$ using $32 \times 32 \times 32$ particles with an initial dust fraction set according to
\begin{equation}
\epsilon(r,0) = \begin{cases}
\epsilon_{0} \left[ 1 - \left( \frac{r}{r_{c}} \right)^{2} \right], & r < r_{c}, \\
0, & \textrm{elsewhere},
\end{cases}
\end{equation}
with $\epsilon_{0} = 0.1$ and $r_{c} = 0.25$. We then evolve the dust diffusion equation, (\ref{eq:depsdth}), discretised according to (\ref{eq:dsdtsph}), while setting the acceleration and thermal energy evolution to zero and assuming $P = \rho$, with the stopping time set to a constant $t_{\rm s} = 0.1$ and the computational timestep set to $\Delta t = 0.05$. Figure~\ref{fig:dustydiffuse} shows the evolution of the dust fraction $\epsilon \equiv \rho_{\rm d}/\rho$ as a function of radius at various times, showing the projection of all particles in the box (points) compared to the exact solution (red lines) at $t=0.0$, 0.1, 0.3, 1.0 and 10.0 (top to bottom). The solution shows a close match to the analytic solution, with agreement to within 0.3 per cent of the analytic solution at all times.

\begin{figure*}
   \centering
   \includegraphics[width=\textwidth]{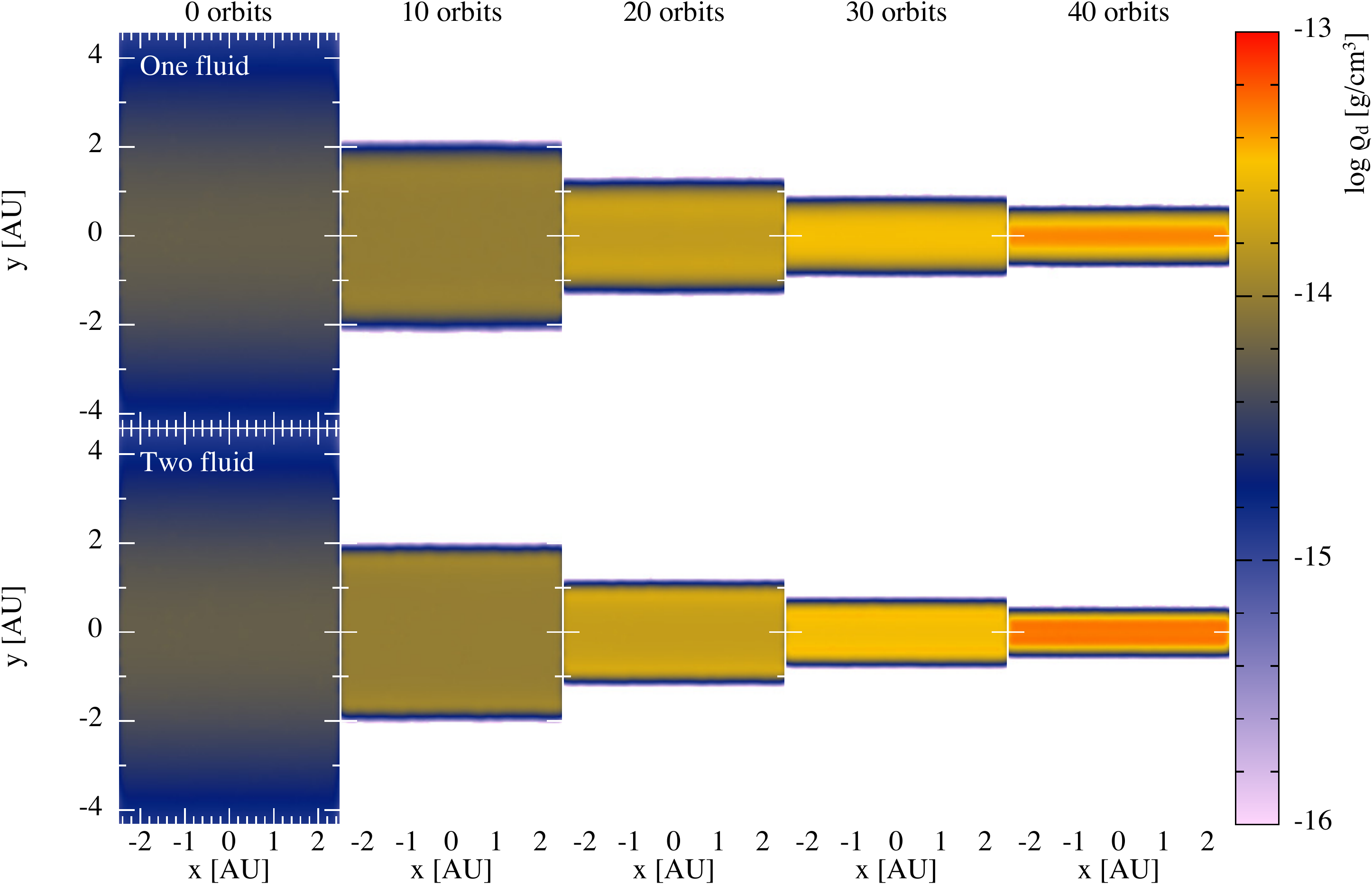}
   \caption{3D version of the dust settling test from \citet{pricelaibe15}, showing the dust density in the `$r$-$z$' plane of a protoplanetary disc. We assume mm-sized grains with a 1\% initial dust-to-gas ratio in a stratified disc atmosphere with $H/R_0 = 0.05$ with $R_0 = 50$~au in (\ref{eq:accsettle}).}
\label{fig:dustsettle}
\end{figure*}

\subsubsection{Dust settling}
 We perform the dust settling test from \citetalias{pricelaibe15} in order to directly compare the \textsc{Phantom} solutions to those produced in \citetalias{pricelaibe15} with the {\sc ndspmhd} code.   To simplify matters we do not consider rotation but simply adapt the 2D problem to 3D by using a thin Cartesian box (as for several of the MHD tests in Section~\ref{sec:mhdtests}). Our setup follows \citetalias{pricelaibe15}, considering a slice of a protoplanetary disc at $R_{0} = 50$ au in the $r-z$ plane (corresponding to our $x$ and $y$ Cartesian coordinates, respectively) with density in the `vertical' direction ($y$) given by
\begin{equation}
\rho(y) = \rho_{0} \exp \left( \frac{-y^{2}}{2H^{2}} \right), \label{eq:rhoz}
\end{equation}
where we choose $H/R_{0} = 0.05$ giving $H = 2.5$ au. We use an isothermal equation of state with sound speed $c_{\rm s} \equiv H \Omega$ where $\Omega \equiv \sqrt{GM/R_{0}^{3}}$, corresponding to an orbital time $t_{\rm orb} \equiv 2\pi/\Omega \approx 353$ yrs. We adopt code units with a distance unit of 10 au, mass in solar masses and time units such that $G=1$, giving an orbital time of $\approx 70.2$ in code units. We apply an external acceleration in the form
\begin{equation}
{\bm a}_{\rm ext} = -\frac{GM y}{\sqrt{{R_{0}^{2} + y^{2}}}},
\label{eq:accsettle}
\end{equation}
where $G=M=1$ in code units.
  
 The particles are placed initially on a close-packed lattice using $32 \times 110 \times 24 = 84~480$ particles in the domain $[x,y,z] \in [\pm 0.25,\pm 3H, \pm\sqrt{3/128}]$. We then use the stretch mapping routine (Section~\ref{sec:stretchmap}) to give the density profile according to (\ref{eq:rhoz}). We set the mid-plane density to $10^{-3}$ in code units, or $\approx 6 \times 10^{-13} $g/cm$^{3}$. The corresponding particle mass in code units is $1.13 \times 10^{-9}$. We use periodic boundaries, with the boundary in the $y$ direction set at $\pm 10H$ to avoid periodicity in the vertical direction. 
 
 Following the procedure in \citetalias{pricelaibe15} we relax the density profile by evolving for 15 orbits with gas only with damping switched on. We then restart the calculation with either i) a dust fraction added to each particle (one-fluid), or ii) a corresponding set of dust particles duplicated from the gas particles (two-fluid). For the dust we assume $1$~mm grains with an Epstein drag prescription, such that the stopping time is given by (\ref{eq:tseps}). Since $\Delta v$ is not available when computing $t_{\rm s}$ with the one-fluid method, we set the factor $f=1$ in (\ref{eq:tseps}) when using this method (this is a valid approximation since by definition $\Delta v$ is small when the one-fluid method is applicable). The dimensionless stopping time $t_{\rm s} \Omega = 8.46 \times 10^{-3}$ initially at the disc midplane. After adding dust we continued the simulation for a further 50 orbits.

 \begin{figure*}
    \centering
    \includegraphics[width=\textwidth]{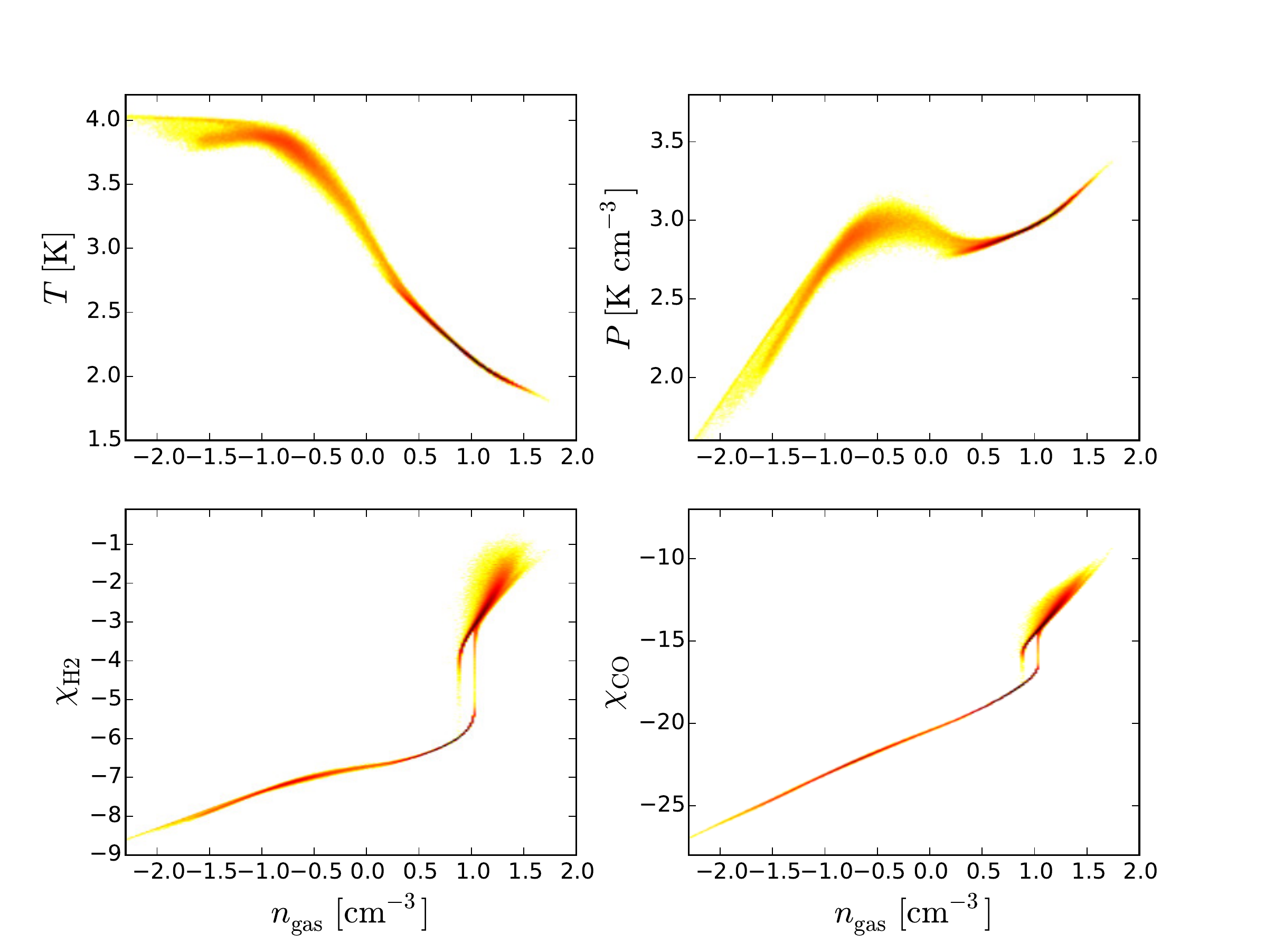}
    \caption{Temperature and pressure profiles resulting from the ISM heating and cooling functions (top) and the abundances of $\rm H_2$ and CO as a function of gas density (bottom). The behaviour of the H$_{2}$ and CO fractions close to $n = 10 \: {\rm cm^{-3}}$ is a consequence of H$_{2}$ self-shielding: in gas which was initially molecular, this remains effective down to lower densities than is the case in gas which was initially atomic. This behaviour is discussed in much greater detail in \citet{2008MNRAS.389.1097D}.}
 \label{fig:coolchem}
 \end{figure*}
 
 Figure~\ref{fig:dustsettle} shows the dust density at intervals of 10 $t_{\rm orb}$, showing the cross section slice through the $z=0$ plane of the 3D box which may be directly compared to the 2D solutions shown in \citetalias{pricelaibe15}. Settling of the dust layer proceeds as expected, with close agreement between the one-fluid (top row) and two-fluid (bottom row) methods, though the two-fluid method is much slower for this test because of the timestep constraint imposed by the stopping time, c.f. (\ref{eq:dtdrag}). The dust resolution is higher in the two-fluid calculation because the set of dust particles follow the dust mass rather than the total mass (for the one-fluid method).



\subsection{ISM cooling and chemistry}
\label{sec:ismtest}
Figure \ref{fig:coolchem} shows the behaviour of the various cooling and chemistry modules used when modelling the ISM. These plots were made from the data from a  simulation of gas rings embedded in a static background potential giving a flat rotation curve \citep{1987gady.book.....B}. Gas is setup in a ring of constant surface density from 5 kpc to 10 kpc in radius, initially at 10~000 K, with a total gas mass of $2\times10^{9}$ M$_\odot$. The top panels show a temperature and pressure profile of all gas particles. The temperature plateaus around 10~000 K and forms a two phase medium visible in the `knee' in the pressure profile, as expected for ISM thermal models \citep{1995ApJ...443..152W}. Much lower temperatures can be reached if the gas is given a higher surface density or if self-gravity is active. In the case of the latter, some energy delivery scheme, or the inclusion of a large number of sink particles, is needed to break apart the cold knots.

The bottom two panels of Figure \ref{fig:coolchem}  show the chemical abundances of $\rm H_2$ and CO. The exact form of the molecular abundance profiles are a function of many variables that are set at run time, with the data in the figure made from the default values. The molecular abundances are strong functions of total gas density, with the CO being a strong function of $\rm H_2$ abundance. If a higher gas mass (e.g. $\times 10$ the value used here) or self-gravity is included then abundances reach a maximum of either 0.5 for $\rm H_2$ or the primordial carbon abundance for CO. See \citet{2008MNRAS.389.1097D} for a detailed discussion of the features of these abundance curves.

\begin{figure*}
\centering
\includegraphics[width=\textwidth]{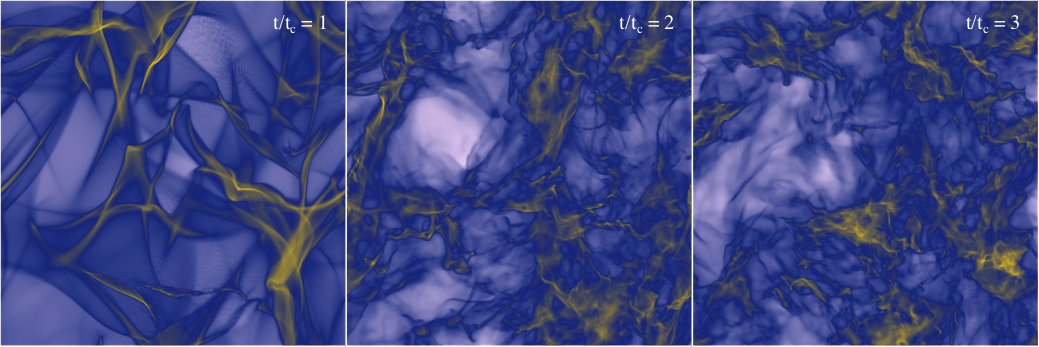}
\caption{Gas column density in \textsc{Phantom} simulations of driven, isothermal, supersonic turbulence at Mach 10, similar to the calculations performed by \citet{pricefederrath10}. We show the numerical solutions at $t = 1$, 2 and 3 crossing times (left to right, respectively). The colour scale is logarithmic between 10$^{-1}$ and 10 in code units. }
\label{fig:turbdens}
\end{figure*}

\begin{figure}
\centering
\includegraphics[width=\columnwidth]{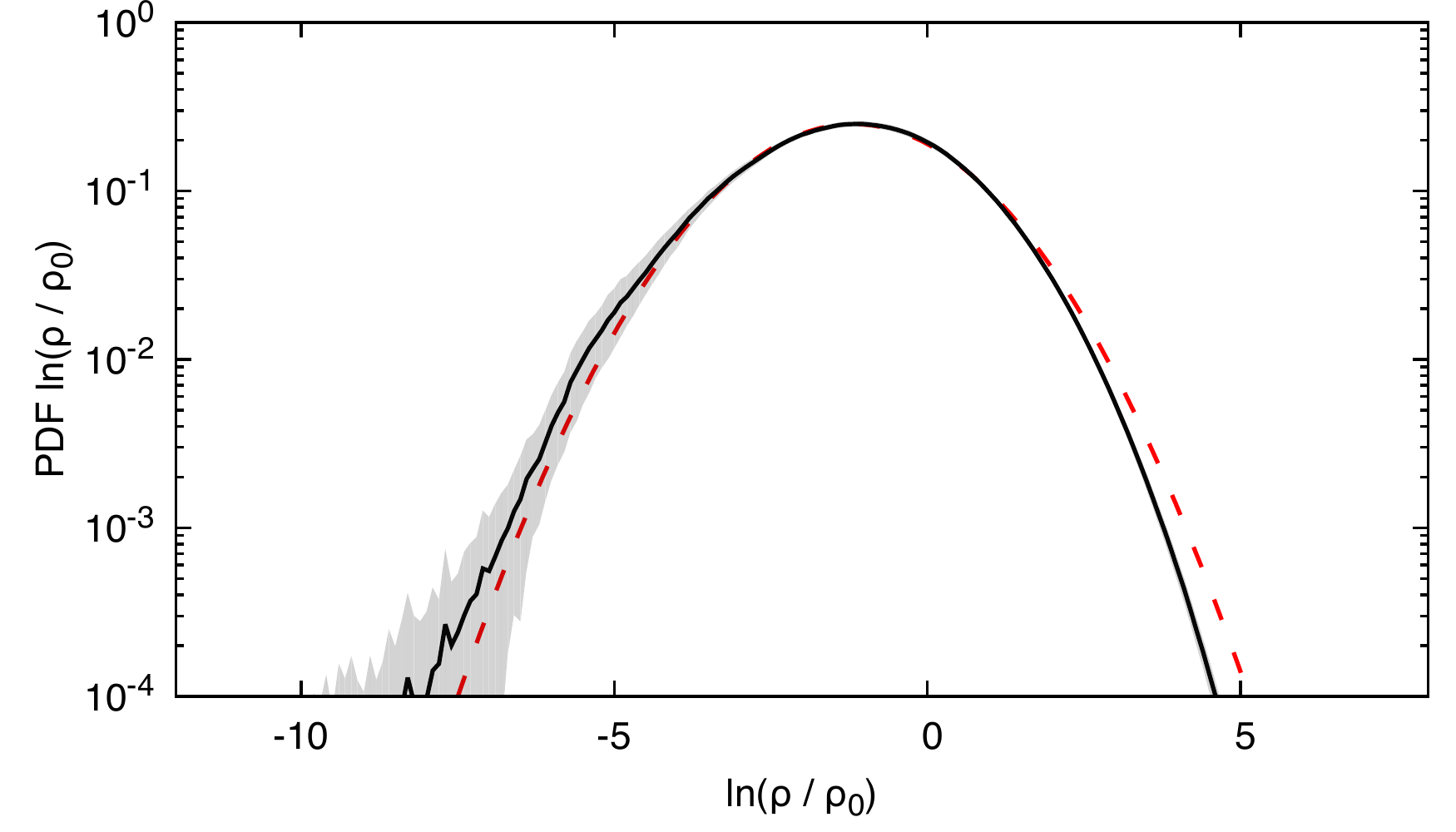}
\caption{Time-averaged PDF of $s = \ln(\rho / \rho_0)$ for supersonic Mach 10 turbulence, with the shaded region representing the standard deviation of the averaging. The PDF is close to a log-normal distribution, shown by the dashed red line.}
\label{fig:turbdenspdf}
\end{figure}

%
%

\section{Example applications}
\label{sec:apps}

 \textsc{Phantom} is already a mature code, in the sense that we have always developed the code with specific applications in mind. In this final section we demonstrate five example applications for which the code is well suited. The setup for each of these applications are provided in the wiki documentation so they can be easily reproduced by the novice user. We also plan to incorporate these examples into an `optimisation suite' to benchmark performance improvements to the code.

\subsection{Supersonic turbulence}
\label{sec:turbulence}
 Our first example application employs the turbulence forcing module described in Section~\ref{sec:turbforcing}. Figure~\ref{fig:turbdens} shows the gas column density in simulations of isothermal supersonic turbulence driven to an rms Mach number of $\mathcal{M} \approx 10$, identical to those performed by \citet{pricefederrath10}. The calculations use $256^3$ particles and were evolved for 10 turbulent crossing times, $t_{\rm c} = L / (2 \mathcal{M})$. This yields a crossing time of $t_{\rm c} = 0.05$ in code units. The gas is isothermal with sound speed $c_{\rm s} = 1$ in code units. The initial density is uniform $\rho_0 = 1$.
 

The gas column density plots in Figure~\ref{fig:turbdens} may be directly compared to the panels in Figure~3 of \citet{pricefederrath10}. Figure~\ref{fig:turbdenspdf} shows the time-averaged probability distribution function (PDF) of $s \equiv \ln(\rho / \rho_0)$. This demonstrates the characteristic signature of isothermal supersonic turbulence, namely the appearance of a log-normal PDF in $s$ \citep[e.g.][]{vazquez-semadeni94,nordlundpadoan99,ogs99}. Indeed, \textsc{Phantom} was used in the study by \citet*{pfb11} to confirm the relationship between the standard deviation and the Mach number in the form
\begin{equation}
\sigma_{\rm s}^{2} = \ln \left( 1 + b^{2} \mathcal{M}^{2} \right),
\end{equation}
where $b=1/3$ for solenoidally driven turbulence, as earlier suggested by \citet{fks08,federrathetal10}.

\begin{figure*}
\centering
\includegraphics[width=0.32\textwidth]{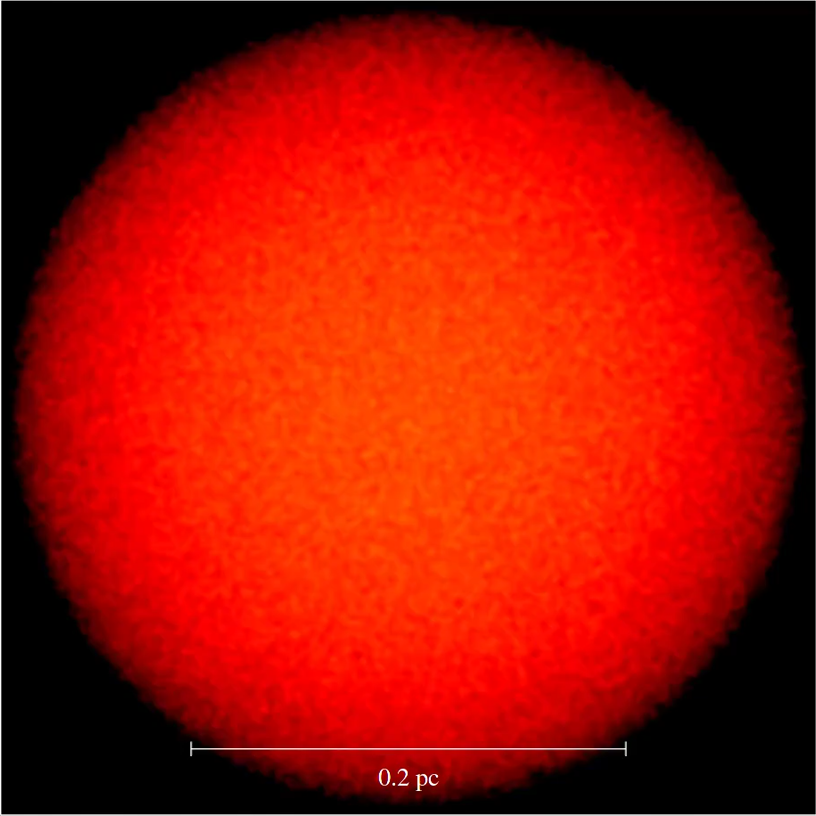}
\includegraphics[width=0.32\textwidth]{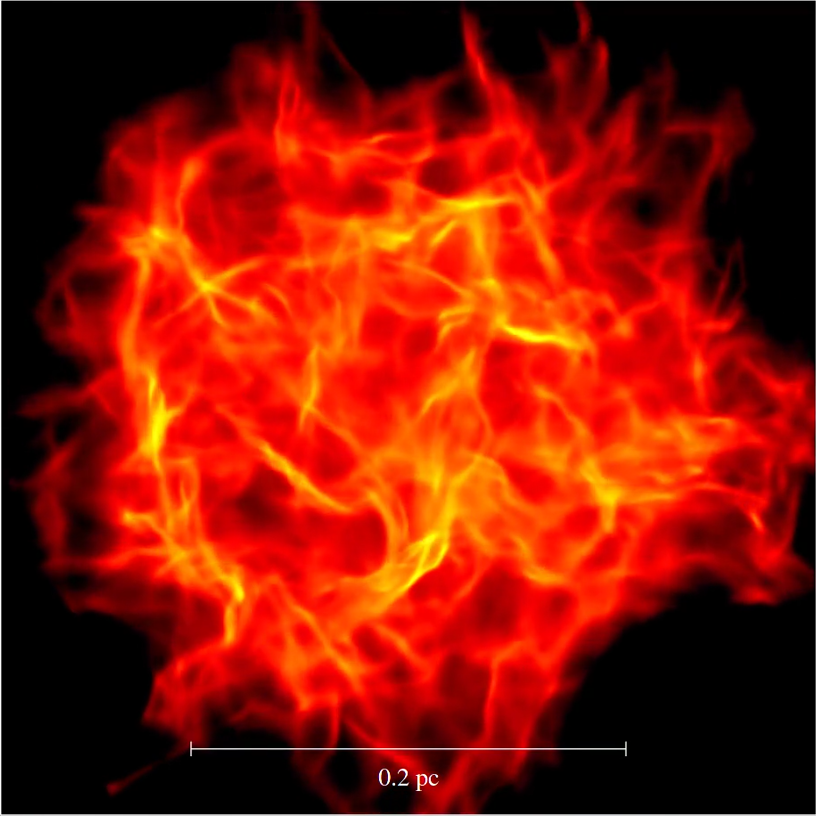}
\includegraphics[width=0.32\textwidth]{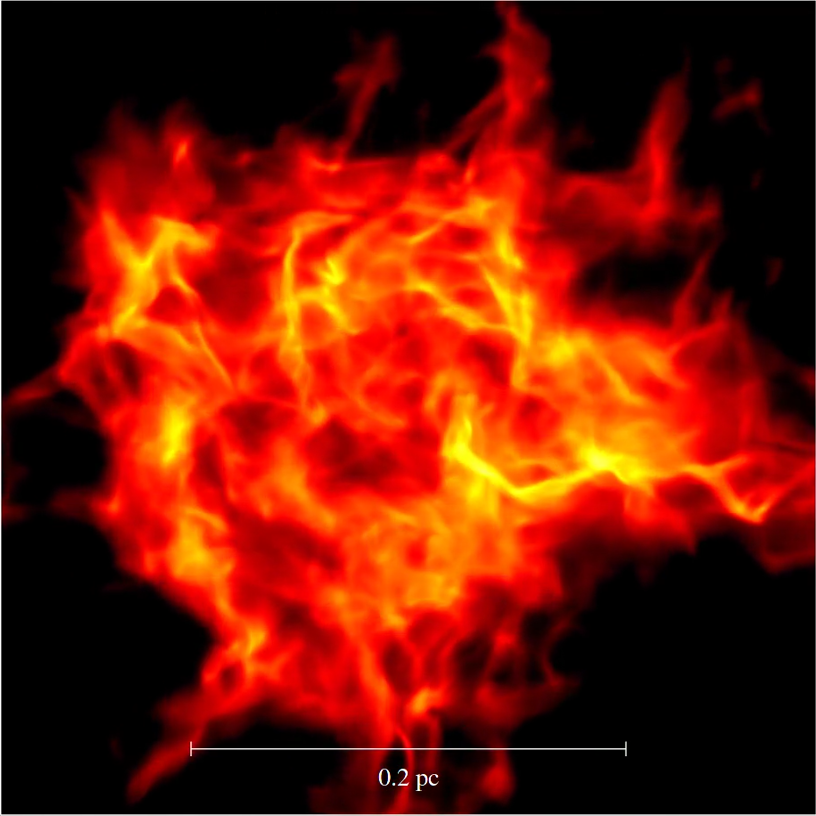}
\includegraphics[width=0.32\textwidth]{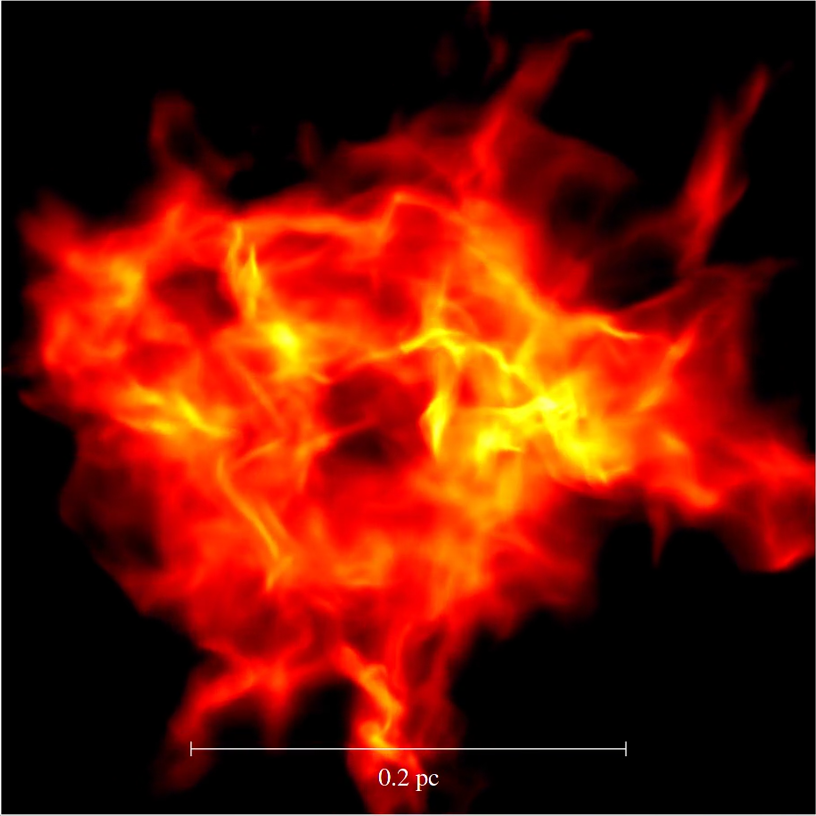}
\includegraphics[width=0.32\textwidth]{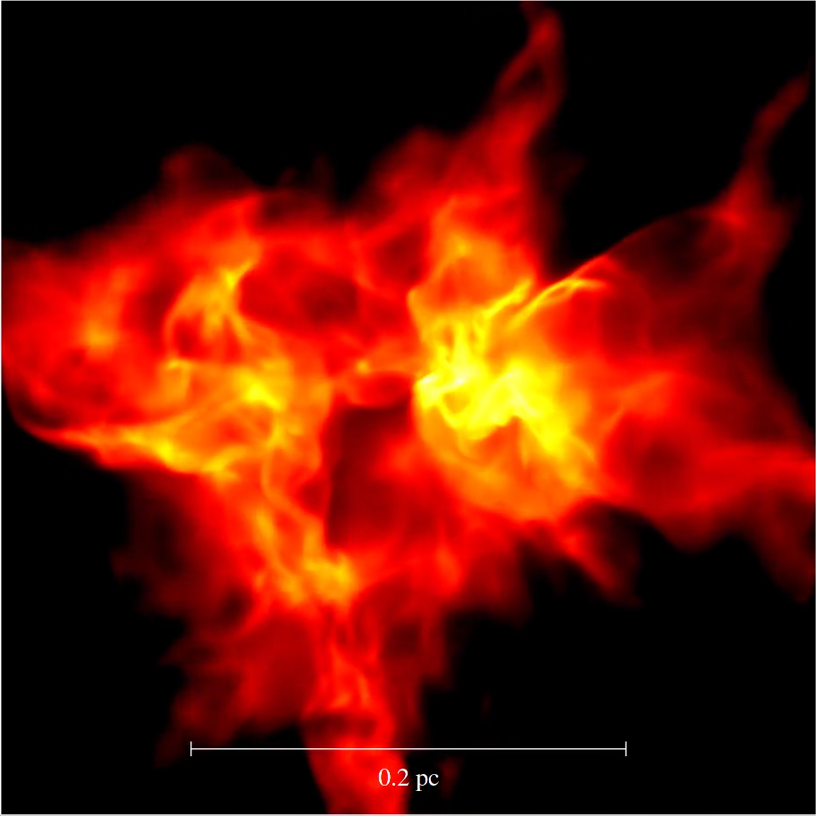}
\includegraphics[width=0.32\textwidth]{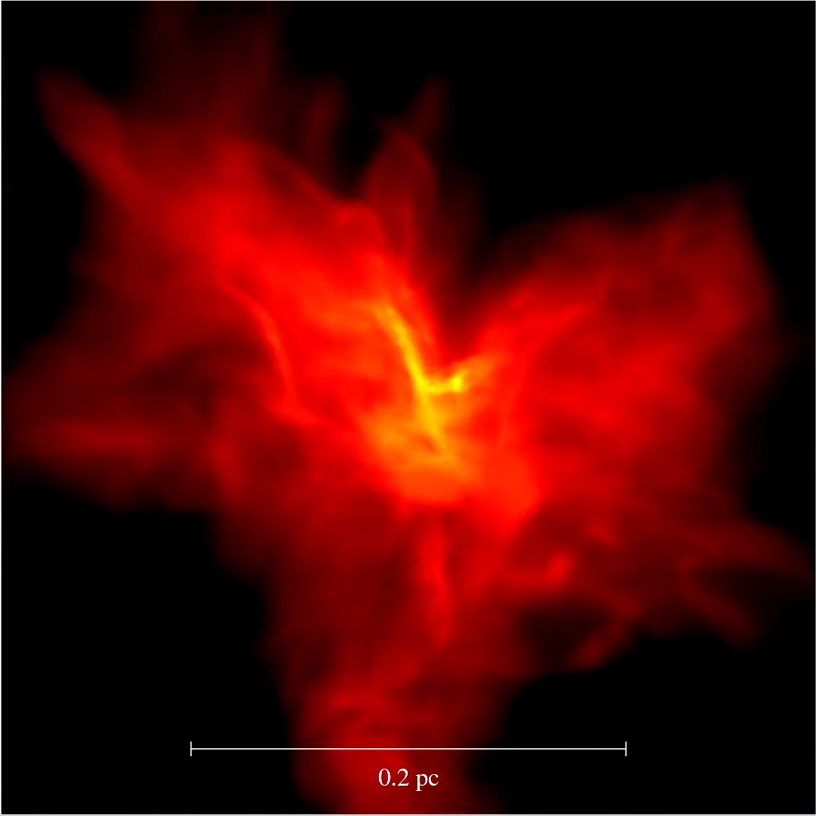}
\includegraphics[width=0.32\textwidth]{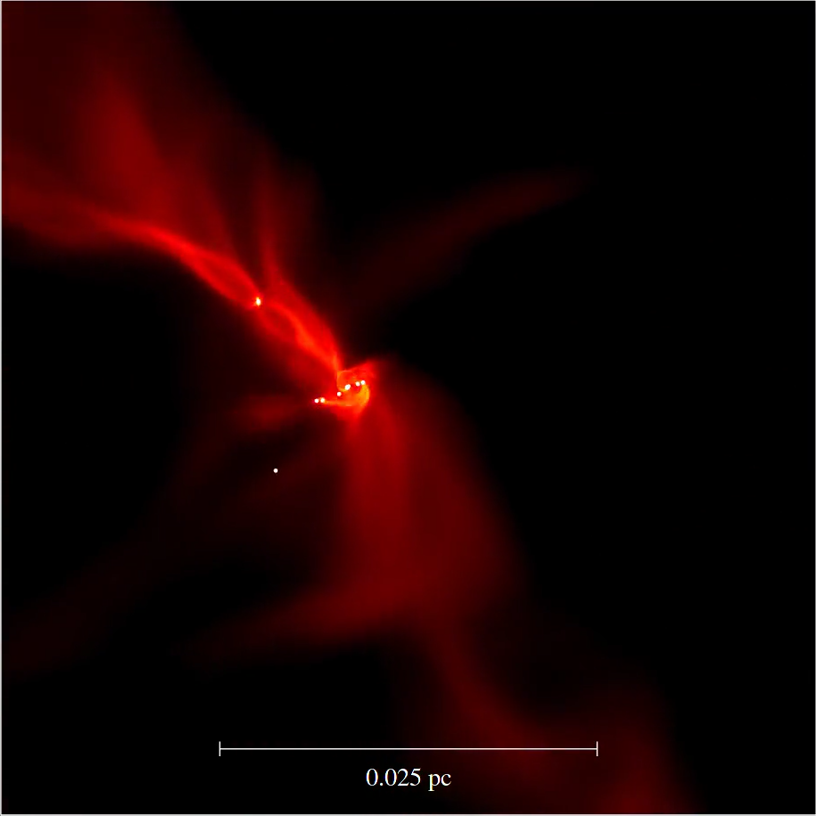}
\includegraphics[width=0.32\textwidth]{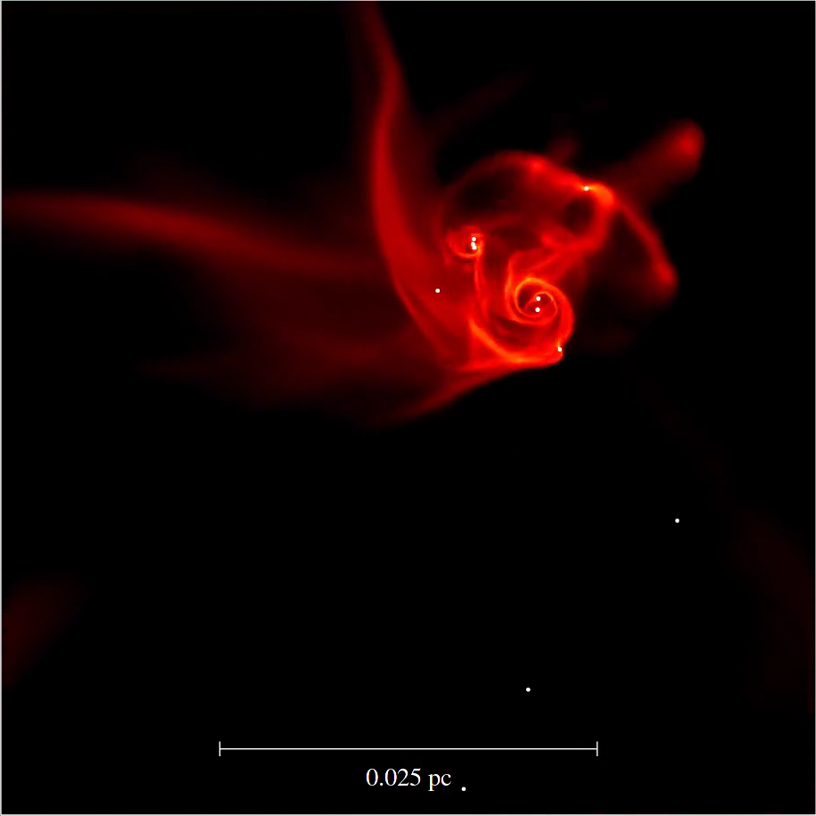}
\includegraphics[width=0.32\textwidth]{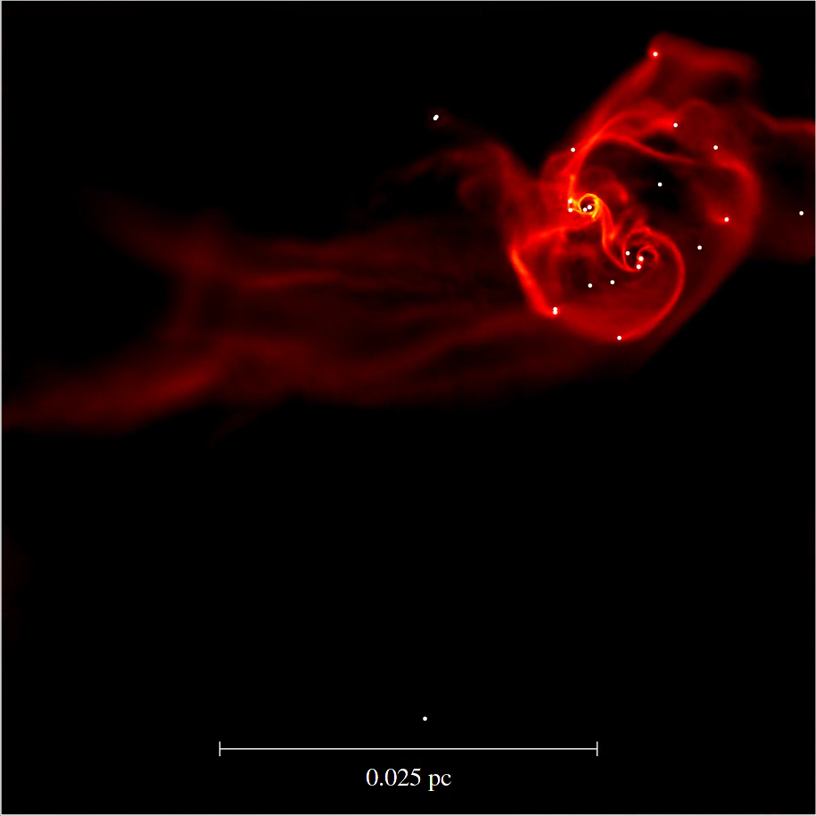}
\caption{Star cluster formation with {\sc Phantom}, showing snapshots of gas column density during the gravitational collapse of a 50 M$_\odot$ molecular cloud core, following \citet{bbb03}. Snapshots are shown every 0.2 t$_{\rm ff}$ (left to right, top to bottom), with the panels after $t > t_{\rm ff}$ zoomed in to show the details of the star formation sequence. As in \citet{bbb03}, we resolve the fragmentation to the opacity limit using a barotropic equation of state.}
\label{fig:cluster}
\end{figure*}

\subsection{Star cluster formation}
 SPH has been used to study star formation since the earliest studies by \citet{gingoldmonaghan82,gingoldmonaghan83,phillips82,phillips86,phillips86a} and \citet{monaghanlattanzio86,monaghanlattanzio91}, even motivating the original development of MHD in SPH by \citet{phillipsmonaghan85}. The study by \citet{bbb03} represented the first simulation of `large scale' star cluster formation, resolved to the opacity limit for fragmentation \citep{rees76,lowlynden-bell76}. This was enabled by the earlier development of sink particles by \citet{bbp95}, allowing star formation simulations to continue beyond the initial collapse \citep{bonnelletal97,batebonnell97}. This heritage is present in \textsc{Phantom} which inherits many of the ideas and algorithms implemented in the {\sc sphng} code.

 Figure~\ref{fig:cluster} shows a series of snapshots taken from a recent application of {\sc Phantom} to star cluster formation by \citet{liptaietal17}. The initial setup follows \citet{bbb03} --- a uniform density sphere of 0.375 pc in diameter with a mass of 50 M$_\odot$. The initial velocity field is purely solenoidal, generated on a 64$^3$ uniform grid in Fourier space to give a power spectrum $P(k) \propto k^{-4}$ consistent with the \citet{larson81} scaling relations, and then linearly interpolated from the grid to the particles. The initial kinetic energy is set to match the gravitational potential energy, ($3/5 GM^2/R$), giving a root mean square Mach number $\approx 6.4$. We set up $3.5 \times 10^6$ particles in the initial sphere placed in a uniform random distribution. We evolve the simulation using a barotropic equation of state $P = K \rho^\gamma$ in the form
\begin{equation}
\frac{P}{\rho} =
\begin{cases}
c^2_{{\rm s}, 0},  & \rho < \rho_1, \\
c^2_{{\rm s}, 0} \left( \frac{\rho}{\rho_1}\right)^{(\gamma_1 - 1)}, & \rho_1 \leq \rho < \rho_2, \\
c^2_{{\rm s}, 0} \left( \frac{\rho_2}{\rho_1}\right)^{(\gamma_1 - 1)} \left( \frac{\rho}{\rho_1}\right)^{(\gamma_2 - 1)},  & \rho_2 \leq \rho < \rho_3, \\
c^2_{{\rm s}, 0} \left( \frac{\rho_2}{\rho_1}\right)^{(\gamma_1 - 1)} \left( \frac{\rho_3}{\rho_2}\right)^{(\gamma_2 - 1)}  \left( \frac{\rho}{\rho_3} \right)^{(\gamma_3 - 1)}, & \rho \geq \rho_3,
\end{cases}
\label{eq:starformationeos}
\end{equation}
where we set the initial sound speed $c_{{\rm s},0} = 2 \times 10^4$ cm s$^{-1}$ and set $[\rho_1,\rho_2,\rho_3] = [10^{-13},10^{-10},10^{-3}]$ g cm$^{-3}$ and $[\gamma_1,\gamma_2,\gamma_3] = [1.4,1.1,5/3]$, as in \citet{bbb03}. We turn on automatic sink particle creation with a threshold density of $10^{-10}$ g cm$^{-3}$, with sink particle accretion radii set to 5~au and particles accreted without checks at 4~au. No sink particles are allowed to be created within 10~au of another existing sink. The calculations satisfy the \citet{bateburkert97} criterion of resolving the minimum Jeans mass in the calculation (known as the opacity limit for fragmentation; \citealt{rees76,lowlynden-bell76}) by at least the number of particles contained within one smoothing sphere.

 \begin{figure*}
\includegraphics[width=\linewidth]{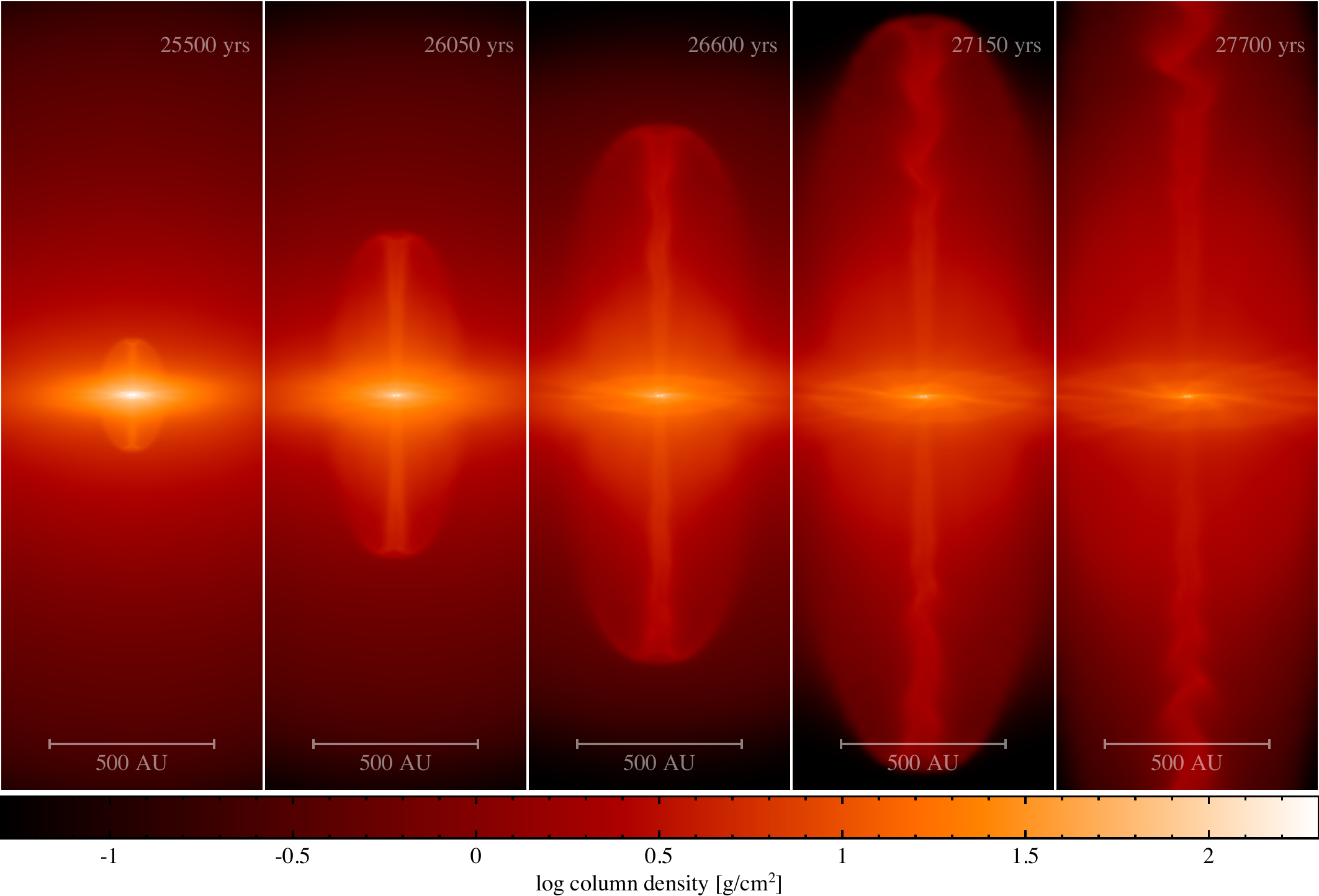}
\caption{Magnetically propelled jet of material bursting out of the first hydrostatic core phase of star formation.}
\label{fig:jetevo}
\end{figure*}

 The snapshots shown in Figure~\ref{fig:cluster} show a similar evolution to the original calculation of \citet{bbb03}. The evolution is not identical since we used a different realisation of the initial turbulent velocity field. A more quantitative comparison can be found in \citet{liptaietal17} where we performed 7 different realisations of the collapse in order to measure a statistically meaningful initial mass function (IMF) from the calculations, finding an IMF in agreement with the one found by \citet{bate09a} in a much larger (500 M$_\odot$) calculation. The IMF produced with a barotropic equation of state does not match the observed local IMF in the Milky Way (e.g.~\citealt{chabrier05}), tending to over-produce low mass stars and brown dwarfs. This is a known artefact of the barotropic equation of state \citep[e.g.][]{matznerlevin05,krumholz06,bate09}, since material around the stars remains cold rather than being heated. It can be fixed by implementing radiative feedback, for example by implementing radiation hydrodynamics in the flux-limited diffusion approximation \citep{whitehousebate04,wbm05,whitehousebate06}. This is not yet implemented in {\sc Phantom} but it is high on the agenda.

\subsection{Magnetic outflows during star formation}
\label{sec:jet}
{\sc Phantom} may also be used to model the formation of individual protostars. We present an example following the initial setup and evolution of \citet{ptb12}. A molecular cloud core with initial density $\rho_0 = 7.4 \times 10^{-18}$ g cm$^{-3}$ is embedded in pressure equilibrium with ambient medium of density $2.5 \times 10^{-19}$ g cm$^{-3}$. The barotropic equation of state given by (\ref{eq:starformationeos}) is used, setting $c_{\rm s} = 2.2 \times 10^{4}$ cm s$^{-1}$. The radius of the core is $4 \times 10^{16}$ cm ($\approx 2700$ au), with the length of the cubic domain spanning $[x,y,z] = \pm 8 \times 10^{16}$ cm. The core is in solid body rotation with angular speed $\Omega = 1.77 \times 10^{-13}$ rad s$^{-1}$. The magnetic field is uniform and aligned with the rotation axis with a mass-to-flux ratio $\mu = 5$, corresponding to $B_0 \approx 163$ $\mu$G. A sink particle is inserted once the gas reaches a density of $10^{-10}$ g cm$^{-3}$, with an accretion radius of 5 au. Thus, this calculation models only the evolution of the first hydrostatic core phase of star formation. The core is composed of $1,004,255$ particles, with $480,033$ particles in the surrounding medium. 

Figure~\ref{fig:jetevo} shows the evolution of a magnetised, collimated bipolar jet of material, similar to that given by \citet{ptb12}. Infalling material is ejected due to the wind up of the toroidal magnetic field. The sink particle is inserted at $t \approx 25~400$ yrs, shortly before the jet begins. The jet continues to be driven while material continues to infall, lasting for several thousand years.

\subsection{Galaxy merger}
 To provide a realistic test of the collisionless $N$-body and SPH implementations, we performed a comparison study where we modelled a galaxy merger, comparing the {\sc Phantom} results with the {\sc Hydra} $N$-body/SPH code (\citealp{CTP95}; \citealp{TC06}). This test requires gravity along with multiple particle types --- gas, stars and dark matter.  Gas interacts hydrodynamically only with itself, and all three particle types interact with each other via gravity (c.f. Table~\ref{tab:types}).

To create a Milky Way-like galaxy, we used {\sc GalactICs} (\citealp{KD95}; \citealp{WD05}; \citealp{WPD08}) to first create a galaxy consisting of a stellar bulge, stellar disc and a dark matter halo.  To create the gas disc, the stellar disc was then duplicated and reflected in the $x=y$ plane to avoid coincidence with the star particles. Ten percent of the total stellar mass was then removed and given to the gas disc.  Although the gas disc initially has a scale height larger than physically motivated, this will quickly relax into a disc that physically resembles the Milky Way.  Next, we added a hot gas halo embedded within the dark matter halo.  The hot gas halo has an observationally motivated $\beta$-profile (e.g. \citealp{CF76}) and a temperature profile given by \citet{KMWSM07}; the mass of the hot gas halo is removed from the dark matter particles to conserve total halo mass.  The mass of each component, as well as particle numbers and particle masses are given in Table~\ref{table:gm:n}.  To model the major merger, the galaxy is duplicated and the two galaxies are placed 70 kpc apart on a parabolic trajectory.  These initial conditions are identical to those used in \citet{wursterthacker13,wursterthacker13a}. To simplify the comparison, there is no star formation recipe, no black holes and no feedback from active galactic nuclei (there are currently no plans to implement cosmological recipes in {\sc Phantom}). Thus, only the SPH and gravity algorithms are being compared.
\begin{center}
\begin{table}
\begin{tabular}{l r r r}
    \hline
 & $M/$M$_\odot$                   & $m/$M$_\odot$                  & $N$ \\
 & (10$^{10}$) &(10$^5$)&     \\
\hline
\hline
Dark matter halo & 89.92    & 89.92 & 100 000 \\
Hot gas halo       &  0.60     &    2.77 &  21 619 \\
Stellar bulge       &  1.34     & 18.10 &   7 407 \\
Stellar disc         &  3.56      & 18.10 &  19 662 \\
Gas disc            &  0.54      &   2.77 &  19 662 \\
\hline
\end{tabular}
\caption{Component breakdown for each galaxy.   For each component, the total mass is $M$, the particle mass is $m$, and the number of particles is $N$.}
\label{table:gm:n} 
\end{table}
\end{center}
  
 \begin{figure}
   \centering
   \includegraphics[height=1.1\textwidth]{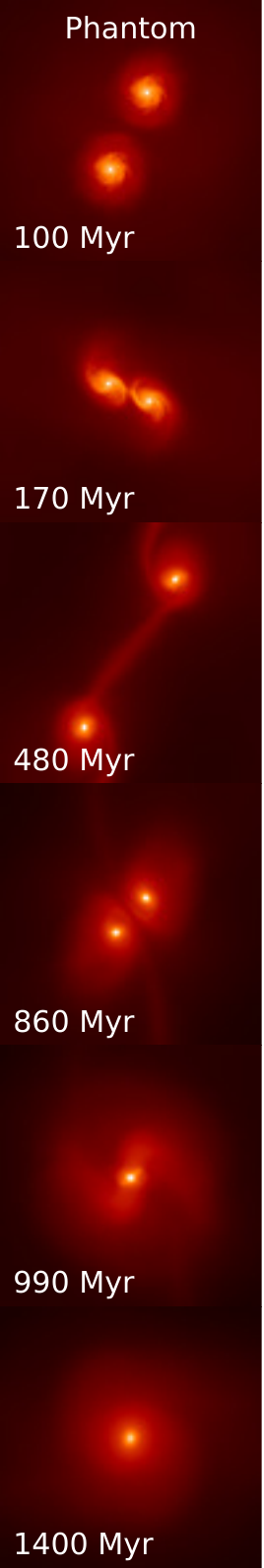} 
   \includegraphics[height=1.1\textwidth]{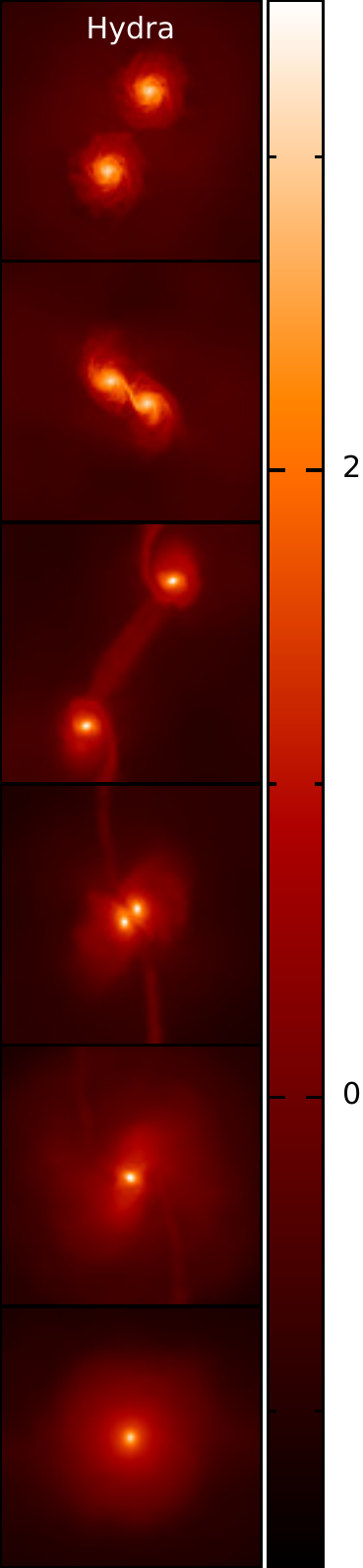} 
   \caption{Evolution of the gas column density in a major merger of two Milky Way-sized galaxies, comparing \textsc{Phantom} to the {\sc Hydra} code. Times shown are from the onset of the simulation, with each frame (100 kpc)$^2$. The colour bar is log (Column density / (M$_\odot$ pc$^{-1}$)).}
\label{fig:gm:den}
\end{figure}
 \begin{figure}
   \centering
   \includegraphics[width=0.45\textwidth]{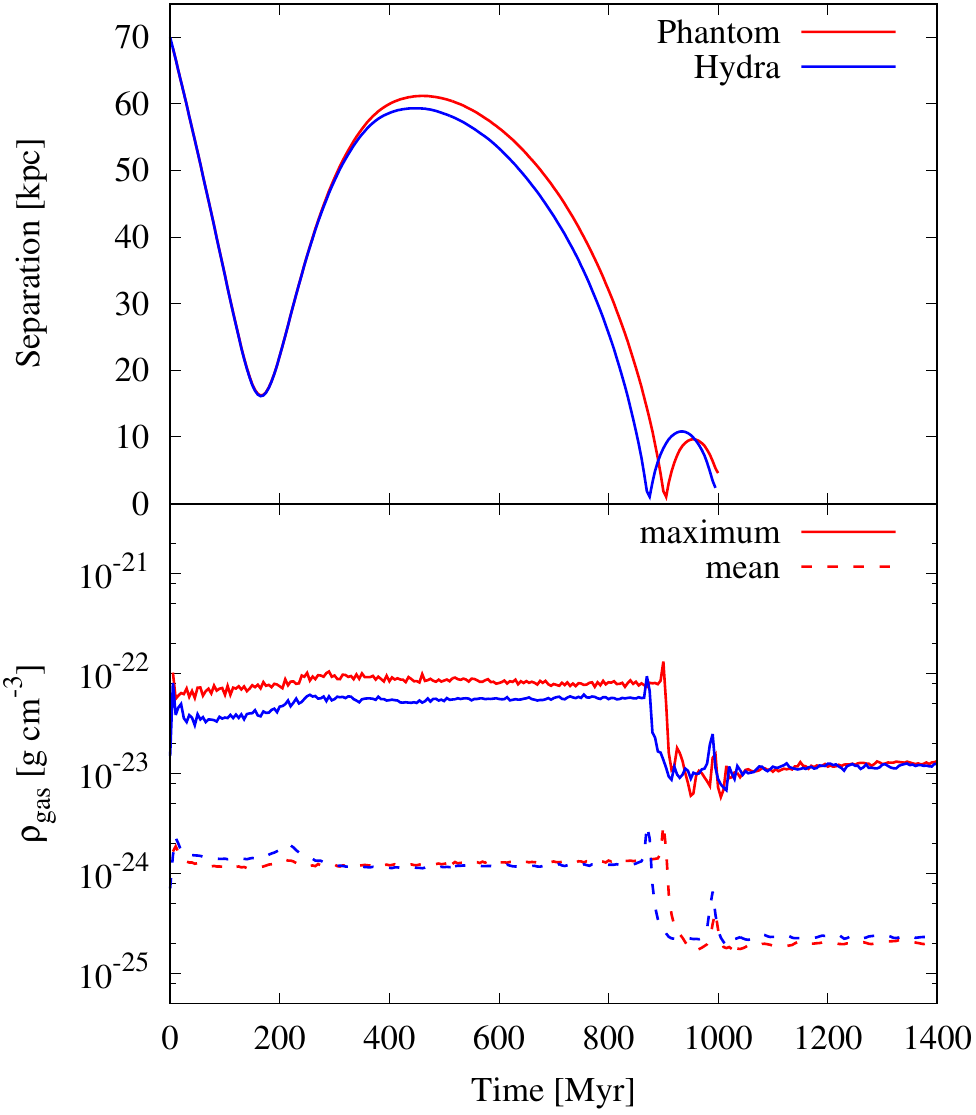}    
   \caption{\emph{Top}: The evolution of the separation of the galaxies, using centre of mass of the star particles that were assigned to each galaxy as a proxy for the galaxy's centre; the stars are sufficiently mixed after 1000~Myr, thus a meaningful separation cannot be calculated.  \emph{Bottom}: The evolution of the maximum (solid) and mean (dashed) gas densities for each model.}
\label{fig:gm:evol}
\end{figure}

 Figure~\ref{fig:gm:den} shows the gas column density evolution from $t=100$ Myr to $t=1.4$ Gyr, comparing {\sc Phantom} (left) to {\sc Hydra} (right), and Figure~\ref{fig:gm:evol} shows the evolution of the separation of the galaxies and the mean and maximum gas density in each model. The evolution of the two galaxy mergers agree qualitatively with one another, with slight differences in the trajectories, evolution times and gas densities between the two codes. Using the centre of mass of the star particles that were assigned to each galaxy as a proxy for the galaxy's centre, the maximum separation at $t \approx 450$ Myr is 59 and 61 kpc for {\sc Hydra} and {\sc Phantom}, respectively.  Second periapsis occurs at 875 and 905 Myr for {\sc Hydra} and \textsc{Phantom}, respectively, which is a difference of 3.4 per cent since the beginning of the simulation.  The maximum gas density is approximately 2 times higher in \textsc{Phantom} prior to the merger, and about 1.2 times higher after the merger; the average gas densities typically differ by less than a factor of 1.2 both before and after the merger.

There are several differences in the algorithms used in {\sc Hydra} compared to \textsc{Phantom}.  The first is the gravity solver.  The long-range gravity in {\sc Hydra} uses an adaptive particle-mesh algorithm \citep{C91}, while \textsc{Phantom} uses a $k$d-tree (c.f. Section \ref{sec:longrangegravity}).  For the short-range gravity, {\sc Hydra} uses a fixed S2 softening length for all particles, where the S2 softening is scaled to an equivalent Plummer softening such that $\epsilon_\text{S2} = 2.34\epsilon_\text{Plummer}$; for this simulation, $\epsilon_i \equiv \epsilon_\text{Plummer} = 300$ pc.  In \textsc{Phantom}, $\epsilon_i = h_i$ for each particle, where $h_i$ is calculated using only the particles of the same type as particle $i$.

A second difference is the treatment of the smoothing length in high density regions.  In {\sc Hydra}, as is common in most galactic and cosmological codes, the smoothing length is limited such that $h_i = \max(h_i,h_\text{min})$, where $h_\text{min} = \epsilon_\text{Plummer}/8 \ (=37.5$ pc).  In \textsc{Phantom}, $h_i$ is always calculated self-consistently and thus has no imposed lower limit.

Finally, \textsc{Phantom} contains an artificial conductivity term (Section~\ref{sec:ac}) that acts to ensure continuous pressure fields across contact discontinuities \citep{price08}.

In \citet{wursterthacker13}, the {\sc Hydra} major merger model was compared to a simulation run using the publicly available version of {\sc Gadget2} (\citealp{syw01}; \citealp{springelhernquist02}).  As here, the comparison was simplified such that only the gravity and SPH solvers were being compared.  They found that the galaxies in each simulation followed similar trajectories and both models reached second periapsis within 0.2 per cent of one another, as measured from the beginning of the simulation.  Note that both {\sc Hydra} and {\sc Gadget2} were both written primarily to solve galactic and cosmological models.

The quantitive difference in results may be attributed to the improved SPH algorithms in \textsc{Phantom} compared to {\sc Hydra}.  The higher density in {\sc Hydra} is consistent with the results in \citet{richardsonetal16}, who found higher densities in {\sc Hydra} compared to the adaptive mesh refinement code {\sc Ramses} \citep{Teyssier2002}. It was determined that this was a result of a combination of the artificial viscosity, $h_\text{min}$ and the suppression of `mixing' (which occurs when no thermal conductivity is applied).  

\subsection{Gap opening in dusty discs}
\begin{figure*}
\centering
\includegraphics[width=0.8\textwidth]{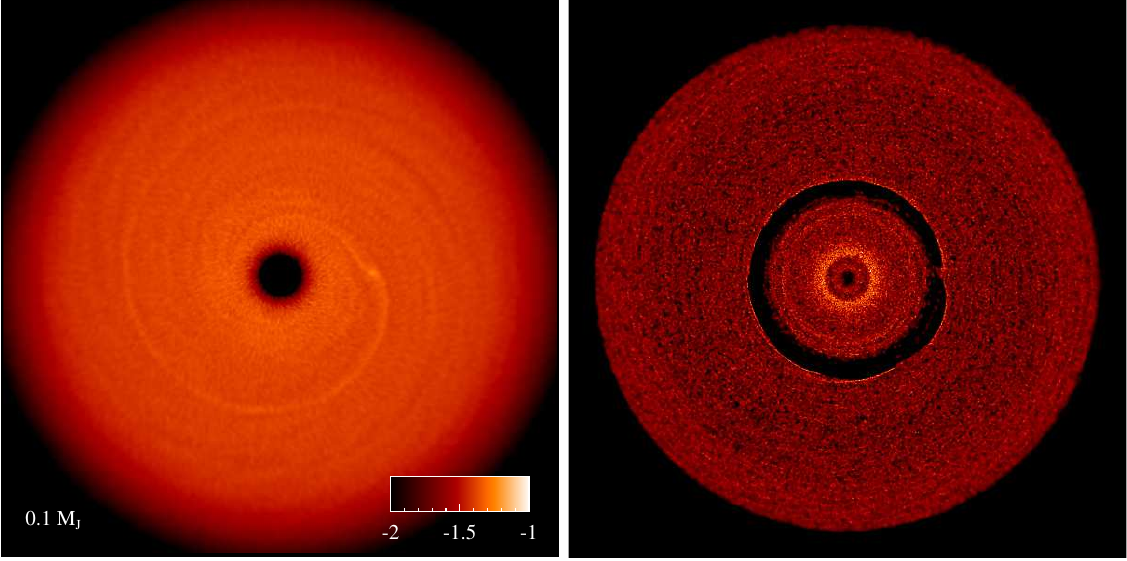}
\includegraphics[width=0.8\textwidth]{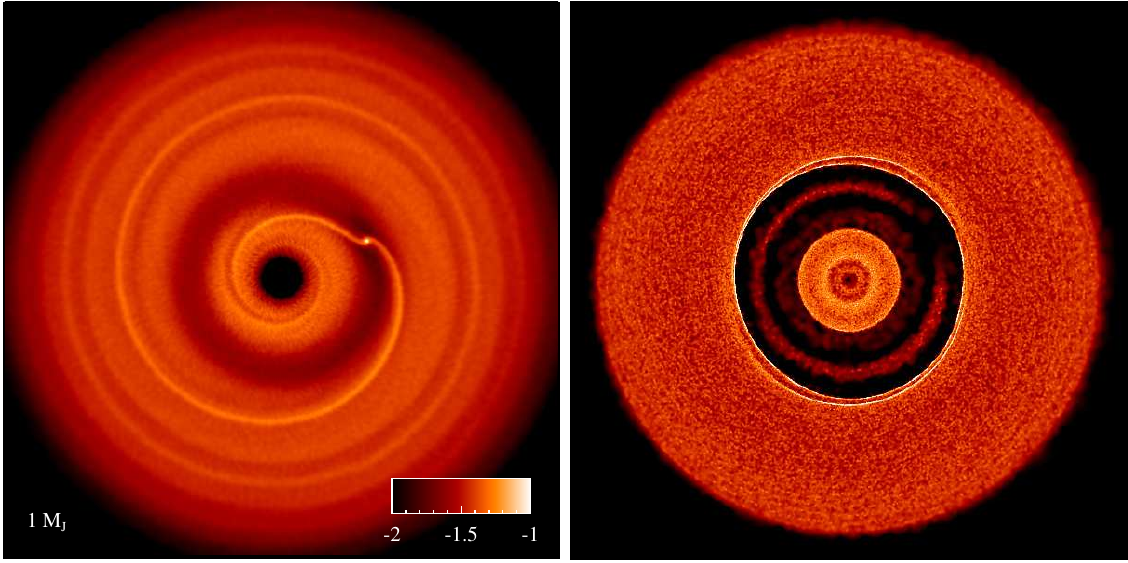}
\caption{Gap opening in dusty protoplanetary discs with {\sc Phantom} (from \citealt{dipierroetal16}), showing surface density in gas (left) and mm dust grains (right) in two simulations of planet-disc interaction with planet masses of 0.1 M$_{\rm Jupiter}$ (top) and 1 M$_{\rm Jupiter}$ (bottom) in orbit around a 1.3 M$_\odot$ star. In the top case a gap is opened only in the dust disc, while in the bottom row the gap is opened in both gas and dust. The colour bar is logarithmic surface density in cgs units.}
\label{fig:dipierro}
\end{figure*}

 Our final example is taken from \citet{dipierroetal16} and builds on our recent studies of dust dynamics in protoplanetary discs with {\sc Phantom} \citep[e.g.][]{dipierroetal15,ragusaetal17}. We perform calculations using the two-fluid approach, setting up a disc with 500~000 gas particles and 100~000 dust particles with $\Sigma \propto r^{-0.1}$ between 1 and 120 au with a total disc mass of $2 \times 10^{-4}$ M$_\odot$. The disc mass is chosen to place the mm dust particles in a regime where the Stokes number is greater than unity. The initial dust-to-gas ratio is 0.01 and we assume a locally isothermal equation of state with $c_{\rm s} \propto r^{-0.35}$, normalised such that $H/R = 0.05$ at 1 au. We use the minimum disc viscosity possible, setting $\alpha_{\rm AV} = 0.1$.
 
 Figure~\ref{fig:dipierro} shows the results of two calculations employing planets of mass 0.1 M$_{\rm Jupiter}$ (top row) and 1.0 M$_{\rm Jupiter}$ (bottom) embedded in a disc around a 1.3 M$_\odot$ star. Left and right panels show gas and dust surface densities, respectively. While the theory of gap opening in gaseous discs is relatively well understood as a competition between the gravitational torque from the planet trying to open a gap and the viscous torques trying to close it \citep[e.g.][]{goldreichtremaine79,goldreichtremaine80}, gap opening in dusty discs is less well understood \citep[see e.g.][]{paardekoopermellema04,paardekoopermellema06}. In \citet{dipierroetal16} we identified two regimes for gap opening in dusty discs where gap opening in the dust disc is either \emph{resisted} or \emph{assisted} by the gas-dust drag. The top row of Figure~\ref{fig:dipierro} demonstrates that low mass planets can carve a gap which is visible only in the \emph{dust} disc, while for high mass planets (bottom row) there is a gap opened in both gas and dust but it is deeper in the dust.
 Moreover, the gap opening mechanism by low mass planets has been further investigated in \citet{dipierrolaibe17}. They derived a grain size-dependent criterion for dust gap opening in discs, an estimate of the location of the outer edge of the dust gap and an estimate of the minimum Stokes number above which low-mass planets are able to carve gaps which appear only in the dust disc. These predictions has been tested against  {\sc Phantom} simulations of planet-disc interaction in a broad range of dusty protoplanetary discs, finding a remarkable agreement between the theoretical model and the numerical experiments.

  Interestingly, our prediction of dust gaps that are not coincident with gas gaps for low mass planets appears to be observed in recent observations of the TW Hya protoplanetary disc, by comparing VLT-SPHERE imaging of the scattered light emission from small dust grains (\citealt{van-boekeletal17}; tracing the gas) to ALMA images of the mm dust emission \citep{andrewsetal16}.

\section{Summary}
\label{sec:summary}
We have outlined the algorithms and physics currently implemented in the \textsc{Phantom} smoothed particle hydrodynamics and magnetohydrodynamics code in the hope that this will prove useful to both users and developers of the code. We have also demonstrated the performance of the code as it currently stands on a series of standard test problems, most with known or analytic solutions. While no code is ever `finished' nor bug-free, it is our hope that the code as it stands will prove useful to the scientific community. Works in progress for future code releases include radiation hydrodynamics, continuing development of the dust algorithms, and an implementation of relativistic hydrodynamics. 

\section*{Acknowledgments}
\textsc{Phantom} is the result of interactions over the years with numerous talented and interesting people. Particular mentions go to Joe Monaghan, Matthew Bate and Stephan Rosswog from whom I (DP) learnt and discussed a great deal about SPH over the years. We also thank Walter Dehnen, James Wadsley, Evghenii Gaburov, Matthieu Viallet and Pedro Gonnet for stimulating interactions. This work and the public release of \textsc{Phantom} was made possible by the award of a 4-year Future Fellowship (FT130100034) to DJP from the Australian Research Council (ARC), as well as funding via Discovery Projects DP130102078 (funding James Wurster and Mark Hutchison) and DP1094585 (which funded Guillaume Laibe and partially funded Terrence Tricco). CN is supported by the Science and Technology Facilities Council (grant number ST/M005917/1). CF gratefully acknowledges funding provided by ARC Discovery Projects (grants~DP150104329 and~DP170100603). SCOG acknowledges support from the Deutsche Forschungsgemeinschaft via SFB 881 (sub-projects B1, B2, B8) and from the  
European Research Council via the ERC Advanced Grant `STARLIGHT' (project number 339177). This work was supported by resources awarded under Astronomy Australia Ltd's merit allocation scheme on the gSTAR and swinSTAR national facilities at Swinburne University of Technology and the Pawsey National Supercomputing Centre. gSTAR and swinSTAR are funded by Swinburne and the Australian Government's Education Investment Fund. We thank Charlene Yang from the Pawsey Supercomputing Centre for particular help and assistance as part of a Pawsey uptake project. We used \textsc{splash} for many of the Figures \citep{price07}. We thank Max Tegmark for his excellent icosahedron module used in various analysis routines. We thank the three referees of this paper for extensive comments on the manuscript.

\bibliography{dan,alex,james}

\begin{appendix}

\section{Details of code implementation}
\begin{figure*}
\begin{center}
\includegraphics[width=\textwidth]{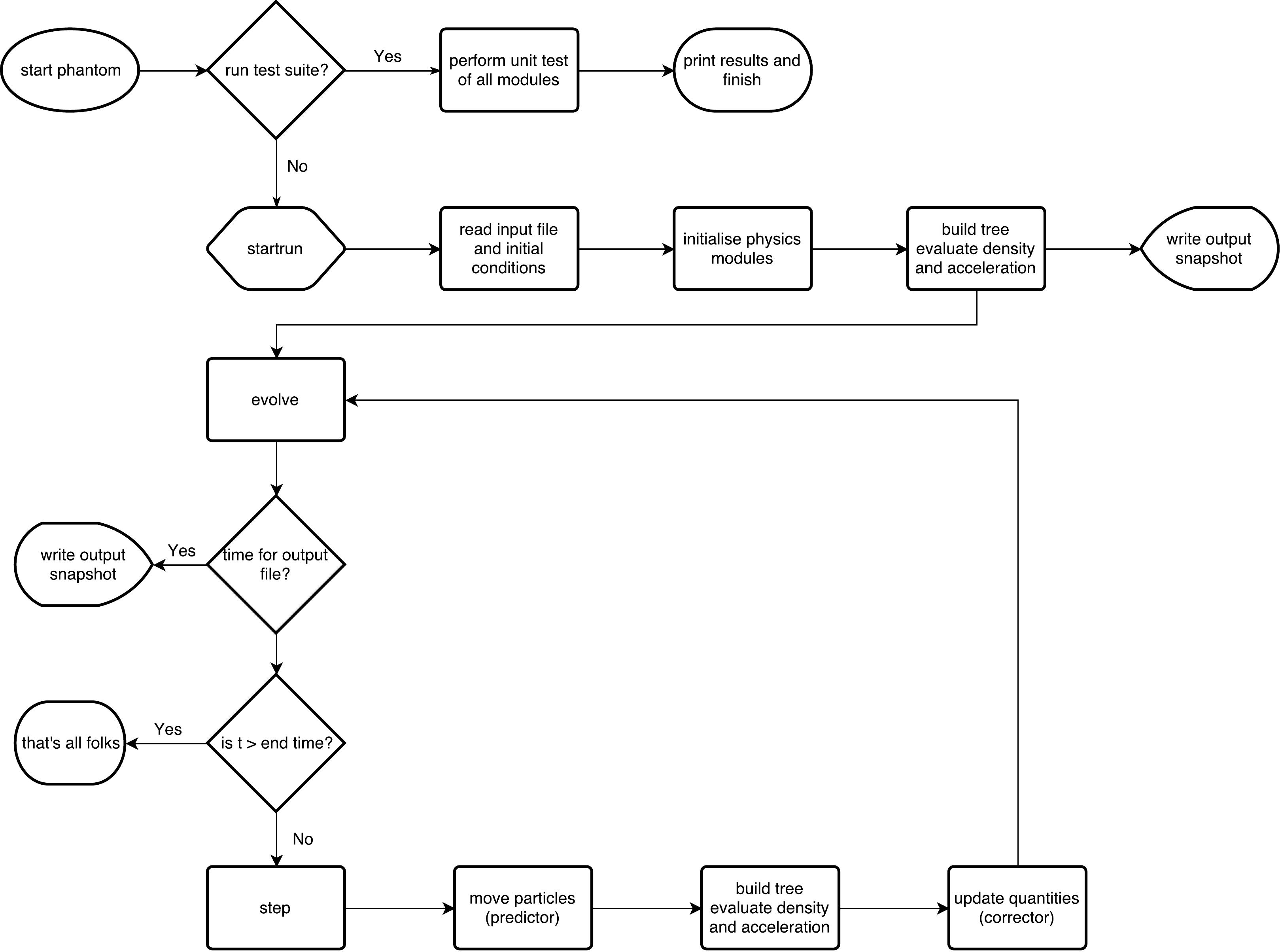}
\caption{Flowchart showing the basic structure of {\sc Phantom}. To the user what appears is a sequence of output files written at discrete time intervals. The core of the code is the timestepping loop, while most of the computational cost is spent building the tree and evaluating density and acceleration by summing over neighbours.}
\label{fig:flowchart}
\end{center}
\end{figure*}

Figure~\ref{fig:flowchart} shows the basic structure of the code. The core of the code is the timestepping loop, while the most time consuming part are the repeated calls to evaluate density and acceleration on the particles via sums over neighbouring particles. Further details of these steps are given in Section~\ref{sec:treecode}.

\subsection{Smoothing kernels}
A \textsc{Python} script distributed with \textsc{Phantom} can be used to generate the code module for alternative smoothing kernels, including symbolic manipulation of piecewise functions using \textsc{sympy} to obtain the relevant functions needed for gravitational force softening (see below). Pre-output modules for the six kernels described in Section~\ref{sec:kdefs} are included in the source code, and the code can be recompiled with any of these replacing the default cubic spline on the command line, e.g. \verb+make KERNEL=quintic+.

  The double hump kernel functions used in the two-fluid dust algorithm (Section~\ref{sec:dragkernel}) can be generated automatically from the corresponding density kernel by the \verb+kernels.py+ script distributed with \textsc{Phantom}, as described in Section~\ref{sec:ksoft}. The pre-generated modules implementing each of the kernels described in Section~\ref{sec:kdefs} hence automatically contain the corresponding double hump kernel function, which is used to compute the drag terms.

\subsection{Particle types}
\label{sec:types}
 Particles can be assigned with a `type' from the list (see Table~\ref{tab:types} in the appendix). The main use of this is to be able to apply different sets of forces to certain particle types (see description for each type). Densities and smoothing lengths are self-consistently computed for \emph{all} of these types except for `dead' particles which are excluded from the tree build and boundary particles whose properties are fixed. However, the kernel interpolations used for these involve only neighbours of the \emph{same type}. Particle masses in {\sc Phantom} are fixed to be equal for all particles of the same type, to avoid problems with unequal mass particles (e.g. \citealt{monaghanprice06}). We use adaptive gravitational force softening for all particle types, both SPH and $N$-body (see Section~\ref{sec:softeningdm}). Sink particles are handled separately to these types, being stored in a separate set of arrays, carry only a fixed softening length which is set to zero by default and compute their mutual gravitational force without approximation (see Section~\ref{sec:sinks}).

\begin{table}
 \setlength{\tabcolsep}{2pt}
 \begin{tabular}{lll}
\hline
\hline
ID & Type & Description \\
\hline
1 & gas & default type, all forces applied \\
2 & dust & drag, external \& gravitational forces \\
3 & boundary & velocity and gas properties fixed \\
4 & star & external and gravitational forces \\
5 & dark matter & same as star, but different mass \\
6 & bulge & same as star, but different mass \\
0 & unknown & usually dead particles \\
\hline
\hline
\end{tabular}
\caption{Particle types in {\sc Phantom}. The density and smoothing length of each type is computed only from neighbours of the same type (c.f. Section~\ref{sec:twofluid}). Sink particles are handled separately in a different set of arrays.}
\label{tab:types}
\end{table}

\subsection{Evaluating density and acceleration}
\label{sec:treecode}

\subsubsection{Tree build}
\label{sec:maketree}
\begin{figure}
\begin{verbatim}
construct_root_node(rootnode,bounds)
push_onto_stack(rootnode,bounds)
number on stack = 1
do while (number on stack > 0)
   pop_off_stack(node,bounds)
   construct_node(node,bounds,boundsl,boundsr)
   if (node was split)
      push_onto_stack (leftchild,boundsl)
      push_onto_stack (rightchild,boundsr)
      number on stack += 2
   endif
enddo
\end{verbatim}
\caption{Pseudo-code for the tree build. The \emph{construct\_node} procedure computes, for a given node, the centre of mass, size, maximum smoothing length, quadrupole moments, and the child and parent pointers and the boundaries of the child nodes.}
\label{fig:treebuild}
\end{figure}

{\sc Phantom} uses a $k$d-tree for nearest neighbour finding and to compute long range gravitational accelerations. The procedure for the tree build is given in Figure~\ref{fig:treebuild}. We use a stack, rather than recursive subroutines, for efficiency and to aid parallelisation (the parallelisation strategy for the tree build is discussed further in Section~\ref{sec:omp-treebuild}). The stack initially contains only the highest level node for each thread (in serial this would be the root node). We loop over all nodes on the stack and call a subroutine to compute the node properties, namely the centre of mass position, the node size, $s$, which is the radius of a sphere containing all the particles centred on the centre of mass, the maximum smoothing length for all the particles contained within the node, pointers to the two node children and the parent node, and, if self-gravity is used, the total mass in the node as well as the quadrupole moment (see Section~\ref{sec:gravity}). The \verb+construct_node+ subroutine also decides whether or not the node should be split (i.e. if the number of particles $> N_{\rm min}$) and returns the indices and boundaries of the resultant child nodes.
 
We access the particles by storing an integer array containing the index of the first particle in each node (\verb+firstinnode+), and using a linked list where each particle stores the index of the next particle in the node (\verb+next+), with an index of zero indicating the end of the list. During the tree build we start with only the root node, so \verb+firstinnode+ is simply the first particle that is not dead or accreted and the \verb+next+ array is filled to contain the next non-dead-or-accreted particle using a simple loop. In the \verb+construct_node+ routine we convert this to a simple list of the particles in each node and use this temporary list to update the linked list when creating the child nodes (i.e., by setting \verb+firstinnode+ to zero for the parent nodes, and filling \verb+firstinnode+ and \verb+next+ for the child nodes based on whether the particle position is to the `left' or the `right' of the bisected parent node).

 The tree structure itself stores 8 quantities without self-gravity, requiring 52 bytes per particle (\verb+x+,\verb+y+,\verb+z+,\verb+size+,\verb+hmax+: 5 x 8-byte double precision; \verb+leftchild+, \verb+rightchild+, \verb+parent+: 3 x 4-byte integer). With self-gravity we store 15 quantities (\verb+mass+ and \verb+quads(6)+, i.e. 7 additional 8-byte doubles) requiring 108 bytes per particle. We implement the node indexing scheme outlined by \citet{gaftonrosswog11} where the tree nodes on each level $l$ are stored starting from $2^{l-1}$, where level $1$ is the `root node' containing all the particles, to avoid the need for thread locking when different sections of the tree are built in parallel. However, the requirement of allocating storage for all leaf nodes on all levels regardless of whether or not they contain particles either limits the maximum extent of the tree or can lead to prohibitive memory requirements, particularly for problems with high dynamic range, such as star formation, where a small fraction of the particles collapse to high density. Hence, we use this indexing scheme only up to a maximum level (\verb+maxlevel_indexed+) which is set such that $2^{\verb+maxlevel_indexed+}$ is less than the maximum number of particles (\verb+maxp+). We do, however, allow the tree build to proceed beyond this level, whereupon the \verb+leftchild+, \verb+rightchild+ and \verb+parent+ indices are used and additional nodes are added in the order that they are created (requiring limited thread locking). 

\subsubsection{Neighbour search}
\begin{figure}
\begin{verbatim}
rcut = size(node) + radkern*hmax(node)
add_to_stack(root node)
number on stack = 1
do while (number on stack > 0)
   nodem = stack(nstack)
   distance = node - nodem
   if (distance < rcut + size(nodem))
      if (node is leaf node)
         ipart = firstinnode(nodem)
         do while(ipart > 0)
            add particle to neighbour list
            nneigh = nneigh + 1
            if (nneigh <= cache size)
               cache positions
            endif
            ipart = next(ipart)
         enddo
      else
         add_to_stack(leftchild)
         add_to_stack(rightchild)
         number on stack += 2
      endif
   endif
enddo
\end{verbatim}
\caption{Pseudo-code for the neighbour search (referred to as the \emph{get\_neigh} routine in Figure~\ref{fig:densitycalc}).}
\label{fig:neighbsearch}
\end{figure}

  The neighbour search for a given `leaf node', $n$, proceeds from the top down. As with the tree build, this is implemented using a stack, which initially contains only the root node. The procedure is summarised in Figure~\ref{fig:neighbsearch}. We loop over the nodes on the stack, checking the criterion
\begin{equation}
r_{nm}^{2} < (s_{n} + s_{m} + R_{\rm kern} h^{n}_{\rm max})^{2}
\end{equation}
where $r_{nm}^{2} \equiv (x_{n} - x_{m})^{2} + (y_{n} - y_{m})^{2} + (z_{n} - z_{m})^{2}$ is the square of the separation between the node centres and $s$ is the node size. Any node $m$ satisfying this criteria, that is not a leaf node, has its children added to the stack and the search continues. If node $m$ is a leaf node, then the list of particles it contains are added to the trial neighbour list and the positions cached. The end product is a list of trial neighbours (\verb+listneigh+), its length \verb+nneigh+ and a cache containing the trial neighbour positions (\verb+xyzcache+) up to some maximum size (12~000 by default, the exact size not being important except that it is negligible compared to the number of particles). Trial neighbours exceeding this number are retrieved directly from memory during the density calculation rather than from the cache. This occurs rarely, but the overflow mechanism allows for the possibility of a few particles with a large number of neighbours, as happens under certain circumstances.

\subsubsection{Density and force calculation}
\label{sec:densityforce}
 Once the neighbour list has been obtained for a given leaf node, we proceed to perform density iterations for each member particle. The neighbours only have to be re-cached if the smoothing length of a given particle exceeds $h_{\rm max}$ for the node, which is sufficiently rare so as not to influence the code performance significantly. In the original version of \textsc{Phantom} (on the \verb+nogravity+ branch) this neighbour cache was re-used immediately for the force calculation but this is no longer the case on the master branch (see Section~\ref{sec:singleloop}). Figure~\ref{fig:densitycalc} summarises the resulting procedure for calculating the density.

\begin{figure}
\begin{verbatim}
!$omp parallel do
do node = 1, number of nodes
   if (node is leaf node)
      call get_neigh(node,listneigh,nneigh)
      i = firstinnode(node)
      do while (i > 0)
         do while not converged
            if (h > hmax(node)) call get_neigh
            do k=1,nneigh
               j = listneigh(k)
               if (n <= cache size)
                  get j position from cache
               else
                  get j position from memory
               endif
               if (actual neighbour)
                  evaluate kernel and dwdh
                  add to density sum
                  add to gradh sum
                  add to div v sum
               endif
            enddo
            update h
            check convergence
         enddo
         i = next(i)
      enddo
   endif
enddo
!$omp end parallel do
\end{verbatim}
\caption{Pseudo-code for the density evaluation in \textsc{Phantom}, showing how Eqs.~\ref{eq:rhosum}, \ref{eq:sphcty} and \ref{eq:omega} are computed. The force evaluation (evaluating Eqs.~\ref{eq:sphmom} and \ref{eq:dudtsph}) is similar except that \emph{get\_neigh} returns neighbours within the kernel radius computed with both $h_{i}$ and $h_{j}$ and there is no need to update/iterate $h$ (see Figure~\ref{fig:forcecalc}).}
\label{fig:densitycalc}
\end{figure}

The corresponding force calculation is given in Figure~\ref{fig:forcecalc}.

\begin{figure}
\begin{verbatim}
!$omp parallel do
do node = 1, number of nodes
   if (node is leaf node)
      call get_neigh(node,listneigh,nneigh,fnode)
      i = firstinnode(node)
      do while (i > 0)
         do n=1,nneigh
            j = listneigh(n)
            if (n <= cache size)
               get j position from cache
            else
               get j position from memory
            endif
            if (dr < Rkern*hi or dr < Rkern*hj)
               evaluate kernel and softening
               add to force sum
            else
               add 1/r^2 forces to force sum
            endif
         enddo
         get i distance from node centre
         expand fnode at position of i
         add long range terms to force sum
         i = next(i)
      enddo
   endif
enddo
!$omp end parallel do
\end{verbatim}
\caption{Pseudo-code for the force calculation, showing how the short and long-range accelerations caused by self-gravity are computed. The quantity \emph{fnode} refers to the long-range gravitational force on the node computed from interaction with distant nodes not satisfying the tree-opening criterion.}
\label{fig:forcecalc}
\end{figure}

\subsubsection{SPH in a single loop}
\label{sec:singleloop}
A key difference in the force calculation compared to the density calculation (Section~\ref{sec:density}) is that computation of the acceleration (Equation~\ref{eq:sphmom}) involves searching for neighbours using both $h_{a}$ and $h_{b}$. One may avoid this requirement, and the need to store various intermediate quantities, by noticing that the $h_{b}$ term can be computed by `giving back' a contribution to one's neighbours. In this way the whole SPH algorithm can be performed in a single giant outer loop, but with multiple loops over the same set of neighbours, following the outline in Figure~\ref{fig:densitycalc}. This also greatly simplifies the neighbour search, since one can simply search for neighbours within a known search radius ($2h_{a}$) without needing to search for neighbours that `might' contribute if their $h_{b}$ is large. Hence a simple fixed grid can be used to find neighbours, as already discussed in Section~\ref{sec:neighb}, and the same neighbour positions can be efficiently cached and re-used (one or more times for the density iterations, and once for the force calculation). This is the original reason we decided to average the dissipation terms as above, since at the end of the density loop one can immediately compute quantities that depend on $\rho_{a}$ (i.e. $P_{a}$ and $q^{a}_{ab}$) and use these to `give back' the $b$ contribution to ones neighbours. This means that the density and force calculations can be done in a single subroutine with effectively only one neighbour call, in principle saving a factor of two in computational cost.

 The two disadvantages to this approach are i) that particles may receive simultaneous updates in a parallel calculation, requiring locking which hurts the overall scalability of the code and ii) that when using individual timesteps only a few particles are updated at any given timestep, but with the simple neighbour search algorithms one is required to loop over all the inactive particles to see if they \emph{might} contribute to an active particle. Hence, although we employed this approach for a number of years, we have now abandoned it for a more traditional approach, where the density and force calculations are done in separate subroutines and the $k$d-tree is used to search for neighbours checking both $h_{a}$ and $h_{b}$ for the force calculation.

\subsection{OpenMP parallelisation}
\label{sec:omp}

\subsubsection{Density and force calculation}
\label{sec:omp-dens}
Shared memory parallelisation of the density and force calculation is described in pseudo-code in Figure~\ref{fig:densitycalc}. The parallelisation is done over the `leaf nodes', each containing around 10 particles. Since the leaf nodes can be done in any order, this can be parallelised with a simple \verb+$omp parallel do+ statement. The neighbour search is performed once for each leaf node, so each thread must store a private cache of the neighbour list. This is not an onerous requirement, but care is required to ensure that sufficient per-thread memory is available. This usually requires setting the \verb+OMP_STACKSIZE+ environment variable at runtime. No thread locking is required during the density or force evaluations (unless the single loop algorithm is employed; see Section~\ref{sec:singleloop}) and the threads can be scheduled at runtime to give the best performance using either dynamic, guided or static scheduling (the default is dynamic). Static scheduling is faster when there are few density contrasts and the work per node is similar, e.g. for subsonic turbulence in a periodic box (c.f.~\citealt{price12a}).

\subsubsection{Parallel tree build}
\label{sec:omp-treebuild}
We use a domain decomposition to parallelise the tree build, similar to \citet{gaftonrosswog11}. That is, we first build the tree nodes as in Figure~\ref{fig:treebuild}, starting from the root-level node and proceeding to its children and so forth, putting each node into a queue until the number of nodes in the queue is equal to the number of \textsc{OpenMP} threads. Since the queue itself is executed in serial, we parallelise the loop over the particles inside the \verb+construct_node+ routine during the construction of each node. Once the queue is built, each thread proceeds to build its own sub-tree independently.

 By default we place each new node into a stack, so the only locking required during the build of each sub-tree is to increment the stack counter. We avoid this by adopting the indexing scheme proposed by \citet{gaftonrosswog11} (discussed in Section~\ref{sec:neighb}) where the levels of the tree are stored contiguously in memory. However, to avoid excessive memory consumption we only use this scheme while $2^{n_{\rm level}} < n_{\rm part}$. For levels deeper than this we revert to using a stack which therefore requires a (small) critical section around the increment of the stack counter.

\subsubsection{Performance}
\begin{figure}
\begin{center}
\includegraphics[width=\columnwidth]{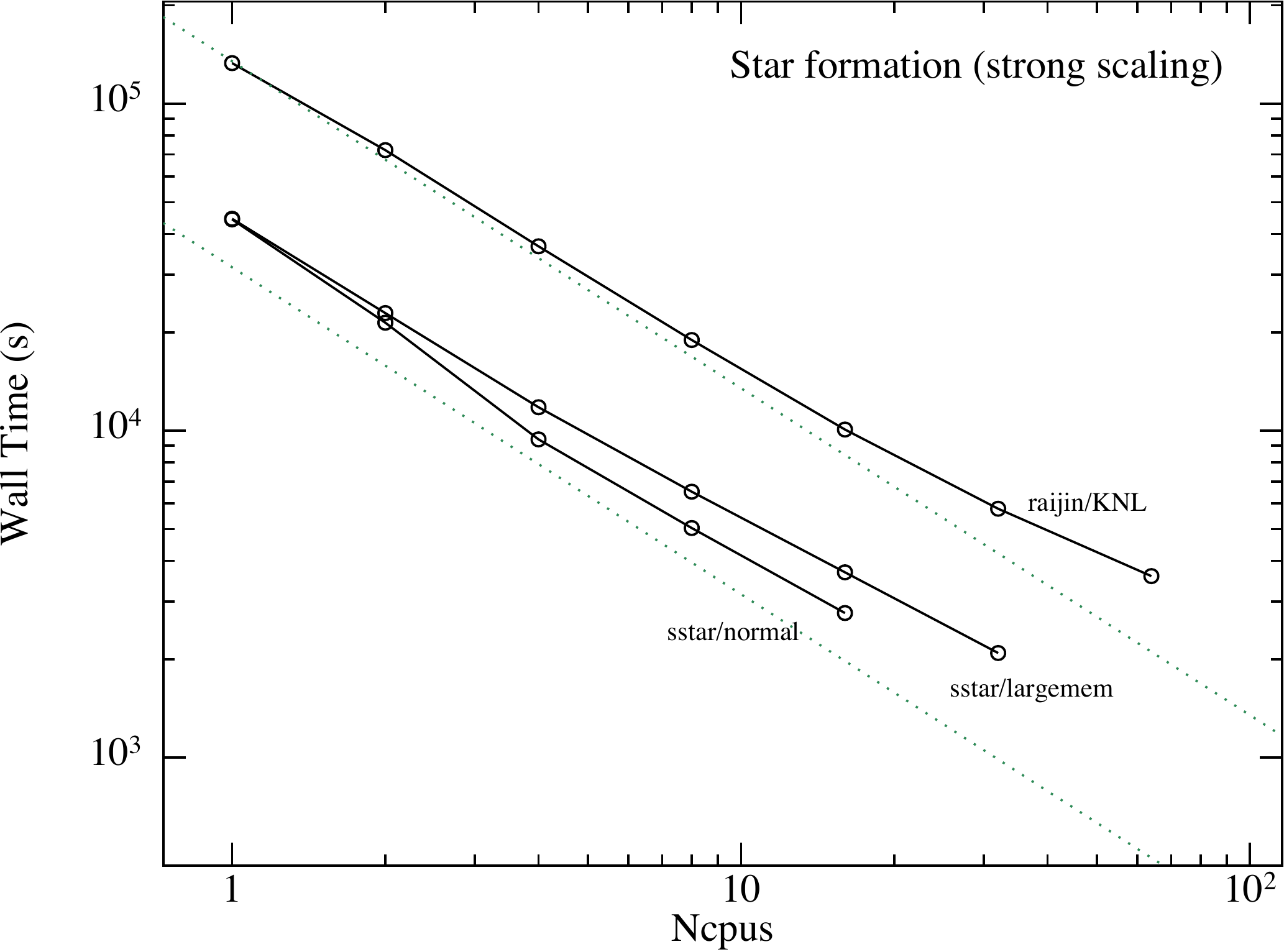}
\caption{Strong scaling results for the pure {\sc OpenMP} code for the magnetised star formation problem, showing wall time as a function of the number of {\sc OpenMP} threads.}
\label{fig:scaling-knl}
\end{center}
\end{figure}

Figure~\ref{fig:scaling-knl} shows strong scaling results for the pure {\sc OpenMP} code. For the scaling tests we wanted to employ a more representative problem than the idealised tests shown in Section~\ref{sec:tests}. With this in mind we tested the scaling using a problem involving the collapse of a molecular cloud core to form a star, as described in Section~\ref{sec:jet}. We used $10^6$ particles in the initial sphere, corresponding to 1.44 million particles in total. We recorded the wall time of each simulation evolved for one free-fall time of the collapsing sphere, corresponding to $t=0.88$ in code units.

\begin{figure}
\begin{center}
\includegraphics[width=\columnwidth]{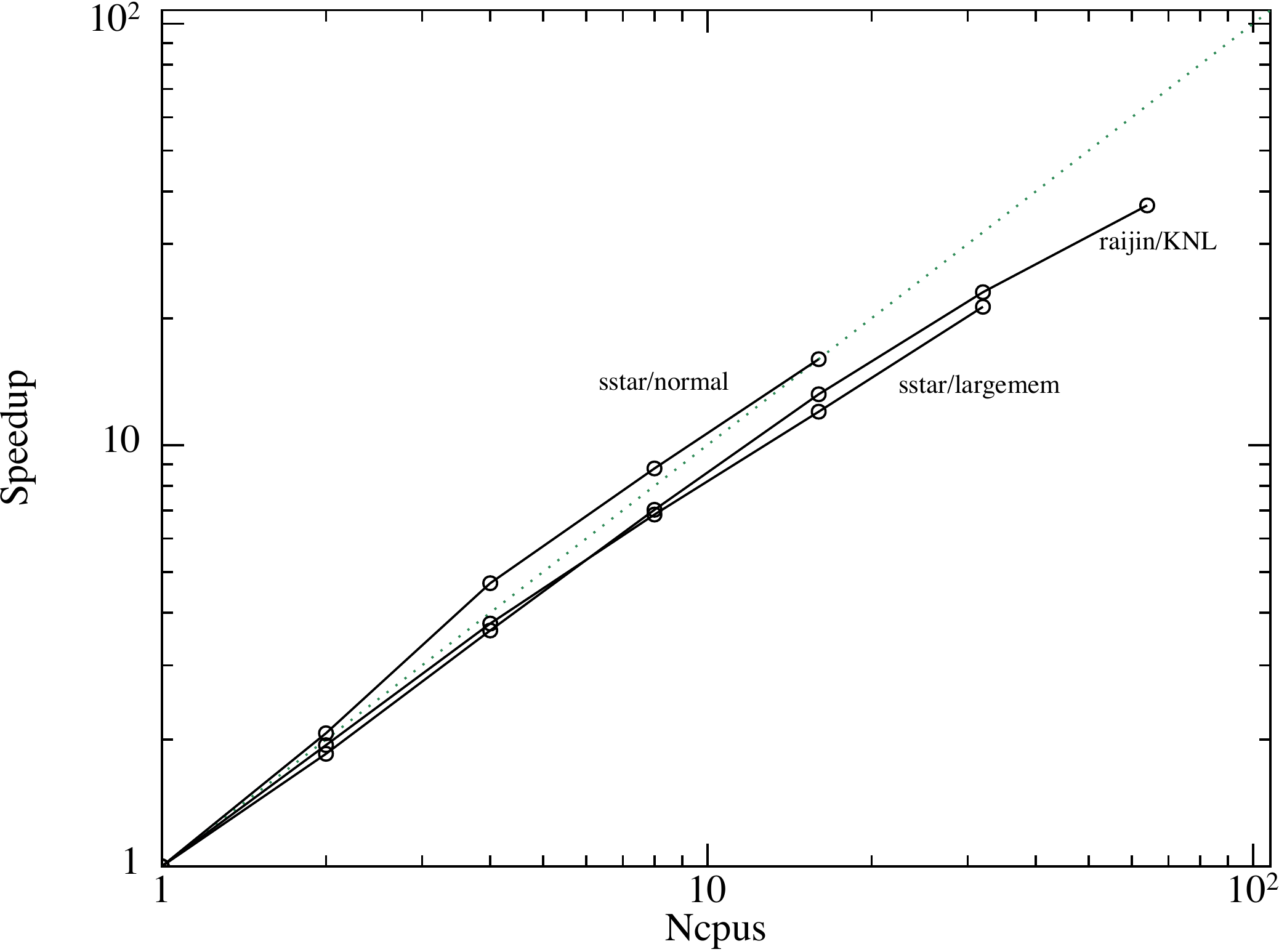}
\caption{Strong scaling results for the pure {\sc OpenMP} code for the magnetised star formation problem, showing speedup as a function of the number of {\sc OpenMP} threads.}
\label{fig:speedup}
\end{center}
\end{figure}

 To show scaling of the {\sc OpenMP} code to a reasonable number of cpus, we performed the test on the Knights Landing$\texttrademark$ (KNL) nodes of the Raijin supercomputer (the main supercomputer of the National Computational Infrastructure in Australia). Each CPU of this machine is an Intel$\textregistered$ Xeon Phi$\texttrademark$ CPU 7230 with a clock speed of 1.30 GHz and a 1024 kB cache size. The results shown in Figure~\ref{fig:scaling-knl} demonstrate strong scaling to 64 CPUs on this architecture. We used the Intel$\textregistered$ Fortran Compiler to compile the code. Timings are also shown for the same calculation performed on two different nodes of the Swinstar supercomputer, namely the `largemem' queue, consisting of up to 32 CPUs using Intel$\textregistered$ Xeon$\texttrademark$ E7-8837 chips running at 2.66 GHz with cache size 24~576 kB, and the `normal' queue, consisting of up to 16 Intel$\textregistered$ Xeon$\texttrademark$ E5-2660 chips running at 2.20 GHz with cache size 20~480 kB. Our shortest wall time is achieved on the largemem queue using 32 CPUs. Figure~\ref{fig:speedup} shows the corresponding speedup (wall time on single cpu divided by wall time on multiple cpus) for each architecture. We find the best scaling on the Swinstar `normal' nodes, which show super-linear scaling and 100\% parallel efficiency on 16 cpus. The parallel efficiency on KNL is 60\% on 64 cpus.

\subsection{MPI parallelisation}
\label{sec:mpi}
 The code has been recently parallelised for distributed memory architecture using the Message Passing Interface (MPI), using a hybrid {\sc mpi}-{\sc OpenMP} implementation. However, {\sc Phantom} does not yet compete with \textsc{gadget} variants in terms of the ability to use large numbers of particles (e.g. the most recent calculations by \citet{bocquetetal16} employed more than $10^{11}$ particles), since almost all of our published simulations to date have been performed with the {\sc OpenMP} code. Nevertheless, we describe the implementation below.
 
 We decompose the domain using the $k$d-tree in a similar manner to the {\sc OpenMP} parallelisation strategy described above. That is, we build a global tree from the top down until the number of tree nodes exceeds the number of MPI threads. Storing a global tree across all MPI threads is not too memory intensive since there are only as many nodes in the global tree as there are MPI threads. Each thread then proceeds to build its own independent subtree. During the neighbour search (performed for each leaf node of the tree) we then flag if a node hits parts of the tree that require remote contributions. If this is the case, we send the information for all active particles contained within the leaf node to the remote processor. Once all nodes have computed density (or force) on their local particles, they proceed to compute the contributions of local particles to the density sums of particles they have received. The results of these partial summations are then passed back to their host processors.
 
 Particles are strictly assigned to a thread by their location in the tree. During the tree build we exchange particles between threads to ensure that all particles are hosted by the thread allocated to their subdomain.
 
  Within each MPI domain the {\sc OpenMP} parallelisation then operates as usual. That is, the {\sc OpenMP} threads decompose the tree further into sub-trees for parallelisation.
 
\subsection{Timestepping algorithm}
\label{sec:indtimestepprac}
When employing individual particle timesteps, we assign particles to bins numbered from zero, where zero would indicate a particle with $\Delta t = \Delta t_{\rm max}$, such that the bin identifier is
 \begin{equation}
i_{{\rm bin}, a} = \max\left\{ {\rm int}\left[\log_{2} \left(\frac{2\Delta t_{\rm max}}{\Delta t_{a}}\right) - \epsilon_{\rm tiny} \right], 0 \right\},
\end{equation}
where $\epsilon_{\rm tiny}$ is a small number to prevent round-off error (equal to the \verb+epsilon+ function in Fortran). The timestep on which the particles move is simply
\begin{equation}
\Delta t = \frac{{\Delta t}_{\rm max}}{2^{n_{\rm max}}},
\end{equation}
where $n_{\rm max}$ is the maximum bin identifier over all the particles. Each timestep increments a counter, $i_{\rm stepfrac}$, which if the timestep hierarchy remains fixed simply counts up to $2^{n_{\rm max}}$. If $n_{\rm max}$ changes after each step, then $i_{\rm stepfrac}$ is adjusted accordingly ($i_{\rm stepfrac} =  i_{\rm stepfrac}/2^{(n_{\rm max, old} - n_{\rm max, new})}$). A particle is active if
\begin{equation}
\mod\left[i_{\rm stepfrac}, 2^{(n_{\rm max} - i_{{\rm bin}, a})} \right] = 0. \label{eq:active}
\end{equation}
Active particles are moved onto a smaller timestep at any time (meaning any time that they are active and hence have their timesteps re-evaluated), but can only move onto a larger timestep if it is synchronised with the next-largest bin, determined by the condition
\begin{equation}
\mod\left[i_{\rm stepfrac}, 2^{(n_{\rm max} - i_{{\rm bin}, a} + 1)} \right] = 0.
\end{equation}
In keeping with the above, particles are only allowed to move to a larger timestep by one bin at any given time. 

 We interpolate the positions of inactive particles by keeping all particles synchronised in time at the beginning and end of the timestep routine. This is achieved by storing an additional variable per particle, $t_{\rm was}$. All particles begin the calculation with $t_{\rm was}$ set to half of their initial timestep, that is
 \begin{equation}
 t_{{\rm was}, a} = \frac12 \left(\frac{\Delta t_{\rm max}}{i_{{\rm bin}, a}}\right).
 \end{equation}
  To be consistent with the RESPA algorithm (Section~\ref{sec:respa}) we first update all particles to their half timestep. These velocities are then used to move the particle positions according to the inner loop of the RESPA algorithm (Equations~\ref{eq:respa1}--\ref{eq:respa4}). We then interpolate all velocities to the current time in order to evaluate the SPH derivative, finishing with the Leapfrog corrector step. Figure~\ref{fig:step} shows the resulting pseudo-code for the entire timestepping procedure.
  
\begin{figure}
\begin{verbatim}
init_step:
 t = 0
 do i=1,n
    twas(i) = 0.5*dt(i)
 enddo

step:
 dt_long = min(dt(1:n))
 do i=1,n
    v(i) = v(i) + (twas(i)-t)*asph(i)
 enddo
 t1 = t + dt_long
 do while (t < t1)
    t = t + dt_short
    do i=1,n
       v(i) = v(i) + 0.5*dt_short*aext(i)
       x(i) = x(i) + dt_short*v(i)
    enddo
    get_external_force(x,aext,dtshort_new)
    get_vdependent_external_force(x,v,aext)
    do i=1,n
       v(i) = v(i) + 0.5*dt_short*aext(i)
    enddo
    dt_short = min(dtshort_new, t1-t)
 enddo
 do i=1,n
    vstar(i) = v(i) + (t-twas(i))*asph(i)
    dtold(i) = dt(i)
 enddo
 get_sph_force(x,vstar,asph,dt)
 do i=1,n
    if (active)
       dt_av = 0.5*(dtold(i)+dt(i))
       v(i) = v(i) + dt_av*asph(i)
       twas(i) = twas(i) + dt_av
    endif
    v(i) = v(i) + (t - twas(i))*asph(i)
    wake_inactive_particles(ibin(i),twas(i),dt_av)
 enddo
\end{verbatim}
\caption{Pseudo-code for the timestepping routine, showing how the interaction between individual timestepping and the RESPA algorithm is implemented. External forces and sink-gas interactions are computed on the fastest timescale $\Delta t_{\rm short} \equiv \Delta t_{\rm ext}$. Additional quantities defined on the particles follow the velocity terms. The variable $t_{\rm was}$ stores the last time the particle was active and is used to interpolate and synchronise the velocities at the beginning and end of each timestep.}
\label{fig:step}
\end{figure}

In particular, the first and last steps in the above (involving \verb+twas+) interpolate the velocity to the current time. All other variables defined on the particles, including the thermal energy, magnetic field and dust fraction, are timestepped following the velocity field.
  
To prevent scenarios were active particles quickly flow into an inactive region (e.g. the Sedov blast wave; see Section~\ref{sec:sedov}), inactive particles can be woken at any time.  On each step, particles who should be woken up will be identified by comparing the \texttt{ibin(i)} of the active particles to \texttt{ibin(j)} of all \texttt{i}'s neighbours, both active and inactive.  If \texttt{ibin(i)+1 > ibin(j)}, then \texttt{j} will be woken up to ensure that its timestep is within a factor of two of its neighbours.  At the end of the step, these particles will have their \texttt{ibin(j)} adjusted as required, and their \texttt{twas(j)} will be reset to the value of a particle with  $i_{{\rm bin}}$ that has perpetually been evolved on that timestep.  Finally, the predicted timestep will be replaced with \texttt{dt\_av=dt\_evolved(j)+0.5dt(j)}, where \texttt{dt\_evolved(j)} is the time between the current time and the time the particle was last active, and \texttt{dt(j)} is the particle's new timestep.


\subsection{Initial conditions: Monte Carlo particle placement}
\label{sec:montecarlo}
For setting up the surface density profile in discs, we use a Monte Carlo particle placement (Section~\ref{sec:icdisc}). This is implemented as follows: We first choose the azimuthal angle as a uniform random deviate $u_{1} \in [\phi_{\rm min},\phi_{\rm max}]$ ($0 \to 2\pi$ by default). We then construct a power-law surface density profile $\Sigma \propto R^{-p}$ using the rejection method, choosing a sequence of random numbers $u_{2} \in [0,f_{\rm max}]$ and iterating until we find a random number that satisfies
\begin{equation}
u_{2} < f,
\end{equation}
where $f \equiv R\Sigma = R^{1 - p}$ and $f_{\rm max} = R_{\rm in}^{1 - p}$ (or $f_{\rm max} = R_{\rm out}^{1 - p}$ if $p \le 1$). Finally, the $z$ position is chosen with a third random number $u_{3} \in [-z_{\rm max}, z_{\rm max}]$ such that $u_{3} < g$, where $g = \exp[-z^{2}/(2H^{2})]$ and $z_{\rm max} = \sqrt{6}H$.

\subsection{Runtime parameters in {\sc Phantom} in relation to this paper}
Table~\ref{tab:params} lists a dictionary of compile-time and runtime parameters used in the code in relation to the notation used in this paper.
\begin{table*}
\begin{tabular}{l|l|l|l}
\hline
\hline
Quantity & Code variable & Description & Reference \\
\hline
$h_{\rm fact}$ & \verb+hfact+ & Ratio of smoothing length to particle spacing & \ref{sec:density} \\
$1/\Omega_{a}$ & \verb+gradh+ & Smoothing length gradient correction term & \ref{sec:density} \\
$\epsilon_h$ & \verb+tolh+ & Tolerance in smoothing length-density iterations & \ref{sec:hrho} \\
$F_{ab}$ & \verb+grkern+ & Scalar part of kernel gradient & \ref{sec:kfunc} \\
 ${\partial W_{ab}(h)}/{\partial h_{a}}$ & \verb+dwdh+ & Derivative of kernel with respect to smoothing length & \ref{sec:kfunc} \\
$K$ &  \verb+polyk+ & Polytropic constant used for barotropic equations of state & \ref{sec:eos} \\
$u_{\rm time}$ &  \verb+utime+ & Code time unit (cgs) & \ref{sec:units} \\
$u_{\rm mass}$ &  \verb+umass+ & Code mass unit (cgs) & \ref{sec:units} \\
$u_{\rm dist}$ &  \verb+udist+ & Code distance unit (cgs) & \ref{sec:units} \\
Eq.~\ref{eq:dudt} & \verb+ipdv_heating+ & Option to turn on/off PdV work term in energy equation & \ref{sec:dudt} \\
$\Lambda_{\rm shock}$ & \verb+ishock_heating+ & Option to turn on/off shock heating in energy equation & \ref{sec:dudt} \\
$\sigma_{\rm decay}$ & \verb+avdecayconst+ & decay constant in artificial viscosity switch & \ref{sec:switches} \\
$\epsilon_{\rm v}$ & \verb+tolv+ & Tolerance on velocity error during timestepping & \ref{sec:timestepping} \\
 $\Delta t_{\rm max}$ & \verb+dtmax+ & Maximum time between output files & \ref{sec:indtimesteps} \\
$R_{\rm acc}$ & \verb+accradius1+ & accretion radius for central potential & \ref{sec:extptmass} \\
$M$ & \verb+binarymassr+ & binary mass ratio for fixed binary potential & \ref{sec:extbinary} \\
$R_{{\rm acc}, 2}$ & \verb+accradius2+ & accretion radius for secondary in fixed binary potential & \ref{sec:extbinary} \\
$k_0$ & \texttt{RadiationPressure} & Radiation pressure in Poynting-Robertson drag & \ref{sec:prdrag} \\
$k_1$ & \texttt{Redshift} & Gravitational redshift in Poynting-Robertson drag & \ref{sec:prdrag} \\
$k_2$ & \texttt{TransverseDrag} & Transverse component of Poynting-Robertson drag & \ref{sec:prdrag} \\
$E_{\rm m}$ & \verb+st_energy+ & Energy in turbulent stirring pattern & \ref{sec:turbforcing} \\
$t_{\rm decay}$ & \verb+st_decay+ & Decay time in turbulent stirring pattern & \ref{sec:turbforcing} \\
$w$ & \verb+st_solweight+ & Solenoidal fraction in turbulent stirring pattern & \ref{sec:turbforcing} \\
$k_{\rm min}$ & \verb+st_stirmin+ & Minimum wavenumber in turbulent stirring pattern & \ref{sec:turbforcing} \\
$k_{\rm max}$ & \verb+st_stirmax+ & Minimum wavenumber in turbulent stirring pattern & \ref{sec:turbforcing} \\
$f_{\rm sol}$ & \verb+st_solweightnorm+ & Solenoidal weighting in turbulent stirring pattern & \ref{sec:turbforcing} \\
$\epsilon$ & \verb+h_soft_sinksink+  & fixed gravitational softening length between sink particles & \ref{sec:sinkaccel} \\ 
$\sigma_{\rm c}$ & \verb+psidecayfac+ & dimensionless ratio of parabolic to hyperbolic $\nabla\cdot\bm{B}$ cleaning & \ref{sec:cleaning} \\
$f_{\rm clean}$ & \verb+overcleanfac+ & multiplier on maximum speed in divergence cleaning & \ref{sec:divbcheck} \\
$\rho_{\rm grain}$ & \verb+graindenscgs+ & intrinsic dust density in cgs units & \ref{sec:epstein} \\
$f_{\rm d}$ & \verb+damp+ & damping parameter for relaxing initial particle distributions  & \ref{sec:damping} \\
\hline
\hline
\end{tabular}
\caption{Various runtime parameters in the code and their relation to this paper}
\label{tab:params}
\end{table*}

\subsection{The \textsc{Phantom} testsuite}
\label{sec:testsuiteapp}
Most numerical codes in astrophysics are tested entirely by their performance on physical problems with known solutions, with solutions that can be compared with other codes and by maintenance of various conservation properties at runtime (see Section~\ref{sec:tests}). We also \emph{unit test} code modules. This allows issues to be identified at a much earlier stage in development. The tests are wrapped into the nightly testsuite. When a bug that escapes the testsuite has been discovered, we have endeavoured to create a unit test to prevent a future recurrence. 

\subsubsection{Unit tests of derivative evaluations}
 The unit test of the density and force calculations checks that various derivatives evaluate to within some tolerance of the expected value. To achieve this, the test sets up $100 \times 100 \times 100$ SPH particles in a periodic, unit cube, and specifies the input variables in terms of known functions. For example, the evaluation of the pressure gradient is checked by setting the thermal energy of each particle according to
\begin{align}
u_{a}(x,y,z) = \frac{1}{2\pi}  & \left[ 3 + \sin \left( \frac{2\pi x}{L_{x}}\right) + \cos \left( \frac{2\pi y}{L_{y}}\right) \right. \nonumber\\
& \left. \phantom{+} + \sin\left(\frac{2\pi z}{L_{z}}\right) \right],
\label{eq:utest}
\end{align}
where $L_{x}$, $L_{y}$ and $L_{z}$ are the length of the domain in each direction and the positions are relative to the edge of the box. We then compute the acceleration according to (\ref{eq:sphmom}) with the artificial viscosity and other terms switched off, assuming an adiabatic equation of state (Equation~\ref{eq:eos}) with $\rho =$~constant. We then test that numerical acceleration on all $10^{6}$ particles is within some tolerance of the analytic pressure gradient expected from (\ref{eq:utest}), namely
\begin{equation}
-\frac{\nabla P_{a}}{\rho_{a}} (x,y,z) = -(\gamma - 1) \nabla u_{a}(x,y,z),
\end{equation}
where typically we use a tolerance of $10^{-3}$ in the relative error $E(x) = \vert x - x_{\rm exact} \vert/x_{\rm exact}$. 

This procedure is repeated for the various derivatives of velocity, including the velocity divergence, curl and all components of the strain tensor. We also test the artificial viscosity terms this way --- checking that they translate correctly according to (\ref{eq:avns}) --- as well as the magnetic field derivatives and time derivatives, magnetic forces, artificial resistivity terms, physical viscosity terms, time derivatives of the dust fraction and the time derivative of the velocity divergence required in the viscosity switch. We perform each of these tests also for the case where derivatives are evaluated on only a subset of the particles, as occurs when individual particle timesteps are employed.

 We additionally check that various conservation properties are maintained. For example, energy conservation for hydrodynamics requires (\ref{eq:dedt}) to be satisfied. Hence, we include a test that checks that this summation is zero to machine precision. Similar tests are performed for magnetic fields, and for subsets of the forces that balance subsets of the thermal energy derivatives (e.g. the artificial viscosity terms).

\subsubsection{Unit tests of sink particles}
 Unit tests for sink particles include i) integrating a sink particle binary for 1000 orbits and checking that this conserves total energy to a relative error of $10^{-6}$ and linear and angular momentum to machine precision; ii) setting up a circumbinary disc of gas particles evolved for a few orbits to check that linear and angular momentum and energy are conserved; iii) checking that circular orbits are correct in the presence of sink-sink softening; iv) checking that accreting a gas particle onto a sink particle conserves linear and angular momentum, and that the resulting centre of mass position, velocity and acceleration are set correctly (c.f. Section~\ref{sec:accrete}); and v) checking that sink particle creation from a uniform sphere of gas particles (Section~\ref{sec:sinkcreate}) succeeds and that the procedure conserves linear and angular momentum.

\subsubsection{Unit tests of external forces}
 For external forces we implement general tests that can be applied to any implemented external potential: i) we check that the acceleration is the gradient of the potential by comparing a finite difference derivative of the potential, $\Phi$, in each direction to the acceleration returned by the external force routine; and ii) we check that the routines to solve matrix equations for velocity dependent forces (e.g. Sections~\ref{sec:lt}--\ref{sec:prdrag}) agree with an iterative solution to the Leapfrog corrector step (Equation~\ref{eq:lfcorrv}).

\subsubsection{Unit tests of neighbour finding routines}
 In order to unit test the neighbour finding modules, (Section~\ref{sec:neighb}), we set up particles in a uniform random distribution with randomly assigned smoothing lengths. We then check that the neighbour list computed with the treecode agrees with a brute-force evaluation of actual neighbours. We also perform several sanity checks: i) that no dead or accreted particles appear in the neighbour list; ii) that all particles can be reached by traversing the tree or link list structure; iii) that nodes tagged as active contain at least one active particle and conversely that iv) inactive cells contain only inactive particles; v) that there is no double counting of neighbours in the neighbour lists and vi) that the cached and uncached neighbour lists are identical. We further check that particle neighbours are found correctly in pathological configurations, e.g. when all particles lie in a one dimensional line along each of the coordinate axes.

\subsubsection{Unit tests of timestepping and periodic boundaries}
 As a simple unit test of both the timestepping and periodic boundaries we set up $50\times50\times 50$ particles in a uniform periodic box with a constant velocity ($v_{x} = v_{y} = v_{z} = 1$) along the box diagonal. We then evolve this for 10 timesteps and check that the density on each particle remains constant and that the acceleration and other time derivatives remain zero to machine precision.
 
\subsubsection{Unit tests of file read/write}
 We check that variables written to the header of the (binary) output files are successfully recovered by the corresponding read routine, and similarly for the particle arrays written to the body of the file. This quickly and easily picks up mistakes made in reading/writing variables from/to the output file.

\subsubsection{Unit tests of kernel module}
We ensure that calls to different kernel routines return the same answer, and check that gradients of the kernel and kernel softening functions returned by the routines are within some small tolerance of a finite difference evaluation of these gradients.

\subsubsection{Unit tests of self-gravity routines}
 In order to unit test the treecode self-gravity computation (Section~\ref{sec:gravity}), we i) check that the Taylor series expansion of the force on each leaf node matches the exact force for a particle placed close to the node centre ii) that the Taylor series expansion of the force and potential around the distant node are within a small ($\sim 10^{-5}$) tolerance of the exact values; iii) that the combined expansion about both the local and distant nodes produces a force within a small tolerance of the exact value and finally iv) that the gravitational force computed on the tree for a uniform sphere of particles is within a small tolerance ($\sim 10^{-3}$) of the force computed by direct summation.
 
\subsubsection{Unit tests of dust physics}
We unit test the dust modules by first performing sanity checks of the dust-gas drag routine --- namely that the transition between Stokes and Epstein drag is continuous and that the initialisation routine completes without error. We then perform a low-resolution {\sc dustybox} test (Section~\ref{sec:dustybox}), checking the solution matches the analytic solution as in Figure~\ref{fig:dustybox}. For one-fluid dust we perform a low resolution version of the dust diffusion test (Section~\ref{sec:dustdiffuse}), checking against the solution at every timestep is within a small tolerance of the analytic solution. Dust mass, gas mass and energy conservation in the one-fluid dust derivatives are also checked automatically.

\subsubsection{Unit tests of non-ideal MHD}
We perform three unit tests on non-ideal MHD.  The first is the wave damping test (see Section~\ref{sec:ni:tests:wave}).  This test also uses the super-timestepping algorithm to verify the diffusive (Eqn~\ref{eq:nimhd:dt} but considering only Ohmic resistivity and ambipolar diffusion) and minimum stable timesteps (Eqn~\ref{eq:dtcour}), but then evolves the system on the smallest of the two timesteps.  The second test is the standing shock test (see Section~\ref{sec:ni:tests:shock}).  This test is also designed as a secondary check on boundary particles.  Both tests used fixed coefficients for the non-ideal terms, and for speed, both tests are performed at much lower resolutions than presented in the paper.  The third test self-consistently calculates the non-ideal coefficients for given gas properties.

\subsubsection{Other unit tests}
 Various other unit tests are employed, including sanity checks of individual timestepping utility routines; checking that the barotropic equation of state is continuous; of the Coriolis and centrifugal force routines; checks of conservation in the generalised Newtonian potential (Section~\ref{sec:gnewton}); of the routine to revise the tree; and of the fast inverse square root routines. 
 
 \subsubsection{Sedov unit test}
As a final ``real'' unit test the code performs a low-resolution ($16^3$) version of the Sedov blast wave test (Section~\ref{sec:sedov}). We check that energy and momentum are conserved at a precision appropriate to the timestepping algorithm (for global timesteps this means momentum conservation to machine precision and energy conservation to $\Delta E/E_{0} < 5 \times 10^{-4}$; for individual timesteps we ensure linear momentum conservation to $2 \times 10^{-4}$ and energy conservation to $2 \times 10^{-2}$.

%
%

\end{appendix}

\label{lastpage}
\end{document}